\documentclass[%
 aip,
 amsmath,amssymb,
 reprint,%
]{revtex4-1}

\usepackage{graphicx}
\usepackage{dcolumn}
\usepackage{bm}
\usepackage{amssymb}
\usepackage{upgreek}
\usepackage{amsmath}
\usepackage{bm}
\usepackage{float}
\usepackage{overpic,multirow,subfigure}
\usepackage{longtable}
\usepackage{xcolor}
\usepackage{listings}

\usepackage[utf8]{inputenc}
\usepackage[T1]{fontenc}
\usepackage{mathptmx,morefloats}

\begin{document}

\preprint{AIP/123-QED}

\title{Turbulent transport and mixing in the multimode narrowband Richtmyer-Meshkov instability}
\thanks{UK Ministry of Defence $\copyright$ Crown Owned Copyright 2019/AWE}%

\author{B. Thornber}
 \email{ben.thornber@sydney.edu.au}
\affiliation{%
School of Aerospace, Mechanical and Mechatronic Engineering,\\ The University of Sydney, Sydney, NSW 2006, Australia
}%
\author{J. Griffond}
\affiliation{%
CEA, DAM, DIF, F-91297 Arpajon, France
}%
\author{P. Bigdelou, I. Boureima, P. Ramaprabhu}
\affiliation{%
Mechanical Engineering and Engineering Science, University of North Carolina--Charlotte, Charlotte, North Carolina 28223,USA
}%
\author{O. Schilling}
\affiliation{%
Lawrence Livermore National Laboratory, Livermore, California 94550,USA
}%
\author{R. J. R. Williams}
\affiliation{%
AWE, Aldermaston, Reading, Berkshire RG7 4PR, United Kingdom
}%

\date{\today}

\begin{abstract}
The mean momentum and heavy mass fraction, turbulent kinetic energy, and heavy mass fraction variance fields, as well as the budgets of their transport equations, are examined at several times during the evolution of a narrowband Richtmyer-Meshkov instability initiated by a Mach $1.84$ shock traversing a perturbed interface separating gases with a density ratio of $3$. The results are computed using the `quarter scale' data from four algorithms presented in the $\theta$-group study of Thornber et al. [Phys. Fluids \textbf{29}, 105107 (2017)]. The present study is inspired by a previous similar study of Rayleigh-Taylor instability and mixing using direct numerical simulation data by Schilling and Mueschke [Phys. Fluids \textbf{22}, 105102 (2010)]. In addition to comparing the predictions of the data from four implicit large-eddy simulation codes, the budgets are used to quantify the relative importance of the terms in the transport equations, and the balance of the terms is employed to infer the numerical dissipation. Terms arising from the compressibility of the flow are examined in particular, i.e., the pressure-dilatation.  The results are useful for validation of large-eddy simulation and Reynolds-averaged modeling of Richtmyer-Meshkov instability.
\end{abstract}

\maketitle

\section{\label{sec:intro}Introduction}

Richtmyer-Meshkov instability (RMI) occurs when a perturbed interface between two fluids of differing properties is impulsively accelerated \cite{Richtmyer1960, Meshkov1969}. This acceleration imparts a vorticity field near the interface, causing any perturbations on the interface to grow in time, regardless of the direction of the impulse relative to the layer. For initial perturbations with amplitude over wavelength $a/\lambda\ll1$, the growth rate of the instability is at first linear \cite{Richtmyer1960}. As the perturbation amplitude becomes nonlinear, the perturbation growth rate decreases, and the shear layers between the peaks and troughs roll up to form vortices. Shear along the interface also triggers smaller scale Kelvin-Helmholtz instabilities, which at later times contribute to the development of a range of vortical structures of differing sizes. For multimode perturbations, this process occurs simultaneously for a range of length-scales, leading to the development of an inhomogeneous turbulent mixing layer at late times. As the deposition of vorticity is impulsive, the layer with linear perturbations evolves through several growth regimes: (i) linear, (ii) nonlinear, (iii) nonlinear transitional (range of length-scales developing), (iv) decaying inhomogeneous variable density turbulent layer. 

This instability occurs in applications ranging from inertial confinement fusion \cite{Clark2016}  and astrophysical flows \cite{Burrows2000}, to augmented mixing in scramjets and jet exhaust plumes \cite{yang1993}. For a comprehensive recent review of research in this field, see Zhou \cite{Zhou2017a,Zhou2017b}.  RMI differs significantly from Kelvin-Helmholtz instability (KHI) and closely related Rayleigh-Taylor instability (RTI) in that the driving is impulsive. With a sufficiently strong impulse, time to develop, and high Reynolds number, linear growth will be followed by a transition to turbulence, where velocity fluctuations decay in time, and the layer grows at a relatively slow rate $\propto t^{\theta}$  with $\theta$ $\approx 0.2$ to $1$ (possible if dominated by linear modes, although a fully turbulent layer is limited to\cite{Barenblatt1983} $\theta \le 2/3$) with a strong dependence on initial conditions \cite{Thornber2010,Elbaz2018,DiStefano2015263,Flippo2016,Krivets2017,Soulard2018,Schilling2010}. As an example $\theta \approx 0.29 $ for the configuration studied in this paper \cite{Thornber2017}, compared to KHI ($\propto t$) and RTI ($\propto t^2$). Despite the relatively small temporal exponent, the growth rate of RMI is sufficiently high that it  significantly impacts the physics of the aforementioned applications, and so there is substantial interest in the accurate modelling of turbulent transport and mixing in RMI. 

In the context of current computational power, turbulent transport models for unsteady Reynolds-Averaged Navier-Stokes (URANS) or Large-Eddy Simulation (LES) are particularly important. URANS methods are in active development and use for design problems, principally due to the complexity of applied computations typically impacted by multiple physical phenomena, complex equations of state, and extreme conditions. Turbulence resolving computations such as Large-Eddy-Simulations require resolution of the full dimensionality of the problem \cite{Clark2016}, whereas for rapid design evaluation and iteration it is desirable to work with a problem reduced to one or two dimensions by appropriate averaging and application of modelling of the full three dimensional effects. This is the key motivation for the development and use of URANS modelling. 

Modelling approaches for decaying, inhomogeneous, variable density turbulent flows where density variations do not satisfy the Boussinesq approximation are particularly complex. The majority of turbulence modelling applied to variable density flows employs a density-weighted, or Favre-averaged approach \cite{Favre1958}. This simplifies the resulting equations describing the evolution of the density-weighted mean compared to Reynolds-averaging. Following the Favre decomposition, two principal additional complexities arise: (i) additional terms associated with density fluctuations and (ii) a change in the definition of the `mean' evolved, perhaps necessitating a modification to the closures of terms from the Reynolds averaged formulation \cite{Chassaing2001}.  URANS modelling approaches applicable to RMI range from single-fluid two-, three-, and four-equation \cite{Dimonte2006,Gauthier1990,Moran2013,Besnard1992,Banerjee2010,Stalsberg2011}, and Reynolds stress \cite{doi:10.1080/14685248.2011.633084,schwarzkopf2016two,Gregoire05,Souffland14},  to two fluid models \cite{Youngs1994,Llor2005,llor_bailly_2003}. There is a strong need for data to validate and further advance these and other models. 

Despite significant advances in facilities and diagnostics \cite{Leinov2009,Malamud2014,mohaghar_carter_pathikonda_ranjan_2019,weber_haehn_oakley_rothamer_bonazza_2014,reese_ames_noble_oakley_rothamer_bonazza_2018,DiStefano2015263,Clark2018,Nagel2017,wan2015,slessor_bond_dimotakis_1998}, experimental data is very challenging to acquire due to the relatively short time scale of RMI in terrestrial facilities, extreme conditions, or uncertainty in the initial perturbations. Such experimental research is critical and permits important validation of individual terms; however, to date it is insufficient to fully determine the required model coefficients for URANS computations. While there is considerable literature examining turbulent transport in shear layers induced by KHI, and some prior studies of compressible RTI \cite{Schilling2010b}, to date there has not been an equivalent study of all individual terms in the turbulent transport equations computed simultaeneously for a turbulent mixing layer developing from a RMI. Thus the objective of the present study is to address a gap in identification of the key mechanisms responsible for variable density effects in the RMI and the terms which must be modelled to accurately represent mean transport.   

Recently, a benchmark multimode, three-dimensional RMI case was computed with multiple independent computational fluid dynamics codes, establishing a well understood `$\theta$-group' dataset \cite{Thornber2017}. A principal goal of that study was to understand and quantify numerical uncertainty in the computation of such flows, and to provide a foundation for basic quantities of interest. The present study computes the  transport equation budgets for the mean momentum, mean heavy fluid mass fraction, heavy fluid mass fraction variance, and specific turbulent kinetic energy from the results of the four codes presented in that paper: Triclade, Turmoil, Flamenco, and Flash. The analysis of this data identifies the terms of greatest importance in the mean transport equations and provides a reliable benchmark for LES and URANS model development and validation. 

The paper is organised as follows. Section \ref{govnum} summarises the numerical methods employed to generate the $\theta$-group dataset, the test case examined, the transport equations examined, and numerical methods employed to compute the terms. Section \ref{results} presents and analyses the turbulent transport budgets for four representative times ranging from just post-shock to a weakly turbulent state. Finally, the key conclusions are summarised in Section \ref{concl}. 

\section{\label{sec:nm} Governing Equations and Computational Details \label{govnum}}

\subsection{Summary of the numerical data employed}

\begin{figure*}\centering
  \begin{overpic}[width=0.7\textwidth]{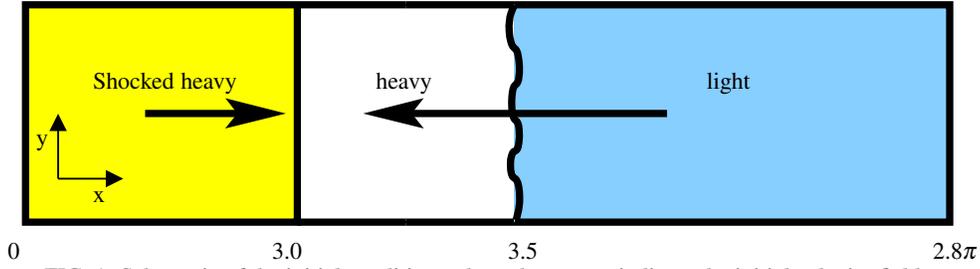}
    \small
    \put(2,9){y}
    \put(8,3){x}
    \put(8,15){Shocked heavy}
    \put(38,15){heavy}
    \put(73,15){light}
    \put(-1,-3){0}
    \put(27,-3){3.0}
    \put(52,-3){3.5}
    \put(97,-3){2.8$\pi$}
\end{overpic}
\caption{Schematic of the initial condition, where the arrows indicate the initial velocity field. \label{rmschematic}}
\end{figure*}

This paper utilises the `quarter scale' data from four algorithms presented in the $\theta$-group study of Thornber {\it et al.} \cite{Thornber2017}. In that paper the full details of the initial conditions, results, and grid convergence studies for each individual algorithm is presented. It established an initialisation to explore the physics of narrowband Richtmyer-Meshkov instability formulated in a way which could be simulated by a wide range of algorithms.  The configuration consisted of a light and a heavy gas, with a density ratio of 3, separated by a perturbed interface. A shock wave of strength Mach $1.84$ impinges on the perturbed layer, triggering Richtmyer-Meshkov instability. The perturbation on the interface is defined using a narrowband power spectrum with constant amplitude between a defined minimum and maximum wavenumber, where modal amplitudes and phases are defined using deterministic random numbers thus generating an exactly reproducible initial condition.

 The initial conditions in the shocked heavy fluid domain $0.0 $ m $<x<3.0$ m are $(\rho,u,p)=(6.375 \,\,\rm{ kg/m}^3,-61.49\,\,\rm{ m/s},400\,\,\rm{ kPa})$. Given a surface perturbation $A(y,z)$, the unshocked heavy fluid lies between $3.0$ m $<x<3.5$ m $+A(y,z)$ with initial properties  $(\rho,u,p)=(3.0\,\,\rm{kg/m}^3,-291.6\,\,\rm{m/s},100\,\,\rm{kPa})$. The third region contains the unshocked light fluid which lies between $3.5$ m $+A(y,z)<x<2.8\pi$ m and is initialised with properties $(\rho,u,p)=(1.0\,\,\rm{kg/m}^3,-291.6\,\,\rm{m/s},100\,\,\rm{kPa})$. The configuration is shown schematically in Fig. \ref{rmschematic}. 

The perturbation power spectrum is constant between length scales of $L/32$ ($\lambda_{min}$) to $L/16$,  where $L$ is the cross-section, and an initial diffuse interface of error function form and thickness $\delta = L/128$ is applied as follows:

\begin{equation}
f_1(y,z)=\frac{1}{2} \,\, {\rm erfc} \left\{\frac{\sqrt{\pi} [x-S(y,z)]}{\delta}\right\},
\end{equation}

\noindent where $f_1$ is the volume fraction of the heavy gas and $S(y,z)=3.5+A(y,z)$. The standard deviation of the perturbation amplitude is $\sigma_0=0.1 \lambda_{min}$ which places the shock-interface interaction initially at the large-amplitude end of the linear regime at the highest wavenumber. The diffuse initial condition ensures that the initial condition was accessible for all numerical schemes, as some algorithms are unstable in the presence of sharp interfaces.  Together, the addition of a diffuse interface and the larger amplitude will impact the initial growth rates, particularly for the highest wavenumbers, which could be expected to be $\approx 20$\% lower than that predicted from linear theory \cite{Duff1962}. Although diffusion is not explicitly represented in the computation, an implicit assumption is that the gases are miscible. Turbulent layers evolving in immiscible fluids may have different growth rates and mixing parameters (see, for example \cite{Dimonte2000,10.1115/1.4038400,PhysRevA.42.7211}).

The heavy and light unshocked gases are at the same temperature in the undisturbed flow, and the specific heats for each component ($C_{v1}$ and $C_{v2}$) are chosen to ensure this. Finally, the problem description is completed by assuming an ideal gas equation of state with a ratio of specific heats $\gamma=5/3$.

The Cartesian computational domain is $L_x\times L_y \times L_z=2.8\pi \times 2\pi \times 2\pi$ m$^3$.  The simulations were run from $t=0$ to $5$ s which is equivalent to a dimensionless time $\tau=246$ (the non-dimensionalisation is detailed in Sec. \ref{sec:nondims}). However, at times later than $2$ s, individual spikes were observed to exit the domain; thus only data up to $2$ s is typically employed. Prior analysis has indicated that by the final time an asymptotic approximately self-similar state is approached--this assumption is examined in more detail here.  Four time instants are examined where the first time is immediately following shock passage ($0.01$ s), the second is shortly after this time where layer mixedness as measured by the mix parameter $\Theta$ has approximately doubled ($0.025$ s), and the third and fourth are approaching a self-similar state ($1.0$ and $2.0$ s). See the discussion in  Sec. \ref{phenom} and  Fig. \ref{sampletimes} for more details.

The configuration was chosen such that the low order statistics for the problem are just converged on the finest grid level using modern compressible algorithms. This introduces a degree of uncertainty on the influence of numerical dissipation on each transport term. This has been addressed in two ways: the first is through a grid convergence study, and the second is by computing all transport terms using data produced from four independent codes/algorithms: Flamenco (Godunov method with fifth order low-Mach-number-corrected MUSCL scheme \cite{Thornber2007c,Garcia2014}); Flash (compressible Euler with the Piecewise Parabolic Method and Monotonized Central limiter, \cite{fryxell2000flash,colella1984piecewise,van1977towards});  Triclade (conservative finite difference using the wave propagation algorithm of Leveque \cite{LeVeque02} with high order accurate corrections \cite{Daru04}) and Turmoil (Lagrange remap method with high-order artificial viscosity \cite{Grinstein2007,Youngs1991,Williams2018}). The boundary conditions are periodic in the homogeneous directions, but the best practise is adopted for each code in the shock-direction which permits the transmission of the reflected rarefaction and transmitted shock through the boundaries with minimal reflections.  The implicit numerical dissipation mechanisms in each of the  algorithms differ substantially. Thus, agreement between the results either implies that for a given quantity (i) numerical dissipation has been minimised or (ii) that numerical dissipation has acted to dissipate the quantity in a statistically similar manner. All simulations have no model for physical viscosity, diffusivity, or conduction.

\subsection{Mean and Fluctuating Components}

This study explores the individual transport terms in the mean momentum and mass fraction, mass fraction variance, and fluctuating kinetic energy equations. A detailed prior study of Rayleigh-Taylor mixing of Schilling and Mueschke \cite{Schilling2010b} presented the required transport equations which are considered here. Referring to the equation numbering in Schilling and Mueschke \cite{Schilling2010b}, this paper examines the transport of mean momentum Eq. (6), mean mass fraction Eqs. (7)--(8), turbulent kinetic energy  Eqs. (14)--(15e), and mass fraction variance Eqs. (19)--(20c). Averaging processes are defined in Eqs. (2)--(3) and accompanying text. In this contribution, the key differences are that the homogeneous directions are $y$ and $z$, and the shock-direction is $x$, with a velocity component $u$, different from the Rayleigh-Taylor case where the principal direction and velocity are labelled $z$ and $w$ respectively. 

The Reynolds average of a field $\phi({\bf x},t)$ over the statistically homogeneous plane is

\begin{equation}
\bar \phi(x,t)=\frac{1}{L_y L_z}\int_0^{L_y}\int_0^{L_z} \phi({\bf x},t)dz dy,
\end{equation}

\noindent which enables a decomposition into mean and fluctuating components, where $\phi({\bf x},t)=\bar \phi(x,t)+\phi'({\bf x},t)$ and the fluctuating component is denoted by a prime. The Favre average is

\begin{equation}
\tilde \phi(x,t)=\frac{\overline {\rho \phi}(x,t)}{\bar\rho (x,t)}=\frac{\int_0^{L_y}\int_0^{L_z} \rho({\bf x},t) \phi({\bf x},t)dz dy}{\int_0^{L_y}\int_0^{L_z} \rho({\bf x},t) dz dy},
\end{equation}

\noindent where $\rho({\bf x},t)$ is density; $\tilde \phi(x,t)$ is defined such that $\phi({\bf x},t)=\tilde \phi(x,t)+\phi''({\bf x},t)$. This averaging reduces the fields to being one dimensional functions of $x$ only. 

The individual terms which may be extracted from a single ILES are listed below, along with the notation used in the figures and analysis. For mean momentum transport:

\begin{multline}
\underbrace{\bar \rho \left(\frac{\partial \tilde u}{\partial t}+u_{trans}\frac{\partial \tilde u}{\partial x}\right)}_{t^{u}}+ 
\underbrace{\bar \rho \left(\tilde {u}-u_{trans}\right)\frac{\partial \tilde u}{\partial x}}_{A^u}= 
\underbrace{-\frac{\partial \bar p}{\partial x}}_{F^u} \\
\underbrace{-\frac{\partial \tau_{11}}{\partial x}}_{R^u}
+ND_u,
\end{multline}

\noindent where $\tau_{11}=\bar {\rho}\widetilde {u''^2}$ where $u$ is the $x$-direction velocity component, $p$ is the pressure, and $ND_u$ represents the effective numerical dissipation or diffusion for this equation. For mean mass fraction transport:

\begin{multline}
\underbrace{\bar \rho \left(\frac{\partial \tilde m_1}{\partial t}+u_{trans}\frac{\partial \tilde m_1}{\partial x}\right)}_{t^{m1}}+
\underbrace{\bar \rho \left(\tilde {u}-u_{trans}\right)\frac{\partial \tilde m_1}{\partial x}}_{A^{m1}}= 
\underbrace{-\frac{\partial \bar {\rho}\widetilde {m_1''u''}}{\partial x}}_{T^{m1}}\\
+ND_{m1},
\end{multline}

\noindent where $m_1$ is the mass fraction of the heavy fluid. For turbulent kinetic energy transport:

\begin{multline}
\underbrace{\bar \rho \left(\frac{\partial \widetilde{E''}}{\partial t}+u_{trans}\frac{\partial \widetilde{E''}}{\partial x}\right)}_{t^{E''}}+
\underbrace{\bar \rho \left(\tilde {u}-u_{trans}\right)\frac{\partial \widetilde{E''}}{\partial x}}_{A^{E''}}= 
\underbrace{-\overline {u''}\frac{\partial \bar p}{\partial x}}_{P_b^{E''}} \\
\underbrace{-\bar {\rho}\widetilde {u_i''u''}\frac{\partial \tilde {u_i}}{\partial x}}_{P_s^{E''}}+
\underbrace{\overline{p'\frac{\partial u_k''}{\partial x_k}}}_{\Pi^{E''}}
\underbrace{-\frac{\partial }{\partial x}\left(\bar \rho \widetilde {E''u''}+\overline{p'u''}\right)}_{T^{E''}}
+ND_{E''},
\end{multline}

\noindent where kinetic energy is $E''=u_i''u_i''/2$. For mass fraction variance transport:

\begin{multline}
\underbrace{\bar \rho \left(\frac{\partial \widetilde{m_1''^2} }{\partial t}+u_{trans}\frac{\partial \widetilde{m_1''^2} }{\partial x}\right)}_{t^{m1''^2}}+
\underbrace{\bar \rho \left(\tilde {u}-u_{trans}\right)\frac{\partial \widetilde{m_1''^2}}{\partial x}}_{A^{m1''^2}}=\\
\underbrace{-2\bar {\rho}\widetilde {m_1''u''}\frac{\partial \tilde m_1}{\partial x}}_{P^{m1''^2}} 
\underbrace{-\frac{\partial \bar \rho \widetilde {m_1''^2 u''}}{\partial x}}_{T^{m1''^2}}
+ND_{m1''}.
\end{multline}

The mean translational velocity $u_{trans}$ of the layer

\begin{equation}
u_{trans}(t)=\frac{\int_0^{L_x} \bar{u}\,\widetilde{m_1}(1-\widetilde{m_1}) \rm{dx}}{\int_0^{L_x} \widetilde{m_1}(1-\widetilde{m_1}) \rm{dx}}
\end{equation}

\noindent accounts for the small non-zero net velocity of the mixing layer in some simulations; for example, in the quarter scale case in Flamenco there is a residual $\tilde {u}\approx 0.5$ m/s attributed to imperfect boundary conditions (shock wave evacuation outside  the computational domain). Each term is computed principally via sixth order central differences, but with biased stencils near the boundaries.

\subsection{Non-Dimensionalisation \label{sec:nondims}}

The non-dimensionalisation follows the analysis outlined in Thornber {\it et al.} \cite{Thornber2017}. These include estimates of the initial growth rate $\dot W_0$ using a power-weighted mean wavelength $\bar \lambda=2\pi/\bar k$ and mean post-shock density $\rho^+_c$, which are defined as

\begin{eqnarray}
\dot W_0&=&0.564\, \sqrt{\frac{7}{12}}\, k_{\rm max}\, At^+\, \sigma^+\, \Delta u,\nonumber\\
\bar k&=&\sqrt{\frac{7}{12}}\,\,k_{\rm max},\,\,\,\rho^+_c=\frac{\rho^+_1+\rho^+_2}{2},
\label{nondims}
\end{eqnarray}

\noindent where the post-shock standard deviation of the perturbation amplitude is $\sigma_0^+=C\sigma_0$, an approximate compression factor \cite{Richtmyer1960} is $C=(1-\Delta u/U_i)$  where $U_i=434$ m/s is the velocity of the incident shock, $\Delta u=291.575$ m/s is the velocity impulse imparted by the shock; post-shock densities were $5.22$ and $1.80$ kg/m$^{3}$ for the heavy and light fluids respectively, giving $\rho_c^+=3.51$ kg/m$^{3}$, $u_c=\dot W_0=12.649$ m/s, and $\bar \lambda=0.2571$ m. The layer centre is defined as the location at which $\tilde m_1 (x,t)=0.5$. 

Three measures of self-similarity may be employed. The first is to examine whether the layer properties at multiple time instants collapse when normalised by characteristic length, time, and mass scale. To examine this, some figures are plotted using axes scaled with the integral width

\begin{equation}
W(t)=\int^{L_x}_0 \bar f_1 (x,t) [1-\bar f_1(x,t)] dx,
\label{intwidtheq}
\end{equation}

\noindent and $\dot W(t)$, the values of which are given in Table \ref{dataresults2}. This Table also gives a second measure, $\log_2 (3W/\bar \lambda)$, which is an indicator of the number of modal generations. Elbaz and Shvarts propose that a mixing layer must achieve $\geq 3$ generations to reach self-similarity \cite{Elbaz2018}. Here approximately $2.29$ generations are reached, an indication that the layer may not be in the fully self-similar regime. A third measure of self-similarity is an integral mixing measure such as the `molecular mixing'

\begin{equation}
\Theta=\frac{\int \overline {f_1 f_2} dx}{\int \bar f_1 \bar f_2 dx}, 
\end{equation}

\noindent which varies by less than $1.5$\% in the last $3 $ s of the Flamenco simulation. The first measure will be explored further within this paper. For the rest of the paper, quantities are plotted against $x/W$, where $x=0$ is the centre of the mixing layer.

\begin{table}
\centering
\caption{Mixing layer width $W$ and growth rate $\dot W$ as a function of time for the code-averaged data for the quarter scale problem.\label{dataresults2}}
\begin{tabular}{ccccc}
$t$ (s) & $\tau$ & $W$ (m) & $\log_2 (3W/\bar \lambda)$  & $\dot W$ (m/s)  \tabularnewline
\hline
\hline
              0.01& 0.49 & 0.07605 &-0.17 & 4.12  \tabularnewline
              0.025&1.23 & 0.1107 & 0.37& 1.436   \tabularnewline
              0.5& 24.6 & 0.235 & 1.46 & 0.122  \tabularnewline
              1& 49.7 &0.2833 & 1.73 & 0.0779  \tabularnewline
              2&98.5 &0.3443  & 2.0& 0.0492  \tabularnewline
              4&197 &0.4195 &2.29 & 0.02956   \tabularnewline
\end{tabular}
\end{table}

\subsection{Grid Convergence \label{conv}}

Grid convergence of the mean flow quantities has been demonstrated in Thornber {\it et al.} \cite{Thornber2017} for a highly resolved simulation where all length scales are four times larger (relative to the domain size). In the current study, higher order statistics are examined; thus it is important to revisit grid convergence to ensure that these metrics are also reasonably converged.  Appendix \ref{convergence} plots all of the individual terms of each transport equation for the code Flamenco at grid resolutions ranging from $180\times 128^2$ to $720 \times 512^2$, in Figs. \ref{ndconv} and \ref{meanconv}. The overall agreement at the two highest grid resolutions for all transport terms is excellent, indicating that the results are reasonably grid converged. 

The spikes of heavy fluid are intense vortical structures of relatively small size (compared to the grid spacing) and as such are the most challenging `large' scale features to converge. Given even small differences in the trajectory at early time, strong interactions of the spike's ring-shaped vortices with other spikes and the mixing layer can cause large differences at late time. This can be seen as a region of relatively lower convergence at $x/W\approx 8$ for the kinetic energy and kinetic energy balances in Figs. \ref{econv} and \ref{etransconv}. Such an intense impact of the spikes on the kinetic energy budget has also been observed in an analysis of the RMI spectral kinetic energy budget \cite{Thornber2012b}.

The levels of noise in each of the quantities is a good indicator of statistical convergence, as no smoothing has been applied. However, the statistical noise does not mask the overall trends for the majority of the results presented. Based on the previous study \cite{Thornber2017}, it is expected that results from Turmoil and Triclade would converge at a similar grid resolution to Flamenco. Flash, however, has historically demonstrated slower convergence compared to the other algorithms, but given the good agreement with the other algorithms presented here, and the fact that Flamenco is reasonably converged at the intermediate grid level (a factor of eight times fewer points than the highest resolution), the Flash results are also expected to be reasonably converged.

\subsection{Note on the Treatment of  Numerical Results}

Compiling the data from each code highlighted numerical and physical issues at different times. Flamenco exhibits post-shock oscillations at the earliest times which are clear in the plots of $\tilde{u}$, Triclade and Turmoil show highly fluctuating acoustic pressure correlations at late times (terms such as $\Pi^{E''}$), and Flash has difficulties at the spike side boundary at late times (visible in the spike side $\widetilde{E''}/u_c^2$ at $t=2$ s for example). An effort has been made here to present all results except where (i) the result is so noisy that it obscures interpretation of the rest of the data or (ii) where it is clearly dominated by a non-physical effect. There are some cases which are marginal (perhaps unphysical), but can fit on the plot without obscuring the data. Where data is deliberately excluded due to such effects, the omission and reason for omission is stated in the figure caption. 

\section{Results \label{results}}

\subsection{Flow Features at Four Time Instants \label{phenom}}

\begin{figure*}
\begin{centering}
\includegraphics[trim={7cm 0. 17cm 0.7cm},clip,height=0.45\textwidth]{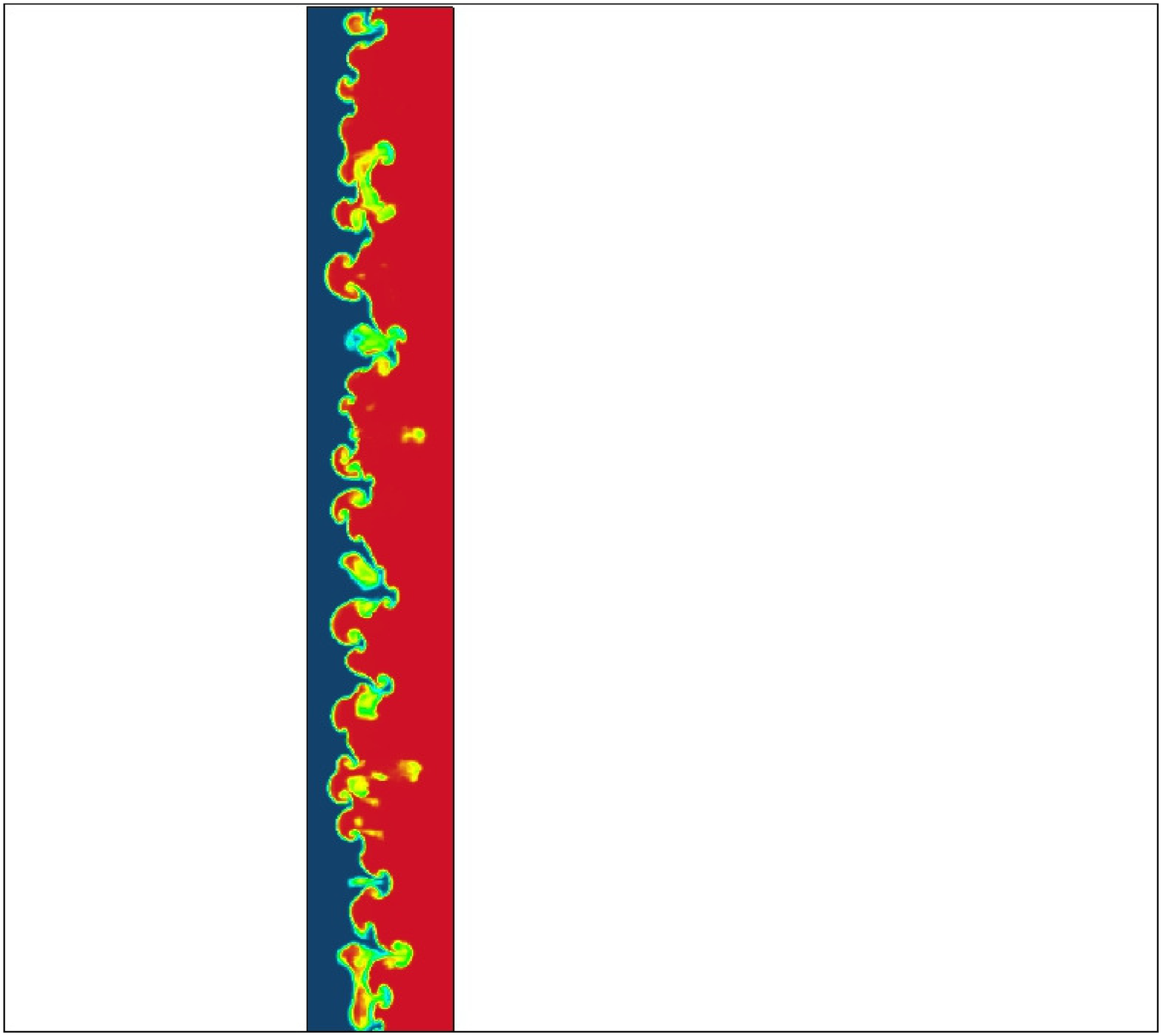}
\includegraphics[trim={5cm 0 12cm 0.7cm},clip,height=0.45\textwidth]{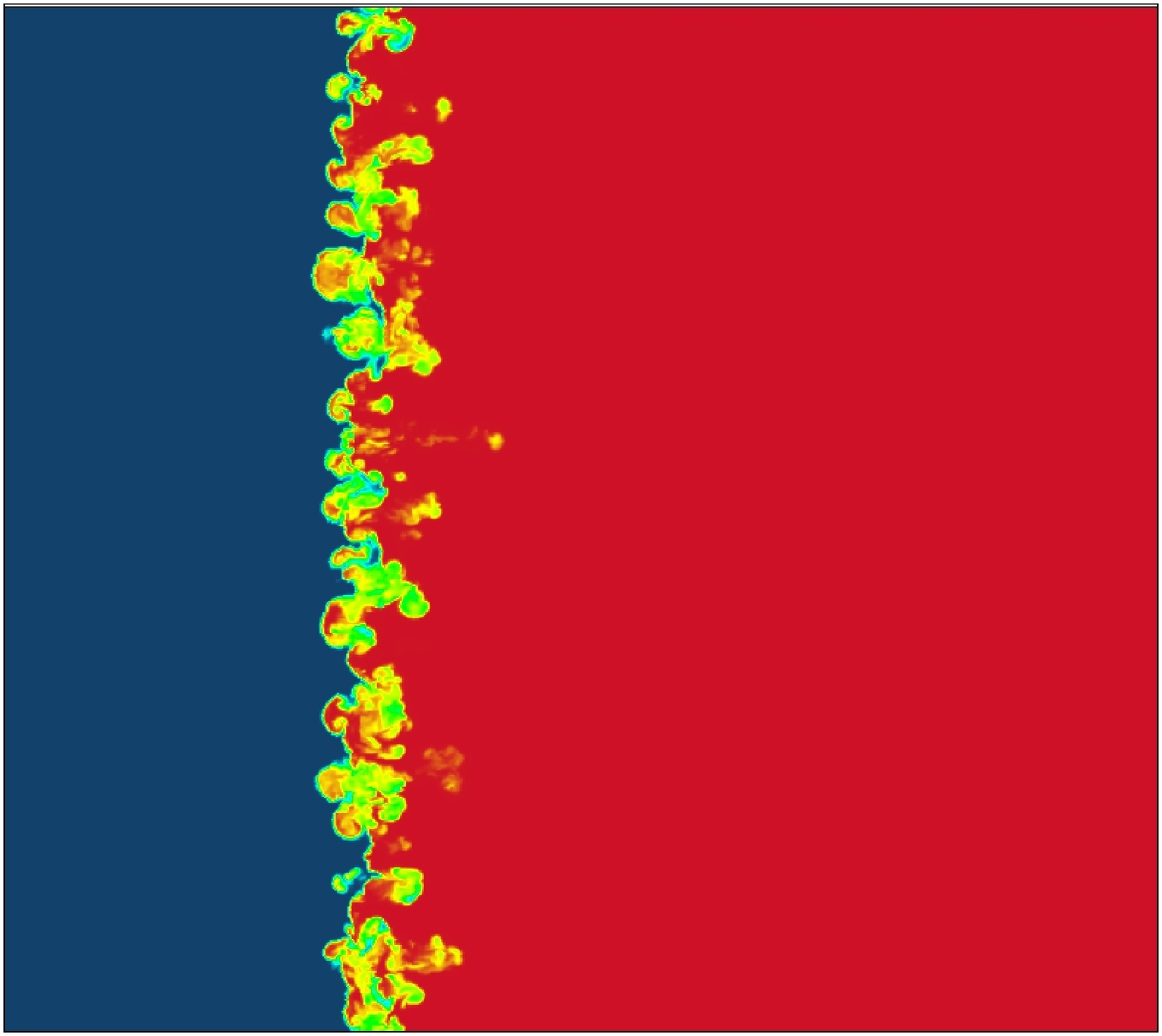}
\includegraphics[trim={5cm 0 7cm 0.7cm},clip,height=0.45\textwidth]{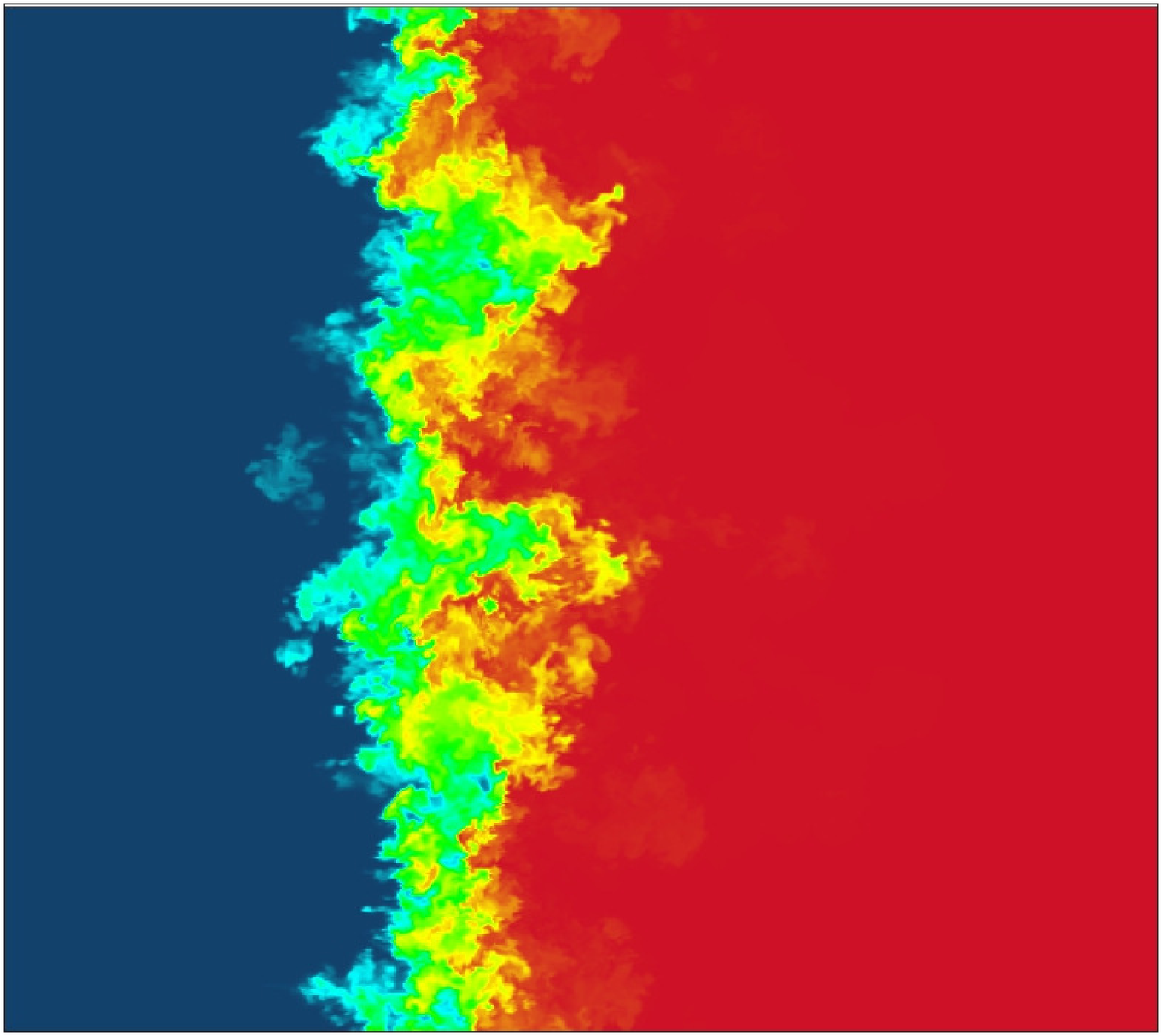}
\includegraphics[trim={5cm 0 4cm 0.7cm},clip,height=0.45\textwidth]{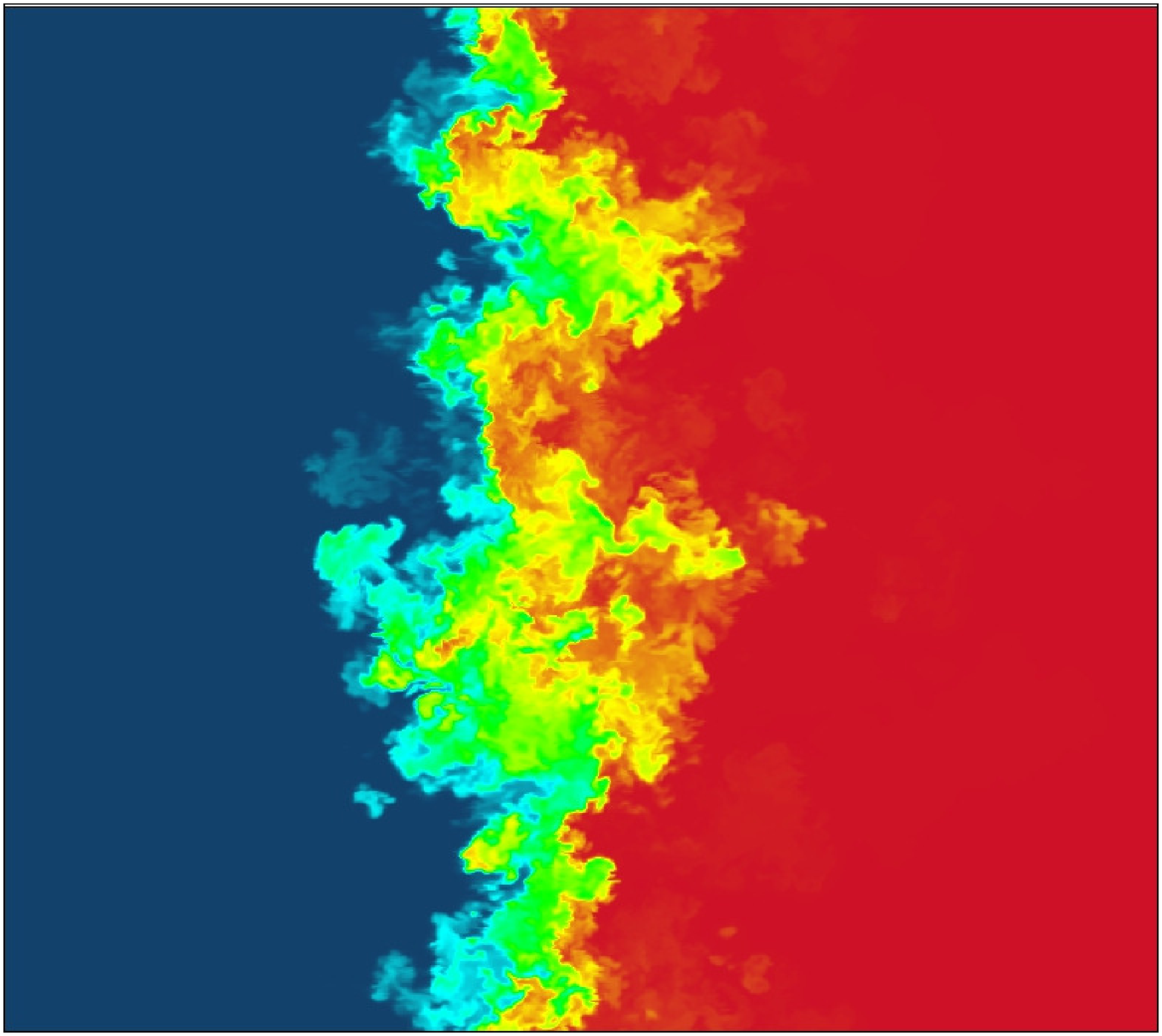}
\caption{Visualisation of slices of volume fraction of the heavy fluid from the Flamenco simulation of the mixing layer at $t=0.01$, $0.025$, $1$, and $2$ s (from left to right). Blue indicates the heavy fluid and red the light fluid.  \label{visslice}} 
\end{centering}
\end{figure*}

\begin{figure*}
\begin{centering}
\includegraphics[width=0.49\textwidth]{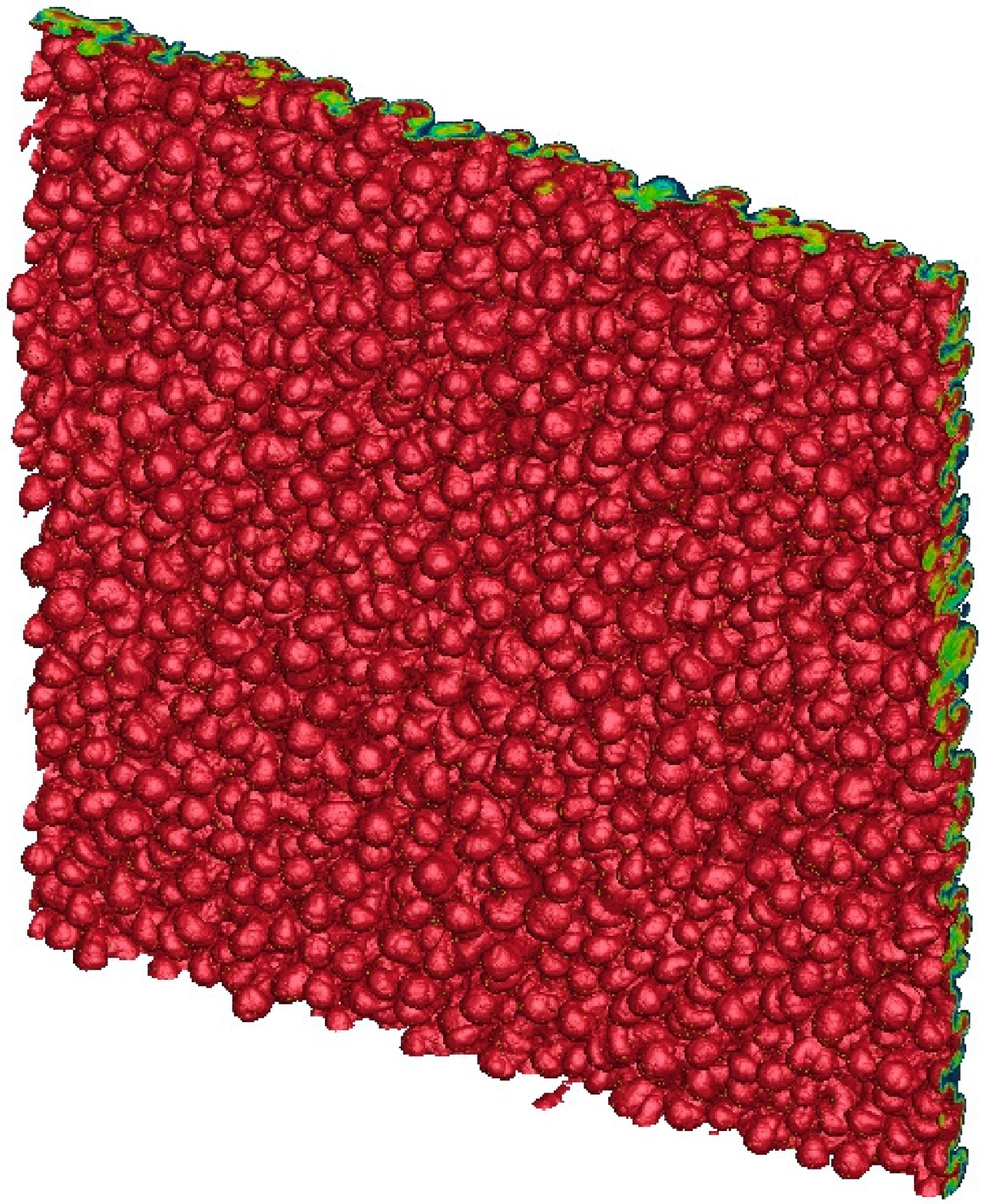}
\includegraphics[width=0.49\textwidth]{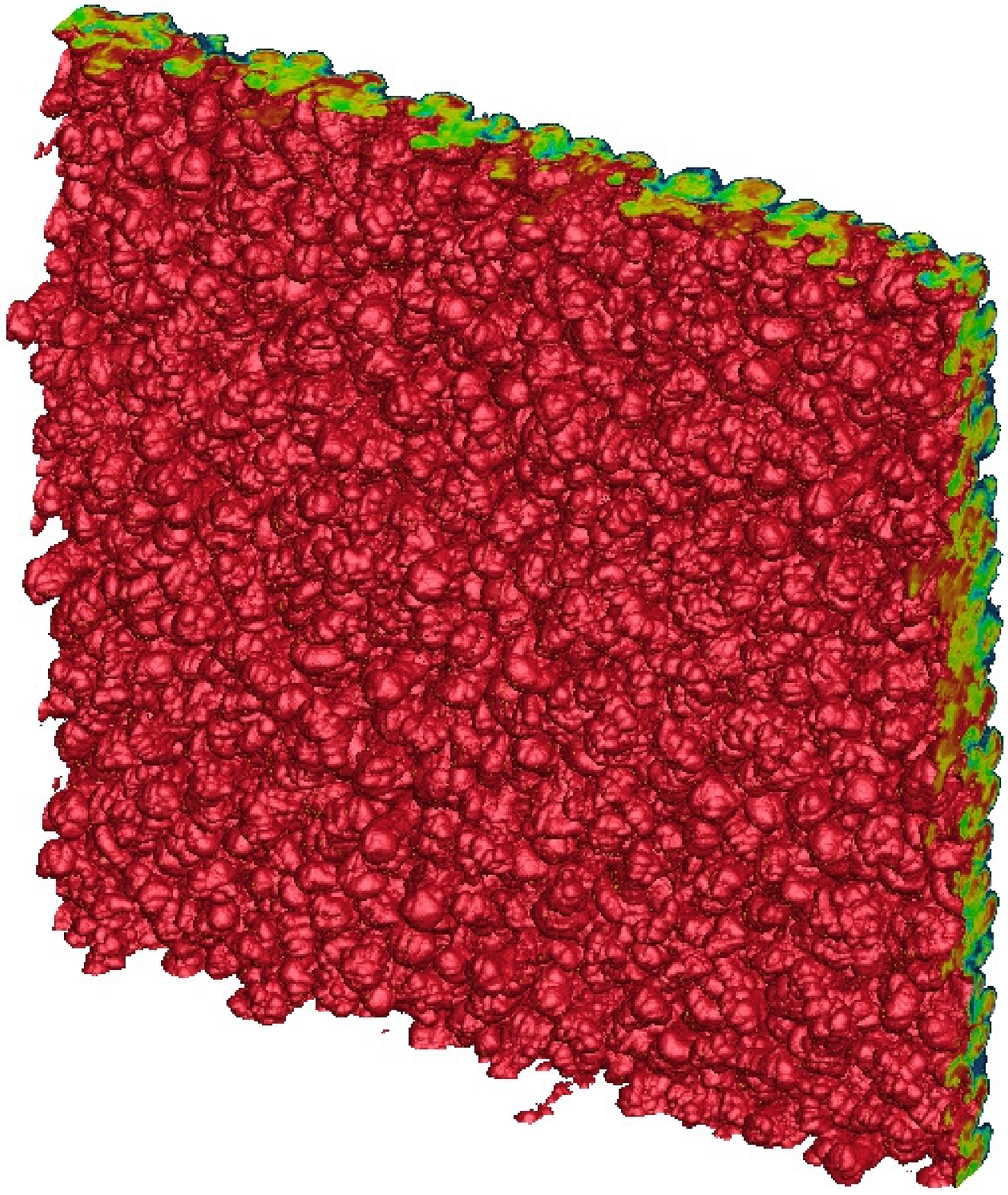}
\includegraphics[width=0.49\textwidth]{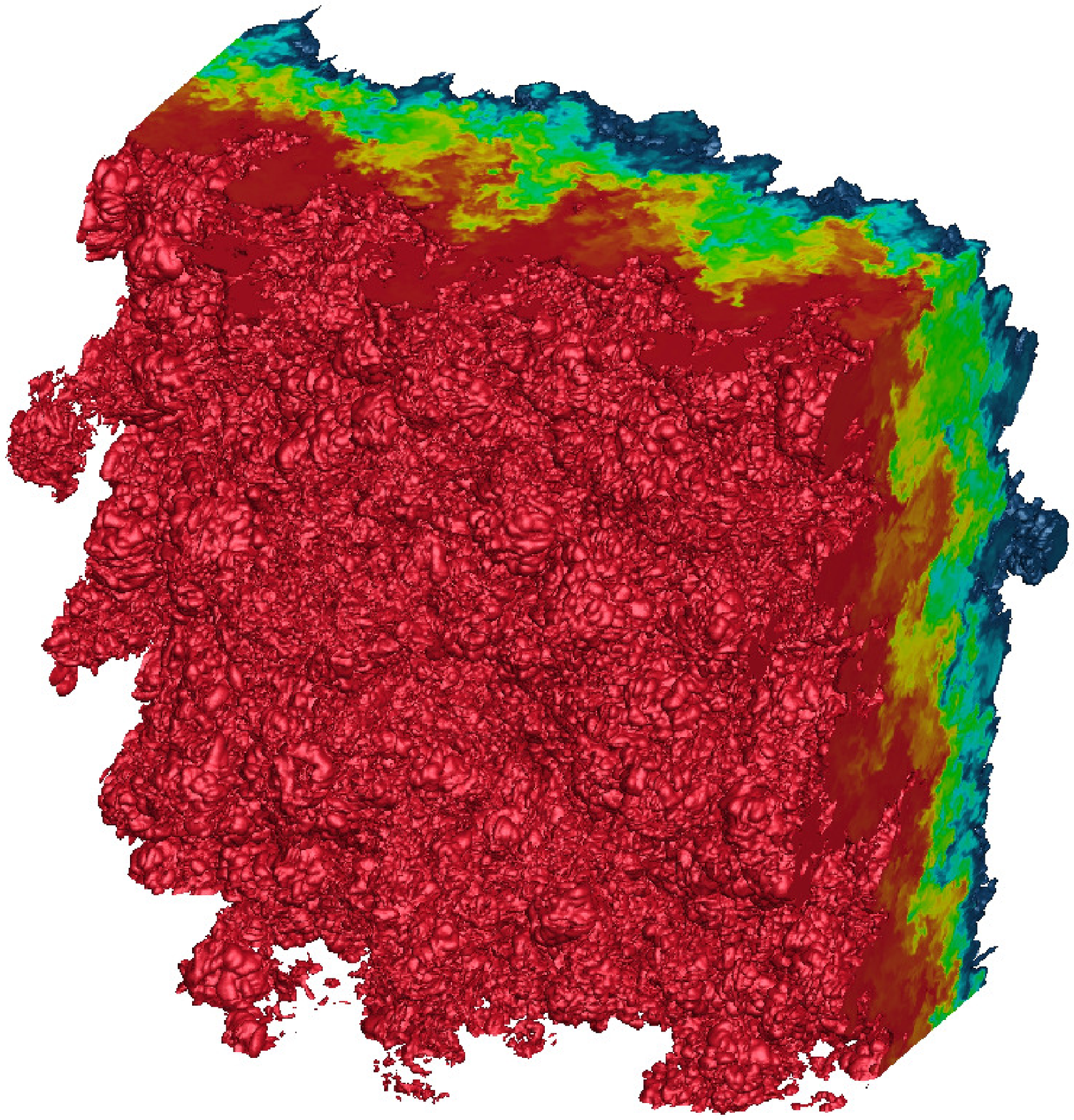}
\includegraphics[trim={7cm 5cm 7cm 7cm},clip,width=0.49\textwidth]{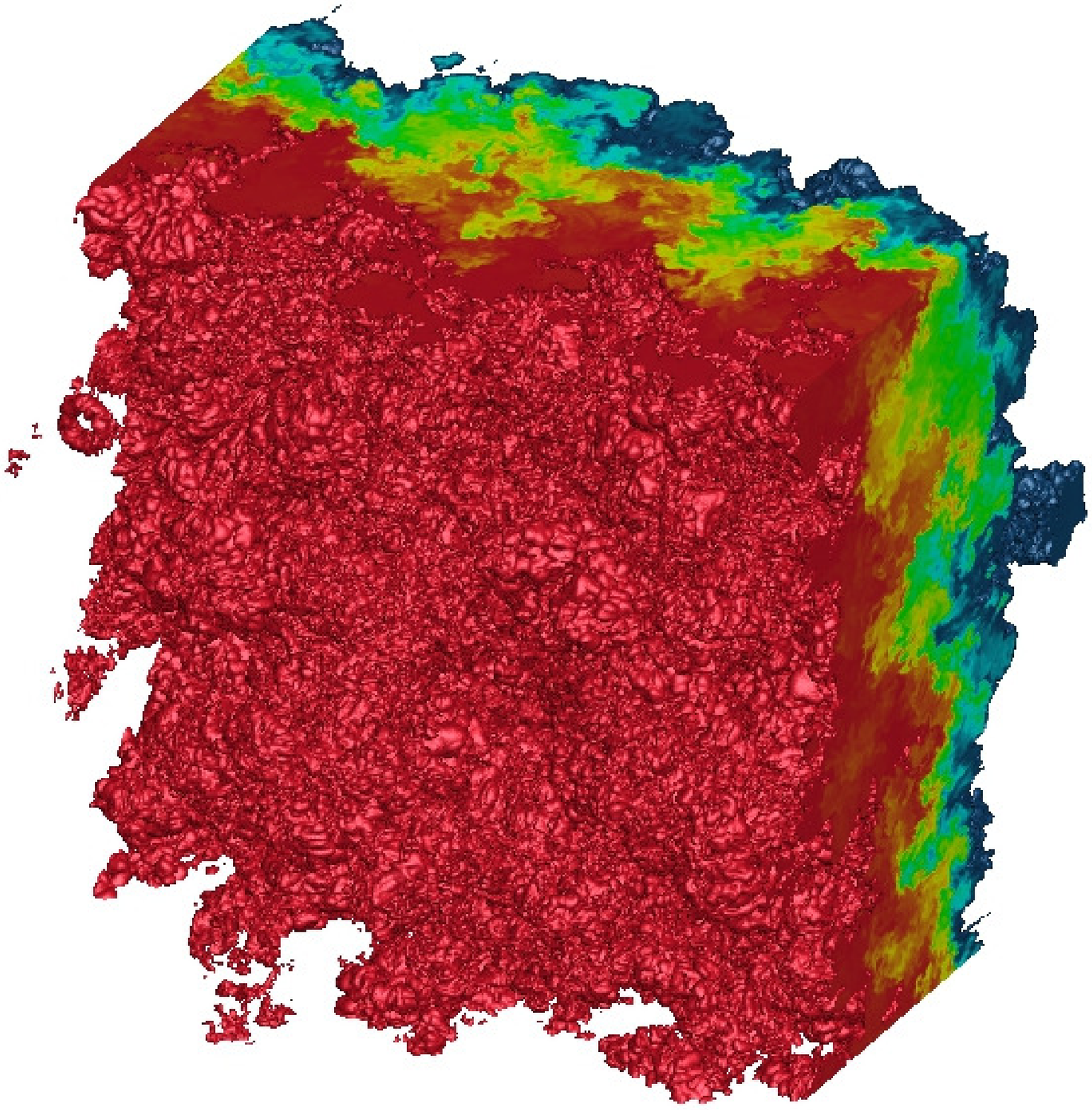}
\caption{Visualisation of volume fractions from the Flamenco simulation of the mixing layer at  $t=0.01$, $0.025$, $1$, and $2$ s, looking at the spike side of the mixing layer. Red indicates the isosurface of heavy fluid volume fraction 0.001 and blue 0.999.  \label{vis}} 
\end{centering}
\end{figure*}

Visualisation of the distribution of the volume fractions within the flow field are shown for the four time instants chosen using slices in the $y$-plane in Fig. \ref{visslice} and three dimensional isosurfaces of volume fraction in Fig. \ref{vis} from the Flamenco code. 

At the initial time $t=0.01$ s, the individual modes at the highest wavenumber have already formed characteristic mushroom shaped bubbles (light fluid penetrating the heavy) and spikes (heavy fluid penetrating the light). As expected at this Atwood number, several spikes have propagated substantially beyond the bulk of the mixing layer. Two of these are visible on the left-most image in Fig. \ref{visslice}, however due to the perspective they are not as cleanly seen in the three dimensional rendering. 

At $t=0.025$ s the layer has begun to transition, where the spikes contain relatively well mixed fluid but the bubble heads still contain some unmixed material. This provides a preliminary indication that turbulent fluctuations are initially stronger on the spike side of the mixing layer. There is visible evidence of the vortex projectiles propagating far from the layer on the spike side, however their intensity and size has reduced. At the two latest times there is a core of well mixed fluid, but the flow is still relatively non-uniform in the homogeneous direction. The layer width grows by $\approx 20$\% between $t=1$ and $2$ s, and low wavenumber modes dominate at $t=2$ s. Strong three-dimensionality and a range of vortex length-scales are apparent at $t=1$ and $2$ s in Fig. \ref{vis}; however, there is still evidence of coherent spike structures such as the ring vortex at the upper left of the three-dimensional visualisation at $t=2$ s.

\begin{figure*}
\begin{centering}
\subfigure[Early Time]{\includegraphics[width=0.49\textwidth]{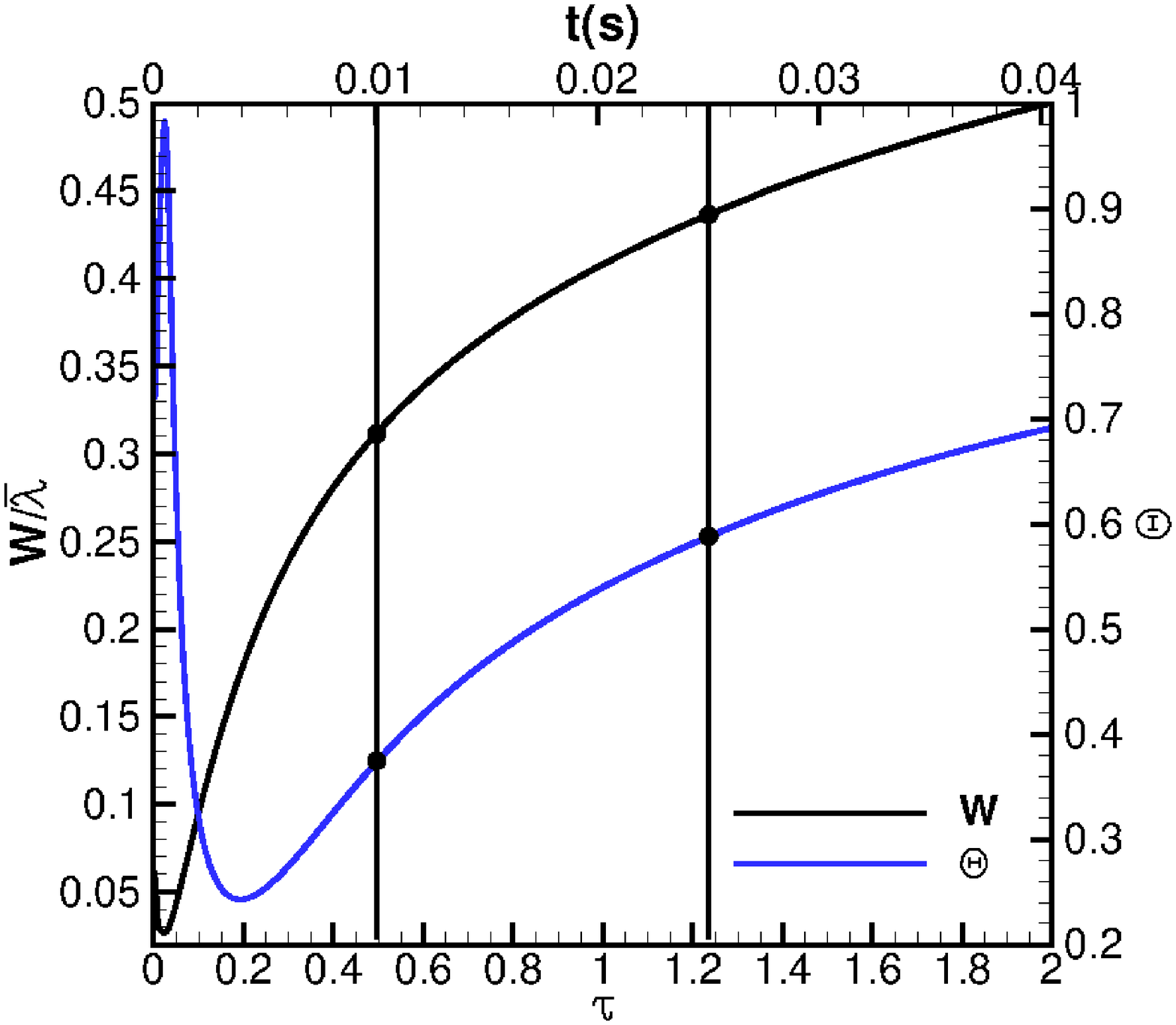}}
\subfigure[Late Time]{\includegraphics[width=0.49\textwidth]{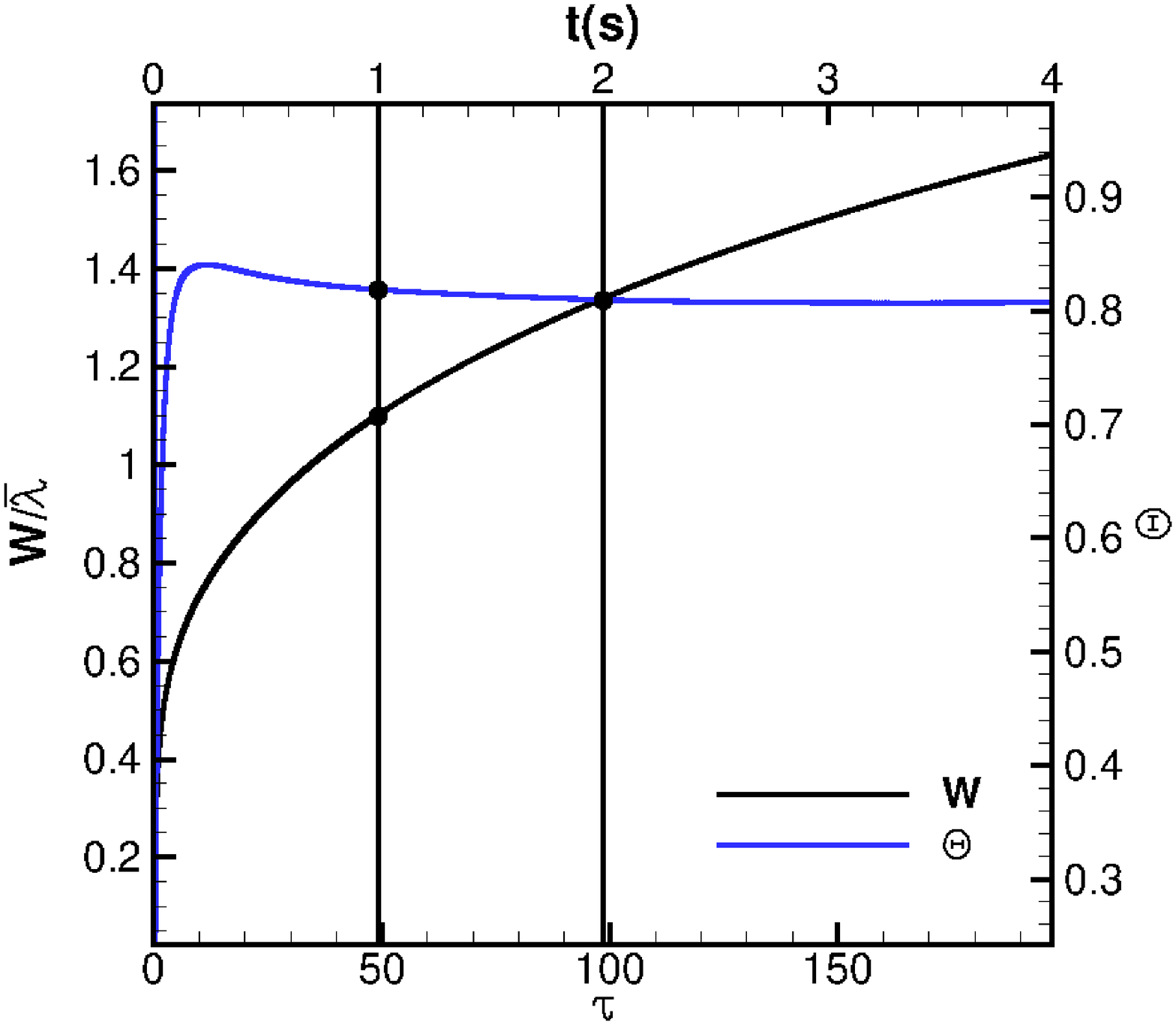}}
\caption{Integral width and molecular mix measure $\Theta$ presented for (a) early and (b) late time, illustrating the times considered with a solid filled circle \label{sampletimes}} 
\end{centering}
\end{figure*}

 Figure \ref{sampletimes} plots the integral width and molecular mixing measure $\Theta$ at early and late times. Superimposed on the figure are the chosen sampling times. The first two sampling times are chosen to be just after the shock passes through the interface and at a short time later. At this stage, the layer is expected to be formed of coherent structures and a wide range of turbulent length scales is not expected. The two latest times are chosen to be at a time where the layer may be approaching self-similarity, but with a compromise between having the widest possible range of turbulent lengthscales and a moderate impact of a lack of statistical accuracy in the homogeneous direction and domain-size constraint \cite{Thornber2016}. From $t=1$ to $4$ s, $\Theta$ changes by only $1.5$\% in Flamenco simulations, an indication that it is approaching a self-similar state (but not yet reached, as discussed earlier).

\subsection{Momentum Transport}

\begin{figure*}
\begin{centering}
\subfigure[\hspace{0.1cm} $t=0.01$ s]{\includegraphics[width=0.49\textwidth]{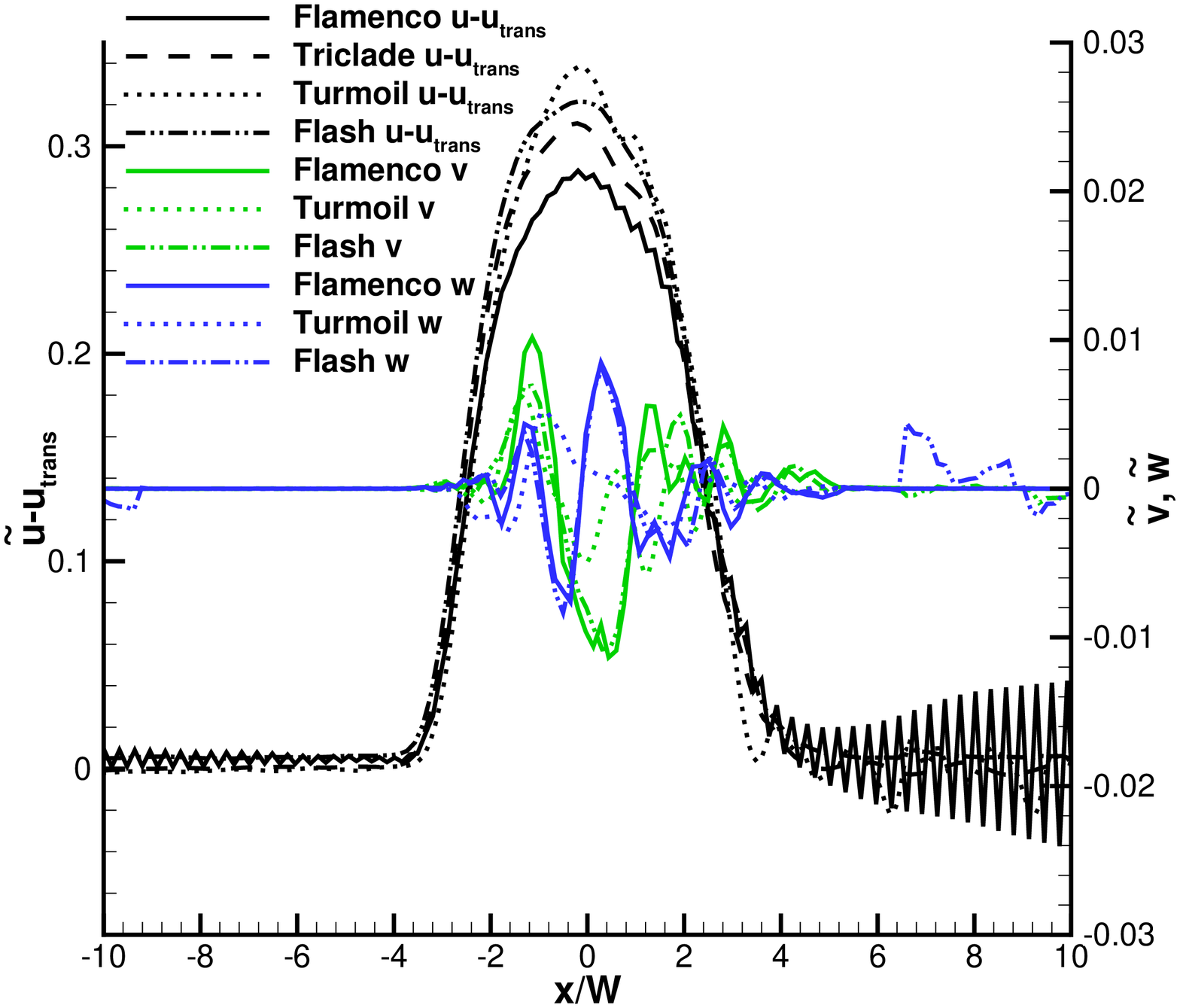}}
\subfigure[\hspace{0.1cm} $t=0.025$ s]{\includegraphics[width=0.49\textwidth]{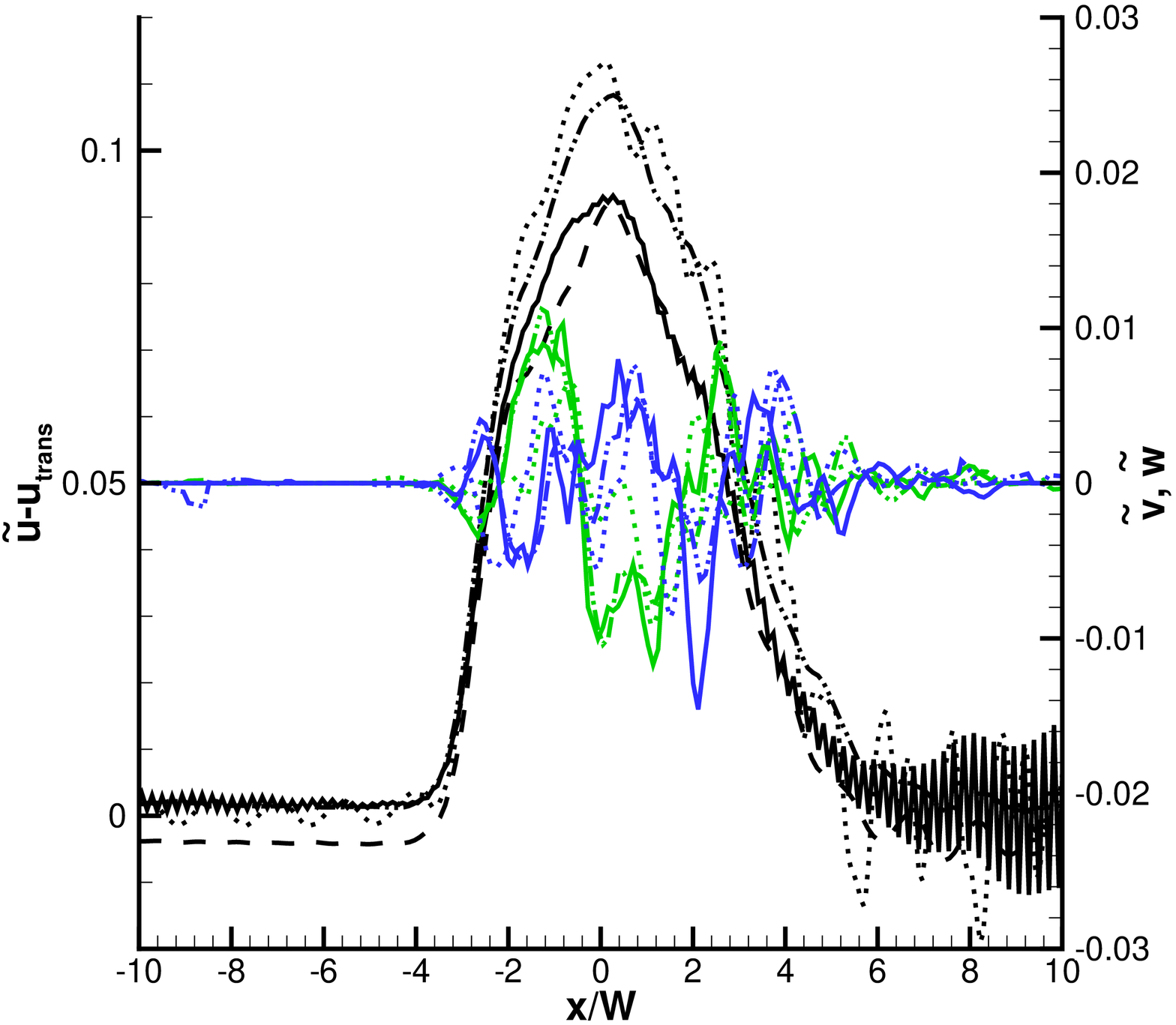}}
\subfigure[\hspace{0.1cm} $t=1$ s]{\includegraphics[width=0.49\textwidth]{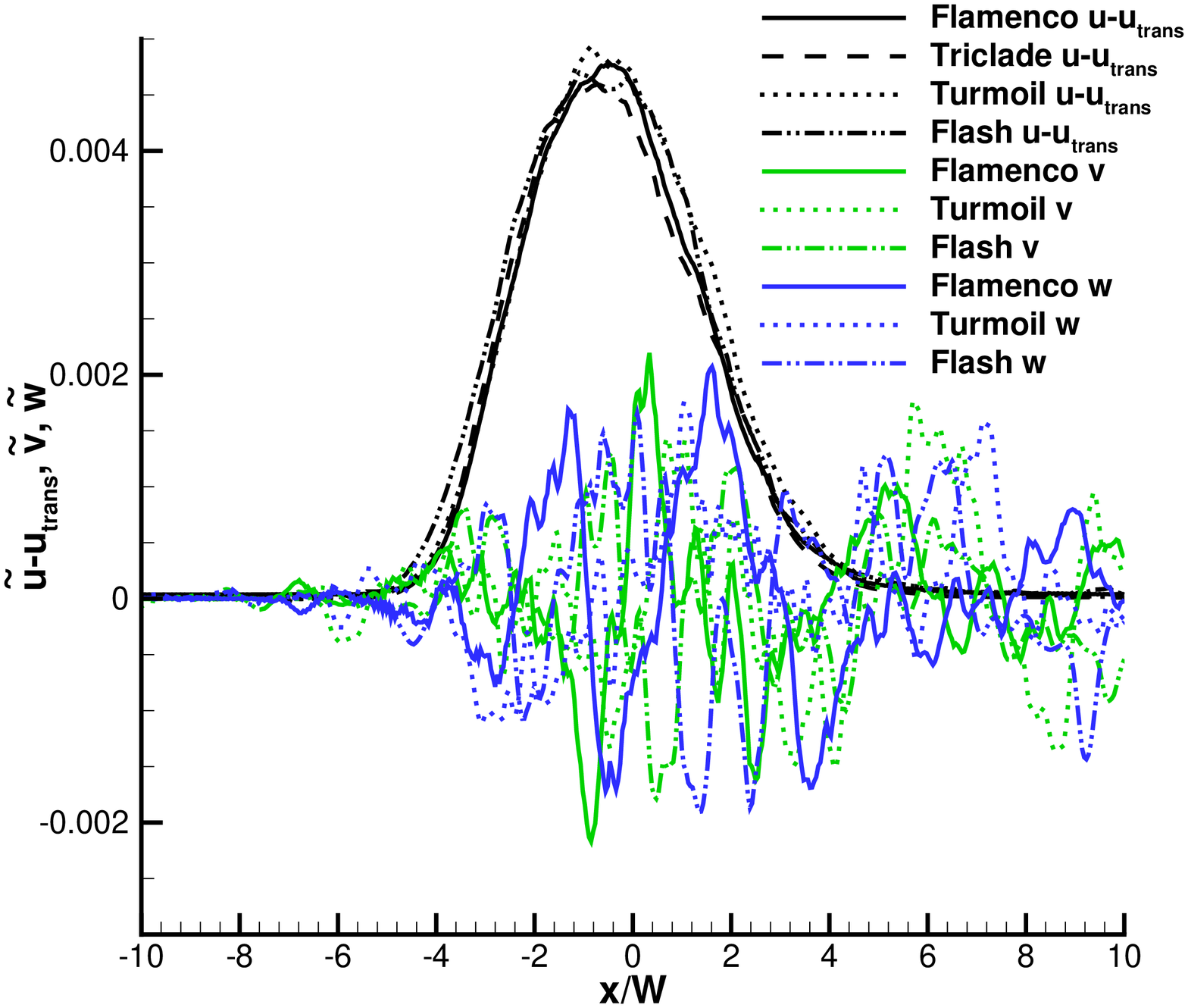}}
\subfigure[\hspace{0.1cm} $t=2$ s]{\includegraphics[width=0.49\textwidth]{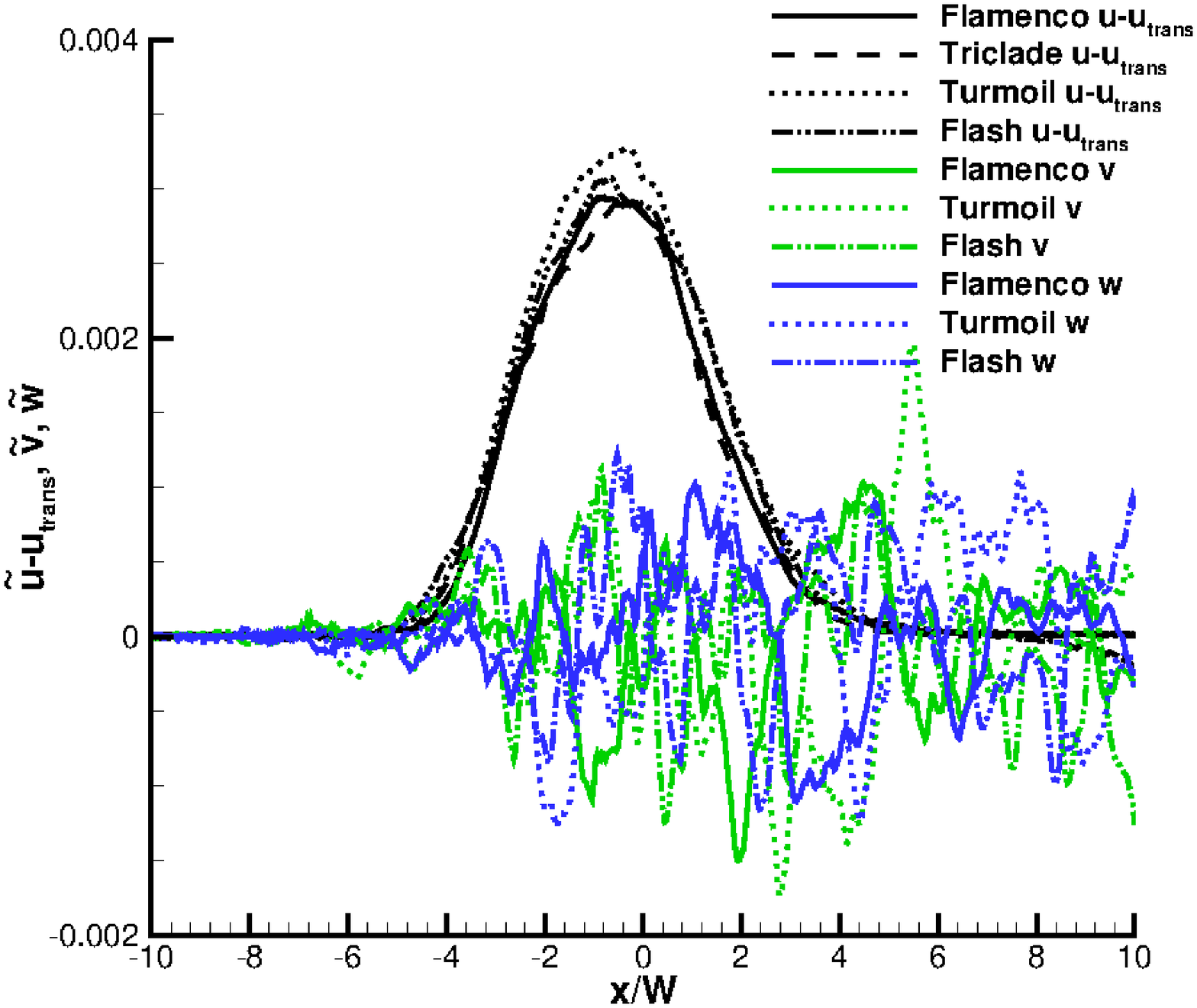}}
\caption{Mean velocity components $\tilde{u}$, $\tilde{v}$, and $\tilde{w}$  at (a) $t=0.01$ s, (b) $0.025$ s, (c) $1$ s, and (d) $2$ s. All terms non-dimensionalised by $u_c$.\label{momtilde}} 
\end{centering}
\end{figure*}

Figure \ref{momtilde} plots the profiles of $\tilde{u}$,$\tilde{v}$, and $\tilde{w}$ normalised by $u_c$ immediately following shock interaction ($0.01$ s), a short time later where mixing is increasing rapidly ($0.025$ s), at the onset of approximate self-similarity ($1.0$ s), and at the latest time ($2.0$ s). The overall agreement for $\tilde{u}$ is very good across the codes, particularly at late time, with the main discrepancies at early times where the peak $\tilde{u}$ varies by 14\% from the average peak value. The Favre-averaged velocity is positive, representing net flux of heavy material into the light material as expected. At early time the profile of $\tilde{u}$ is asymmetric, with a larger gradient on the bubble side 
($x<0$)
and more gradual decrease on the spike side
($x>0$);
 however, at late times it is reasonably symmetric and the peak velocity is at the centreline (defined by mean mass fraction of one half). Asymmetries are expected at this moderate Atwood number. 

The Favre-averaged in-plane velocities $\tilde{v}$ and $\tilde{w}$ are relatively large compared with $\tilde{u}$ at late times, where it could be expected that these quantities should average to zero in the limit of an infinite domain. As the domain size is finite, these components may take on values roughly equivalent to the noise observed in the evaluation of $\tilde{u}$. This is clearly not the case as the profile of $\tilde{u}$ is relatively smooth. The large magnitude of $\tilde{v}$ and $\tilde{w}$ also extends well into the spikes ($x/W>4$), where the plane-averaged heavy mass fraction is $<1$\%, and these fluctuations are associated with the propagation of `vortex projectiles' exiting the mixing layer. As this is a turbulent mixing layer, it is not surprising to see differences in the locations of the peaks and troughs for the in-plane components between the algorithms, especially given that the mean $\tilde{v}$ and $\tilde{w}$ are $0.6$\% of the peak values at a given $x$ location. 

The relatively large magnitude is most likely due to the mean gradient of $\tilde{u}$ in the $x$ direction being constrained to be close to zero due to approximate incompressibility and the corresponding divergence-free constraint. Due to numerical miscibility, the current simulations will not satisfy $\nabla \cdot {\bf u}=0$, as $\nabla \cdot {\bf u}=-D \,\ln \rho/Dt$, where $D(.)/Dt$ is the material derivative. Although the flow at late times is at low Mach number and nearly incompressible, the advected density changes due to numerical mixing.  The periodic boundary conditions in the $y$ and $z$ directions do not constrain the net flow in that direction, thus leading to the observed profiles. 

\begin{figure}
\begin{centering}
\includegraphics[width=0.49\textwidth]{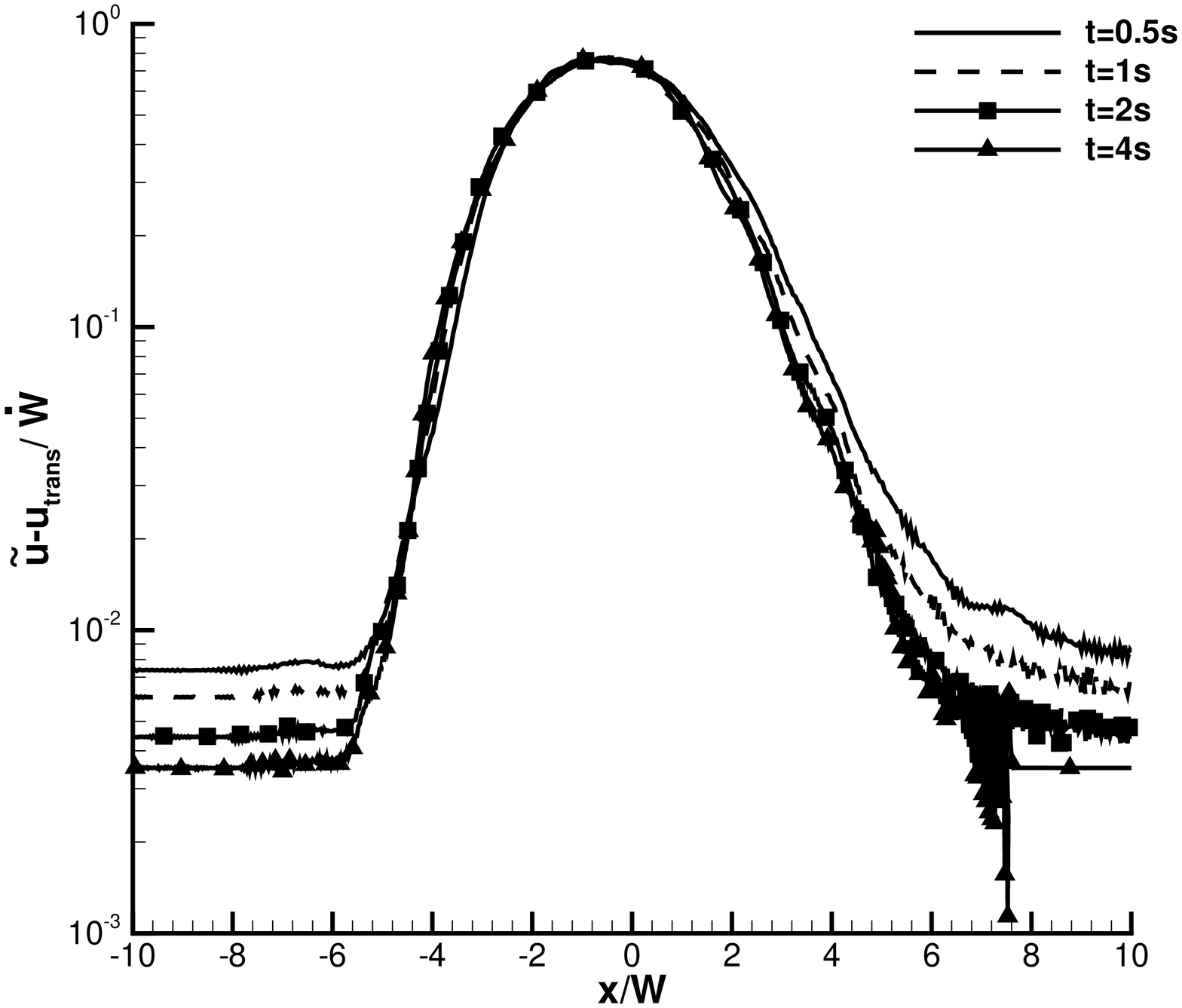}
\caption{Mean velocity component $\tilde{u}$ at $t=0.5$, $1$, $2$, and $4$ s (Flamenco results only) non-dimensionalised by $W(t)$ and $\dot{W}(t)$.  \label{ssmom}}
\end{centering}
\end{figure}

Late time self-similarity is examined in Fig. \ref{ssmom}, where the Flamenco data has been non-dimensionalised by the integral width $W(t)$ and time rate of change of the integral width $\dot {W}(t)$. An additional point at $t=4$ s has been added, where at that time spikes have left the computational domain, but the bulk of the mixing layer may be assumed to be reasonably well resolved. The momentum profiles collapse well for the bulk of the layer under the chosen scalings for $t>1$ s, however the spike side is taking longer to achieve approximate self-similarity.

\begin{figure*}
\begin{centering}
\subfigure[\hspace{0.1cm} $t=0.01$ s]{\includegraphics[width=0.49\textwidth]{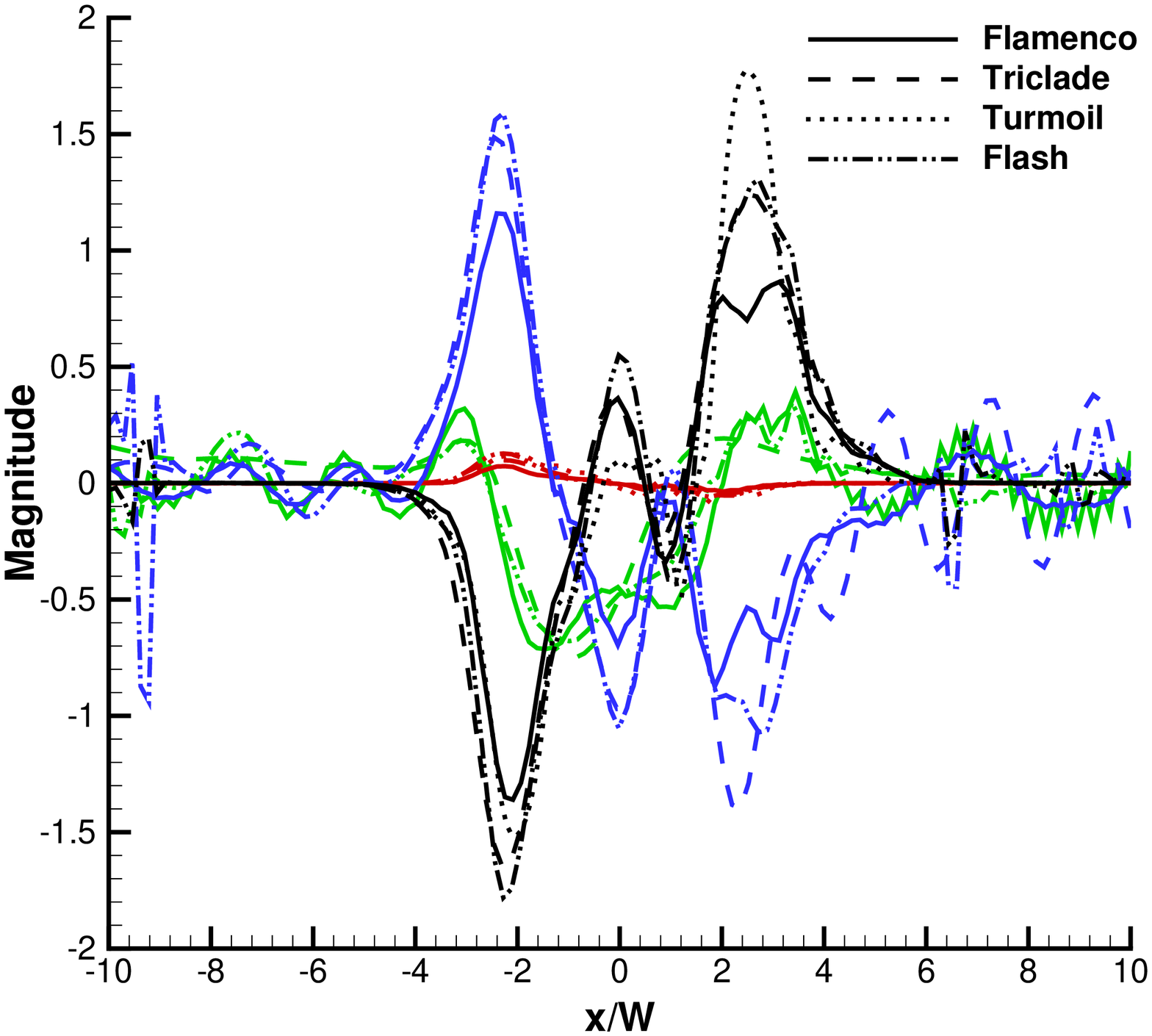}}
\subfigure[\hspace{0.1cm} $t=0.025$ s]{\includegraphics[width=0.49\textwidth]{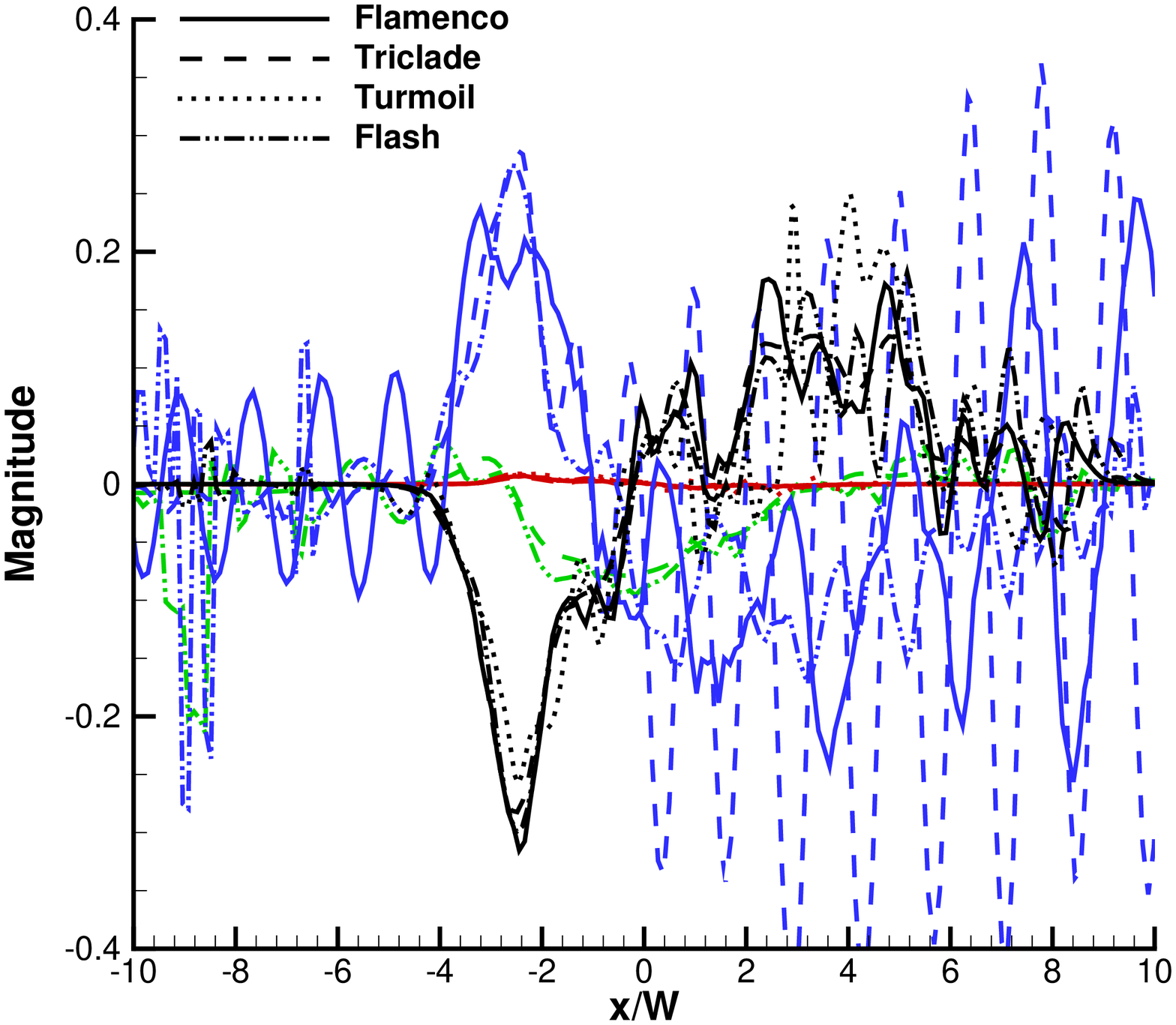}}
\subfigure[\hspace{0.1cm} $t=1$ s]{\includegraphics[width=0.49\textwidth]{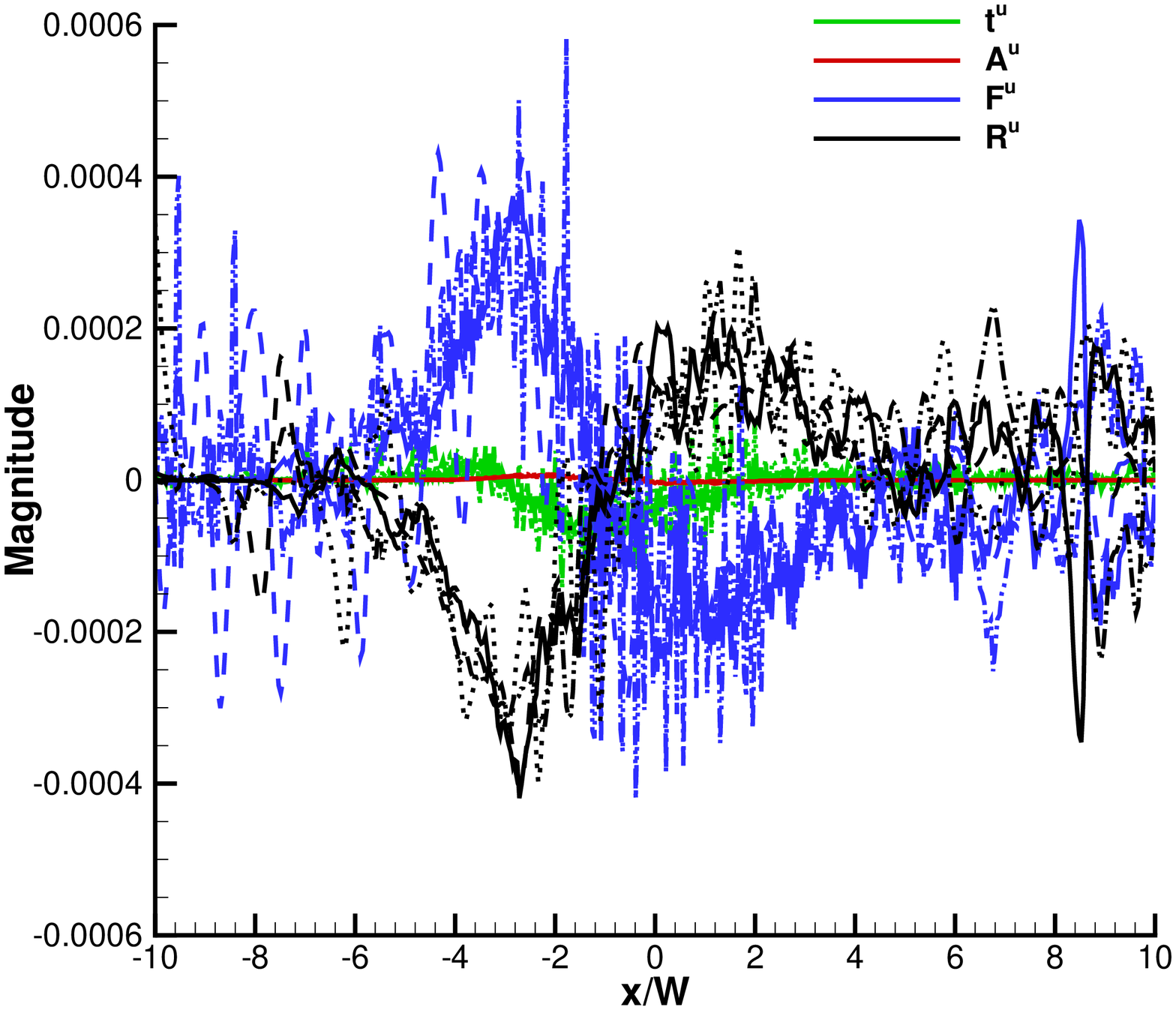}}
\subfigure[\hspace{0.1cm} $t=2$ s]{\includegraphics[width=0.49\textwidth]{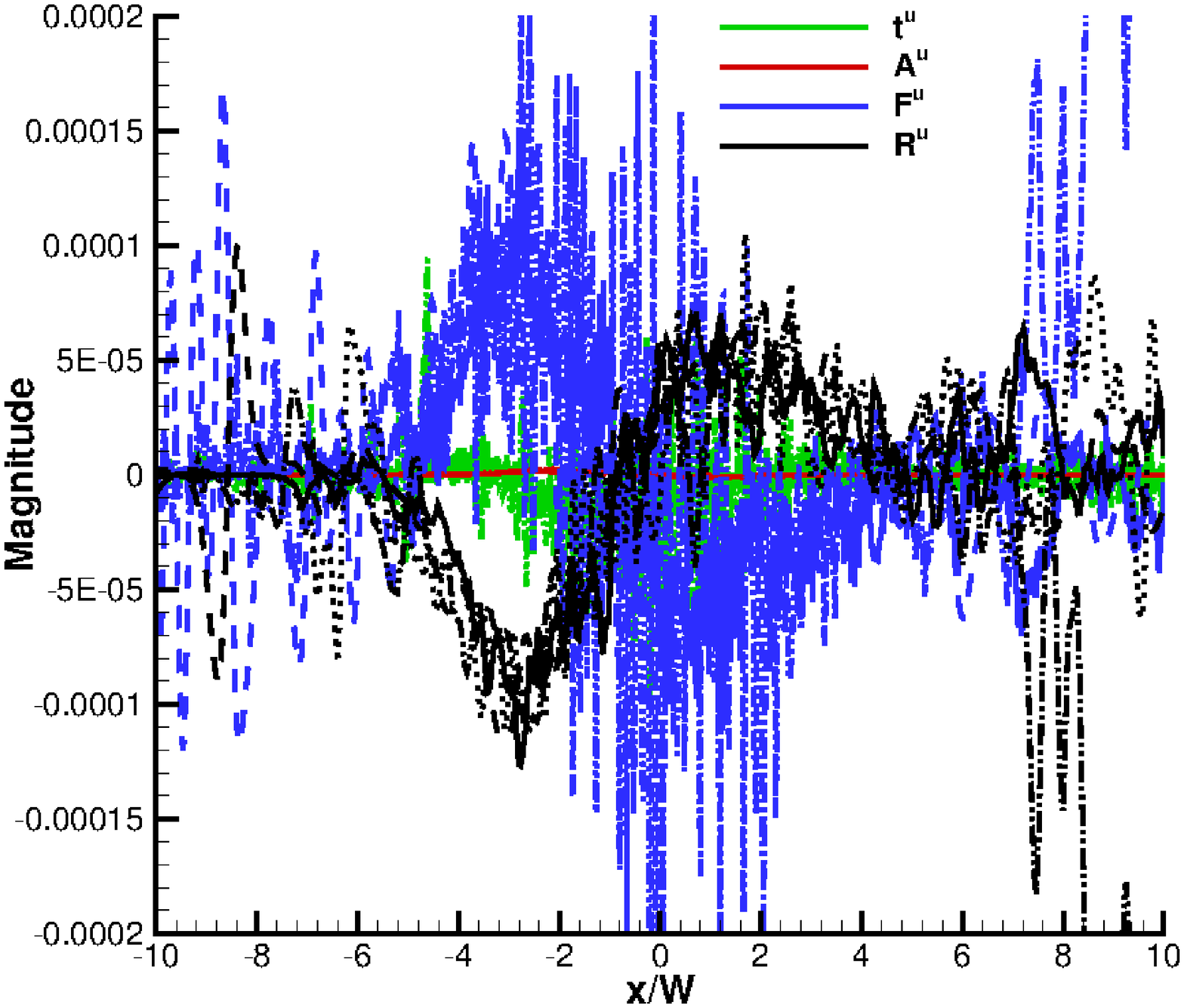}}
\caption{Individual terms in the $\tilde{u}$ transport equation at (a) $t=0.01$ s, (b) $0.025$ s, (c) $1$ s, and (d) $2$ s. Flamenco time derivative at $t=0.025$ s is not shown as it is dominated by the dissipation of the oscillations (see Fig. \ref{momflam}), and Turmoil $F^u$ at $t=0.025$, $1$, and $2$ s are not shown due to large fluctuations which mask the other curves (see Fig. \ref{momsmooth} for smoothed $F^u$). All terms non-dimensionalised by $\rho_c u_c^2/\bar \lambda$. \label{mom}}
\end{centering}
\end{figure*}

\begin{figure*}
\begin{centering}
\subfigure[\hspace{0.1cm} $t=0.01$ s]{\includegraphics[width=0.49\textwidth]{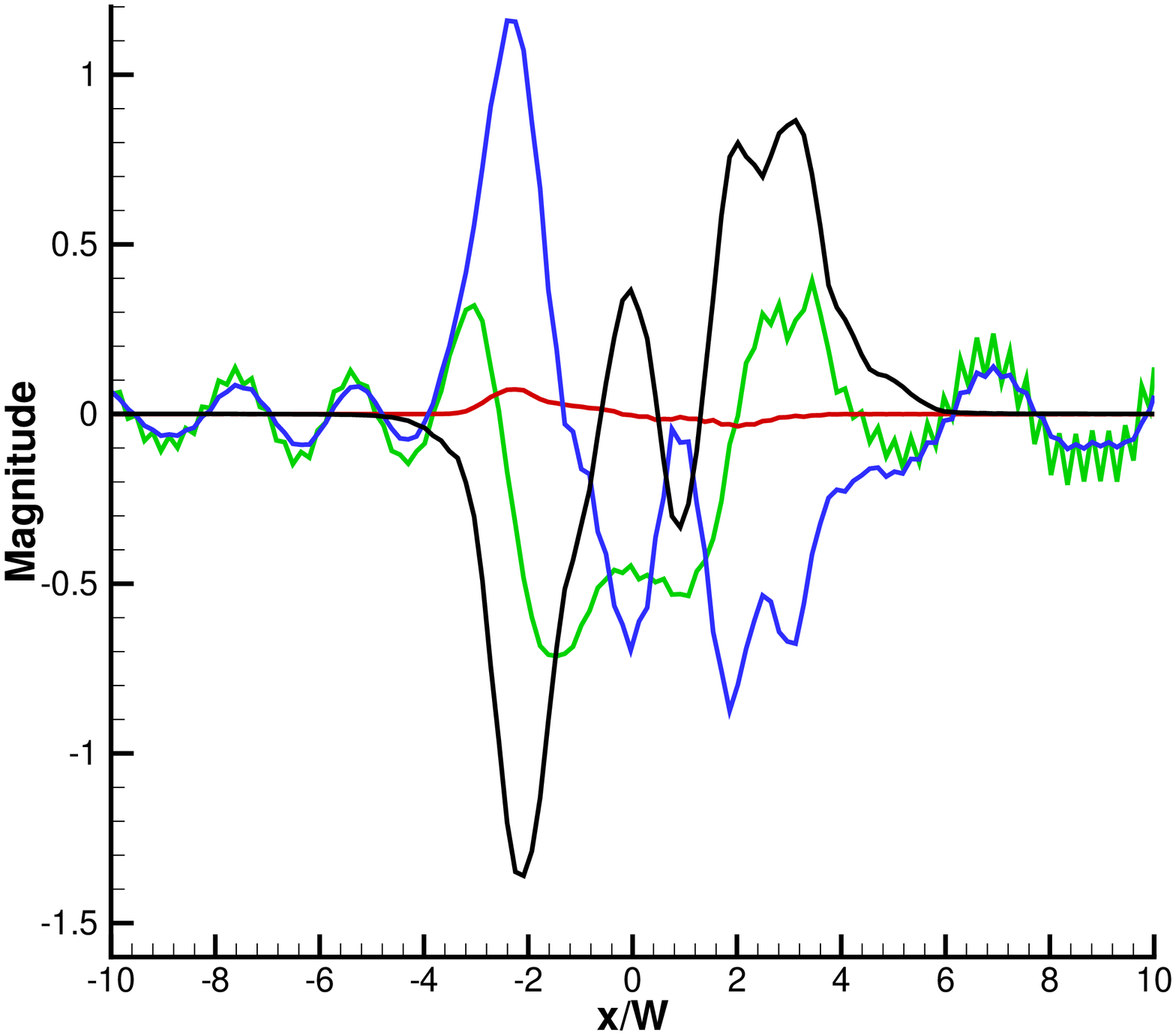}}
\subfigure[\hspace{0.1cm} $t=0.025$ s]{\includegraphics[width=0.49\textwidth]{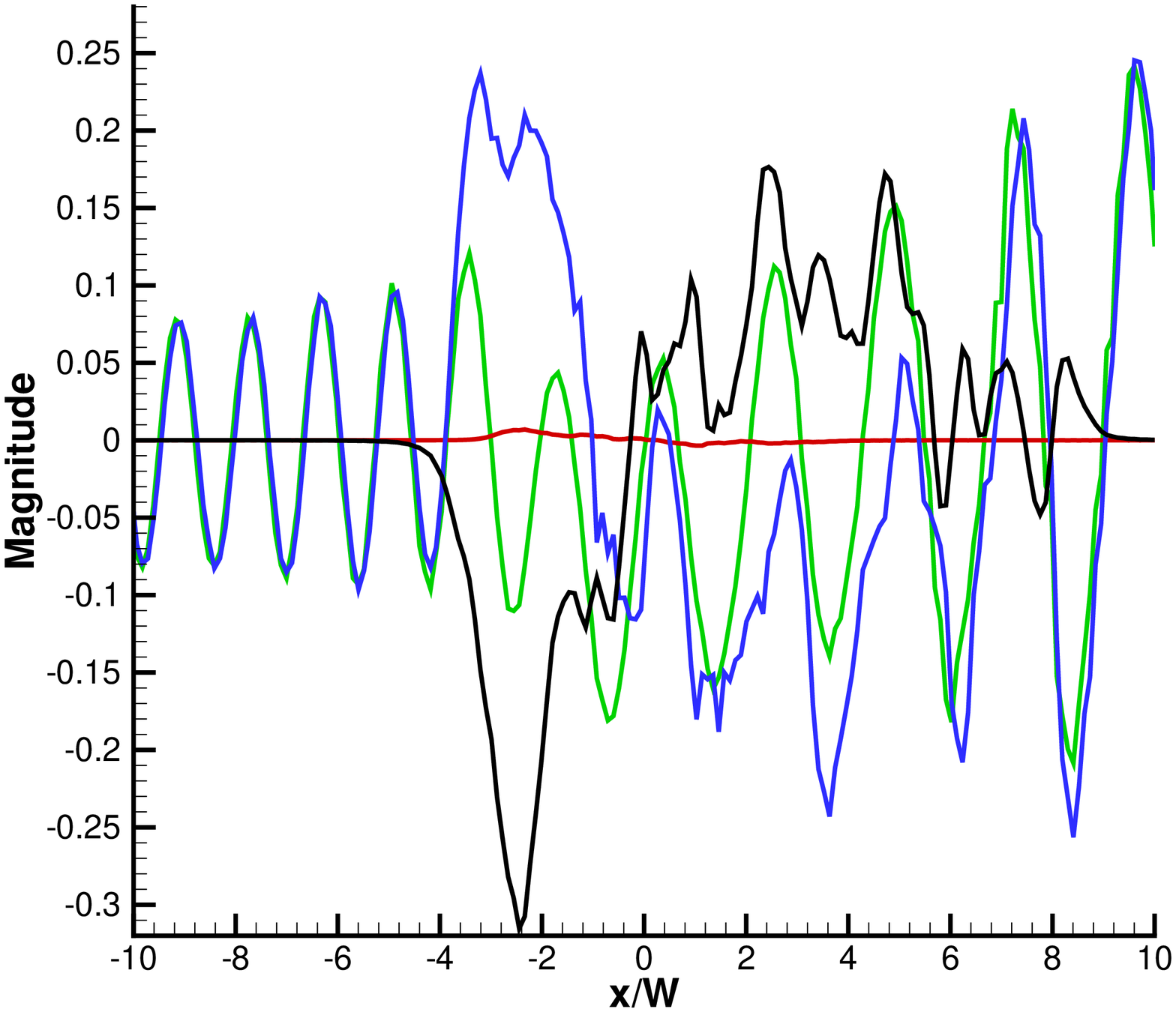}}
\subfigure[\hspace{0.1cm} $t=1$ s]{\includegraphics[width=0.49\textwidth]{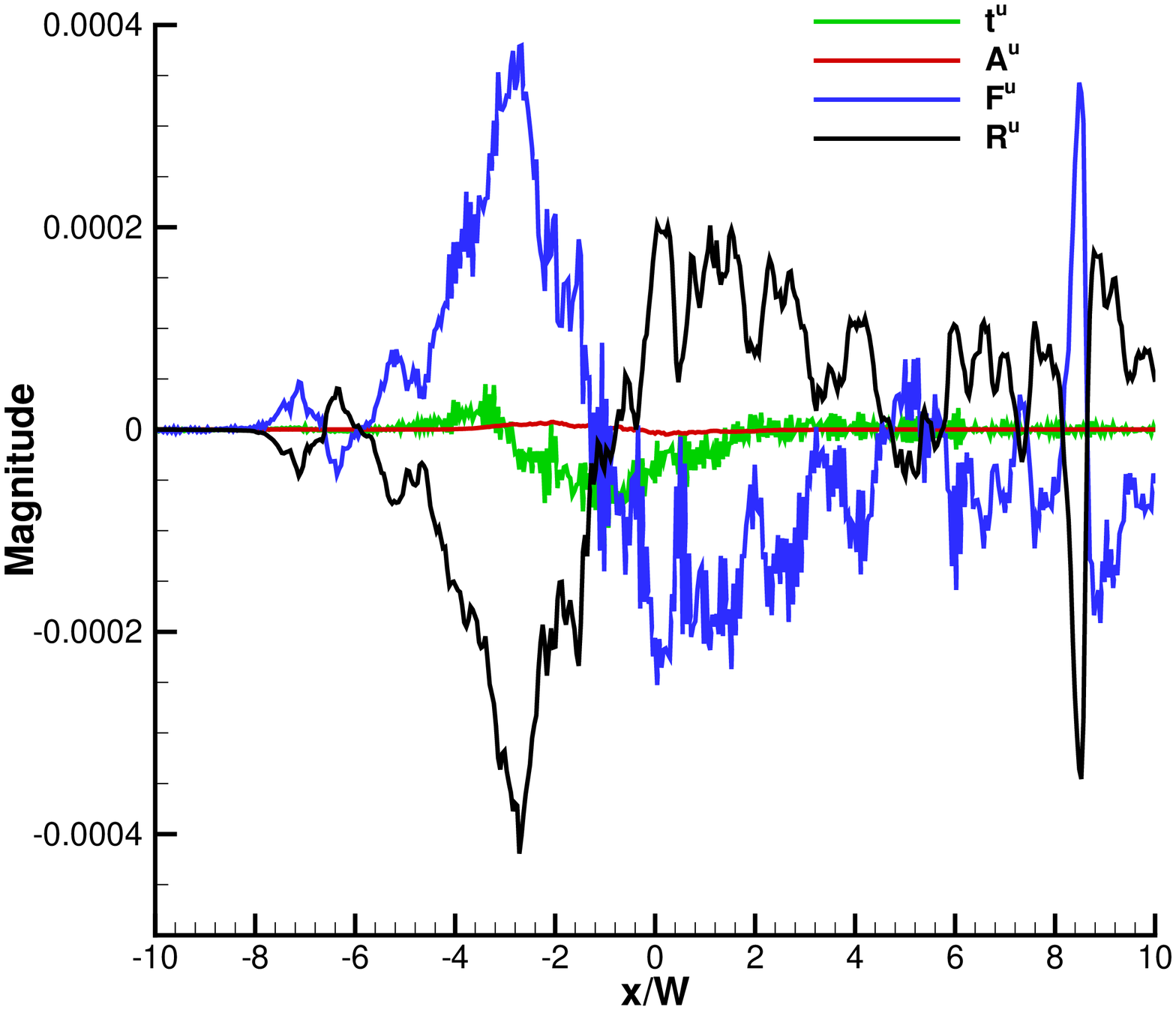}}
\subfigure[\hspace{0.1cm} $t=2$ s]{\includegraphics[width=0.49\textwidth]{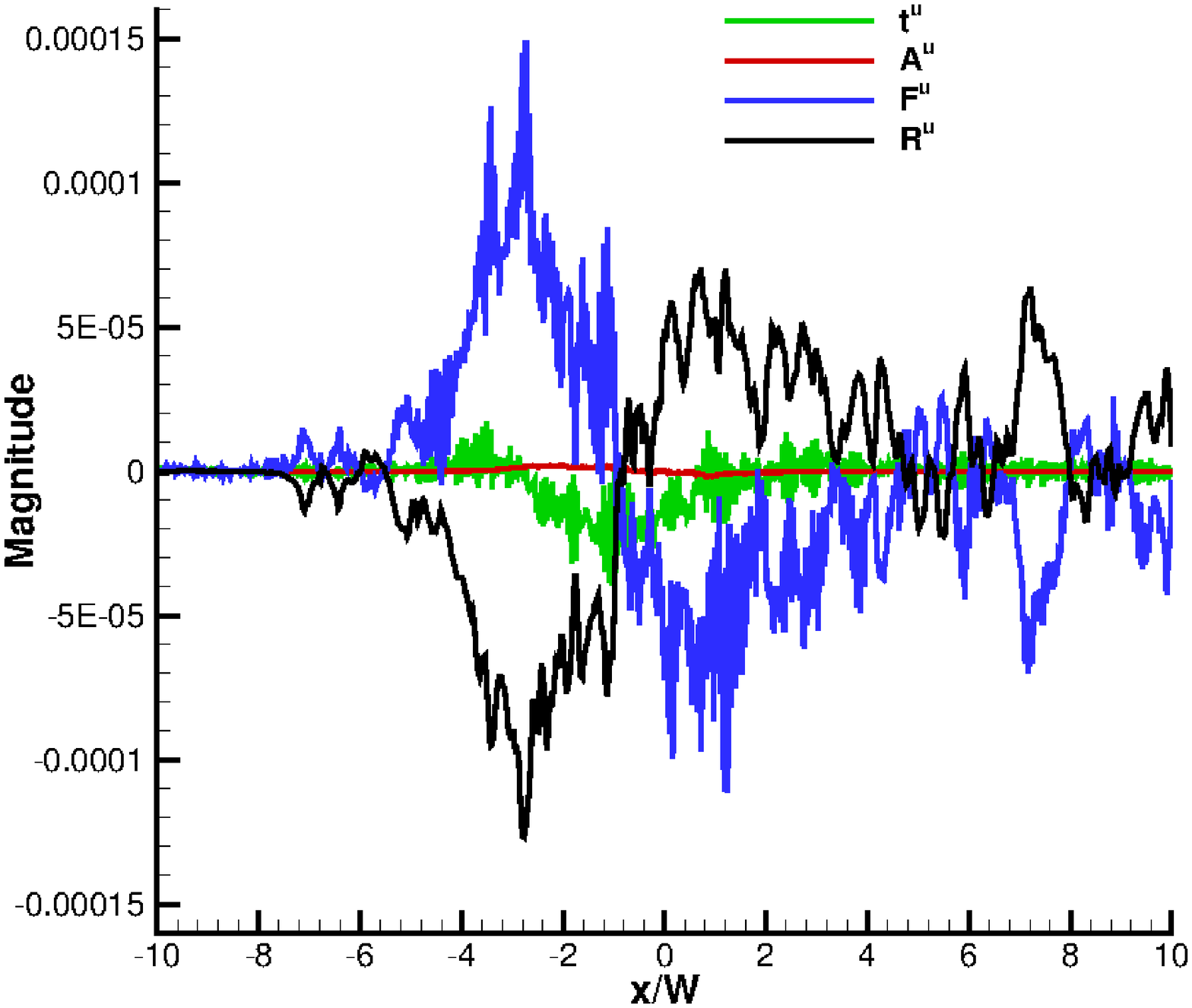}}
\caption{Individual terms in the $\tilde{u}$ transport equation at (a) $t=0.01$ s, (b) $0.025$ s, (c) $1$ s, and (d) $2$ s computed from the Flamenco results. All terms non-dimensionalised by $\rho_c u_c^2/\bar \lambda$.  \label{momflam}}
\end{centering}
\end{figure*}

\begin{figure*}
\begin{centering}
\subfigure[\hspace{0.1cm} $t=1$ s]{\includegraphics[width=0.49\textwidth]{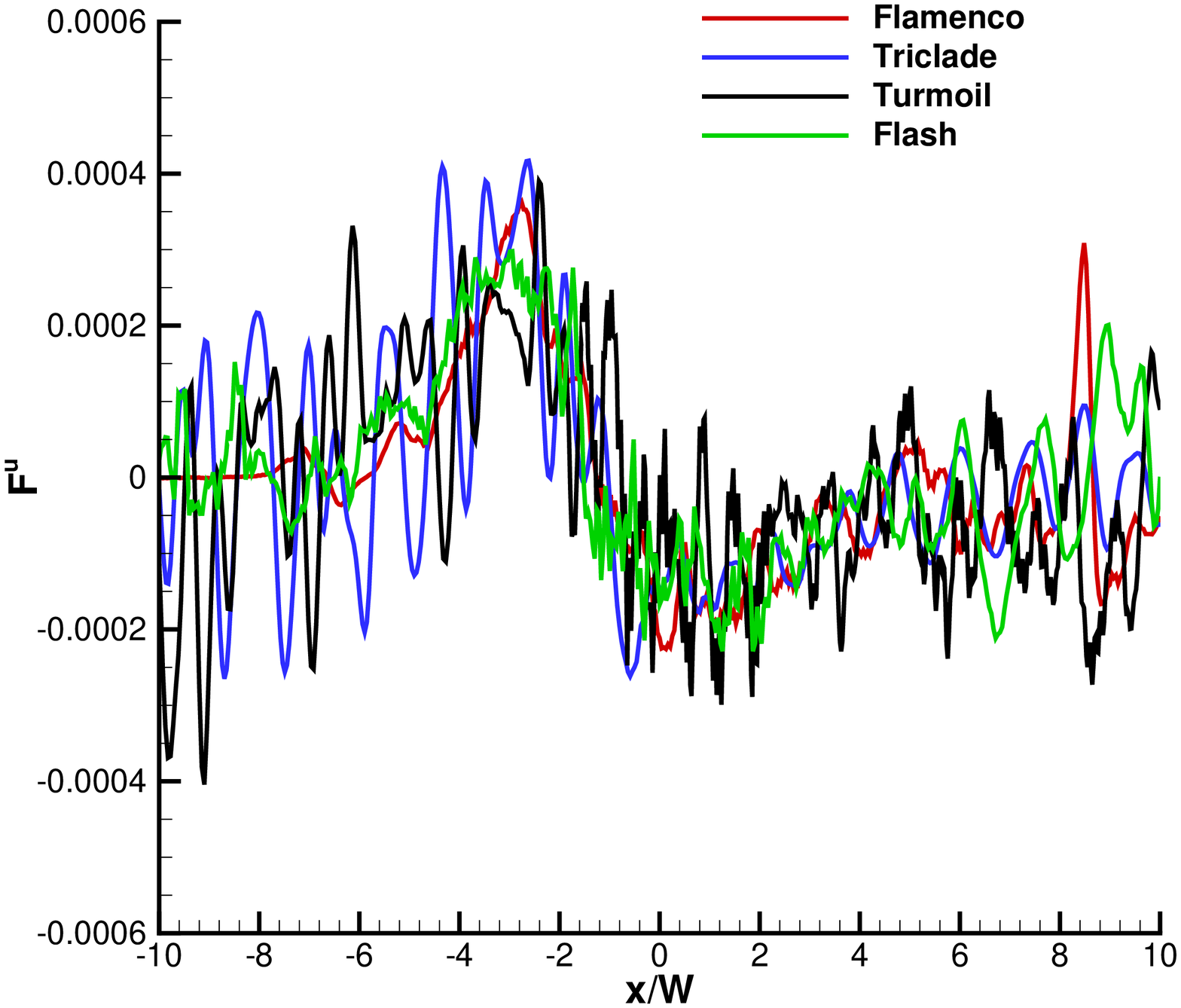}}
\subfigure[\hspace{0.1cm} $t=2$ s]{\includegraphics[width=0.49\textwidth]{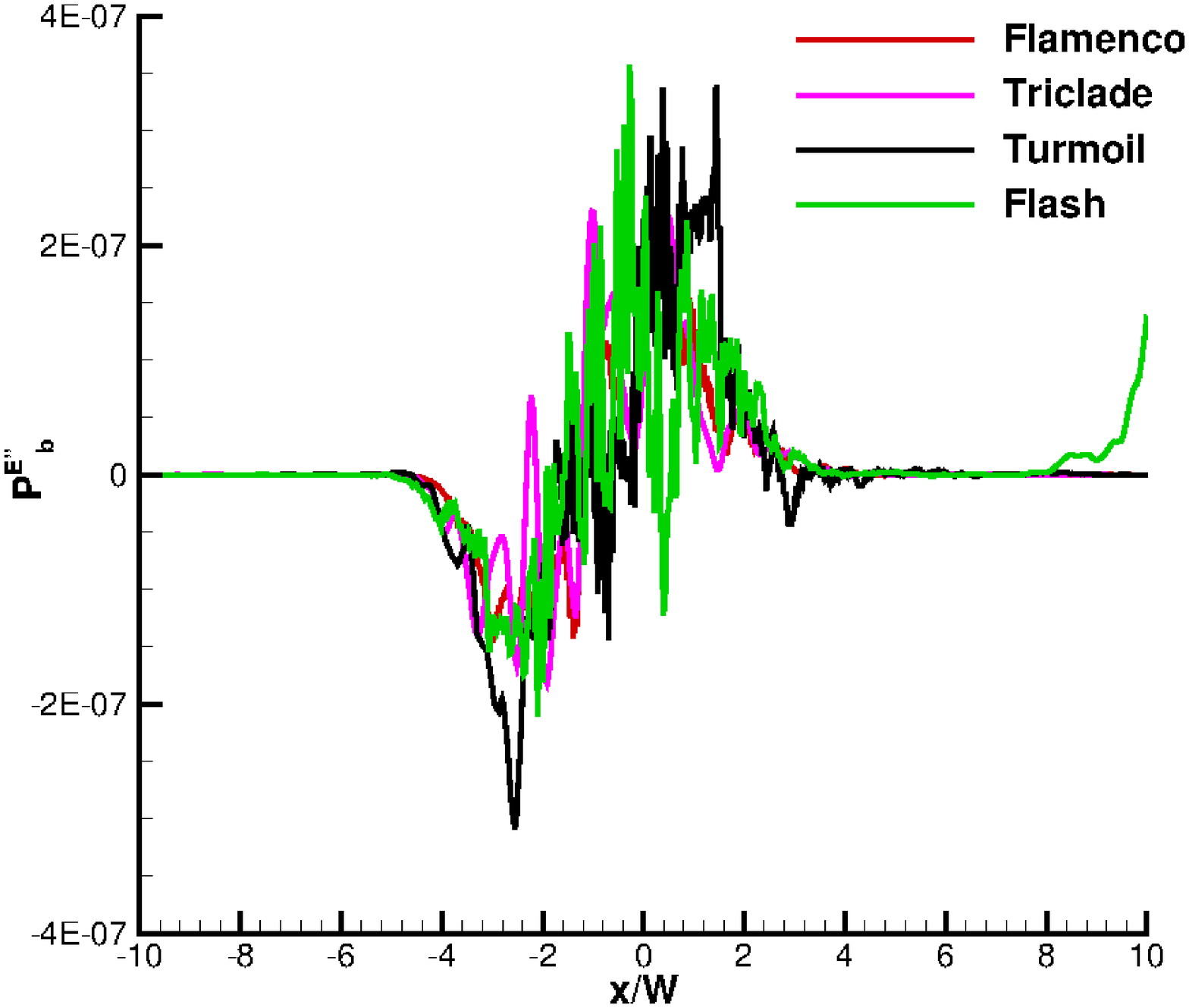}}
\caption{Smoothed $F^u$ term at (a) $t=1$ s and (b) $2$ s. All terms non-dimensionalised by $\rho_c u_c^2/\bar \lambda$. \label{momsmooth}}
\end{centering}
\end{figure*}

Figure \ref{mom} shows the individual components of the mean momentum transport equation for all codes, Fig. \ref{momflam} shows the same time instants but only the Flamenco results so that the relative magnitudes of the individual terms may be more clearly discerned, and Fig. \ref{momsmooth} plots the $F^u$ term alone, where the data has been smoothed by a simple five point moving average. The codes agree well for all terms except for the pressure gradient term $F^u$, which shows good agreement on the bubble side of the mixing layer but relatively poor agreement on the spike side. Although the bulk of the mixing layer is contained in $-4 \le x/W \le 4$, there is a substantial contribution to the mean momentum equation at $x/W > 4$ which may be attributed to highly energetic ejected spikes, features which combine both very high velocities and low pressure cores in a localised feature. Such features have been observed in previous analysis of spectral energy transfer in the RMI \cite{Thornber2012} and RTI \cite{Cook2002}. 

As has been shown for a small Atwood number Rayleigh-Taylor instability \cite{Schilling2010}, the time rate of change of mean momentum $t^u$ is mostly generated as the remainder of two larger terms, the mean pressure gradient $F^u$ and the Reynolds stresses $R^u$, where $F^u\approx -R^u$ at all times. At early time both $R^u$ and $F^u$ are approximately anti-symmetric across the centreline, but at later times the bubble side amplitudes are larger and closer to the core of the mixing layer than the spike side amplitudes, as the gradient of the Reynolds stress $\tau_{11}$ is much sharper on the bubble than the spike side (see the section on kinetic energy transport in Sec. \ref{ketrans}).  Mean advection is nearly negligible at all times, which is expected as the mean layer position is approximately stationary with respect to the chosen reference frame.

\begin{figure}
\begin{centering}
\includegraphics[width=0.49\textwidth]{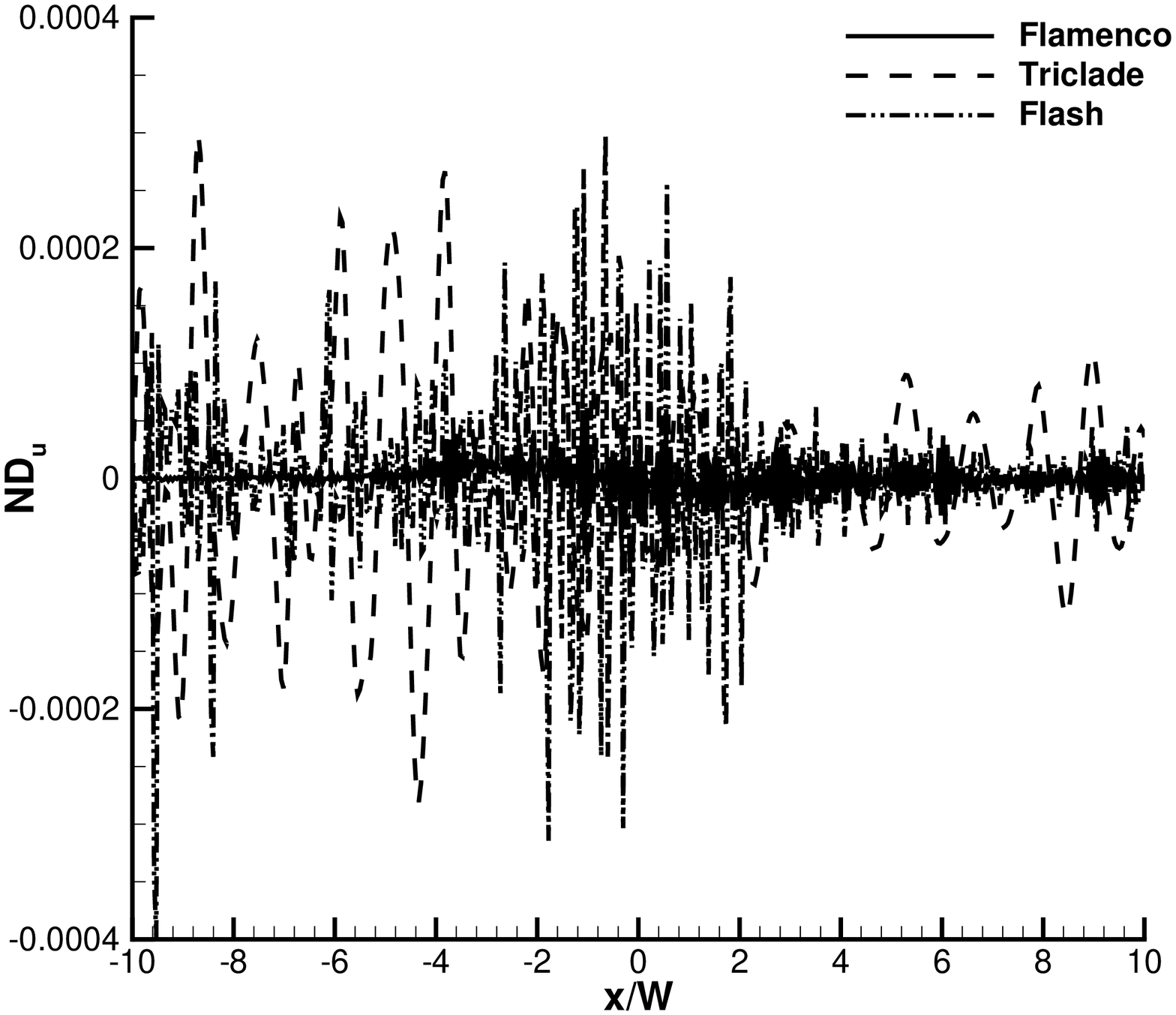}
\caption{Non-dimensional numerical dissipation estimated by the residual in the mean momentum transport balance at $t=1$ s.  \label{momnd}}
\end{centering}
\end{figure}

The mean momentum transport equation has contributions from numerical dissipation $ND_{u}$ representing the under-resolved components of the Reynolds stresses. As the grid resolution increases, the magnitude of these implicitly modelled effects should decrease. Figure \ref{momnd} plots the non-dimensional numerical dissipation term for Flash, Flamenco, and Triclade, computed as the residual of the mean momentum equation.  Although the overall magnitude is similar to the other terms in the momentum equation, the magnitude is largely due to the presence of noise in the computation of the individual physical terms as can be seen clearly in Fig. \ref{mom}. Examining only the Flamenco results (which have relatively low noise), shows that the peak magnitude is less than $3.5\times 10^{-5}$, which is ten times lower than the dominant physical terms, and the rms magnitude defined over $-5 < x/W < 5$ is $7.6\times 10^{-6}$. A similar conclusion can be reached if a simple moving average is applied to the results. It should be noted however that as the net change of mean momentum is a result of the difference between the two largest physical terms $F^u$ and $R^u$, the numerical dissipation represents approximately one quarter of the magnitude of the overall rate of change of $\tilde{u}$ and is not negligible. 

\subsection{Mass Fraction Transport}

\begin{figure*}
\begin{centering}
\subfigure[\hspace{0.1cm} $t=0.01$ s]{\includegraphics[width=0.49\textwidth]{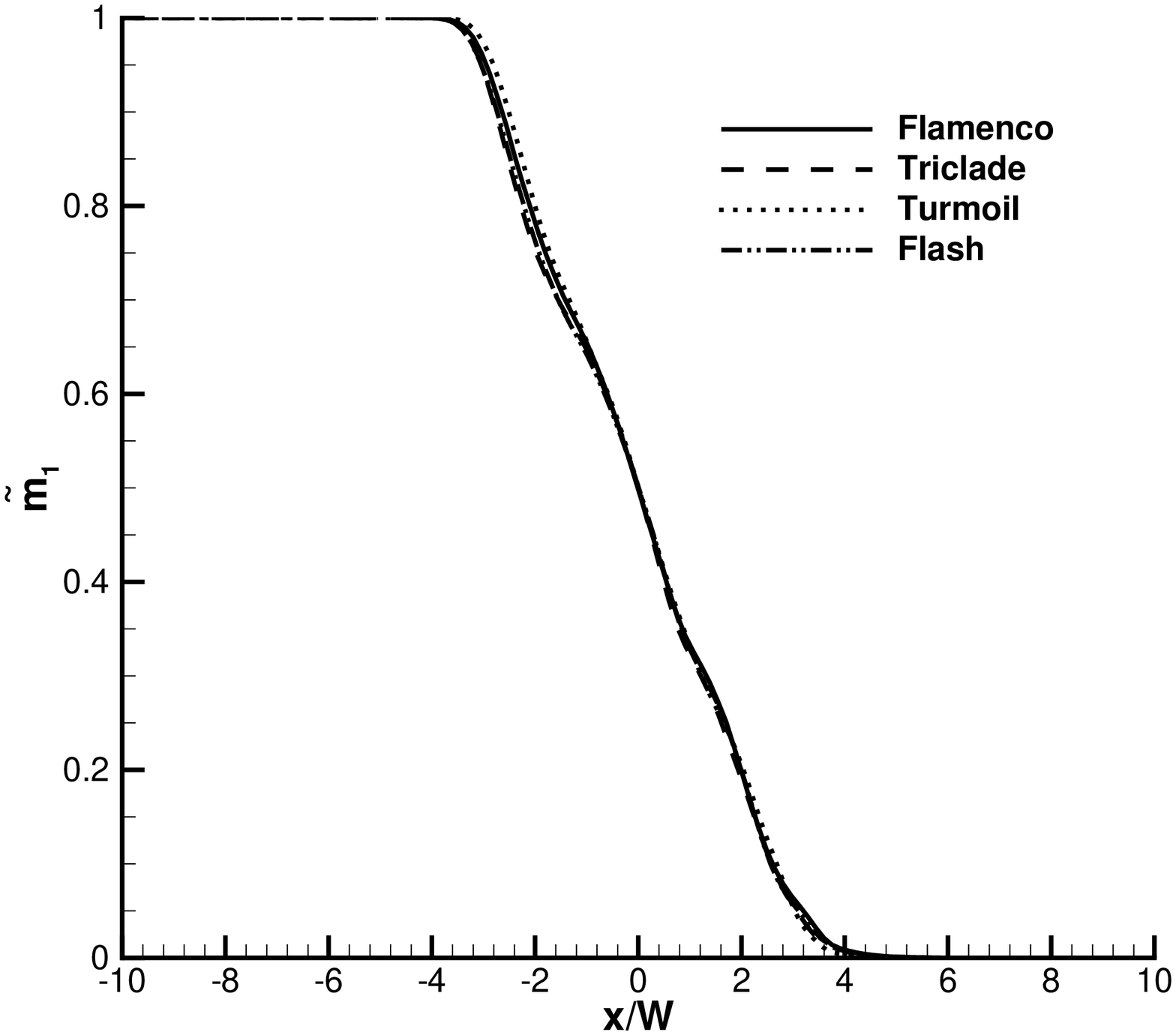}}
\subfigure[\hspace{0.1cm} $t=0.025$ s]{\includegraphics[width=0.49\textwidth]{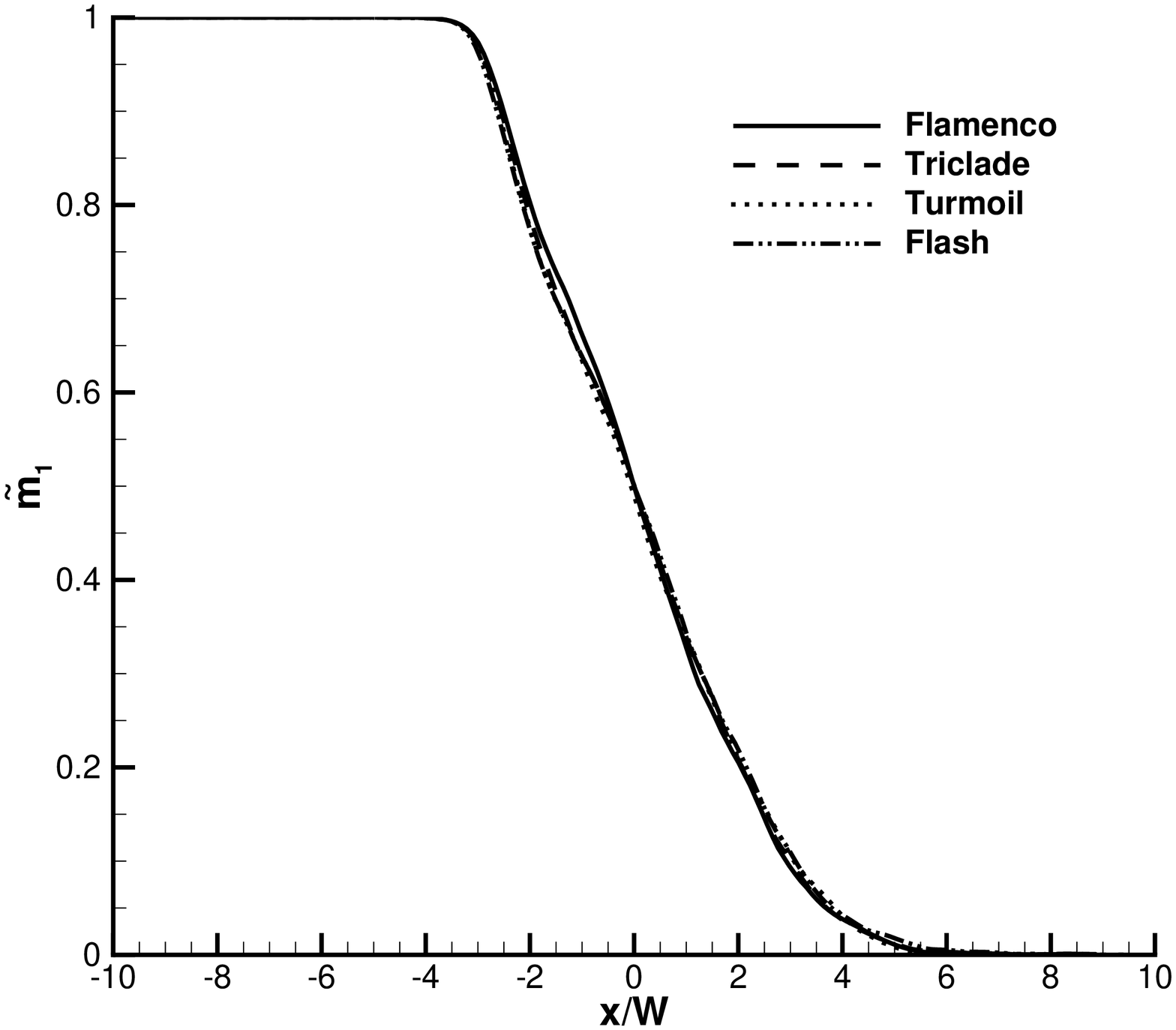}}
\subfigure[\hspace{0.1cm} $t=1$ s]{\includegraphics[width=0.49\textwidth]{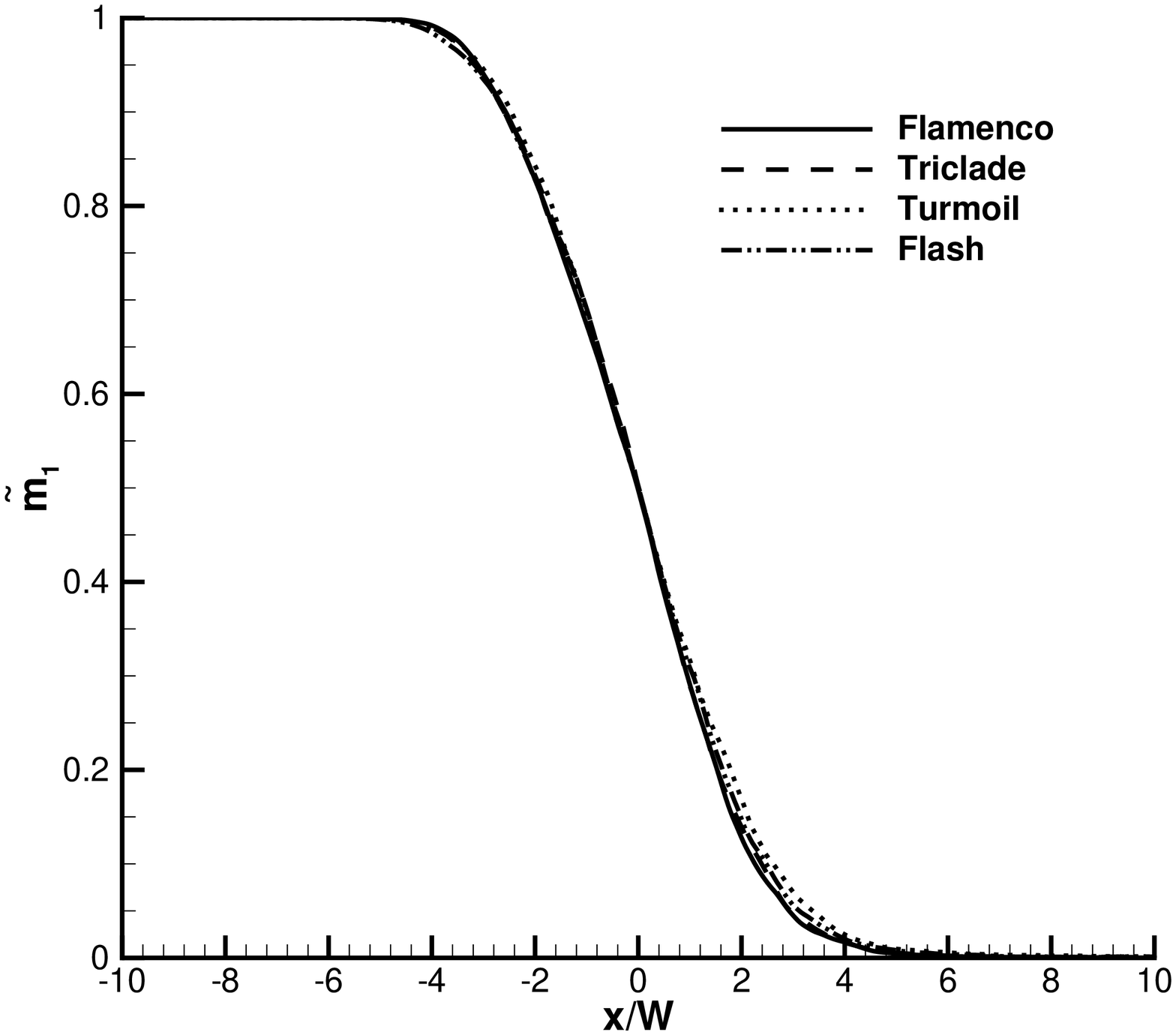}}
\subfigure[\hspace{0.1cm} $t=2$ s]{\includegraphics[width=0.49\textwidth]{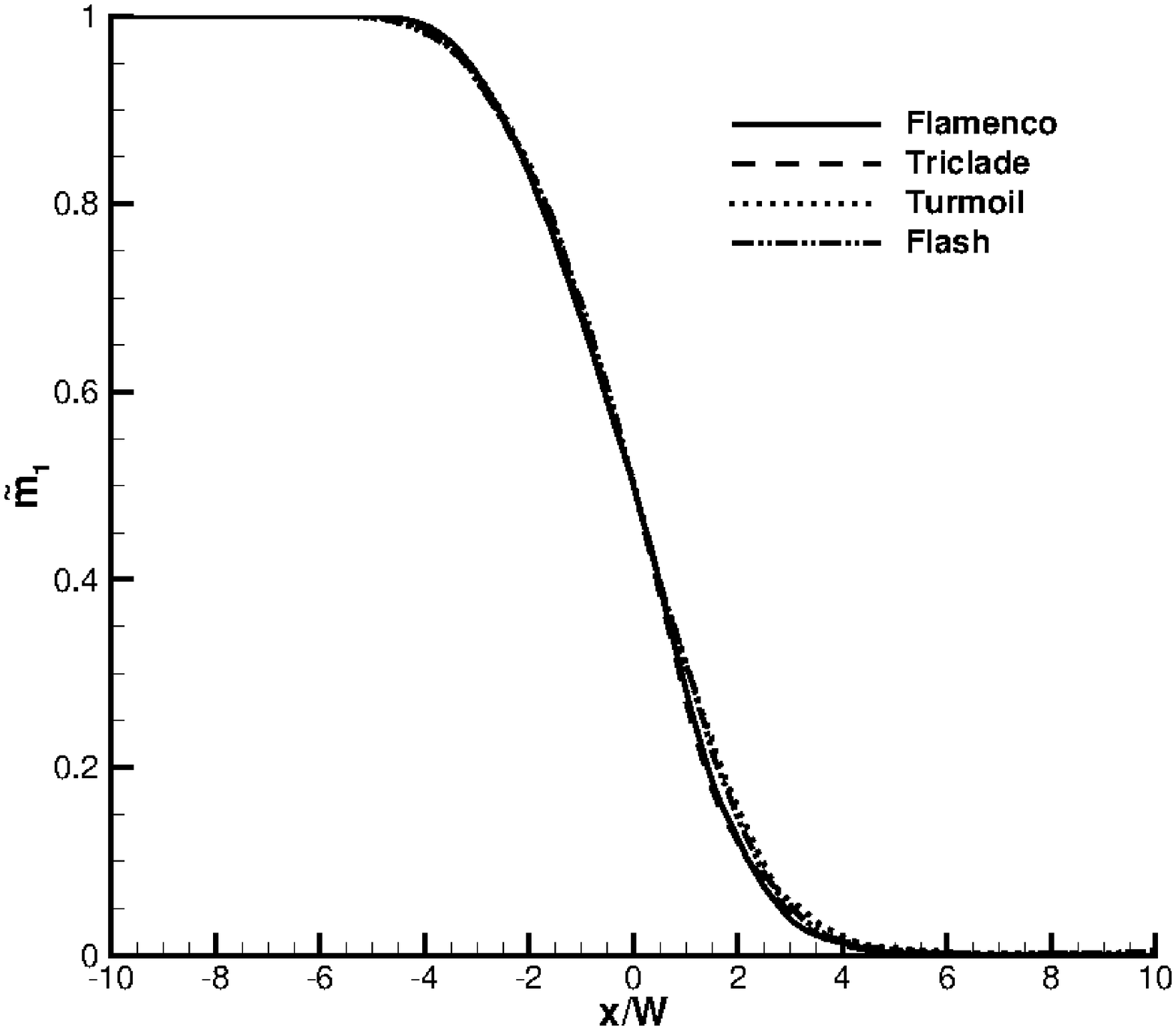}}
\caption{Mean mass fraction $\widetilde{m_1}$ at (a) $t=0.01$ s, (b) $0.025$ s, (c) $1$ s, and (d) $2$ s. \label{mftilde}}
\end{centering}
\end{figure*}

\begin{figure*}
\begin{centering}
\includegraphics[width=0.49\textwidth]{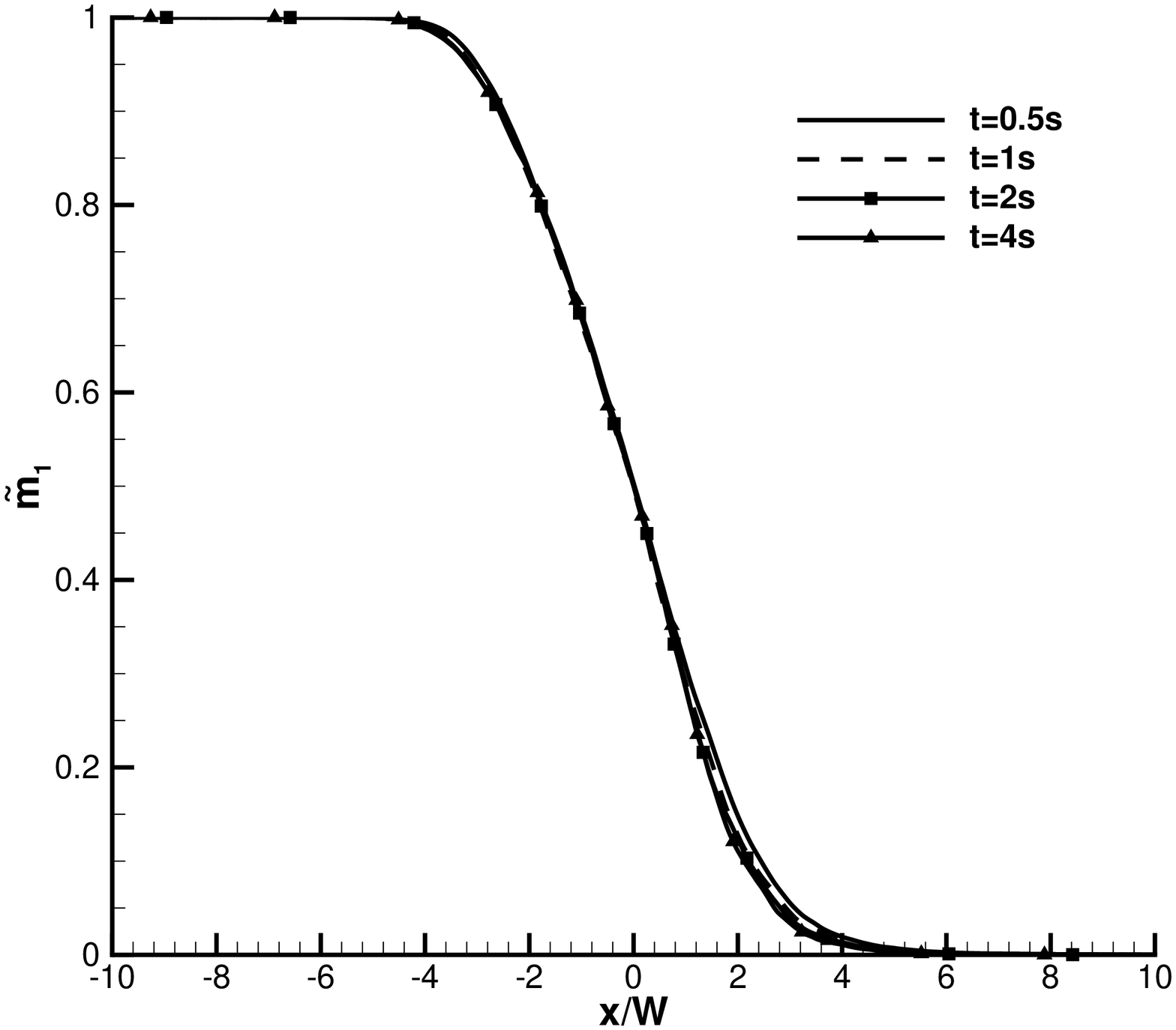}
\includegraphics[width=0.49\textwidth]{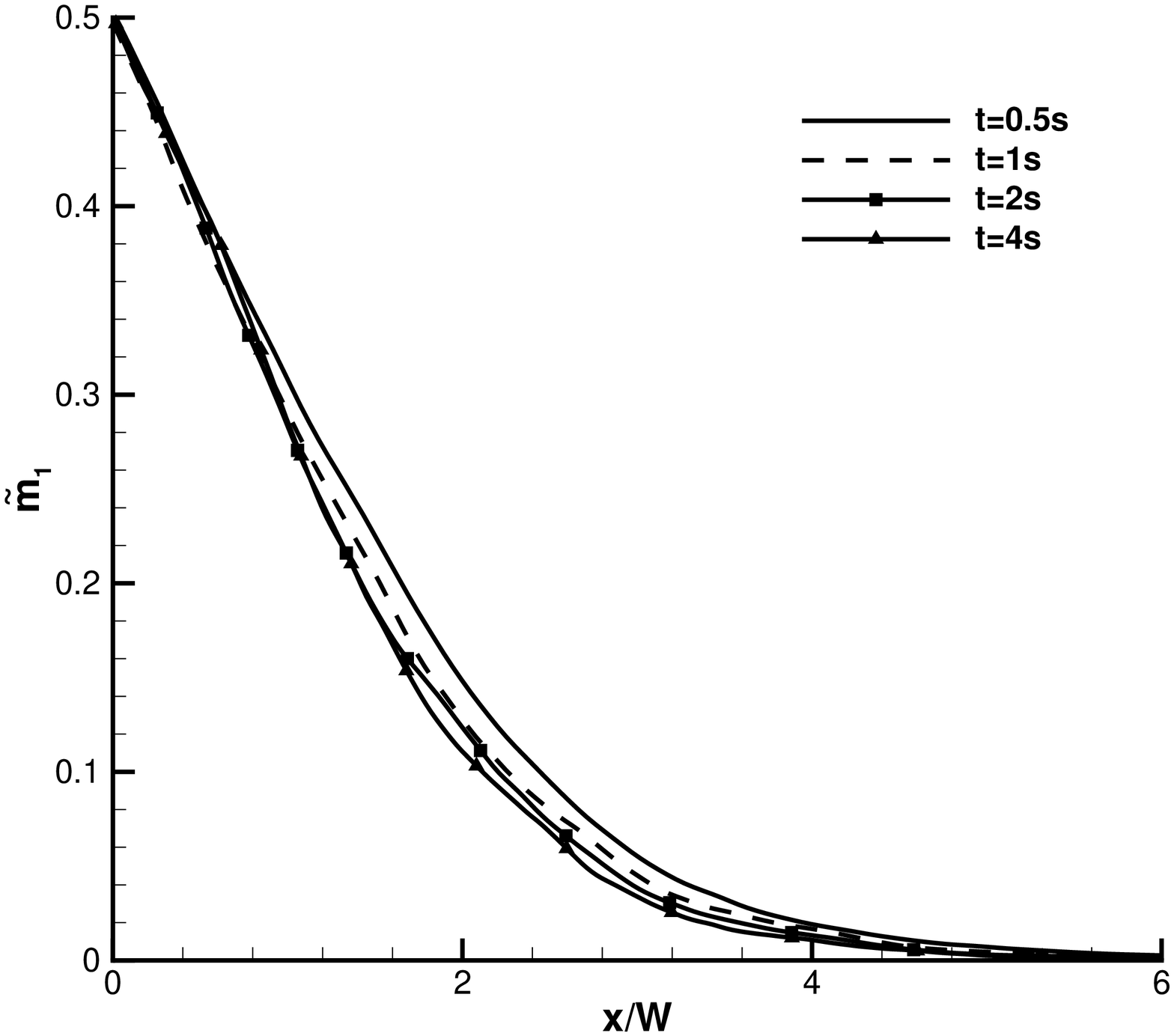}
\caption{Mean mass fraction $\widetilde{m_1}$ at $t=0.5$, $1$, $2$, and $4$ s (right), and a close up of the spike side (left) (Flamenco results only).  \label{mfss}}
\end{centering}
\end{figure*}

Figure \ref{mftilde} presents the profile of the mean heavy fluid mass fraction $\tilde m_1$. The profile is approximately linear, and at late time exhibits no discernable kinks indicating a region of well mixed flow, with very small mean diffusive transport. All codes are in good agreement barring some late time differences on the spike side. Figure \ref{mfss} shows the scaled $\tilde m_1$ profile from Flamenco at four times, demonstrating an excellent collapse for the three latest times on the bubble side in particular. The figure also shows a close-up of the spike side, where the size of the spikes relative to the mixing layer is still decreasing at late times, a further indication that a fully self-similar state is not yet attained. This is consistent with breakup and dissipation of these strong vortical structures.

\begin{figure*}
\begin{centering}
\subfigure[\hspace{0.1cm} $t=0.01$ s]{\includegraphics[width=0.49\textwidth]{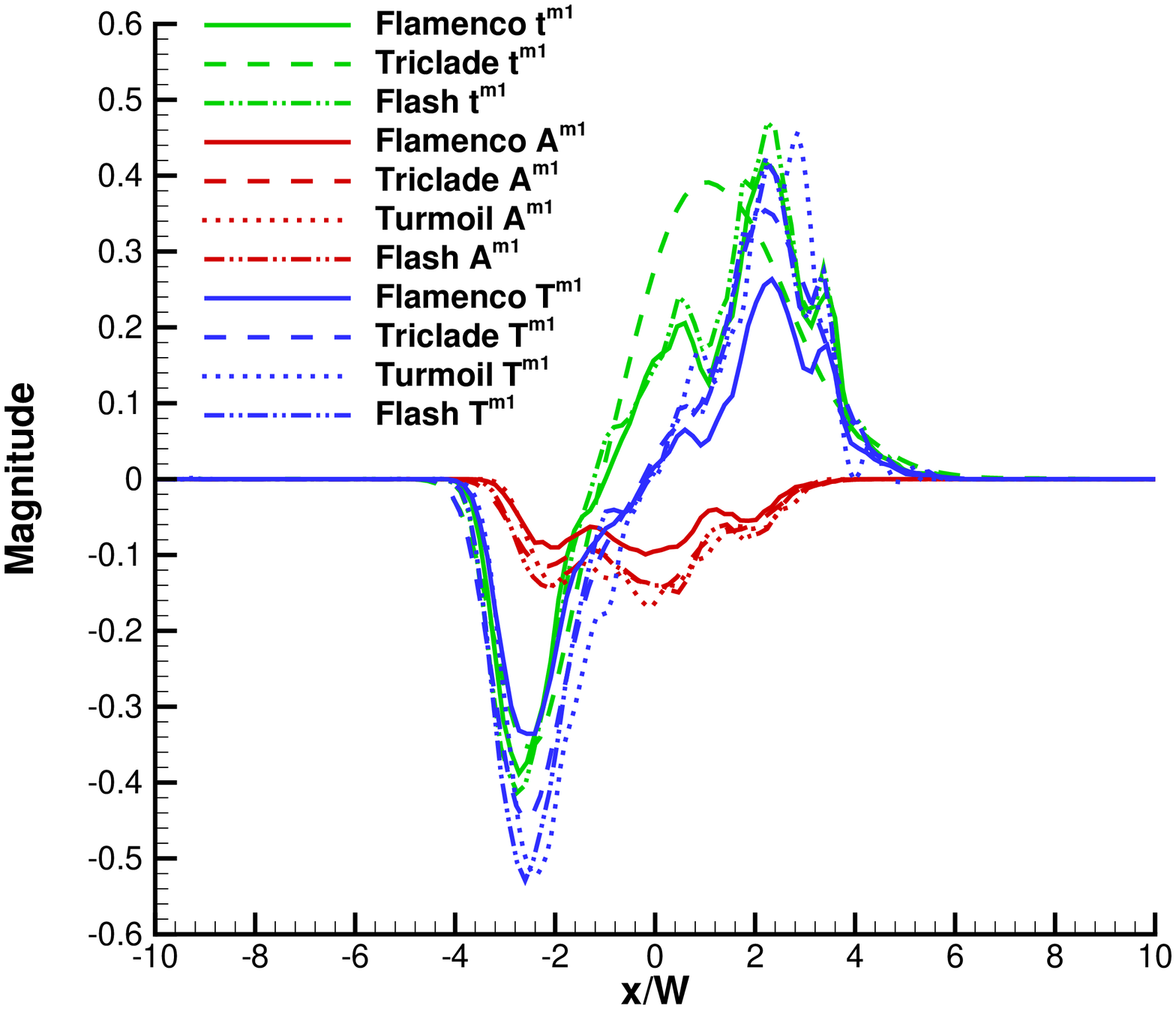}}
\subfigure[\hspace{0.1cm} $t=0.025$ s]{\includegraphics[width=0.49\textwidth]{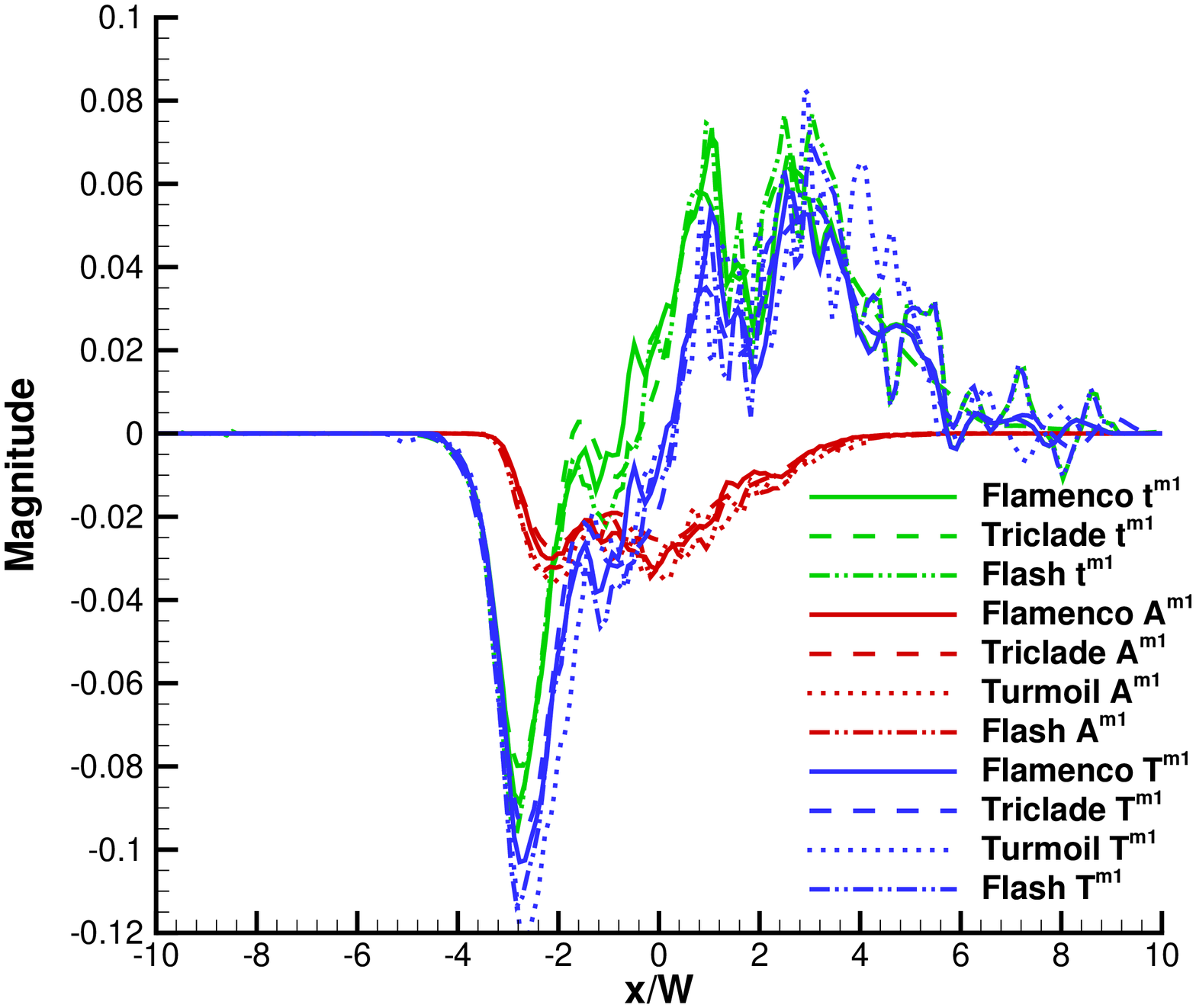}}
\subfigure[\hspace{0.1cm} $t=1$ s]{\includegraphics[width=0.49\textwidth]{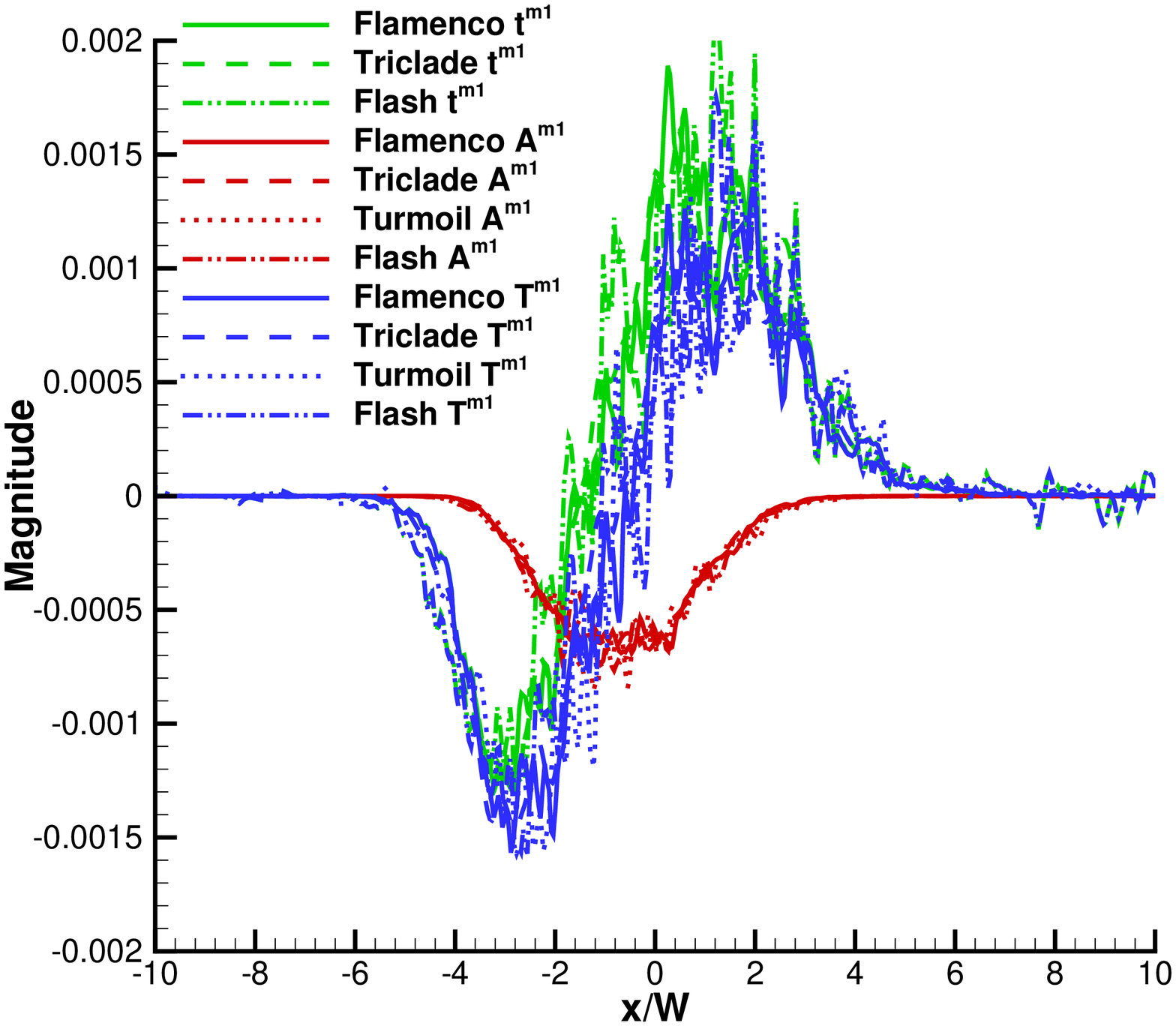}}
\subfigure[\hspace{0.1cm} $t=2$ s]{\includegraphics[width=0.49\textwidth]{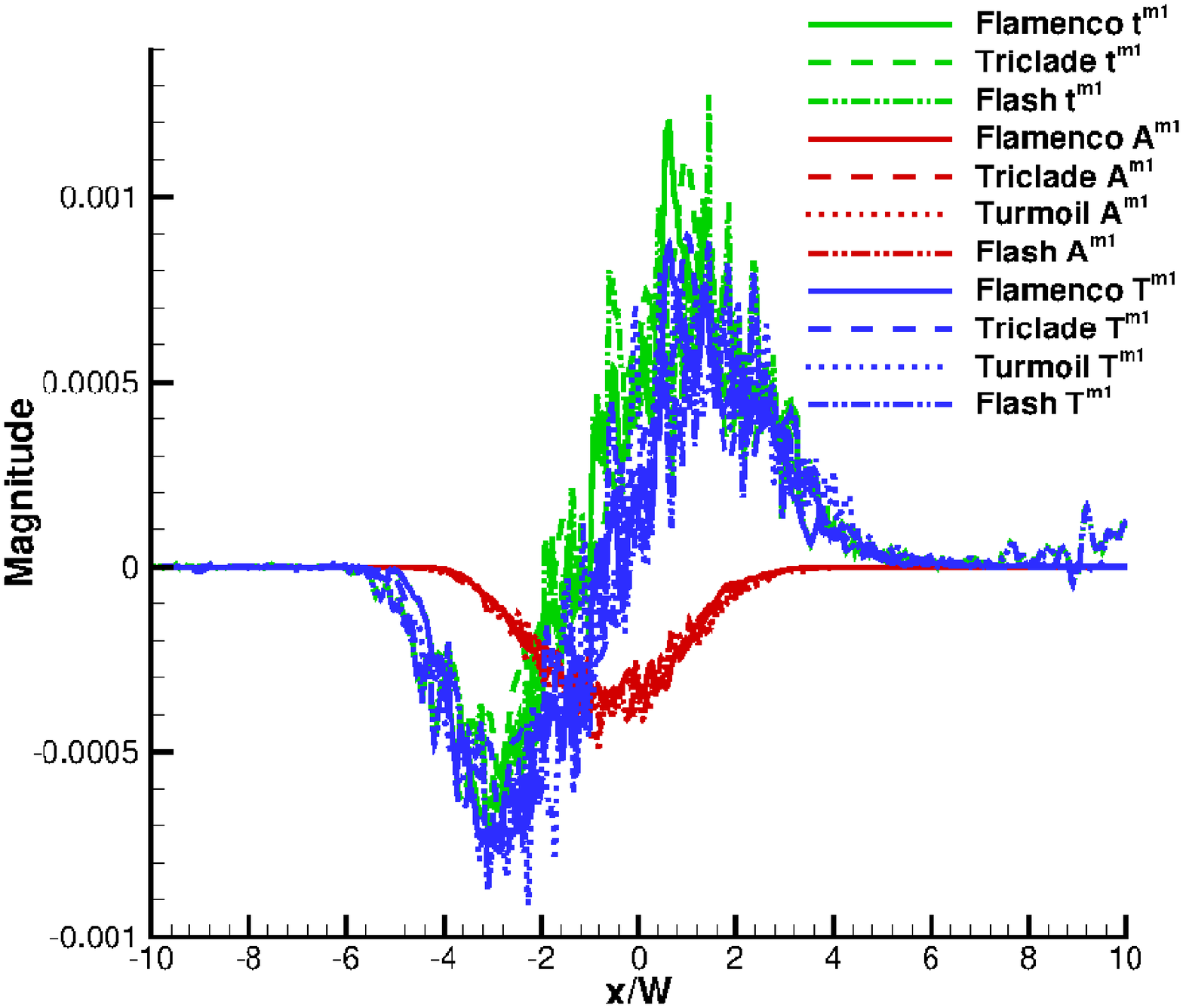}}
\caption{Individual terms in the $\widetilde{m_1}$ transport equation at (a) $t=0.01$ s, (b) $0.025$ s, (c) $1$ s, and (d) $2$ s.  All terms non-dimensionalised by $\rho_c u_c/\bar \lambda$.  \label{mf}}
\end{centering}
\end{figure*}

Figure \ref{mf} plots all terms in the $\tilde m_1$ transport equation, where all code results collapse well. Turbulent transport $T^{m1}$ is approximately anti-symmetric about the centre of the mixing layer and dominates for $x/W<0$, however due to mean advection $A^{m1}$ being negative throughout the layer, the time rate of change $t^{m1}$ dominates for $x/W>0$. Mean advection $A^{m1}$ is non-negligible, symmetric at the two latest times, and negative throughout the layer due to the net transport from the heavy to light side, but is less than half the magnitude of turbulent transport. As a result, the time variation of $\tilde m_1$ largely follows $T^{m1}$ but is further skewed towards the heavy side due to the contribution from $A^{m1}$.

\begin{figure}
\begin{centering}
\includegraphics[width=0.49\textwidth]{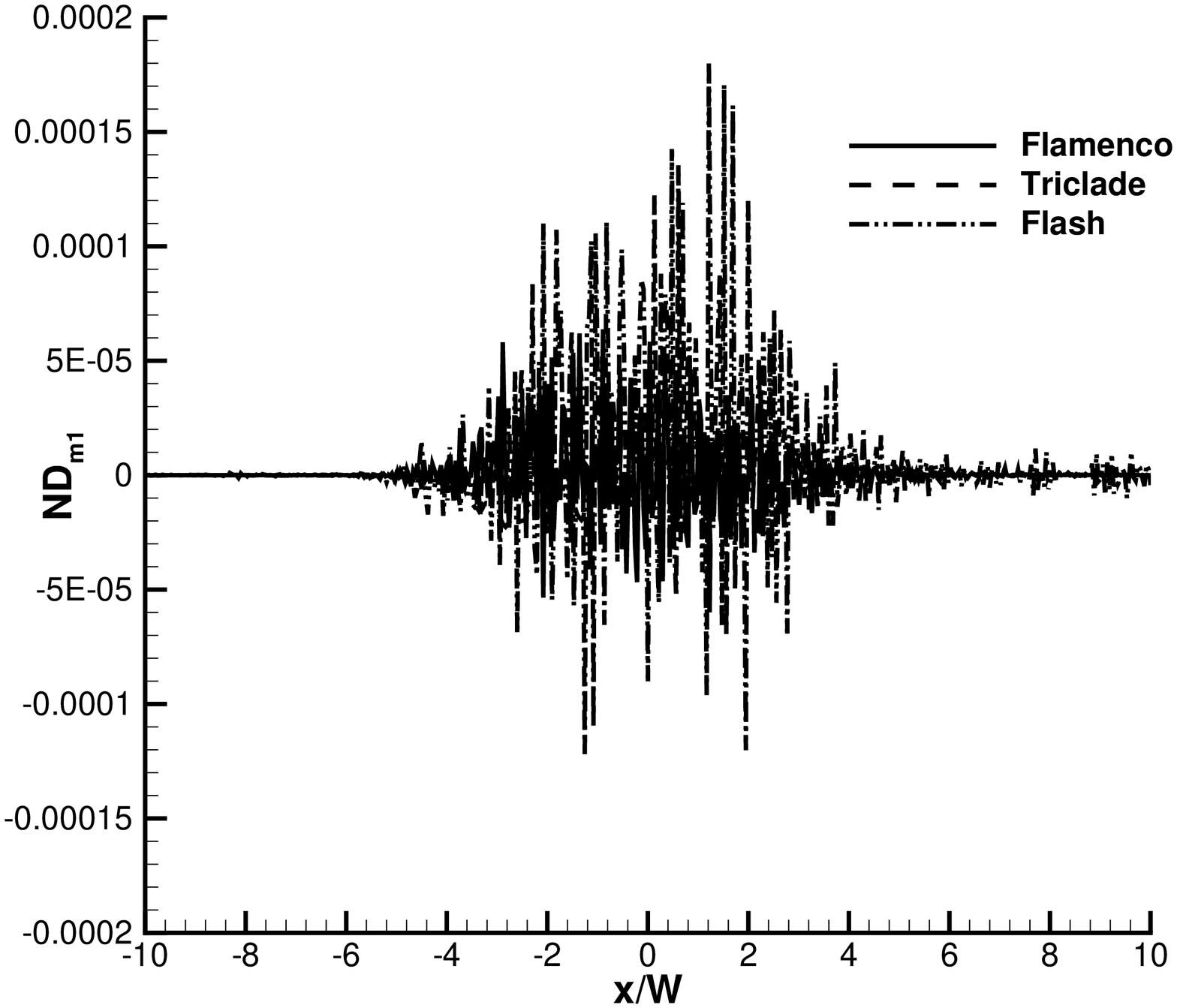}
\caption{Non-dimensional numerical dissipation estimated by the residual in the $\tilde m_1$ turbulent transport balance at $t=1$ s.  \label{mfdiss}}
\end{centering}
\end{figure}

Similar to the mean momentum equation, the results presented here also include numerical dissipation $ND_{m1}$, which is estimated as the remainder in the governing equations and shown in Fig. \ref{mfdiss} for $t=1$ s. Again the result is quite noisy, and the peak of $9\times 10^{-5}$ is at least one order of magnitude lower than the peak of the dominant physical terms. There is no discernable structure to the numerical dissipation profile, exhibiting as symmetric noise about zero.

\subsection{Mass Fraction Variance Transport}

\begin{figure*}
\begin{centering}
\subfigure[\hspace{0.1cm} $t=0.01$ s]{\includegraphics[width=0.49\textwidth]{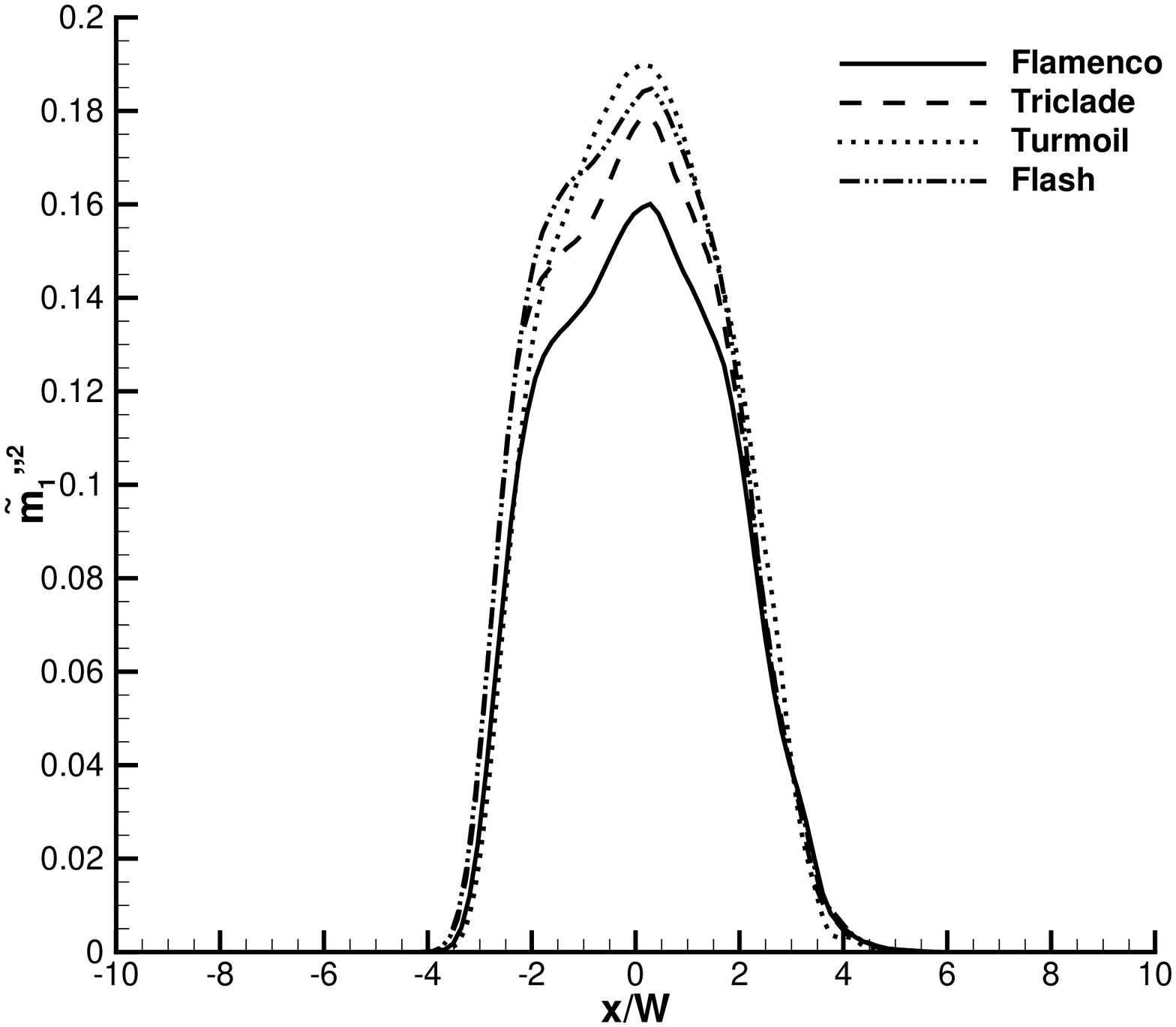}}
\subfigure[\hspace{0.1cm} $t=0.025$ s]{\includegraphics[width=0.49\textwidth]{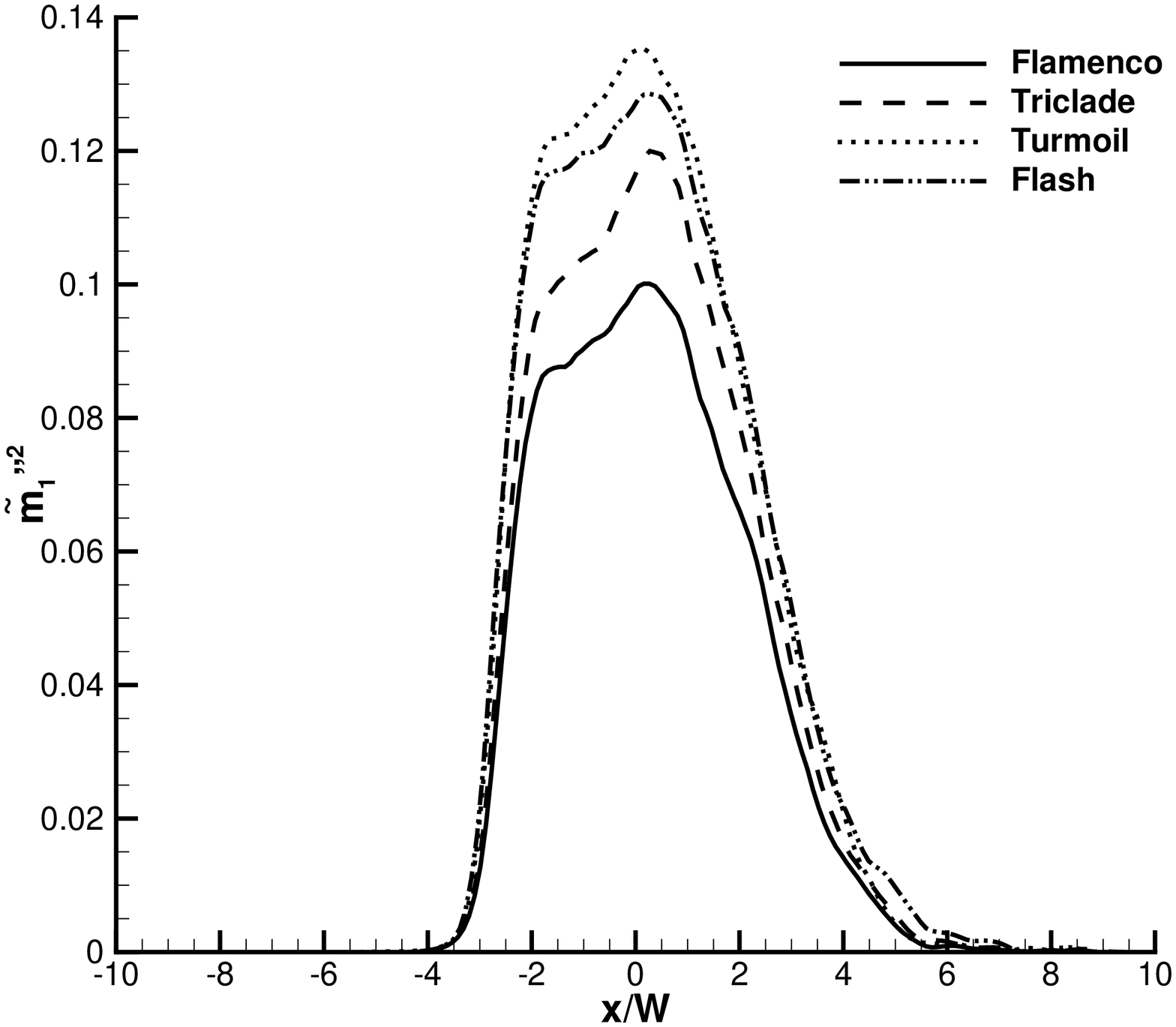}}
\subfigure[\hspace{0.1cm} $t=1$ s]{\includegraphics[width=0.49\textwidth]{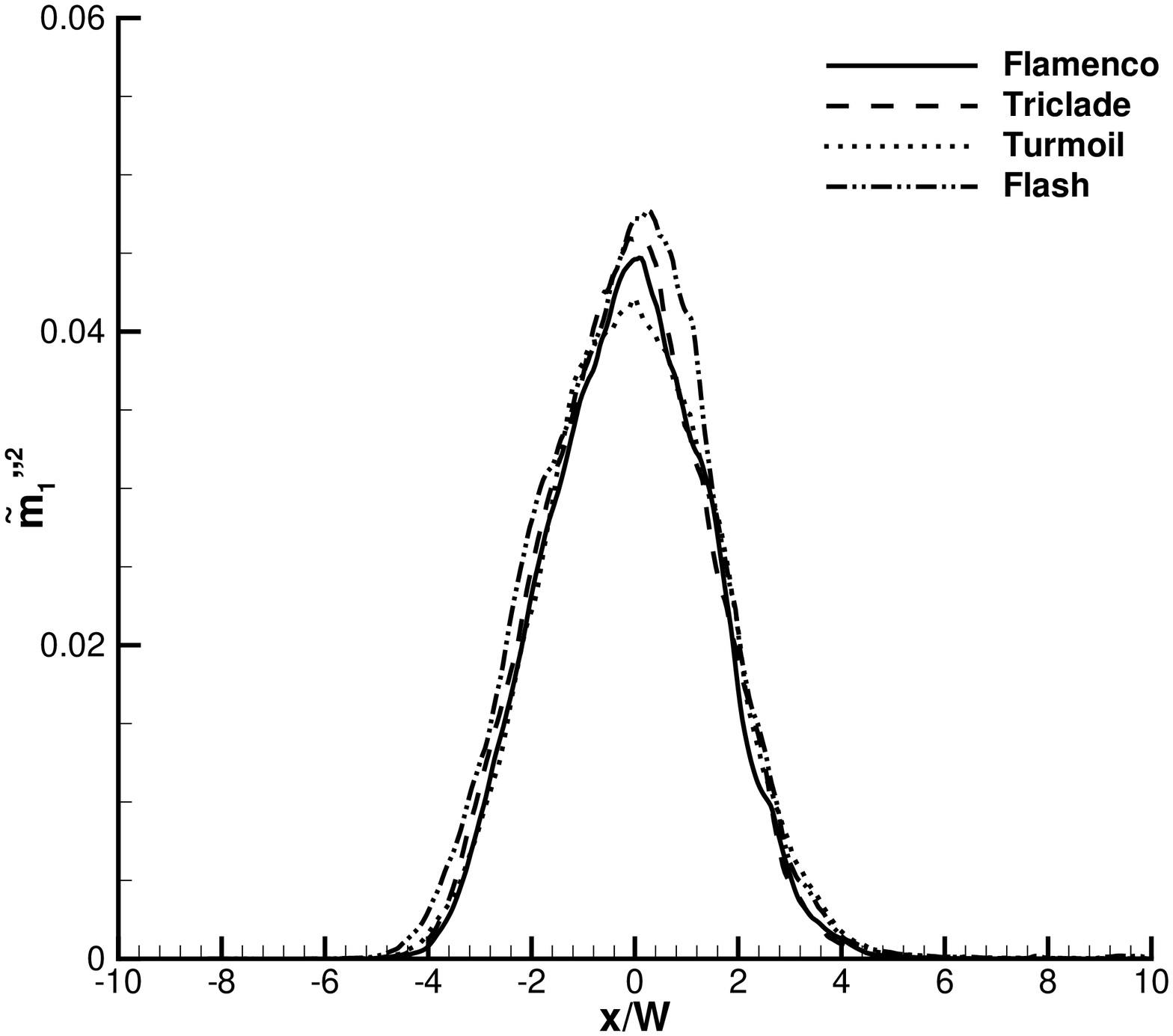}}
\subfigure[\hspace{0.1cm} $t=2$ s]{\includegraphics[width=0.49\textwidth]{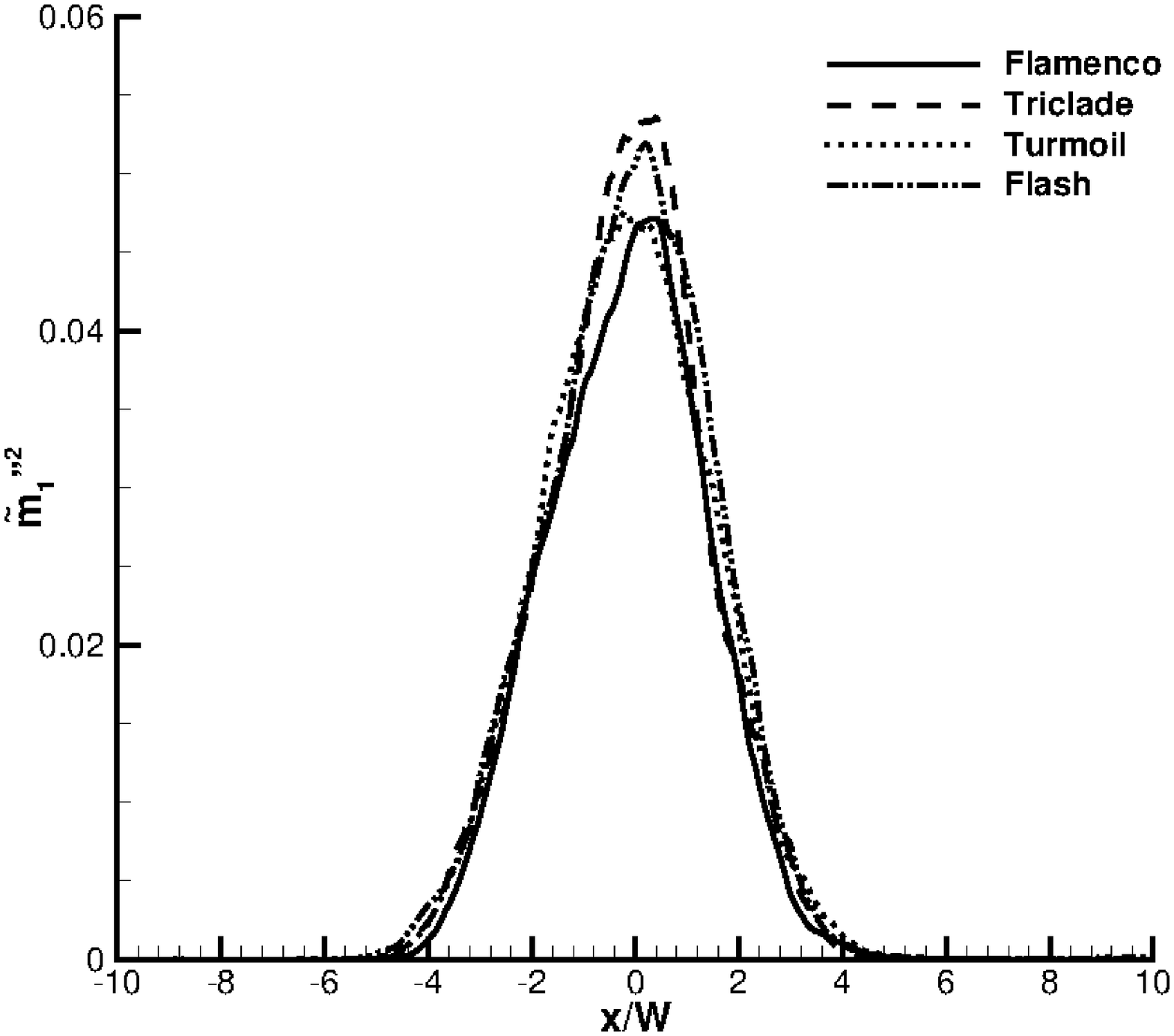}}
\caption{Mass fraction variance $\widetilde{m_1''^2}$ at (a) $t=0.01$ s, (b) $0.025$ s, (c) $1$ s, and (d) $2$ s.  \label{mfvtilde}}
\end{centering}
\end{figure*}

Profiles of the heavy species mass fraction variance $\widetilde{m_1''^2}$ are shown in Fig. \ref{mfvtilde} for four representative times. At early times, the profiles peak moderately on the spike side of the mixing layer and are relatively symmetric. There is a substantial difference in the peak mass fraction variance from each code, bounded by Flamenco as the lowest and Turmoil as the highest. This is expected as at early times the layer is in a state of transition, where the initially smooth density variation across the interface is being sharpened to a grid-scale discontinuity by shear flows driven by the growth of RMI. The minimum thickness of the layer is a function of the dissipation within each code, hence the increased uncertainty. At the latest two times, the mean gradients within the layer are smoother and the codes are in better agreement, although the peaks are still only in agreement to within $\pm$ 10\%. As time progresses, the variance decreases towards a self-similar profile as stirring reduces the gradients through mixing.

\begin{figure}
\begin{centering}
\includegraphics[width=0.49\textwidth]{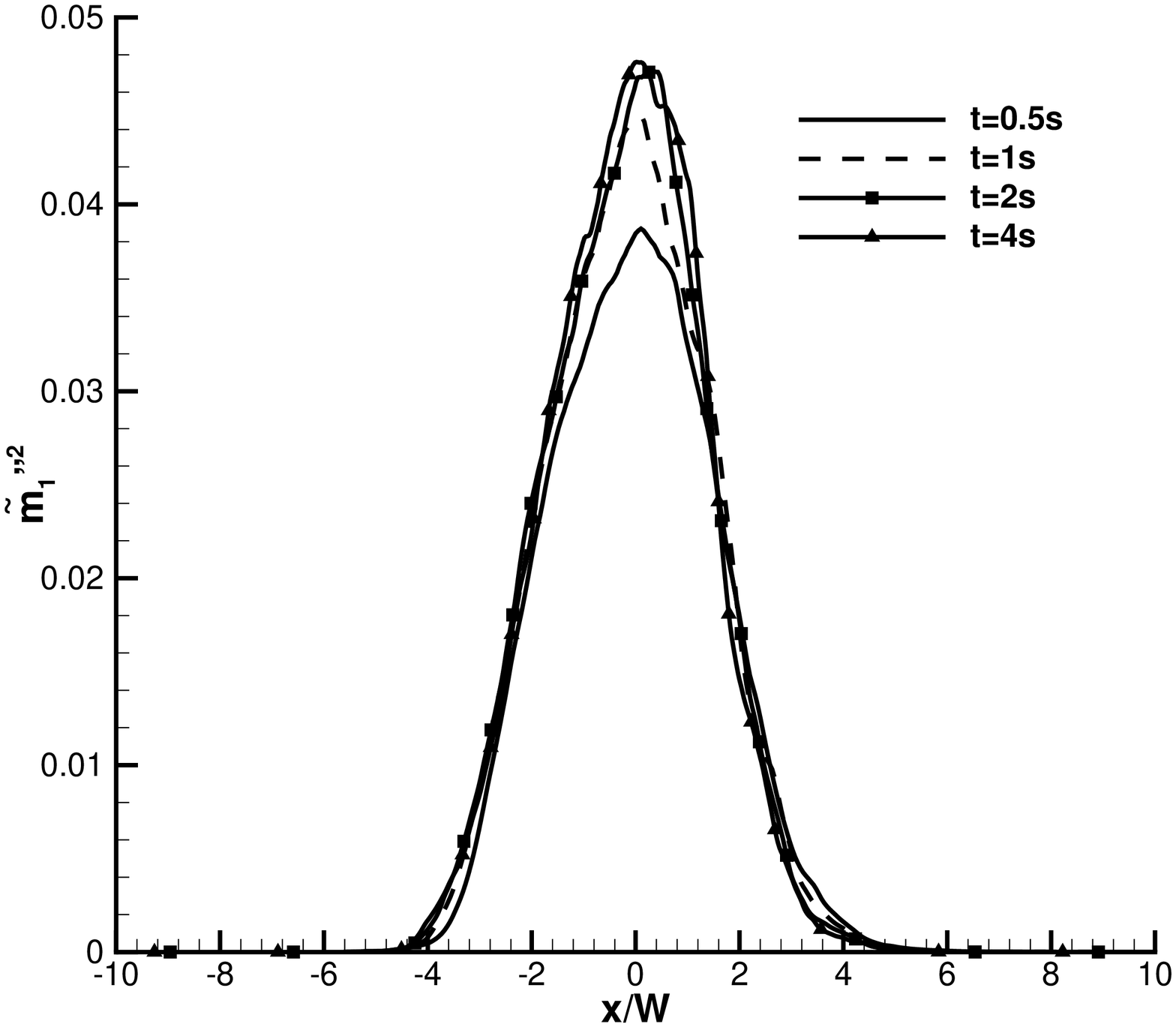}
\caption{Mass fraction variance $\widetilde{m_1''^2}$ at $t=0.5$, $1$, $2$, and $4$ s (Flamenco results only).  \label{mfvss}}
\end{centering}
\end{figure}

Figure \ref{mfvss} shows the collapse of the  $\widetilde{m_1''^2}$ profiles at several times. Except for the earliest `transitional' time of $0.5$ s, the profiles at the other three times collapse well, having a similar peak value of $\approx 0.046$. Mass fraction variance gradients are steeper on the spike side of the layer compared to the bubble side. 

\begin{figure*}
\begin{centering}
\subfigure[\hspace{0.1cm} $t=0.01$ s]{\includegraphics[width=0.49\textwidth]{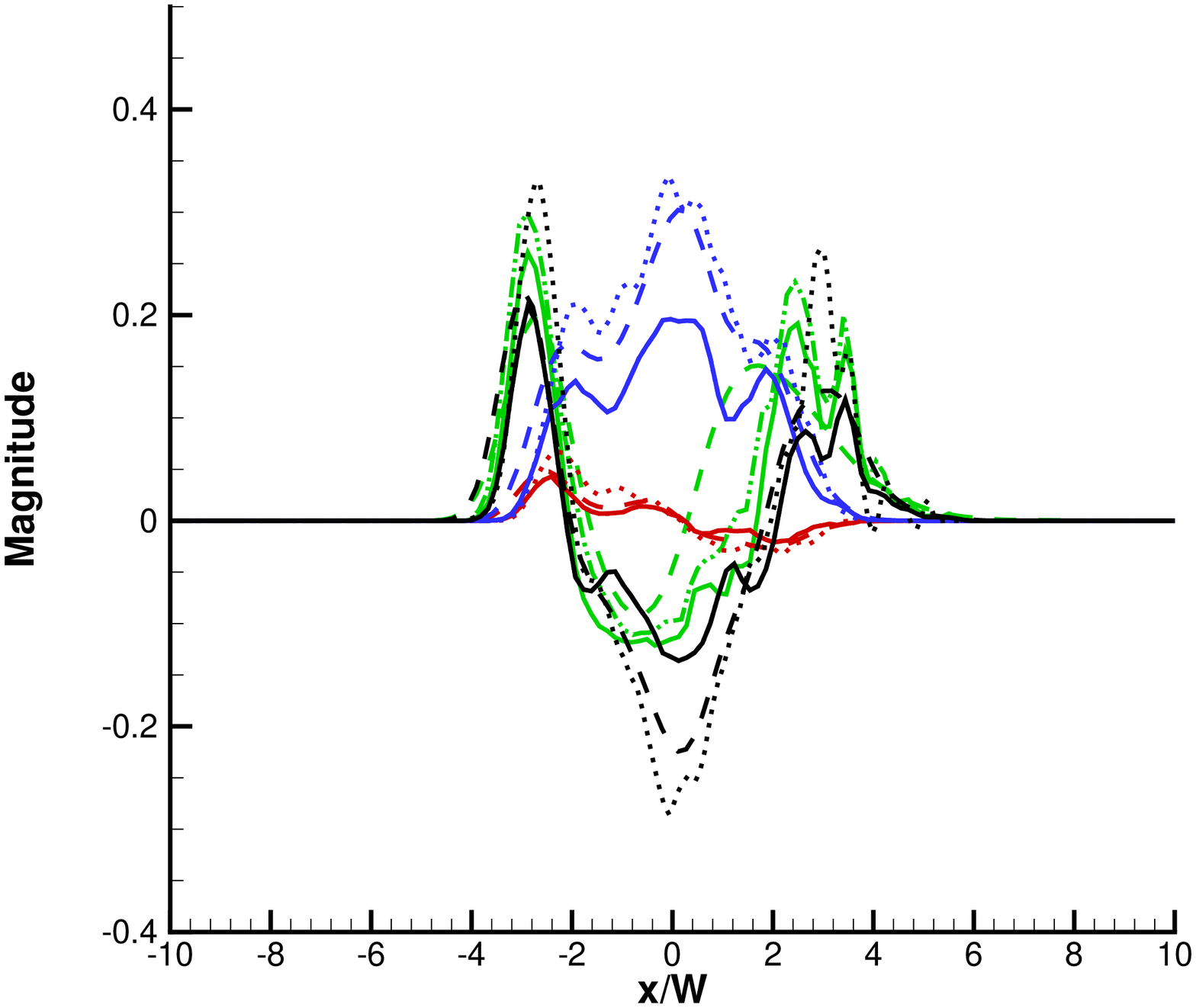}}
\subfigure[\hspace{0.1cm} $t=0.025$ s]{\includegraphics[width=0.49\textwidth]{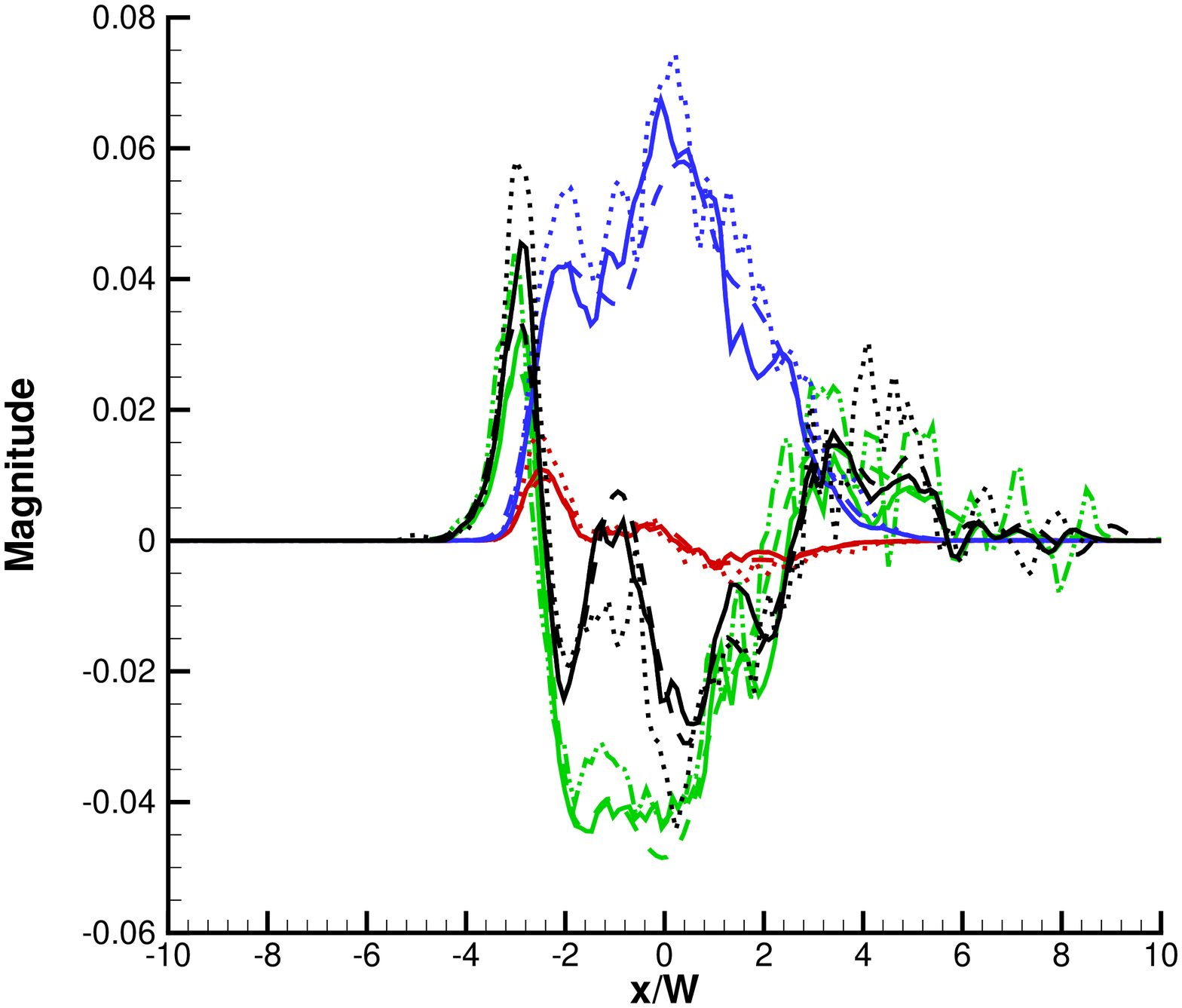}}
\subfigure[\hspace{0.1cm} $t=1$ s]{\includegraphics[width=0.49\textwidth]{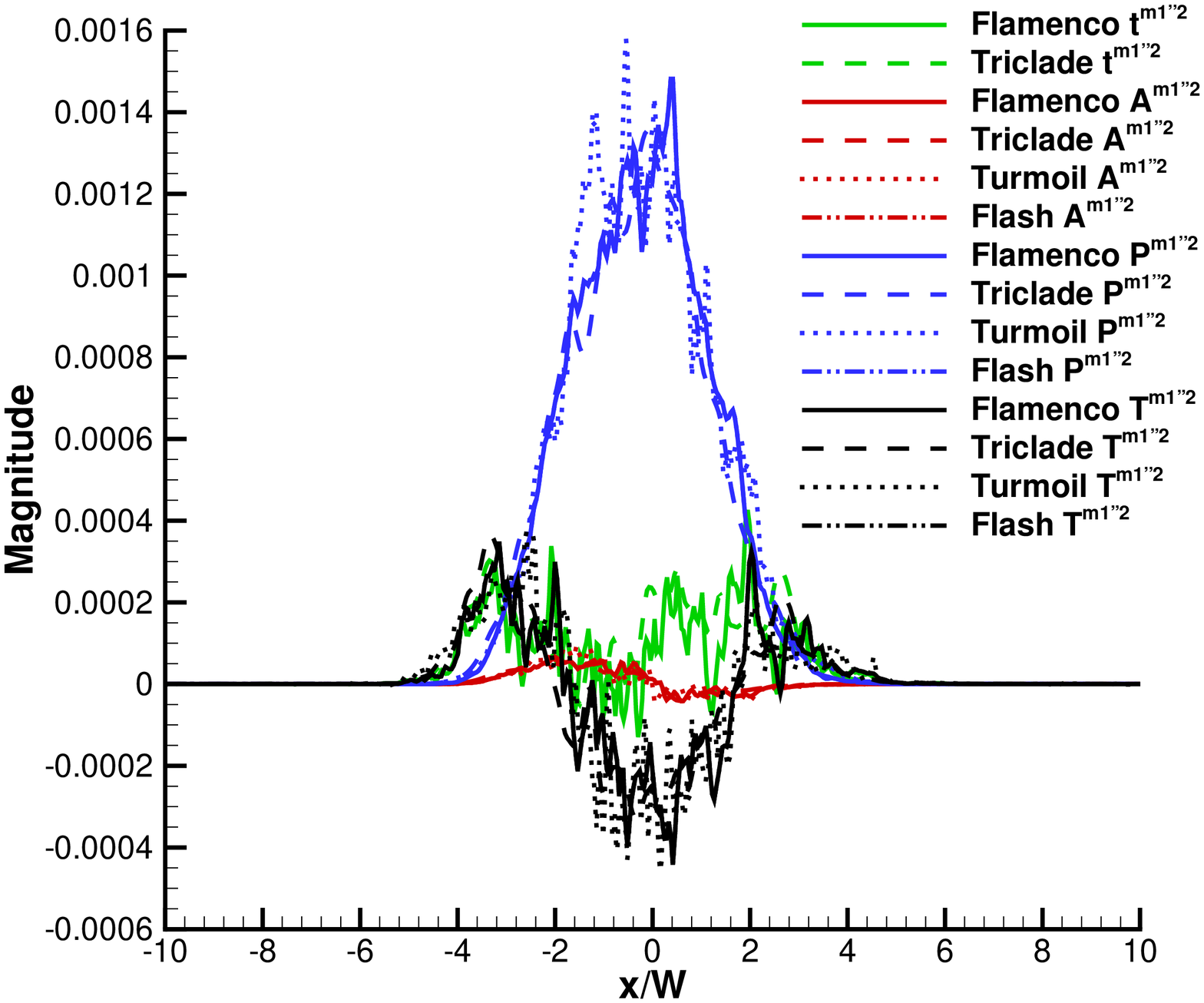}}
\subfigure[\hspace{0.1cm} $t=2$ s]{\includegraphics[width=0.49\textwidth]{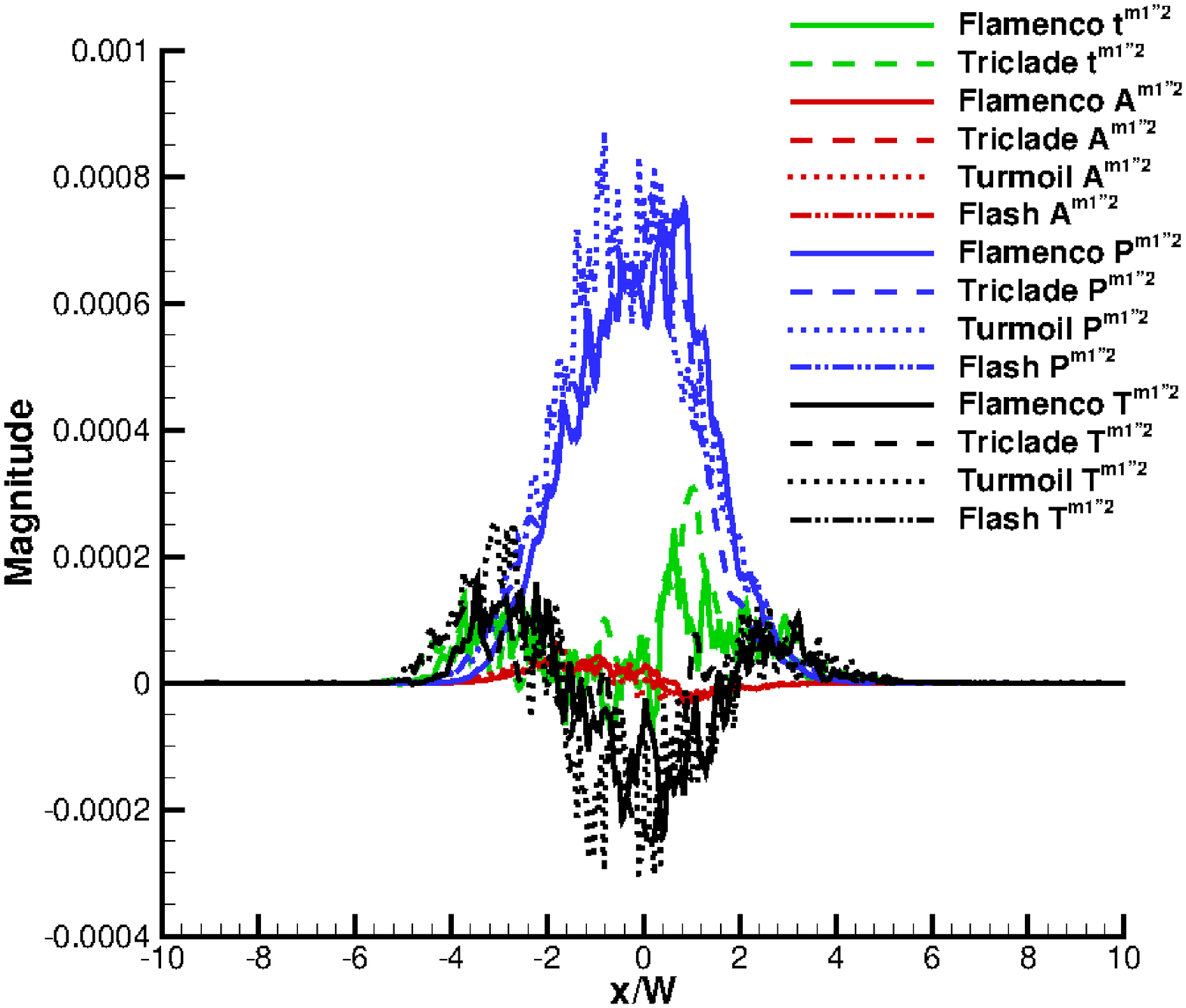}}
\caption{Individual terms in the $\widetilde{m_1''^2}$ transport equation at (a) $t=0.01$ s, (b) $0.025$ s, (c) $1$ s, and (d) $2$ s.  All terms non-dimensionalised by $\rho_c u_c/\bar \lambda$. Flash time derivatives are not shown at $1$ and $2$ s as they follow the same trends as Flamenco and Triclade but contain substantially higher noise, which otherwise masks the underlying trend.  \label{mfv}}
\end{centering}
\end{figure*}

Individual terms in the transport equation for  $\widetilde{m''_1}$ are plotted in Fig. \ref{mfv} for all codes. At the two earliest times, all terms but mean advection are important. At the latest time, mean production dominates all other physically resolved terms by a factor of three, although it must be noted that mean numerical dissipation is of a similar magnitude. Turbulent transport increases variance at the bubble and spike fronts, but reduces it within the core. Although this also occurs at the later times, it is dominated by mean production which is asymmetrically weighted towards the bubble side of the layer. 

\begin{figure*}
\begin{centering}
\includegraphics[width=0.49\textwidth]{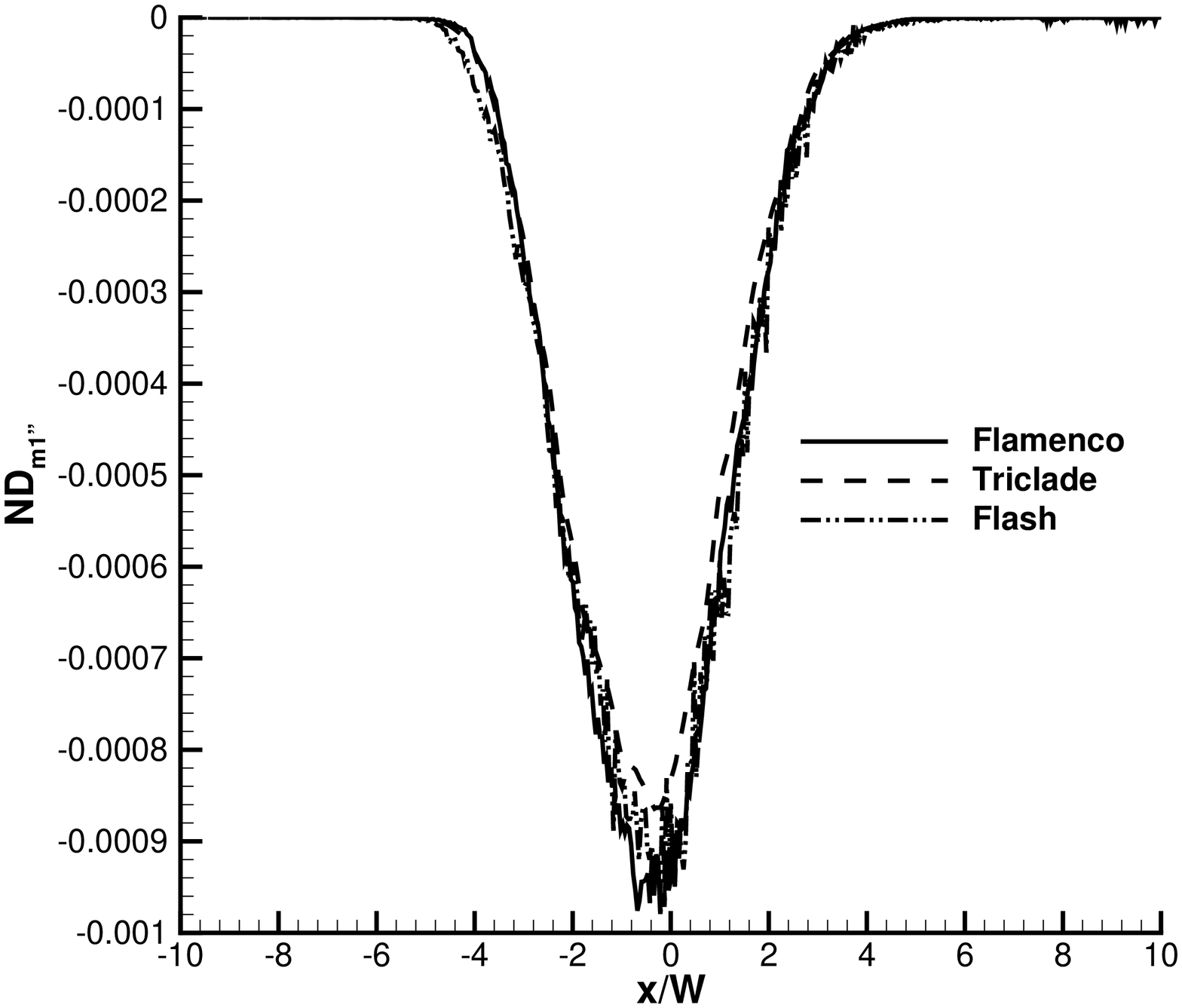}
\includegraphics[width=0.49\textwidth]{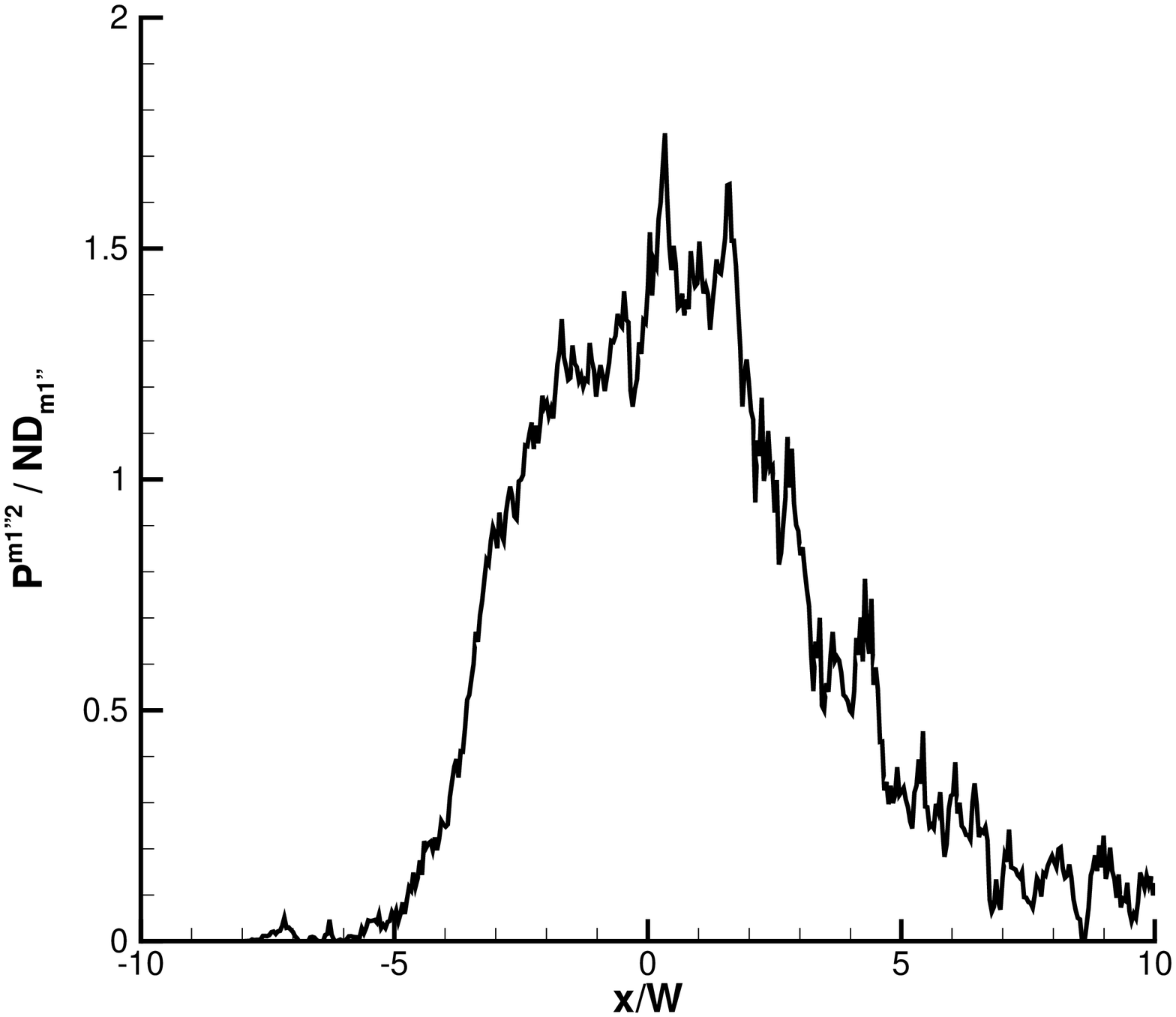}
\caption{Non-dimensional numerical dissipation estimated by the residual in the mass fraction variance transport balance at $t=1$ s (left), and the production rate over dissipation rate (right).  \label{mfvdiss}}
\end{centering}
\end{figure*}

The numerical dissipation $ND_{m1''}$ has been computed for $t=1$ s and is plotted in Fig. \ref{mfvdiss} for Flamenco, Flash and Triclade, alongside the production over dissipation for Flamenco. The agreement between the codes is excellent, particularly given the different algorithmic approaches. As described above, numerical dissipation is of similar magnitude to the largest physical terms, and is purely dissipative. The ratio of production to dissipation peaks at $\approx 1.5$, somewhat larger than the Rayleigh-Taylor instability value of $1.25$ reported by Schilling and Mueschke \cite{Schilling2010b} and takes a volume weighted average of $0.92$ over the region $-5 < x/W < 5$.
    
\subsection{Turbulent Kinetic Energy Transport \label{ketrans}}

\begin{figure*}
\begin{centering}
\subfigure[\hspace{0.1cm} $t=0.01$ s]{\includegraphics[width=0.49\textwidth]{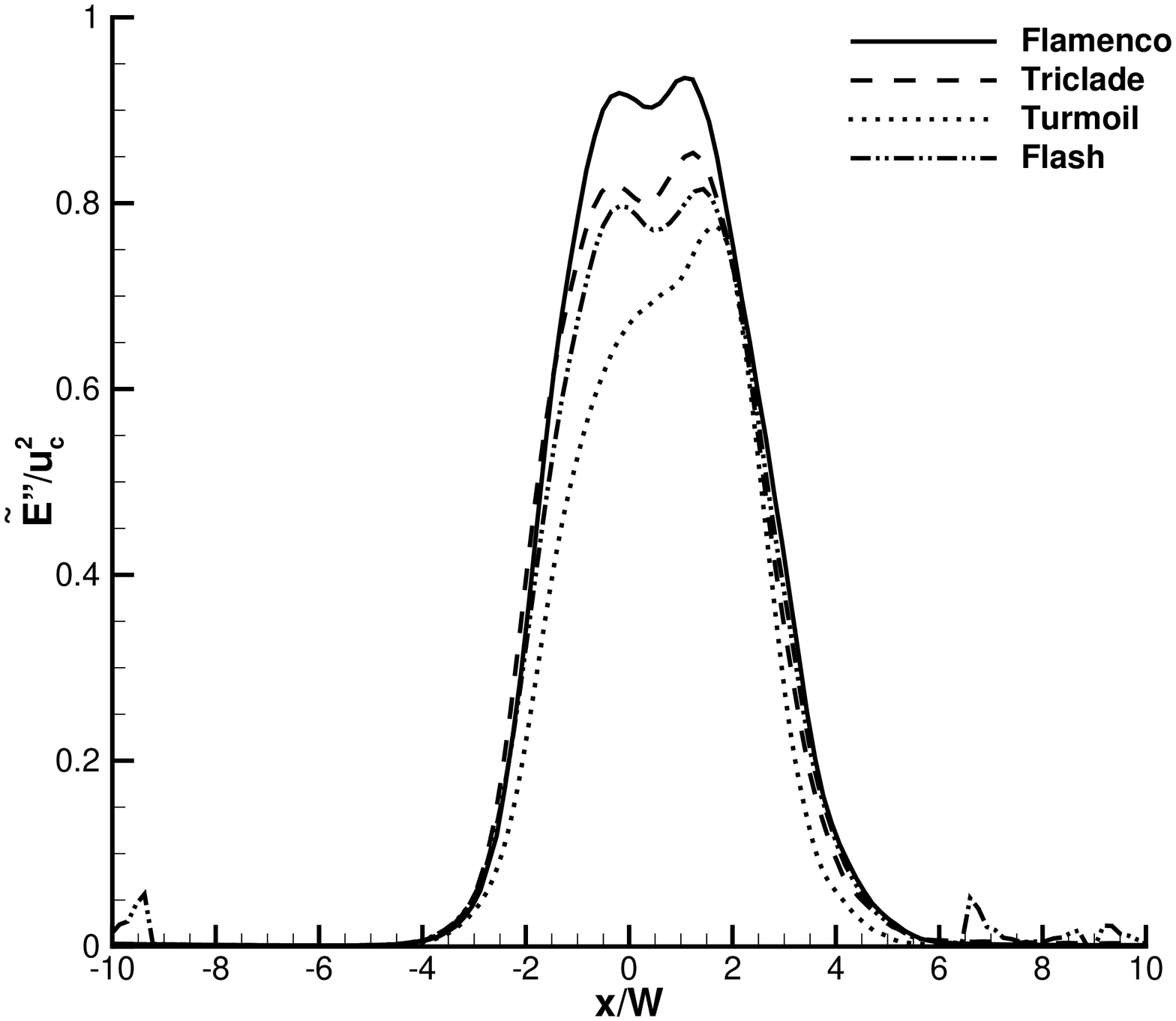}}
\subfigure[\hspace{0.1cm} $t=0.025$ s]{\includegraphics[width=0.49\textwidth]{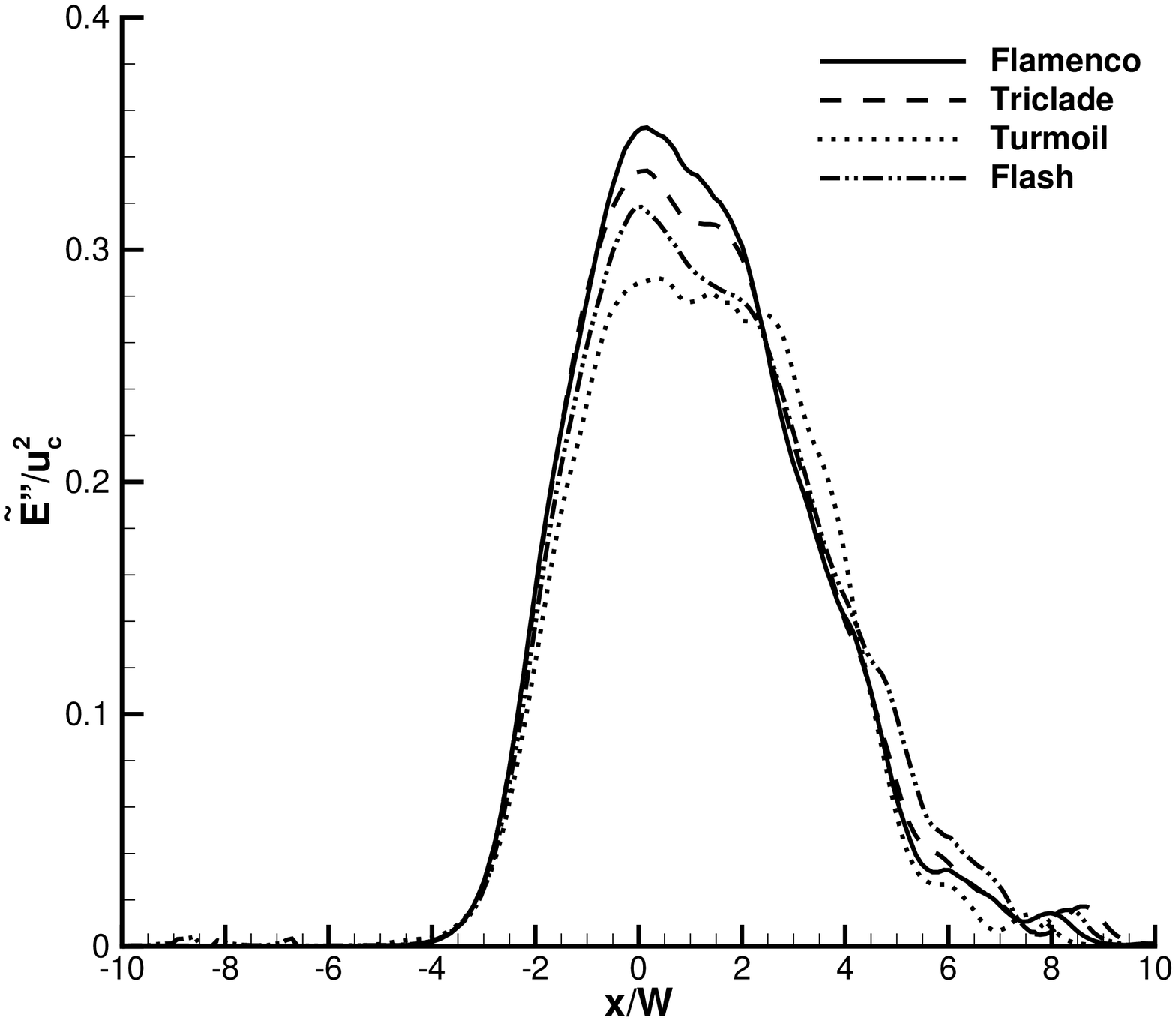}}
\subfigure[\hspace{0.1cm} $t=1$ s]{\includegraphics[width=0.49\textwidth]{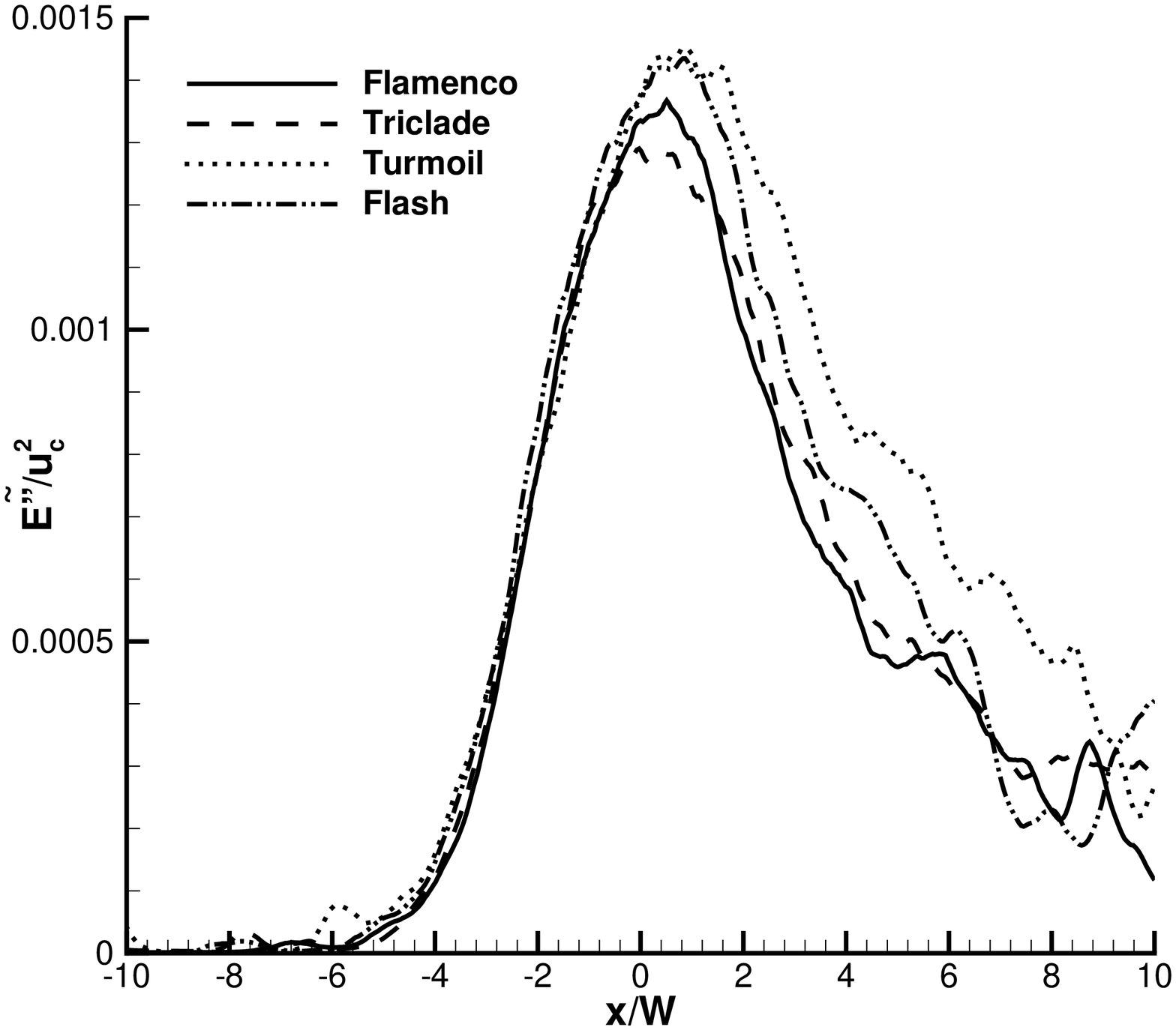}}
\subfigure[\hspace{0.1cm} $t=2$ s]{\includegraphics[width=0.49\textwidth]{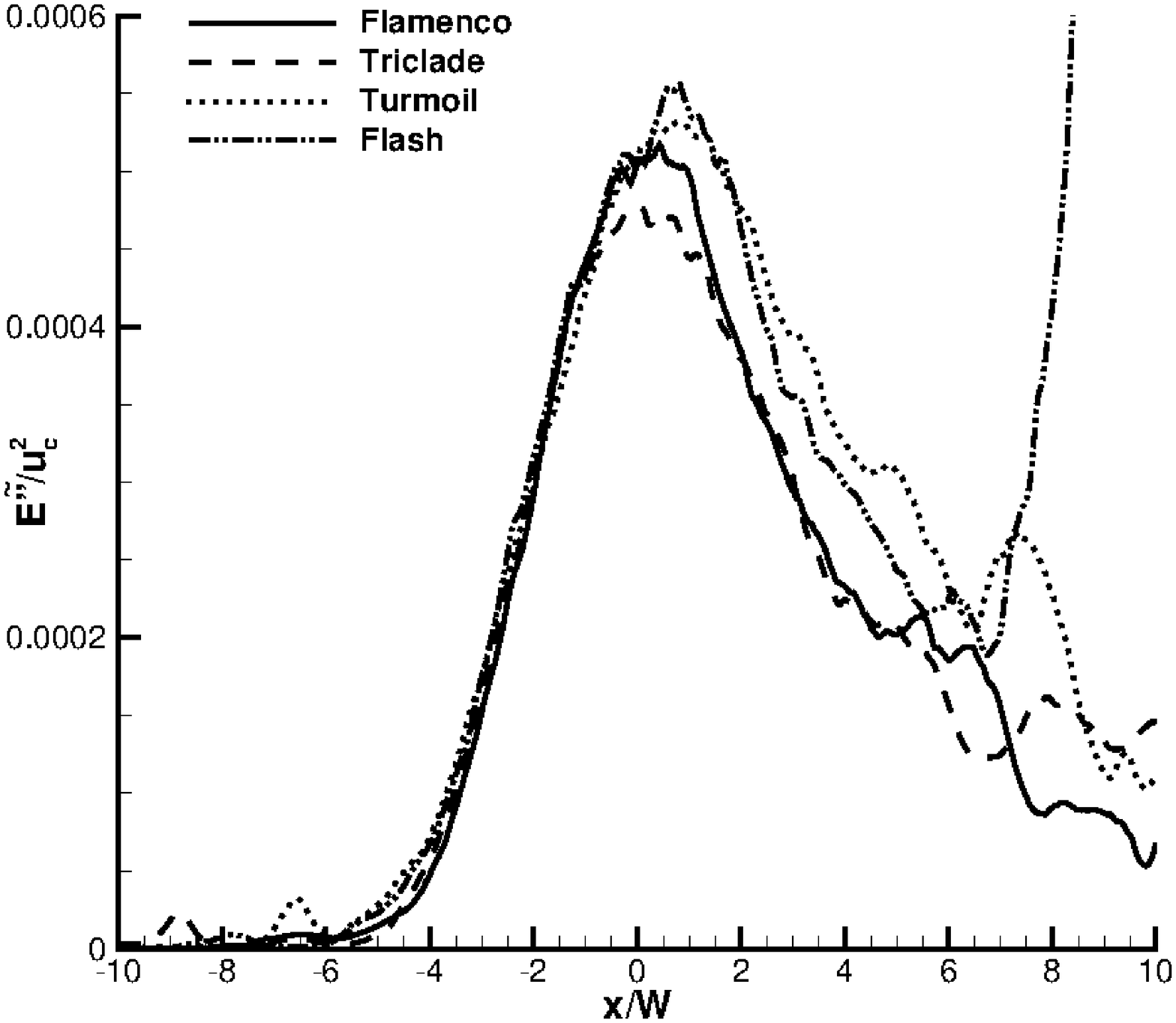}}
\caption{Normalised turbulent kinetic energy $\widetilde{E''}/u_c^2$ at (a) $t=0.01$ s, (b) $0.025$ s, (c) $1$ s, and (d) $2$ s.  \label{ketilde}}
\end{centering}
\end{figure*}

The dimensionless turbulent kinetic energy $\widetilde {E''}$ is shown in Fig. \ref{ketilde} for four times. At the earliest time all codes show a `double peak' structure, where the peaks are focused on the bubble and spike heads respectively. The highly energetic spikes persist at late times, causing a broadening of the profile into the light fluid ($x/W>0$). The bulk of the turbulent kinetic energy content in the bubbles is restricted to the region $x/W>-6$, and increases steeply on the bubble side. This is consistent with the visualisations in Fig. \ref{visslice} which show an earlier onset of mixing on the spike side (an indicator of stirring by a range of length scales) than on the bubble side. The advection of spikes away from the layer at early time smooths the spatial contribution of the spike to the turbulent kinetic energy profiles, and turbulent transport acts to merge the early `double bump' profile into a single peak. The peak is slightly biased to the spike side of the mixing layer at late time, where most of the kinetic energy is located. For example, at $t=1$ s and $x/W=5.5$ there is substantial $\widetilde {E''}$ even though $\tilde m_1 <0.1$\%. At late times the profile is asymmetric. In comparing the codes, there are consistent peak positions in Triclade, Turmoil, and Flamenco. Flash shows large values on the spike side at the latest time, which is attributed to unphysical interactions with the boundaries.

\begin{figure}
\begin{centering}
\includegraphics[width=0.49\textwidth]{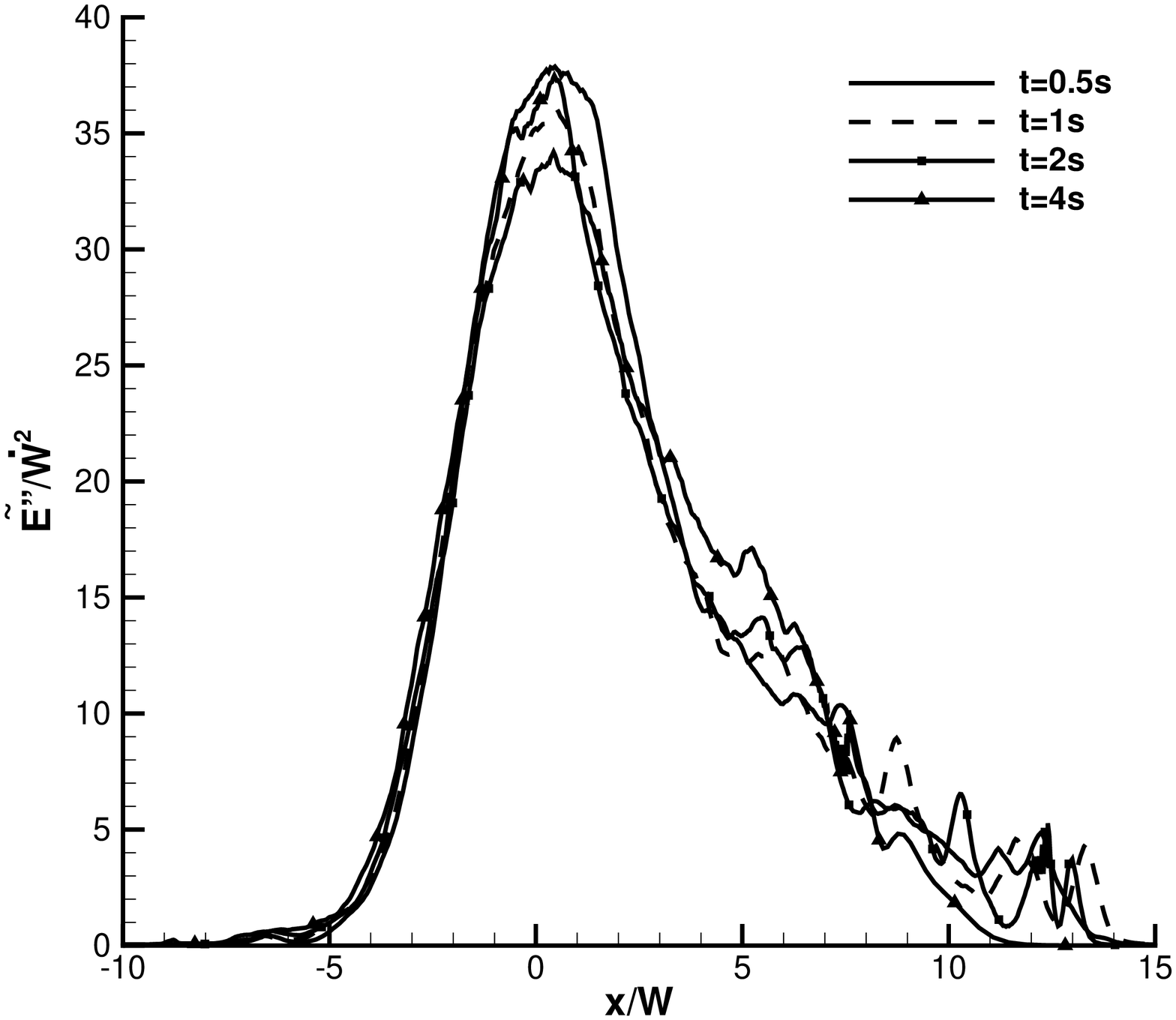}
\caption{Normalised turbulent kinetic energy $\widetilde{E''}/u_c^2$ at $t=0.5$ s, $1$, $2$, and $4$ s (Flamenco results only).  \label{kess}}
\end{centering}
\end{figure}

The scaling of the profiles of  $\widetilde {E''}$ with the mixing layer width and its growth rate is plotted in Fig. \ref{kess}. Overall the collapse is excellent for the core of the layer, but notable differences can be seen in the prediction of the peaks in $\widetilde {E''}$ in the spikes. As observed previously, the approximate vortex ring structures do not scale with the same power law as the rest of the mixing layer, and generate isolated peaks and troughs in the $\widetilde {E''}$ profile which are not consistent from one time to the next when scaled using the integral layer properties. However, the overall shape of the profile is in good agreement given the factor of $8$ reduction in dimensional kinetic energy between the first time and the last shown in Fig. \ref{kess}.

\begin{figure*}
\begin{centering}
\subfigure[\hspace{0.1cm} $t=0.01$ s]{\includegraphics[width=0.49\textwidth]{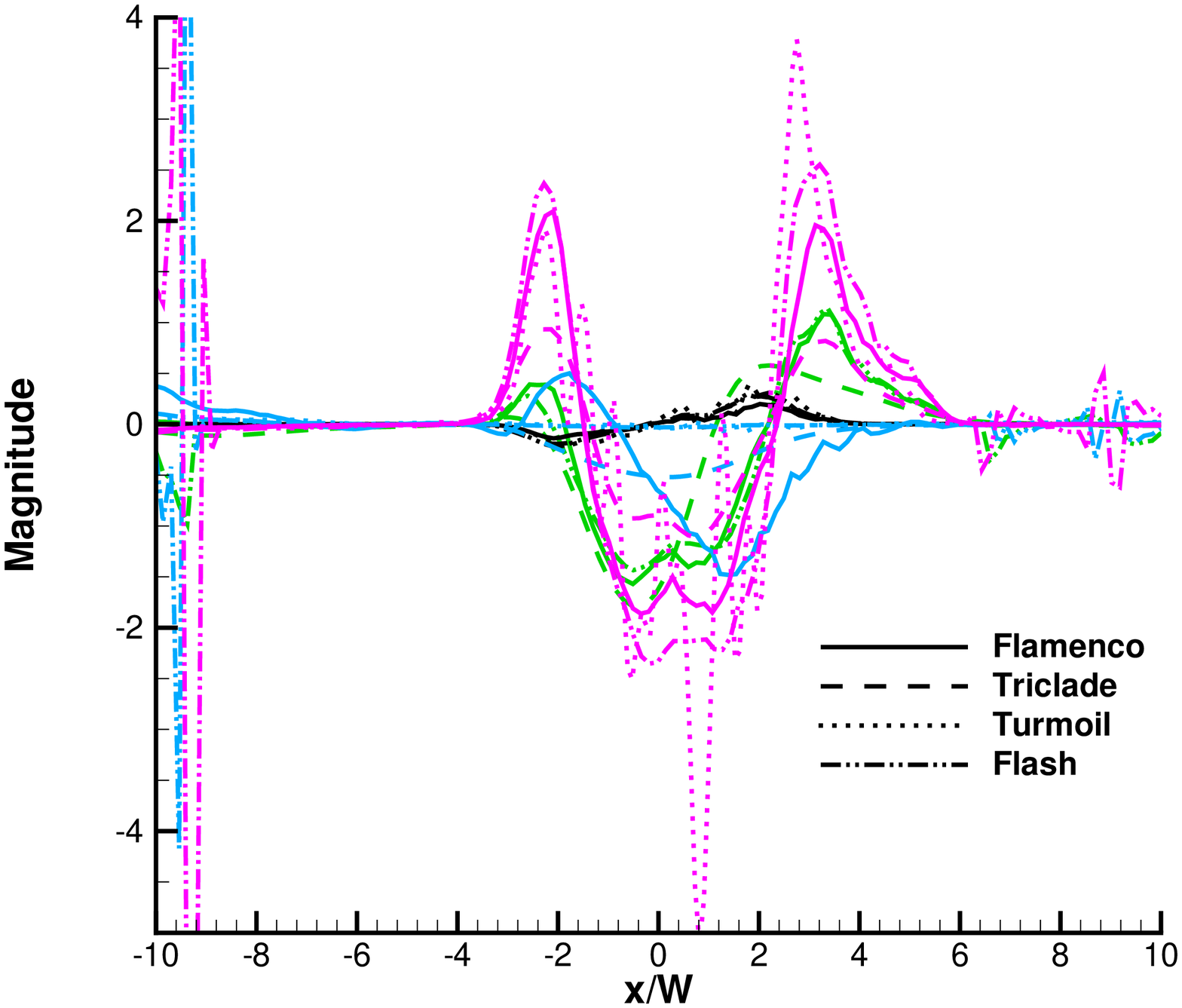}}
\subfigure[\hspace{0.1cm} $t=0.025$ s]{\includegraphics[width=0.49\textwidth]{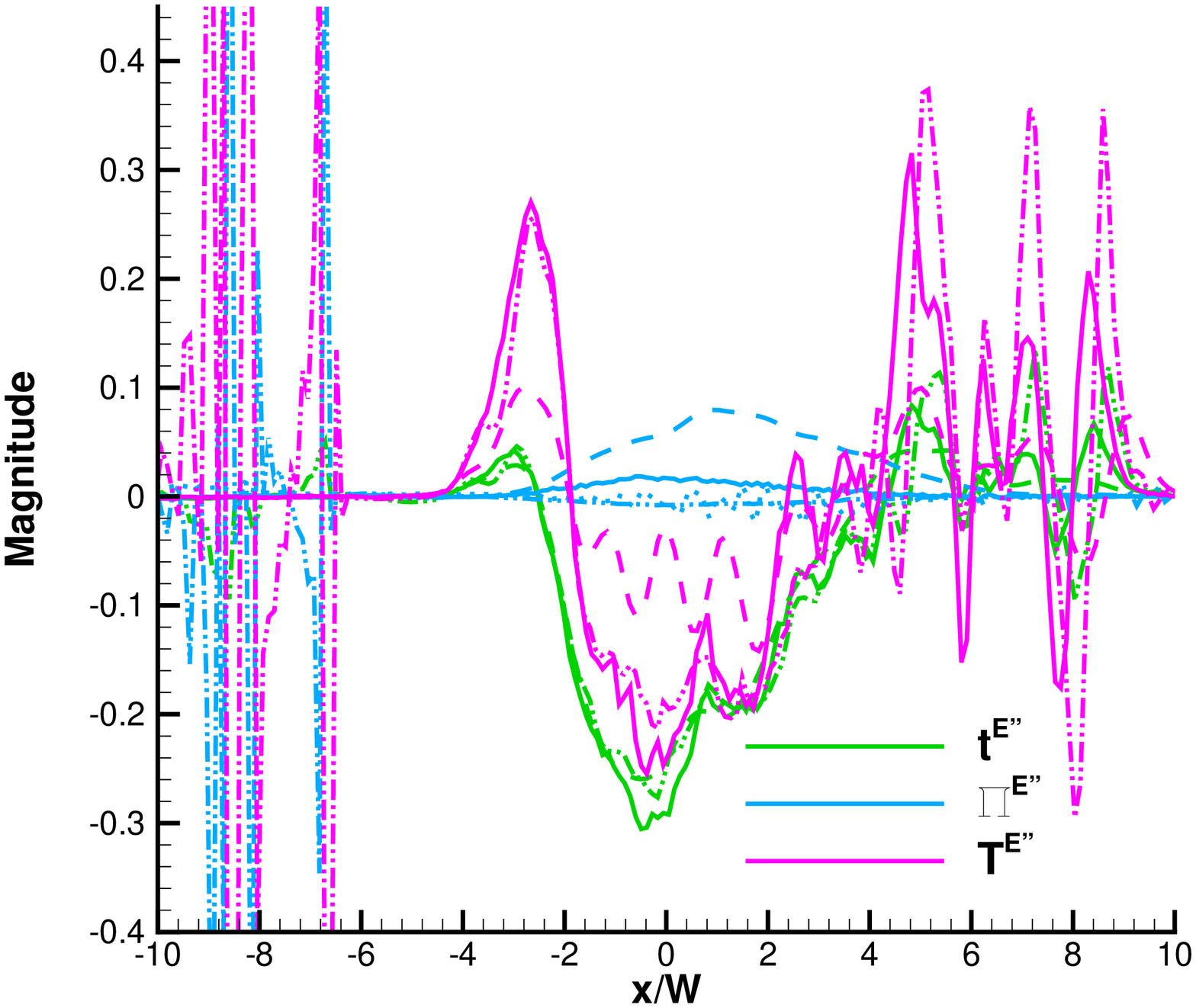}}
\subfigure[\hspace{0.1cm} $t=1$ s]{\includegraphics[width=0.49\textwidth]{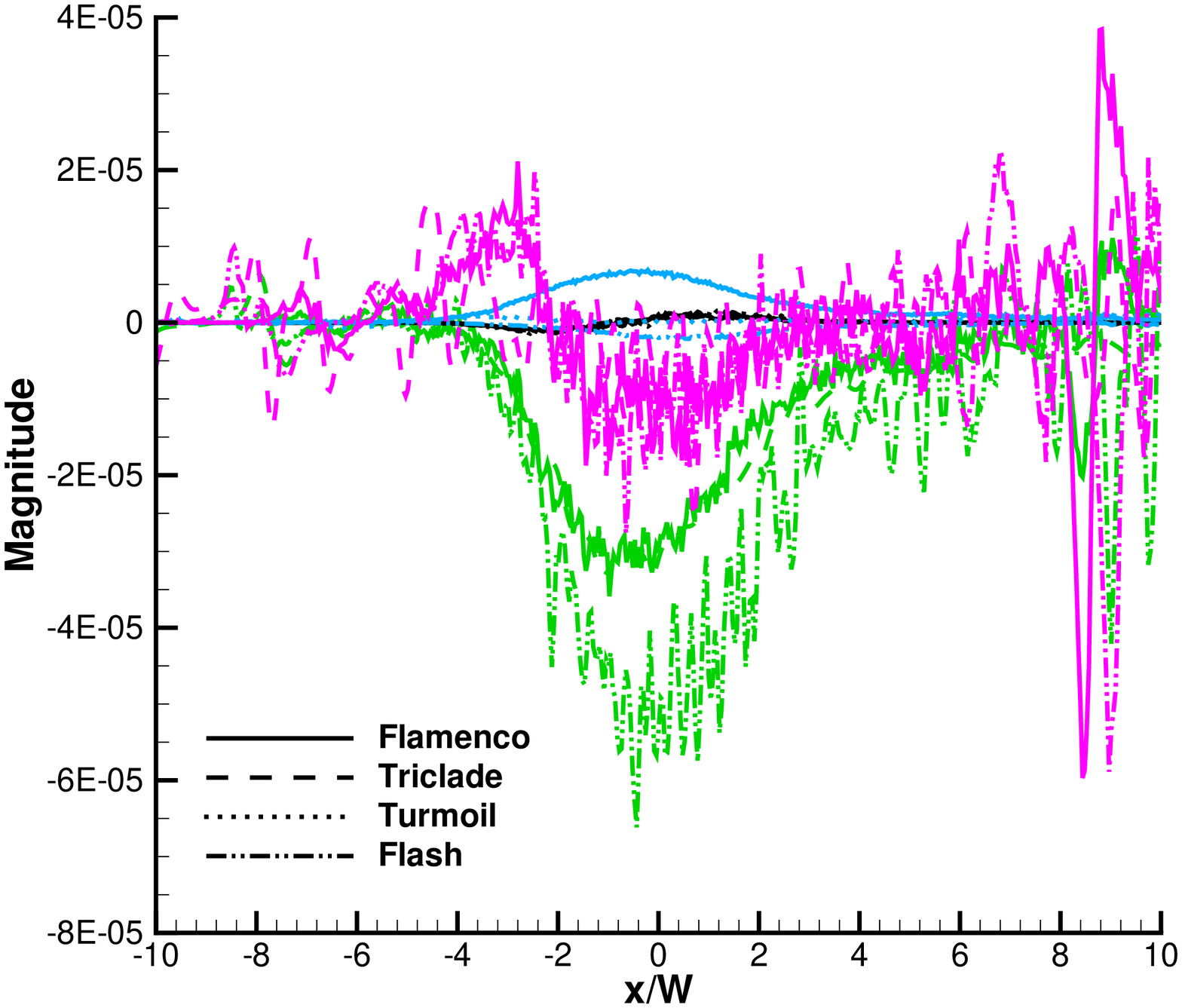}}
\subfigure[\hspace{0.1cm} $t=2$ s]{\includegraphics[width=0.49\textwidth]{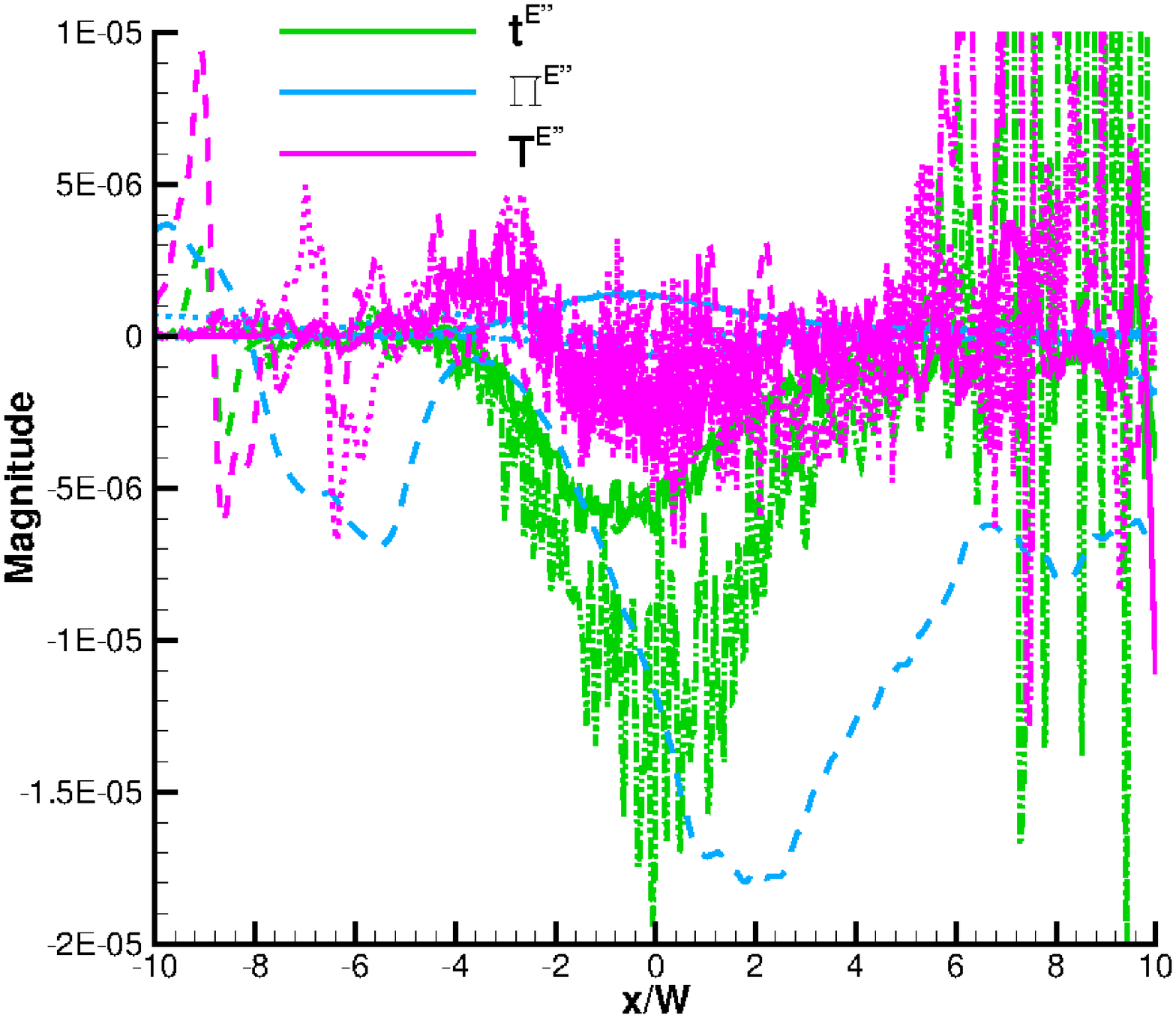}}
\caption{Large magnitude individual terms in the $\widetilde{E''}$ transport equation at (a) $t=0.01$ s, (b) $0.025$ s, (c) $1$ s, and (d) $2$ s. All terms non-dimensionalised by $\rho_c u_c^3/\bar \lambda$. Note that $t=0.025$ s does not include $T^{E''}$ for Turmoil, $t=1$ s  does not include $T^{E''}$ for Turmoil and $\Pi^{E''}$ for Triclade. Turmoil data is not shown as fluctuations mask the other lines; see Fig. \ref{kesmooth} for full data for $T^{E''}$.   \label{ke1}}
\end{centering}
\end{figure*}

\begin{figure*}
\begin{centering}
\subfigure[\hspace{0.1cm} $t=0.01$ s]{\includegraphics[width=0.49\textwidth]{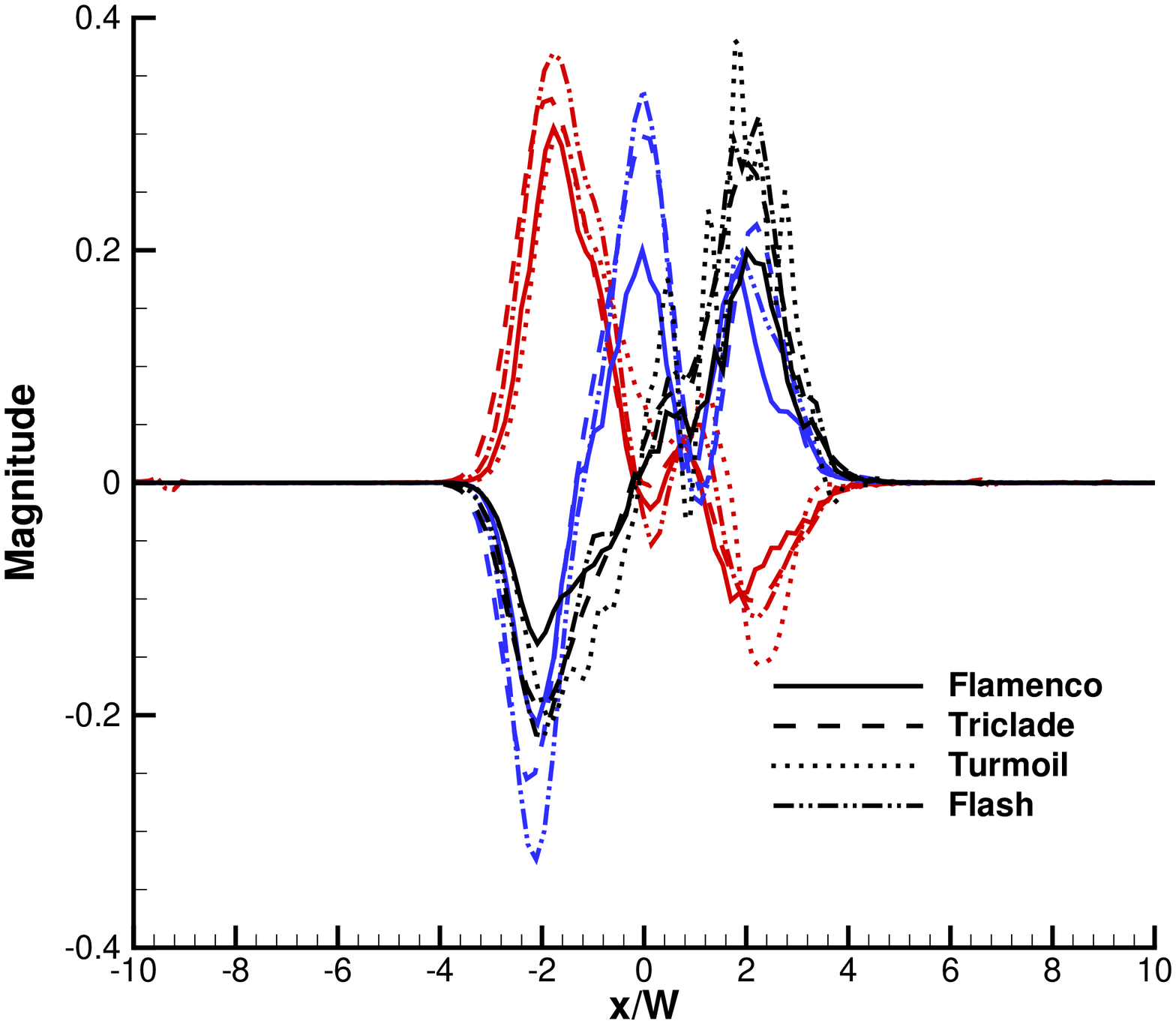}}
\subfigure[\hspace{0.1cm} $t=0.025$ s]{\includegraphics[width=0.49\textwidth]{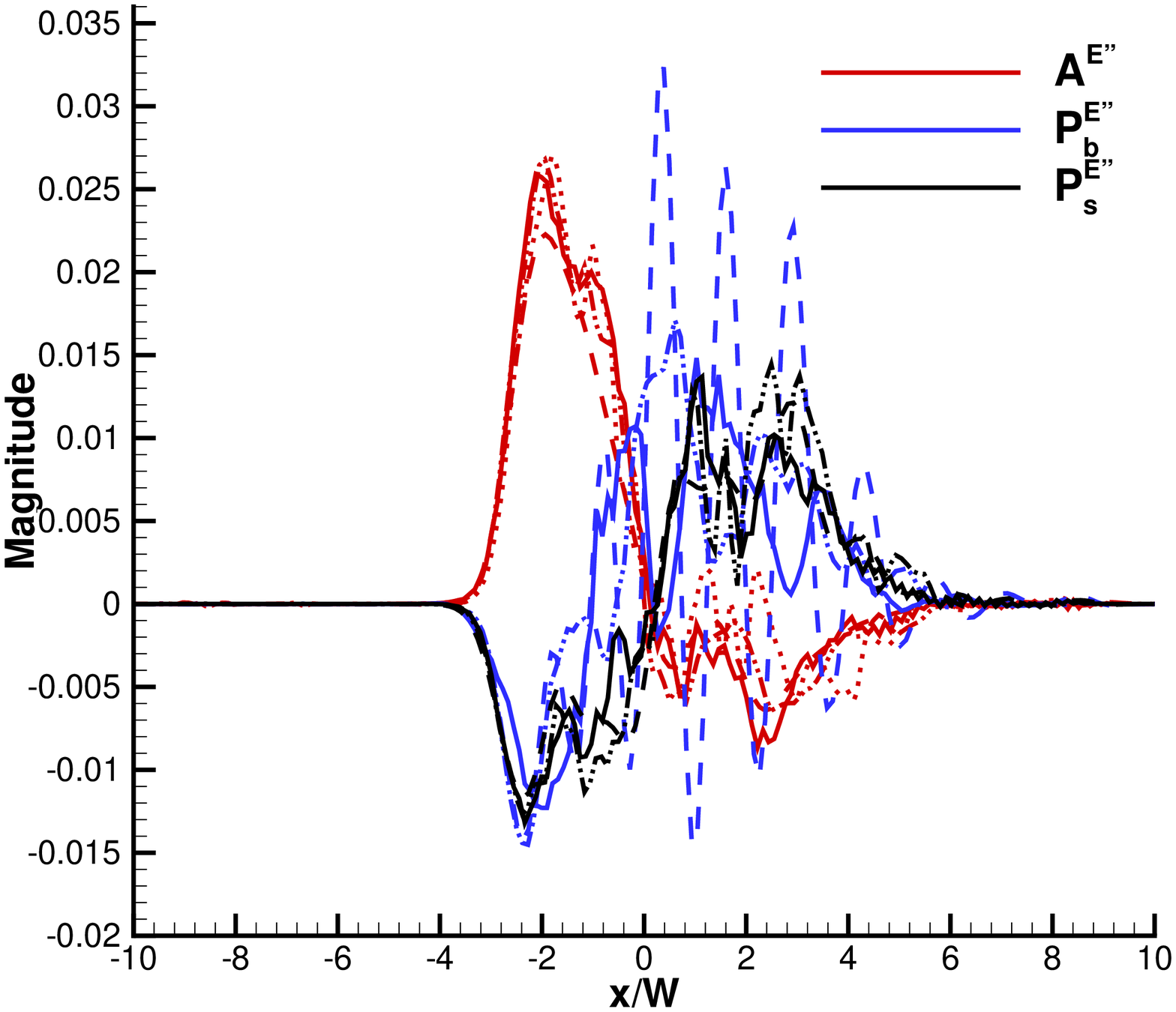}}
\subfigure[\hspace{0.1cm} $t=1$ s]{\includegraphics[width=0.49\textwidth]{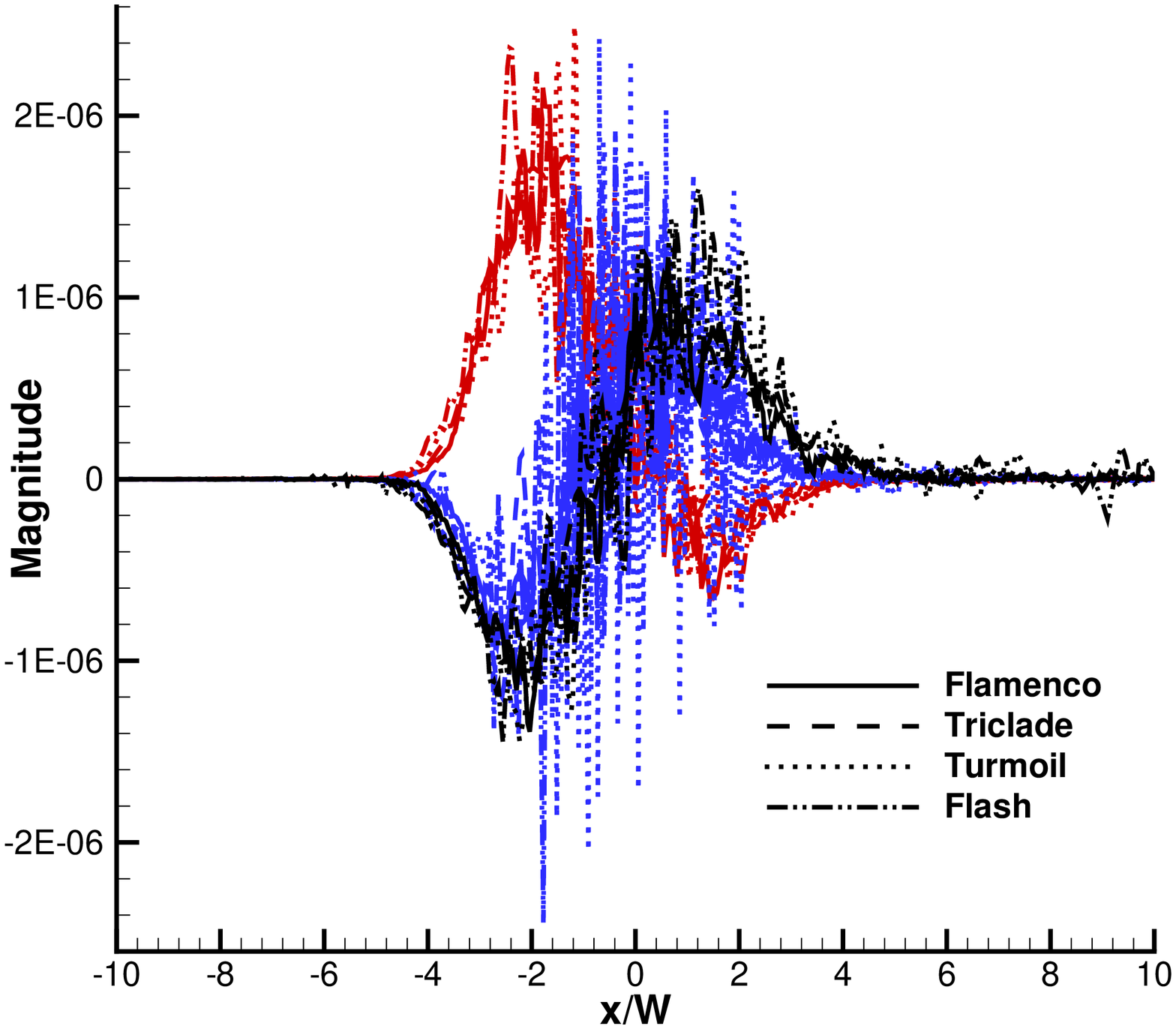}}
\subfigure[\hspace{0.1cm} $t=2$ s]{\includegraphics[width=0.49\textwidth]{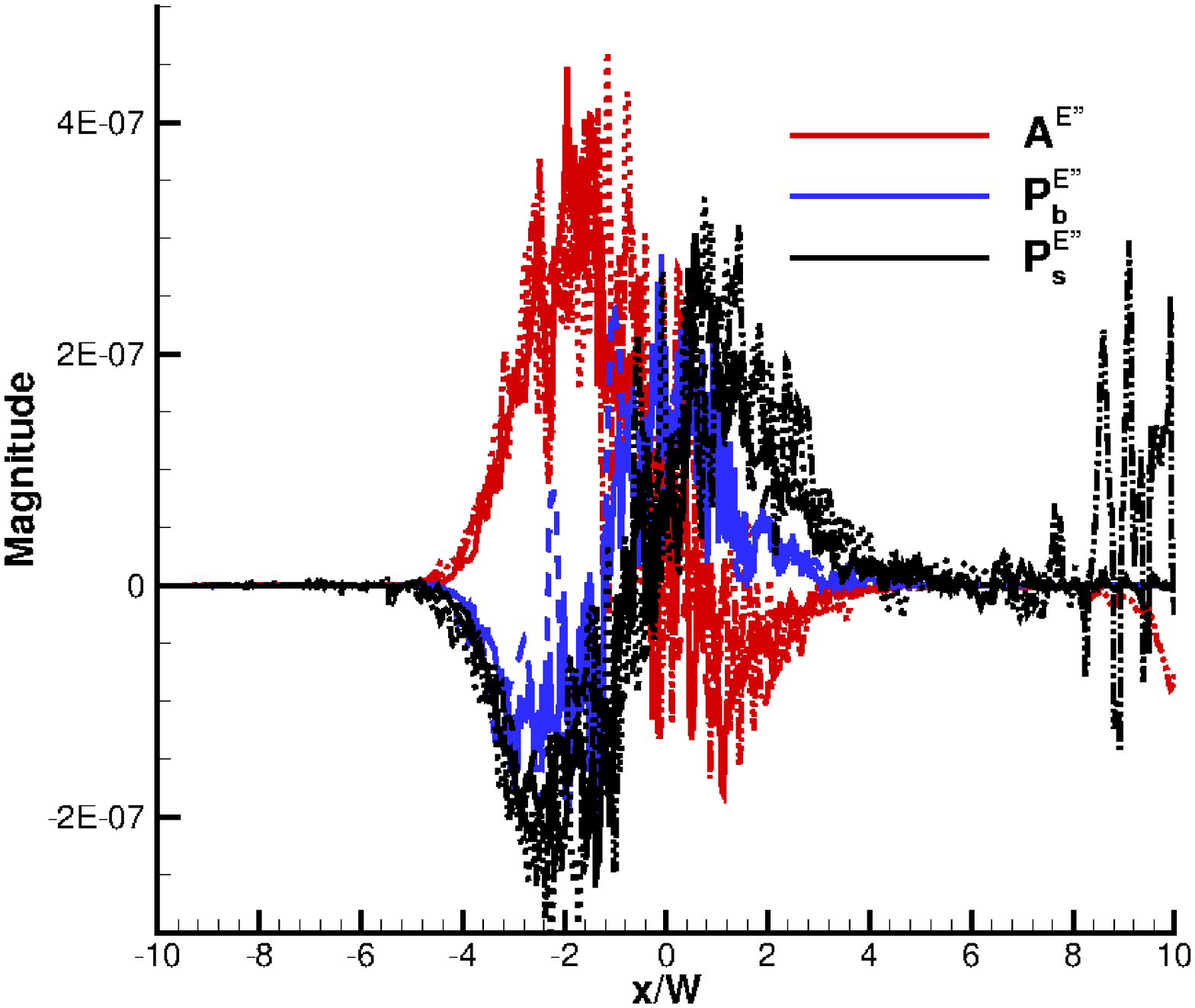}}
\caption{Small magnitude individual terms in the $\widetilde{E''}$ transport equation at (a) $t=0.01$ s, (b) $0.025$ s, (c) $1$ s, and (d) $2$ s. All terms are non-dimensionalised by $\rho_c u_c^3/\bar \lambda$. Note that the $t=0.01$, $t=0.025$, and $2$ s does not include Turmoil $P_b^{E''}$, $0.025$ s also does not include Turmoil $P_s^{E''}$, and $t=2$ s does not show the Flash $P_b^{E''}$ term. Turmoil data is not shown as fluctuations mask the other curves; see Fig. \ref{kesmooth} for full data for $P^{E''}_b$.  \label{ke2}}
\end{centering}
\end{figure*}

\begin{figure*}
\begin{centering}
\subfigure[\hspace{0.1cm} $t=0.01$ s]{\includegraphics[width=0.49\textwidth]{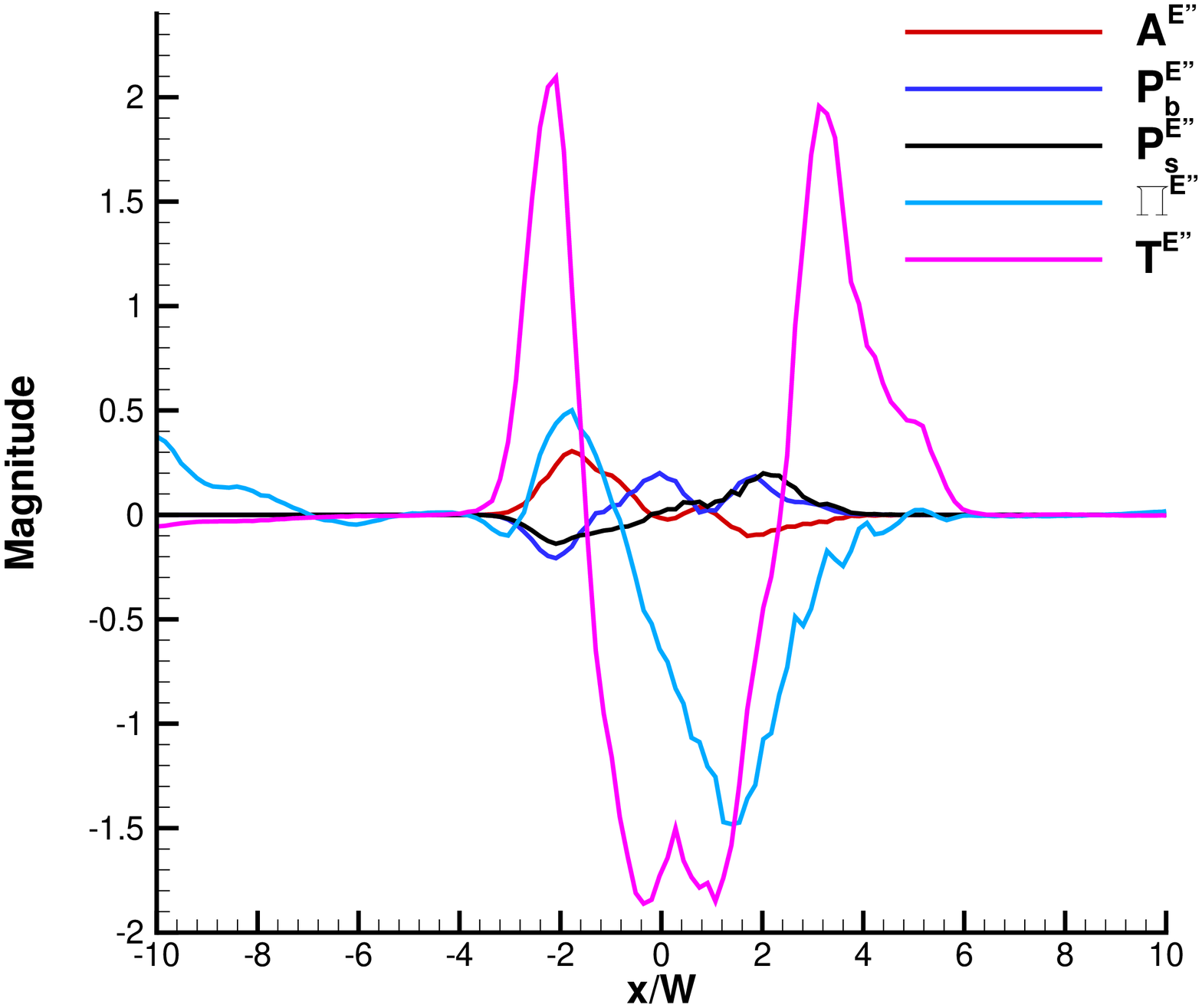}}
\subfigure[\hspace{0.1cm} $t=0.025$ s]{\includegraphics[width=0.49\textwidth]{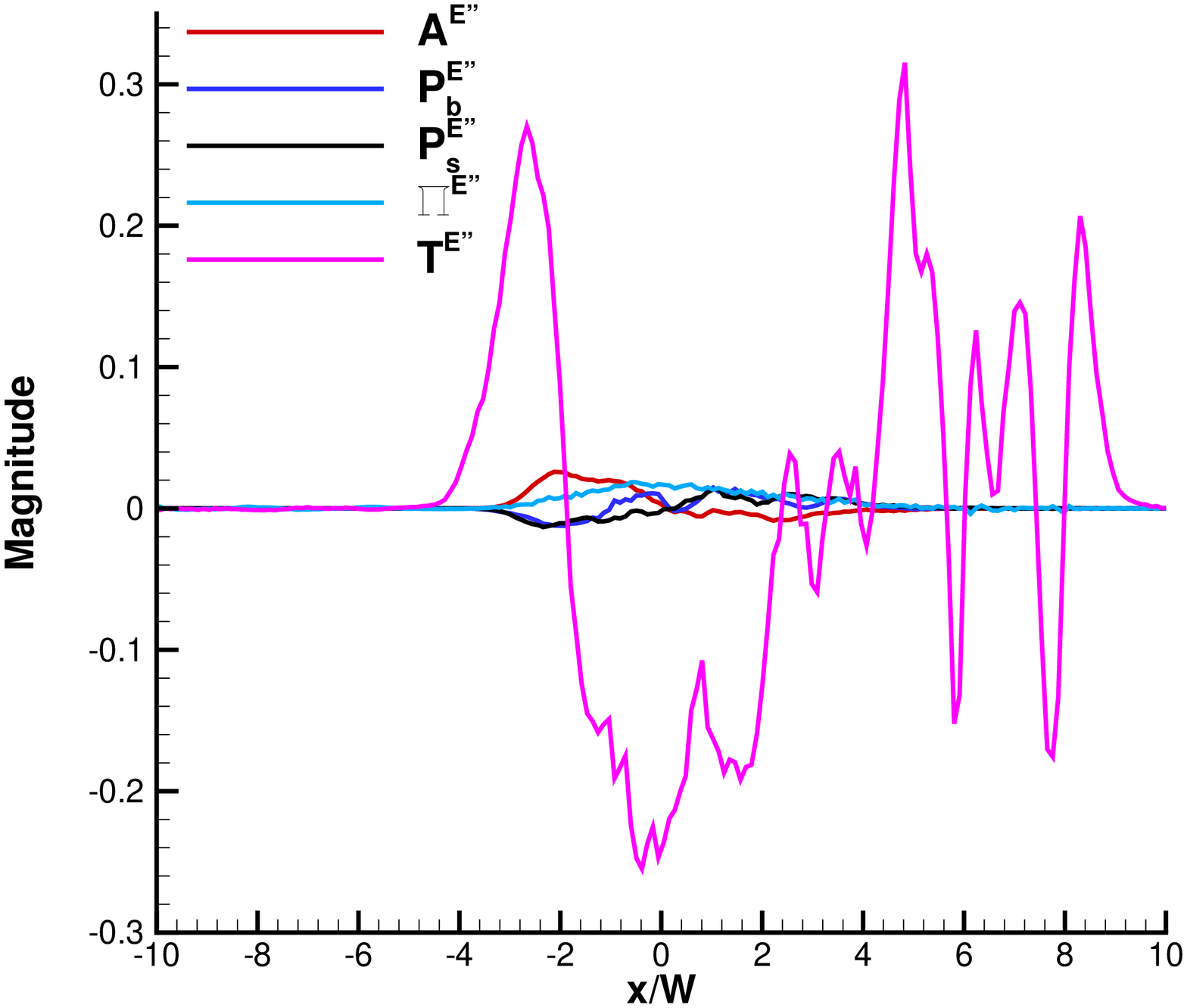}}
\subfigure[\hspace{0.1cm} $t=1$ s]{\includegraphics[width=0.49\textwidth]{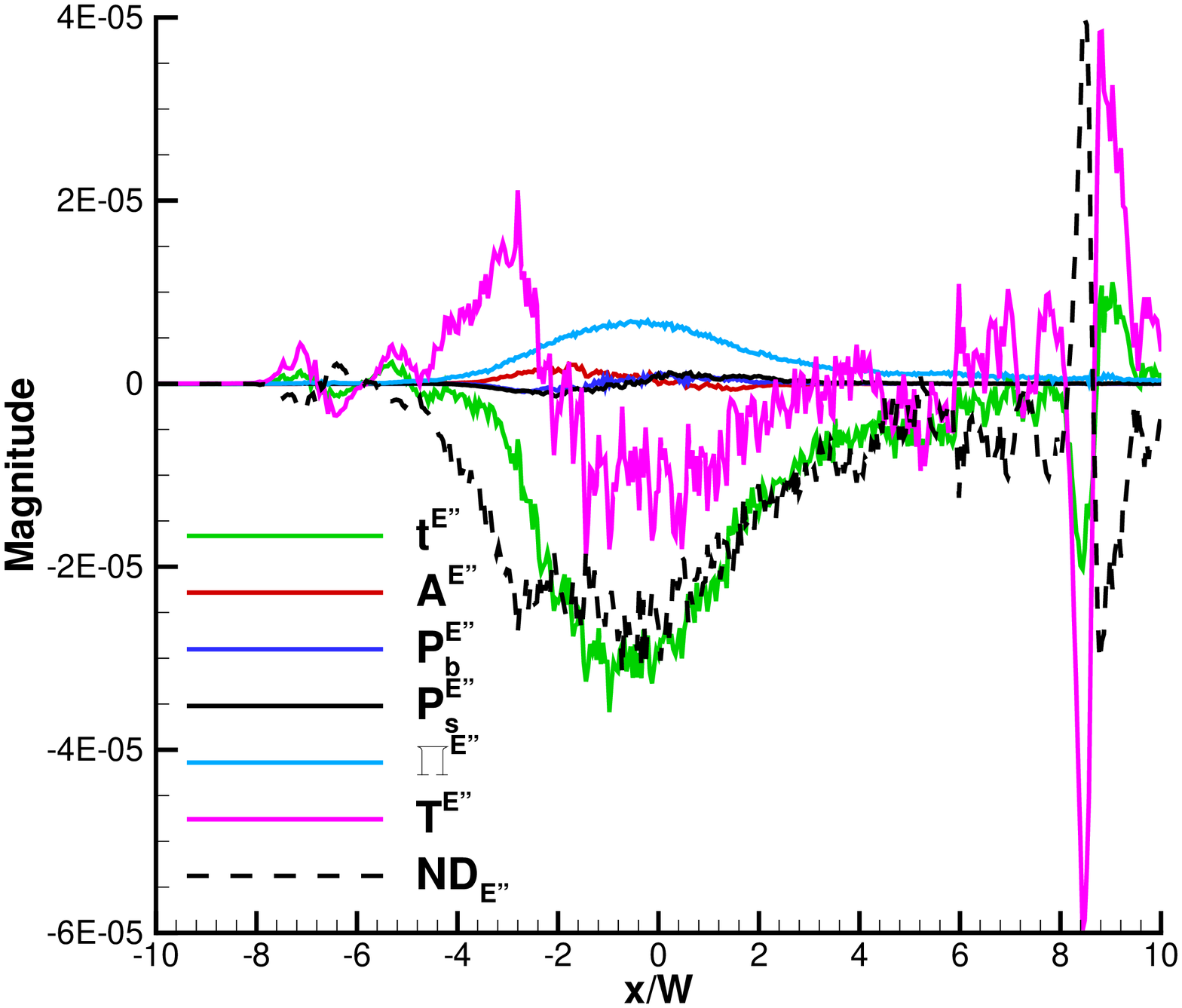}}
\subfigure[\hspace{0.1cm} $t=2$ s]{\includegraphics[width=0.49\textwidth]{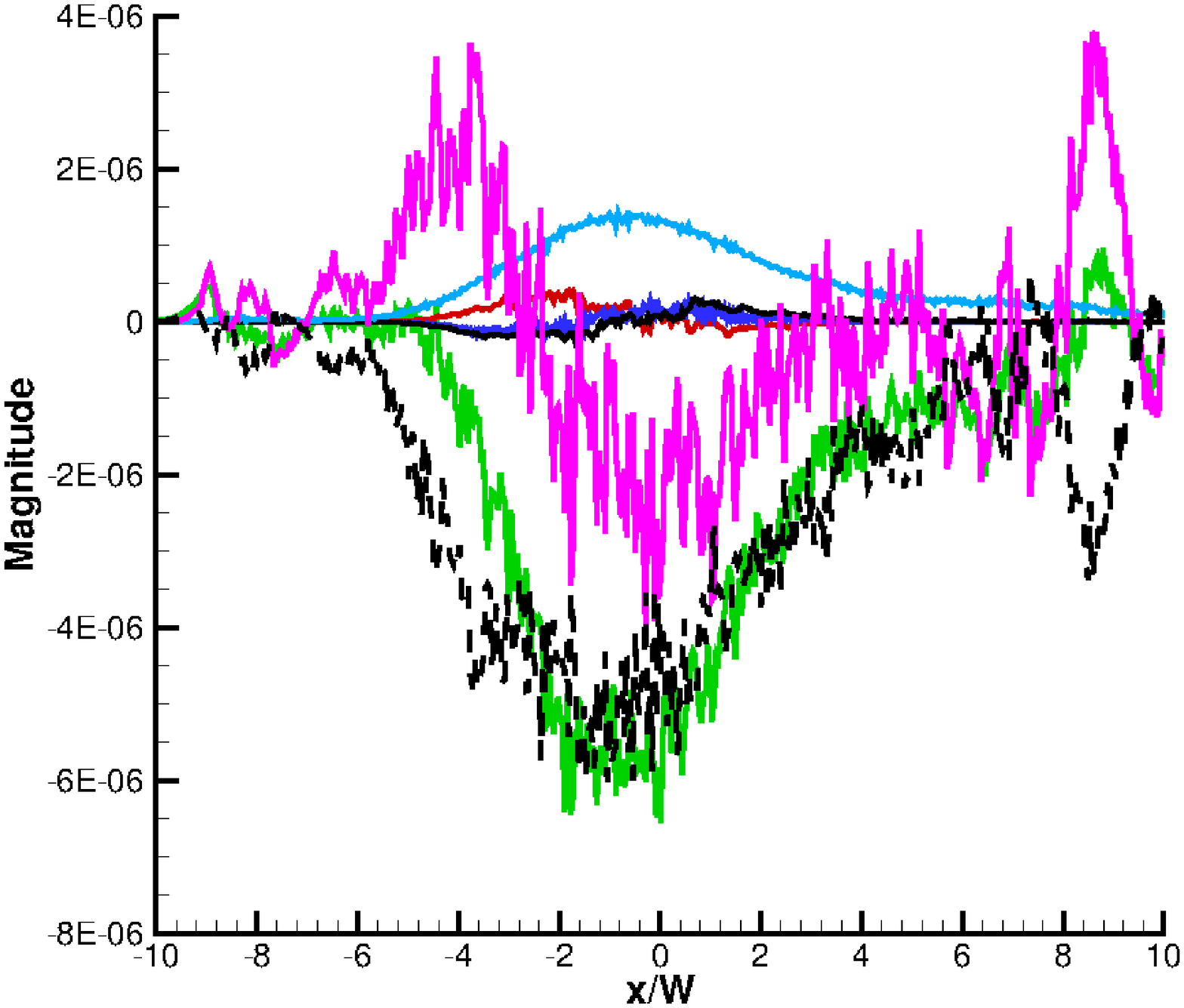}}
\caption{Individual terms in the $\widetilde{E''}$ transport equation at (a) $t=0.01$ s, (b) $0.025$ s, (c) $1$ s, and (d) $2$ s computed from the Flamenco results. All terms non-dimensionalised by $\rho_c u_c^3/\bar \lambda$.  \label{keflam}}
\end{centering}
\end{figure*}

\begin{figure*}
\begin{centering}
\subfigure[\hspace{0.1cm} $P_b^{E''}$ at $t=1$ s]{\includegraphics[width=0.49\textwidth]{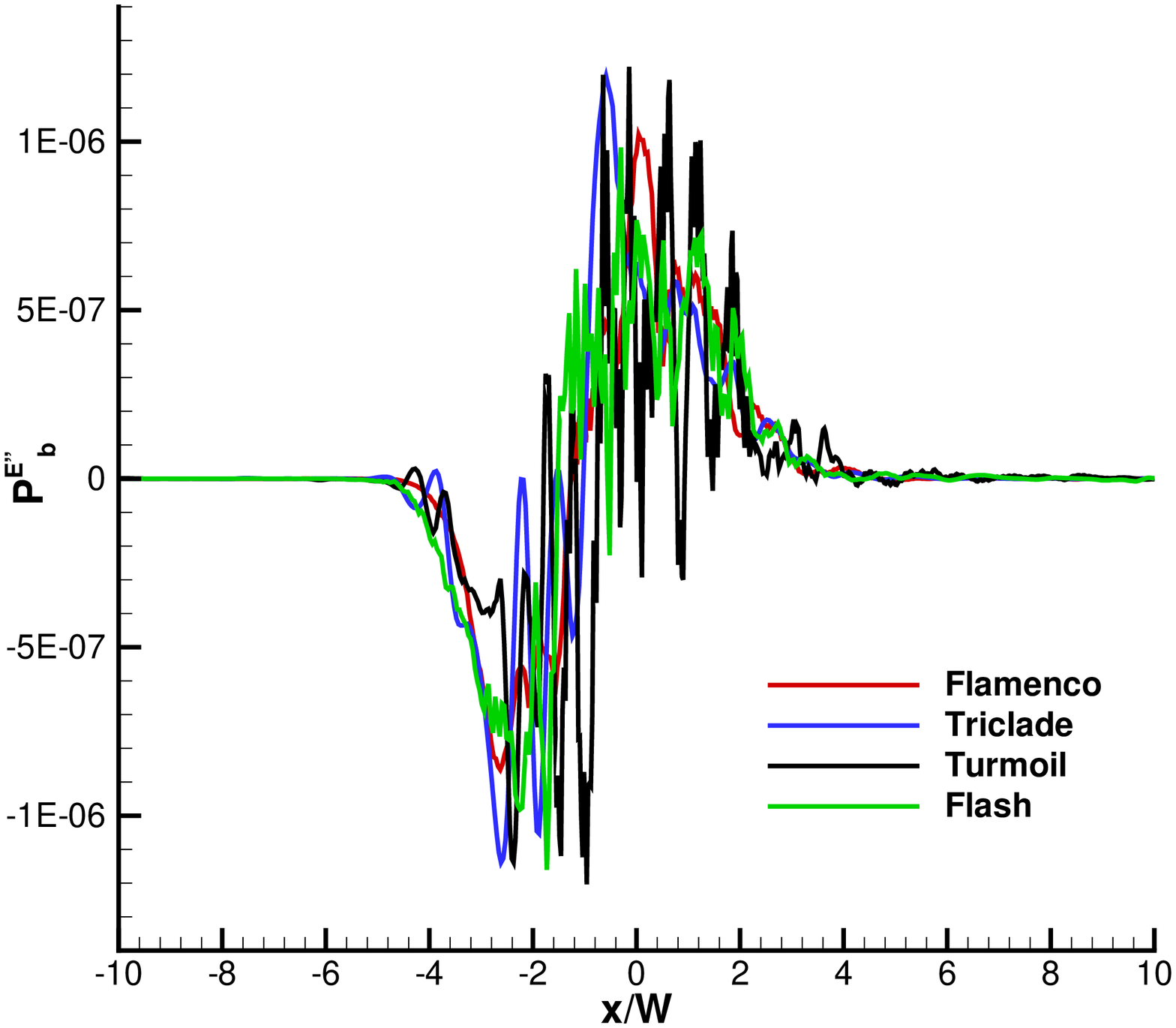}}
\subfigure[\hspace{0.1cm} $P_b^{E''}$ at $t=2$ s]{\includegraphics[width=0.49\textwidth]{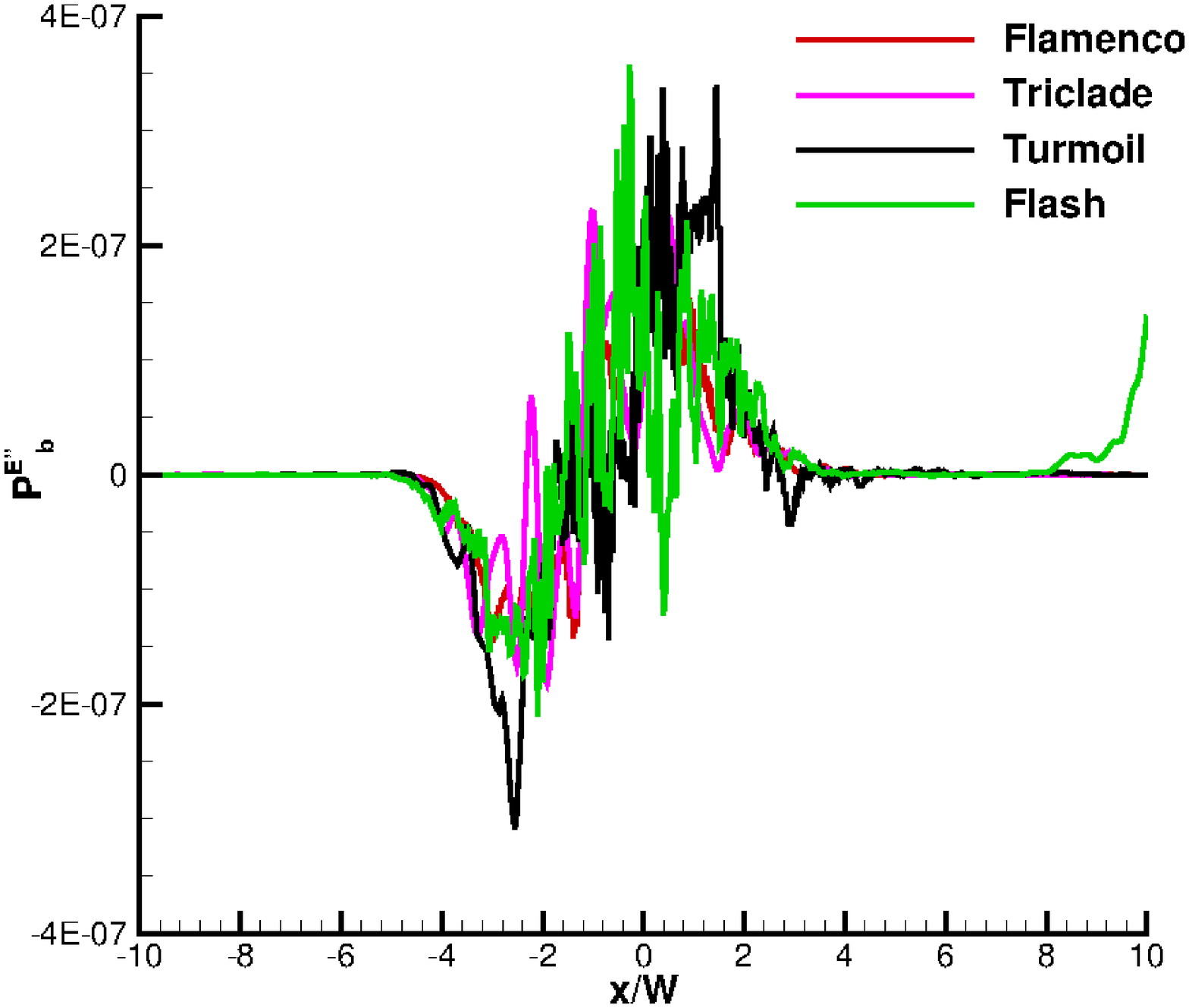}}
\subfigure[\hspace{0.1cm} $T^{E''}$  at $t=1$ s]{\includegraphics[width=0.49\textwidth]{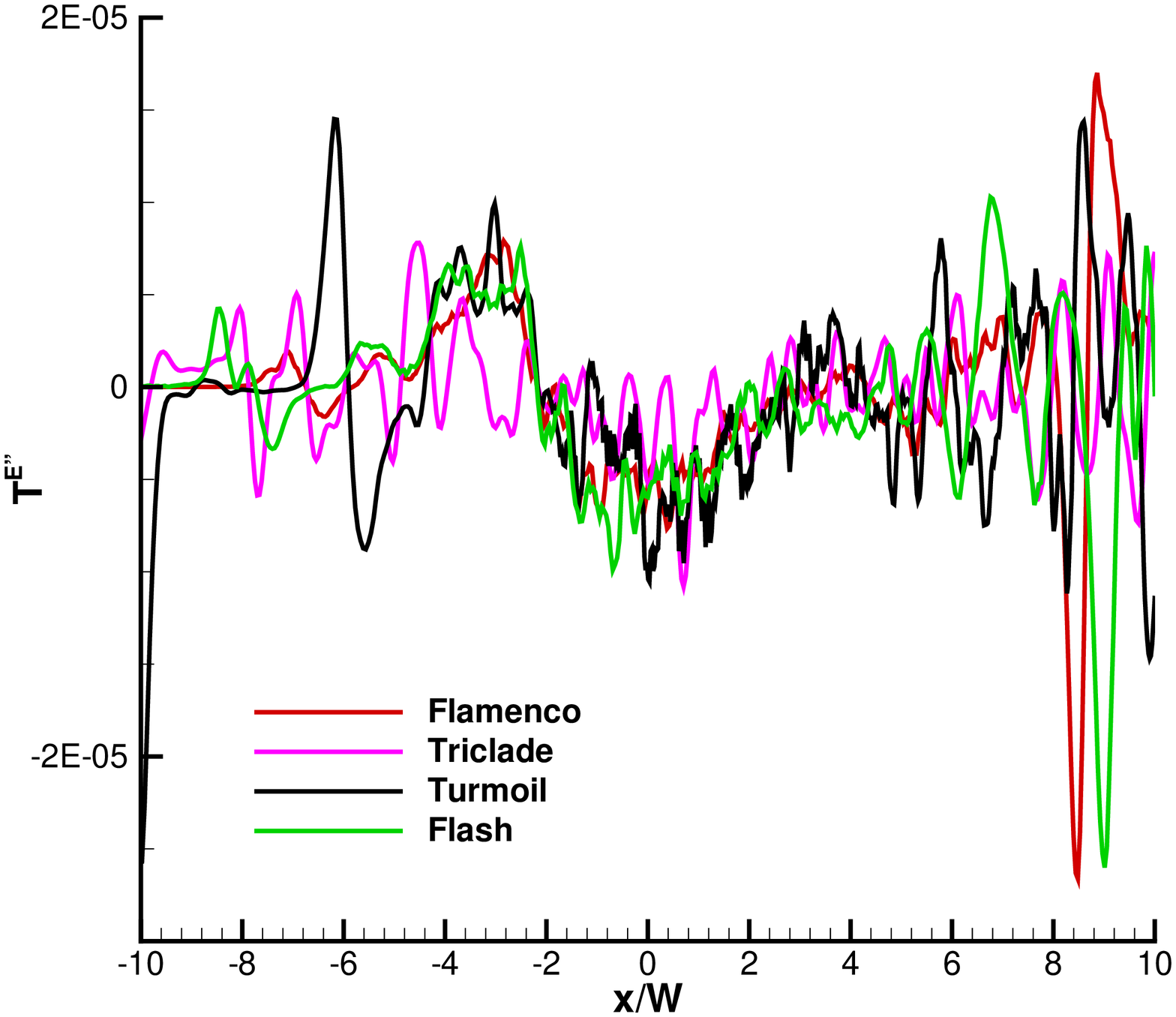}}
\subfigure[\hspace{0.1cm} $T^{E''}$  at $t=2$ s]{\includegraphics[width=0.49\textwidth]{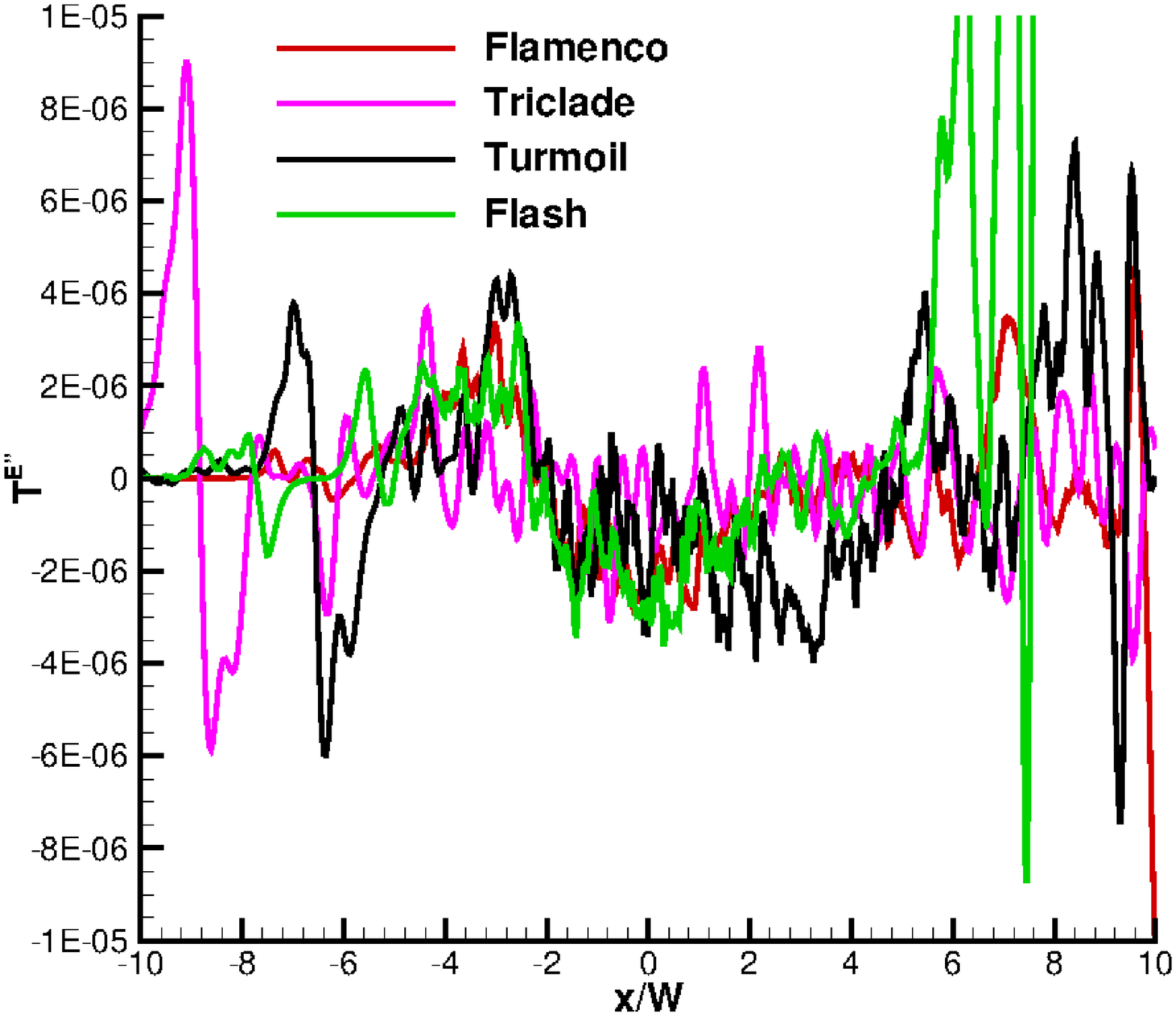}}
\caption{Smoothed $P_b^{E''}$ (top row, (a) and (b) ) and $T^{E''}$ (bottom row (c) and (d))  plotted for all codes at  $1$ s (left (a) and (c)) and $2$ s (right (b) and (d)). All terms non-dimensionalised by $\rho_c u_c^3/\bar \lambda$.   \label{kesmooth}}
\end{centering}
\end{figure*}

There are six terms in the kinetic energy budget, of which the time variation $t^{E''}$, pressure-dilatation correlation $\Pi^{E''}$, and turbulent transport  $T^{E''}$ terms are relatively large, and mean advection $A^{E''}$, buoyancy production $P_b^{E''}$, and shear production $P_s^{E''}$ are relatively small. The large terms are plotted together in Fig. \ref{ke1}, and the small terms in Fig. \ref{ke2}. To aid a clear interpretation of the relative magnitude of each of the terms, the full budget is plotted for the Flamenco results only in Fig. \ref{keflam}. Finally,  $P_b^{E''}$ and $T^{E''}$ have strong fluctuations, thus Fig. \ref{kesmooth} plots those two terms independently at $t=1$ and $2$ s where a five point moving average has been applied to smooth the data. 

First, a note on some numerical issues in computing specific terms. The overall level of fluctuations in the results is substantially larger than in the transport equations for $\tilde m_1$ and $\tilde u$, which is partly expected as $\widetilde{E''}$ is a second order quantity; however, there were also specific difficulties in the computation of terms involving correlations of pressure fluctuations. This led to a large variation in results from code to code for the pressure-dilatation $\Pi^{E''}$ and turbulent transport $T^{E''}$ terms. 

Several different stencils were employed to compute the pressure-dilatation term, ranging from second to sixth order, to explore the convergence of the numerical results. For Flamenco, it was noted that both the magnitude and sign changed for the pressure-dilatation correlation from second order estimation of derivatives to sixth order. It is encouraging that the third and fourth order stencils are converging towards the sixth order result.

Now consider the individual terms in Fig. \ref{ke1}. At the first and second times, turbulent transport $T^{E''}$ forms two symmetric peaks at the bubble and spike fronts, indictating a transport/redistribution of turbulent kinetic energy from the core of the layer to the edges. Turbulent transport is the largest term contributing to the time rate of change of $\widetilde{E''}$. The peak in $T^{E''}$ at the bubble side remains at later time; however, the peak on the spike side advects away from the layer, becoming flatter and indicating a net transport of $\widetilde{E''}$ from the spike to bubble side. 

The pressure-dilatation term is symmetric, peaking at the centre of the layer, and increases  $\widetilde{E''}$. As pointed out in Schilling and Mueschke \cite{Schilling2010b}, although the Mach number is low, mean dilatation is non-negligible due to mixing of the fluids: thus, $\nabla \cdot {\bf u}=-\nabla \cdot(D \nabla \ln\rho)$, with $D$  an effective numerical diffusion the form of which may be more complex than Fickian. 

Both Turmoil and Triclade results exhibited large fluctuations in pressure fluctuation-based terms which precluded the extraction of clear trends from their data. This is due to the treatment of acoustic waves in both of those algorithms, which both dissipate acoustic fluctuations less strongly than Flamenco and Flash. Acoustic waves add a `fast' component to the kinetic energy budget, but are a mostly isentropic process, and thus should not impact the overall time rate of change of $\widetilde{E''}$. Figure \ref{ketilde} shows that this is the case: the peak kinetic energies have reduced by a factor of $1000$ over the simulation time, yet all codes remain in good agreement; thus, net dissipation must be of a similar order of magnitude for all algorithms and very likely to be scaling relative to a physical (rather than numerical) dissipation rate. This division of the pressure-dilatation into compressible and incompressible components is in agreement with previous observations for homogeneous decaying turbulence and shear layers \cite{Sarkar1992,ristorcelli_1997,Chassaing2001}, where it was noted that although the fast component has a large magnitude, its global impact is small compared with the `incompressible'  correlation.

\begin{figure*}
\begin{centering}
\subfigure[\hspace{0.1cm} Flamenco $\sqrt{\langle p'^2\rangle} (t)$ for $t=0.01$ (top line), $0.1$, $0.2$ ... $0.9$ and $1.0$ s (bottom line).]{\includegraphics[width=0.49\textwidth]{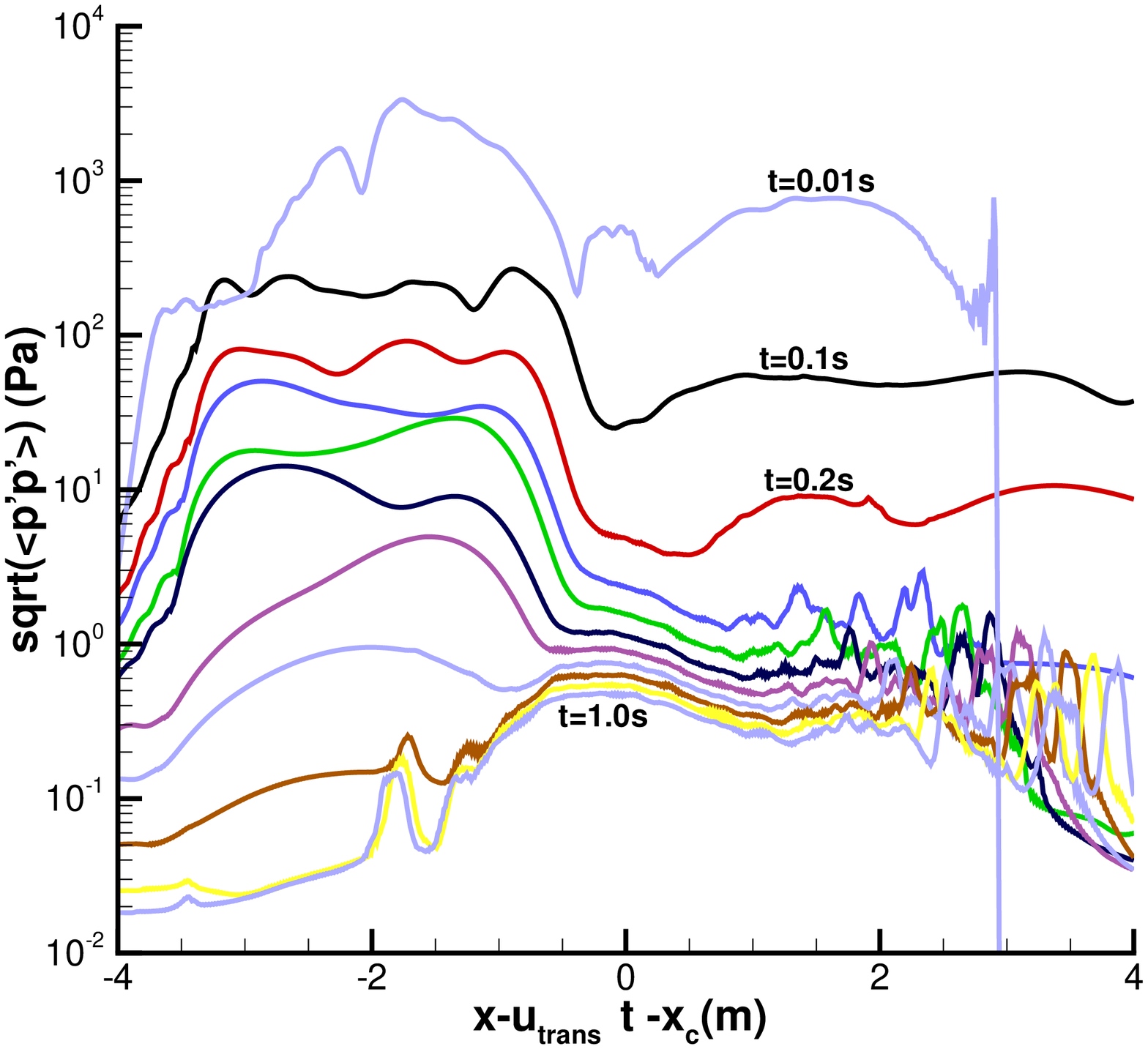}}
\subfigure[\hspace{0.1cm} $\sqrt{\langle p'^2\rangle}$ averaged in the heavy fluid. \label{pdpdhf}]{\includegraphics[width=0.49\textwidth]{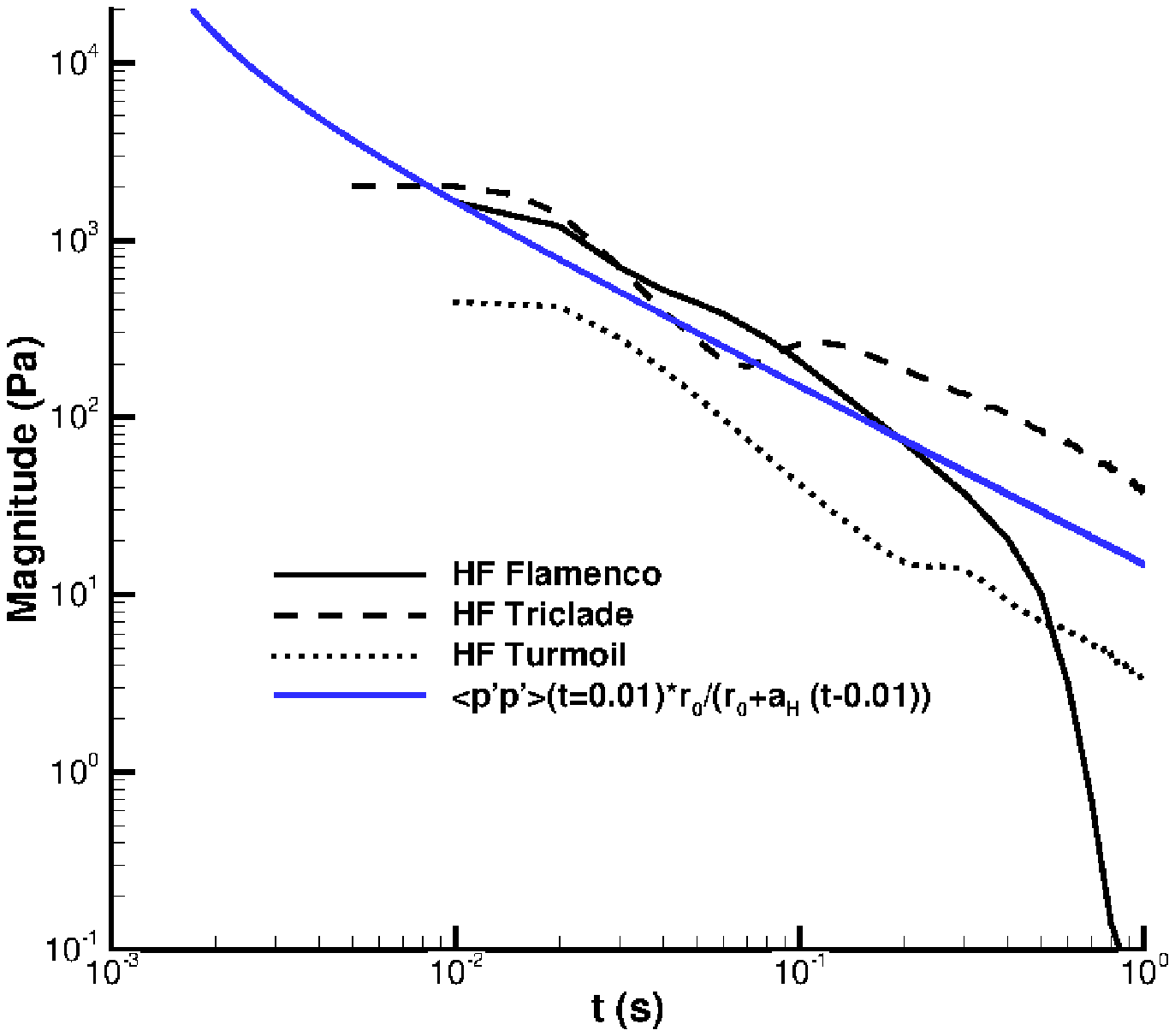}}
\subfigure[\hspace{0.1cm} $\sqrt{\langle p'^2\rangle}$ sampled at the layer centre. \label{pdpdc}]{\includegraphics[width=0.49\textwidth]{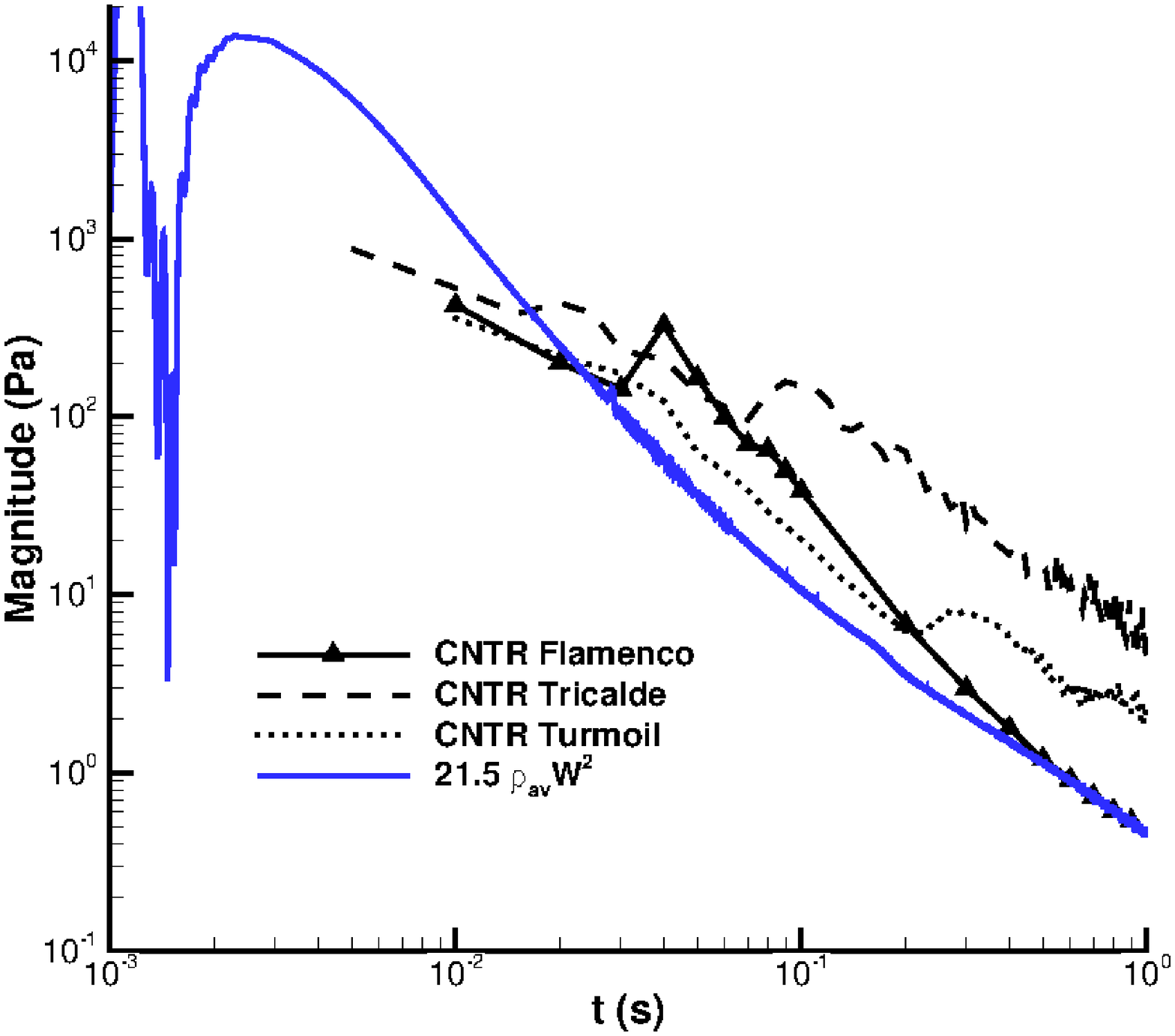}}
\subfigure[\hspace{0.1cm} $\sqrt{\langle p'^2\rangle}$ averaged in the light fluid. \label{pdpdlf}]{\includegraphics[width=0.49\textwidth]{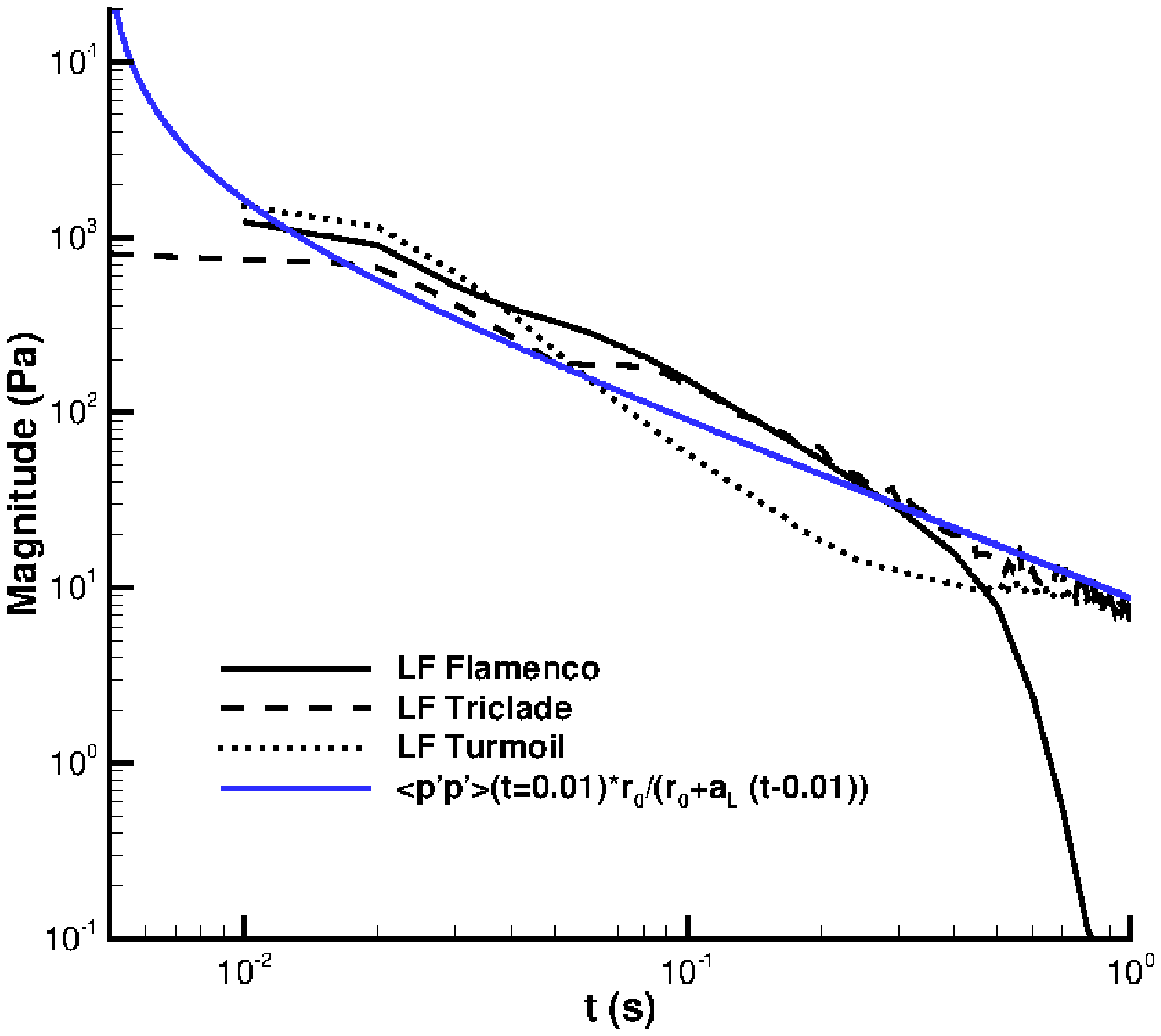}}
\caption{The variation of (a) $\sqrt{\langle p'^2\rangle} (t)$ as a function of time and space in Flamenco, and as a function of time in (b) the heavy fluid,  (c) the light fluid, and (d) the layer centre for Flamenco, Triclade, and Turmoil.   \label{pfluc}}
\end{centering}
\end{figure*}

Figure \ref{pfluc} plots the spatial distribution of the rms pressure variance $\sqrt{\langle p'^2\rangle}$ at several times using Flamenco, and the temporal variation at the centre of the mixing layer
 [defined by $x$ such that $\tilde m_1(x,t)=0.5$]
and the integral average in the heavy bulk fluid 
[$x$ such that $\tilde m_1(x,t)>0.99$] 
and light fluid
[$x$ such that $\tilde m_1(x,t)<0.01$]. At late time, the pressure fluctuations within the mixing layer in Flamenco clearly follow an incompressible scaling $\approx 21.5 \rho_{av} \dot W^2$ with $\rho_{av}=(\rho_H^++\rho_L^+)/2$. This is a reasonable result noting that the integral width is substantially smaller than the actual width of the layer which leads to the large constant of proportionality. However, for Triclade and Turmoil the pressure fluctuations are substantially larger.

Focusing on the fluctuations in the pure heavy fluid in Fig. \ref{pdpdhf} and pure light fluid in Fig. \ref{pdpdlf}, the fluctuations are  modelled as a simple spherical source deposited at shock-interaction, where given conservation of acoustic energy the pressure fluctuations should scale with $\sqrt{\langle p'^2\rangle} (t)\approx \sqrt{\langle p'^2\rangle} (t_0) r_0(t_0)/[r_0(t_0)+a_{H,L} t)]$. Here $a_H$ and $a_L$ are the speed of sound in the shocked heavy/light fluid respectively, $t_0$ is a sampling time measured from shock interaction and the initial acoustic radius is modelled as $r_0(t_0)=\bar \lambda/2+a_H t_0$. 

The overall agreement is very good from $t=0.01$ to $0.1$ s between Flamenco, Triclade, and Turmoil, despite the complexities of the acoustic waves and the influence of the exiting rarefaction at early times ($t<0.1$ s). At later times, numerical dissipation in Flamenco reduces the magnitude of the acoustic waves in the heavy and light fluids, to below the pressure fluctuations produced by turbulent motion in the layer. This is not seen in Turmoil and Triclade, which preserve acoustic fluctuations better. One dimensional tests indicate that Turmoil damps acoustic waves less strongly than Flamenco, which is consistent with this result, especially noting that by $t=1$ s the wave has travelled through the cross-section more than $400$ times and so numerical dissipation has had ample opportunity to cause the observed differences. This explains the discrepancy between the three codes in the layer centre, where pressure fluctuations in the heavy fluid penetrate the layer and are at a larger magnitude than pressure fluctuations generated through vortical motion alone.

At even later times acoustic pressure fluctuations will dominate the vortical fluctuations by orders of magnitude. This also explains the difficulty in obtaining agreement amongst the codes for transport terms including pressure gradients, and correlations including pressure fluctuations and gradients. Despite this, the impact on the overall transport balance is small.

The time derivative $t^{E''}$ is plotted in Fig. \ref{ke1} and shows an increase in turbulent kinetic energy at both the bubble and spikes at the earliest times. However, at $t=1$ and $t=2$ s the time rate of change is negative throughout the core of the mixing layer, with positive variations only at the isolated vortex `projectiles'. Note that the overall balance is impacted by dissipation provided by the implicit numerical model, which is discussed at the end of this section.

Examining the smaller terms plotted in Fig. \ref{ke2}, mean advection $A^{E''}$ shows a transport of kinetic energy from the bubble to spike side of the layer. Both buoyancy and shear production are negative on the bubble side and positive on the spike side, and all three terms are only non-zero within the core of the mixing layer. Some oscillations in  $P_s^{E''}$ may be seen at $x/W>8$ at $t=2$ s for Flash, due to the unphysical interaction of the spikes with the boundary condition.

The relative magnitude of all six terms can be seen clearly in Fig. \ref{keflam} for the Flamenco results only. It is clear that the time evolution of $\widetilde{E''}$ is dominated by two physical effects, the turbulent transport 
and the pressure-dilatation correlation, and by numerical dissipation with a mean magnitude of the same order as the resolved physical terms.

\begin{figure}
\begin{centering}
\includegraphics[width=0.49\textwidth]{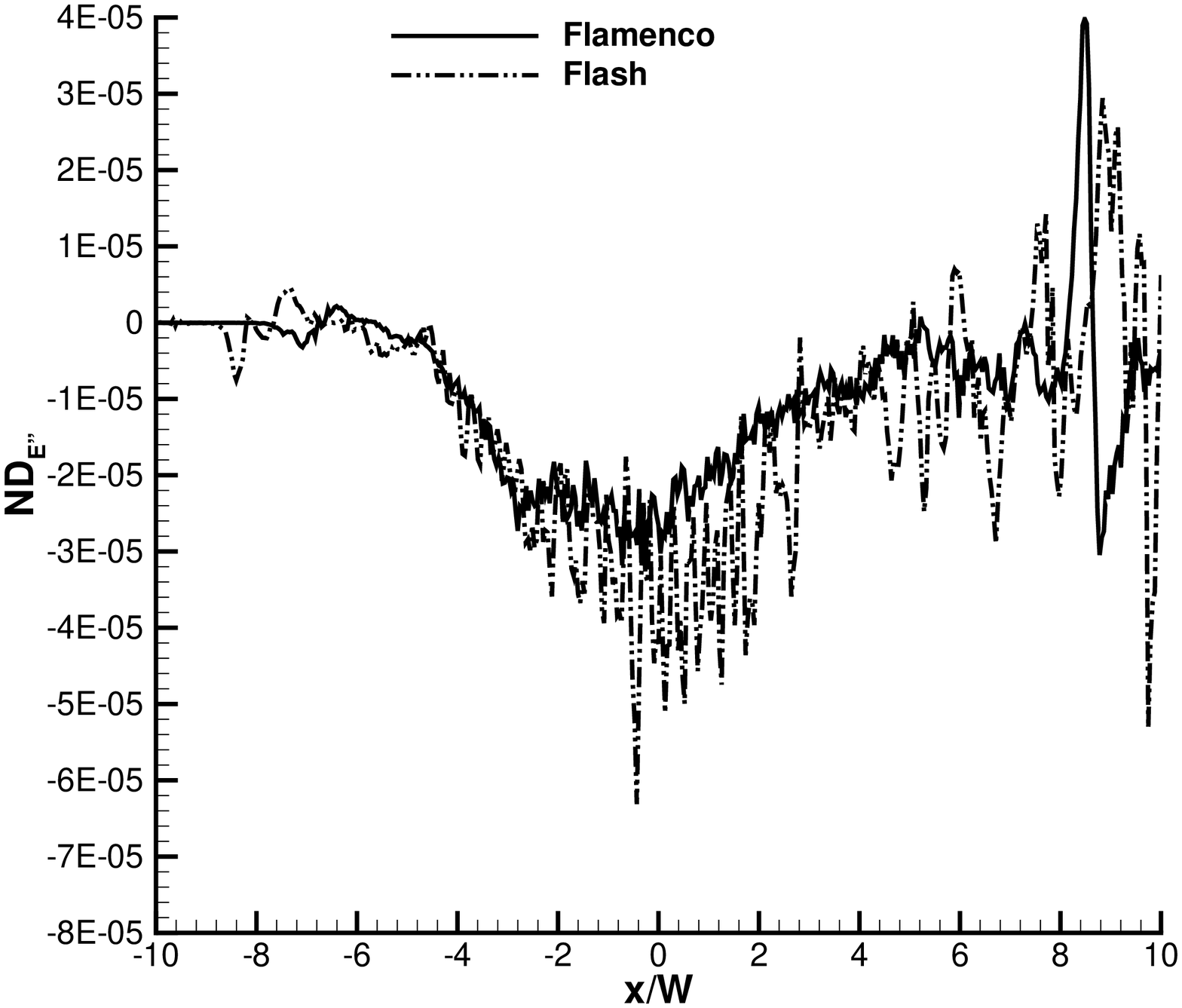}
\caption{Non-dimensional numerical dissipation estimated by the residual in the mean kinetic energy transport balance at $t=1$ s.  \label{kediss}}
\end{centering}
\end{figure}

The numerical dissipation $ND_{E''}$ has been computed for $t=1$ s and is plotted in Fig. \ref{kediss}. This figure compares the numerical dissipation computed for Flamenco and Flash, where the Triclade results are not shown as they are dominated by the fluctuations in the pressure-dilatation term due to the acoustic field. As mentioned in the previous paragraph, the numerical dissipation of turbulent kinetic energy is of a similar order to the dominant terms in the transport balance. This large order of magnitude is to be expected. Turbulent kinetic energy is conserved until it reaches the viscous scales, where it is dissipated. In an Implicit LES, turbulent kinetic energy is dissipated by numerical dissipation as it reaches the grid scale. The mechanism of turbulent dissipation is highly code-dependent, with some algorithms having dissipation in the remap phase for low Mach number flows  (e.g. Turmoil), and others using upwind schemes to dissipate on the individual wave strengths (e.g. Flamenco, Flash, Triclade). 

However, Implicit LES relies on the concept that the actual dissipation rate is driven by the largest scales of turbulence, and independent of the detailed mechanism of dissipation as long as the dissipation does not impact the large scales. Thus, when a sufficiently wide range of large scales are resolved by a numerical algorithm, the flow would evolve in a statistically identical  manner regardless of whether the smallest scales of the flow are resolved or not. The word `statistical' is used since turbulent flows are very sensitive to small perturbations, hence an exact match is not expected. 

That the four codes used in the current study have such good agreement in the turbulent kinetic energy profiles at a very late time, where peak values have reduced by a factor of 1000, indicates that the decay rates are very similar from algorithm to algorithm. Thus, although only results from two codes are plotted in Fig. \ref{kediss}, there is reasonable confidence that a similar magnitude would be observed for the other two codes. Finally, the grid convergence of $ND_{E''}$ for Flamenco as shown in Fig. \ref{ndconv} also indicates that the simulations have achieved the necessary scale separation such that the total dissipation rate is determined by processes at resolved scales, rather than by numerical dissipation.

\section{Conclusions \label{concl}}

An analysis of turbulent transport in an inhomogeneous compressible turbulent mixing layer induced by Richtmyer-Meshkov instability has been presented. The analysis used data from four independent ILES codes which were applied to the quarter-scale problem presented in the $\theta$-group collaboration \cite{Thornber2017}. All terms in the high-Reynolds number (inviscid) limit of the Reynolds-averaged transport equations for momentum, species mass fraction, turbulent kinetic energy, and mass fraction variance were computed in the inhomogeneous direction. Numerical dissipation provided an implicit subgrid model in all algorithms, which was computed from the balance of the transport equations. The objective of this study was to provide insight into the dominant physical mechanisms influencing turbulent transport, and to facilitate the future development and validation of reduced order modelling of those quantities. The inclusion of four numerical methods also provided an understanding of the confidence in predicting each of the terms, and to compare and contrast the impact of numerical dissipation. 

For the momentum transport equation, there is high confidence in the mean momentum in the turbulent stage, with codes agreeing to within $\approx \pm 4$\% in the shock-direction. The time rate of change of momentum is dominated by the balance of the pressure gradient and Reynolds stresses, and numerical dissipation is very low. The pressure gradient term was particularly noisy due to acoustic modes. Advecting spikes which escape the mixing layer cause large variations in the individual transport terms, the position and magnitude of which were not easily captured numerically. The results here may also be impacted by the small sample size of such spikes which escape the developing mixing layer at an early time. The individual transport terms have a spread of $\approx \pm 10$\% due to statistical errors and acoustic fluctuations, but the agreement between codes is good and could be employed to validate URANS and LES modelling approaches. 

For mass fraction transport, the mean profiles are nearly identical at all times; thus, there is a high level of confidence. The time variation of mass fraction distribution largely follows the turbulent transport term, moderated by mean advection, and numerical dissipation is small. Here there is very good agreement between all algorithms,  and thus a high level of confidence in the interpretation of the dominant terms. The mass fraction variance shows a higher variation at early times, but again agreement is very good when the layer transitions to turbulence ($\approx \pm 5$\%). All terms in the transport equation are important, but mean advection is lower by a factor of approximately two. Numerical dissipation is purely negative as expected, and is of a similar magnitude to the largest resolved terms. All terms considered are in sufficient agreement to be useful for validation of lower order modelling results.

The analysis of the turbulent kinetic energy transport equation was more challenging. The largest error in turbulent kinetic energy occurred at the earliest, non-turbulent times ($\approx \pm 10$\%), but with excellent agreement at later times ($\approx \pm 3$\%). Temporal evolution is dominated by turbulent transport and numerical dissipation (implicit subgrid dissipation), with pressure-dilatation the third largest. The computed pressure-dilatation term was strongly impacted by acoustic modes, rendering the confidence in the results for that term relatively low compared to the rest of the data, given the numerically challenging problem of evolving sound waves for relatively long periods. Despite this uncertainty, the impact on the overall development of kinetic energy in the layer is concluded to be low, as evidenced by the excellent agreement in late time kinetic energy for algorithms which have substantially different mean pressure fluctuations at late time.

\section{Acknowledgements}

The authors greatly acknowledge Dr. Karnig Mikaelian who suggested the original `$\theta$-group' collaboration at the Fourteenth International Workshop on the Physics of Compressible Turbulent Mixing in 2014. This research was supported under Australian Research Council's Discovery Projects funding scheme (project number DP150101059). The lead author would like to acknowledge the computational resources at the National Computational Infrastructure through the National Computational Merit Allocation Scheme which were employed for all Flamenco cases presented here. Flash was developed by the DOE-sponsored ASC/Alliance Center for Astrophysical Thermonuclear Flashes at the University of Chicago. This work was performed under the auspices of the U.S. Department of Energy by Lawrence Livermore National Laboratory under Contract DE-AC52-07NA27344.

\appendix
\section{Grid Convergence \label{convergence}}

\begin{figure*}
\begin{centering}
\subfigure[\hspace{0.1cm} Numerical dissipation in the $\tilde{u}$  transport equation]{\includegraphics[width=0.49\textwidth]{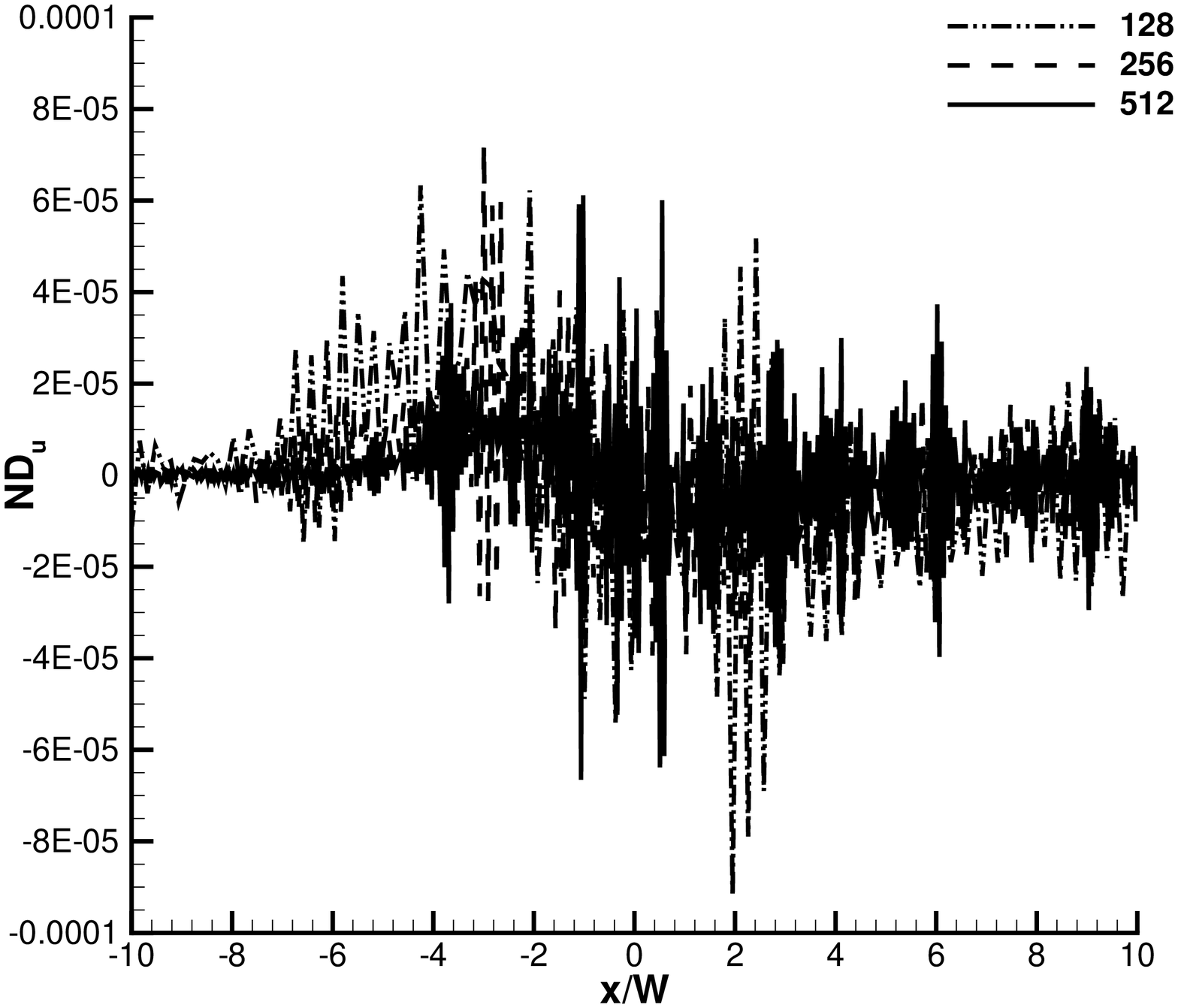}}
\subfigure[\hspace{0.1cm} Numerical dissipation in the $\widetilde{m_1}$  transport equation]{\includegraphics[width=0.49\textwidth]{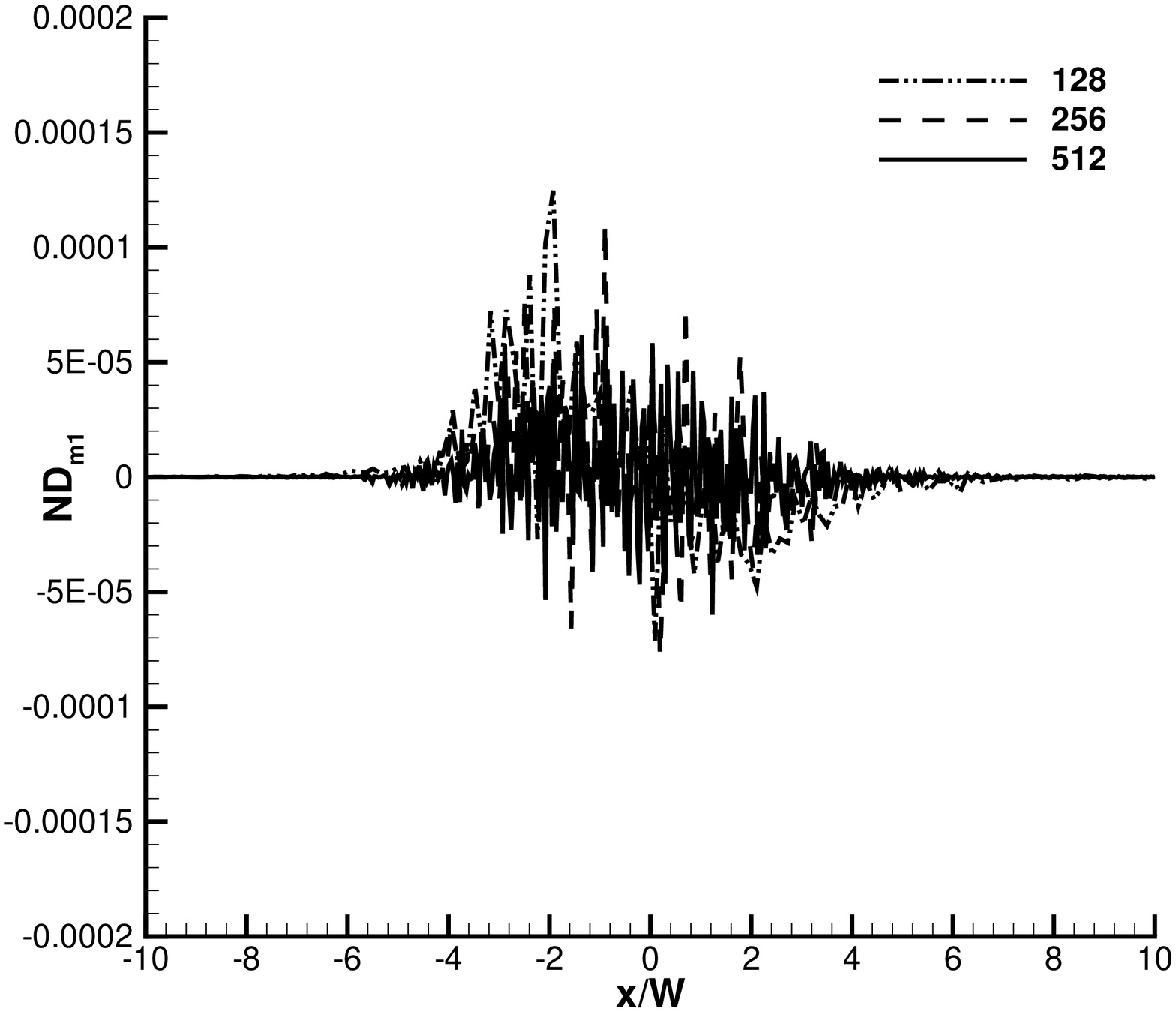}}
\subfigure[\hspace{0.1cm} Numerical dissipation in the $\widetilde{E''}$   transport equation]{\includegraphics[width=0.49\textwidth]{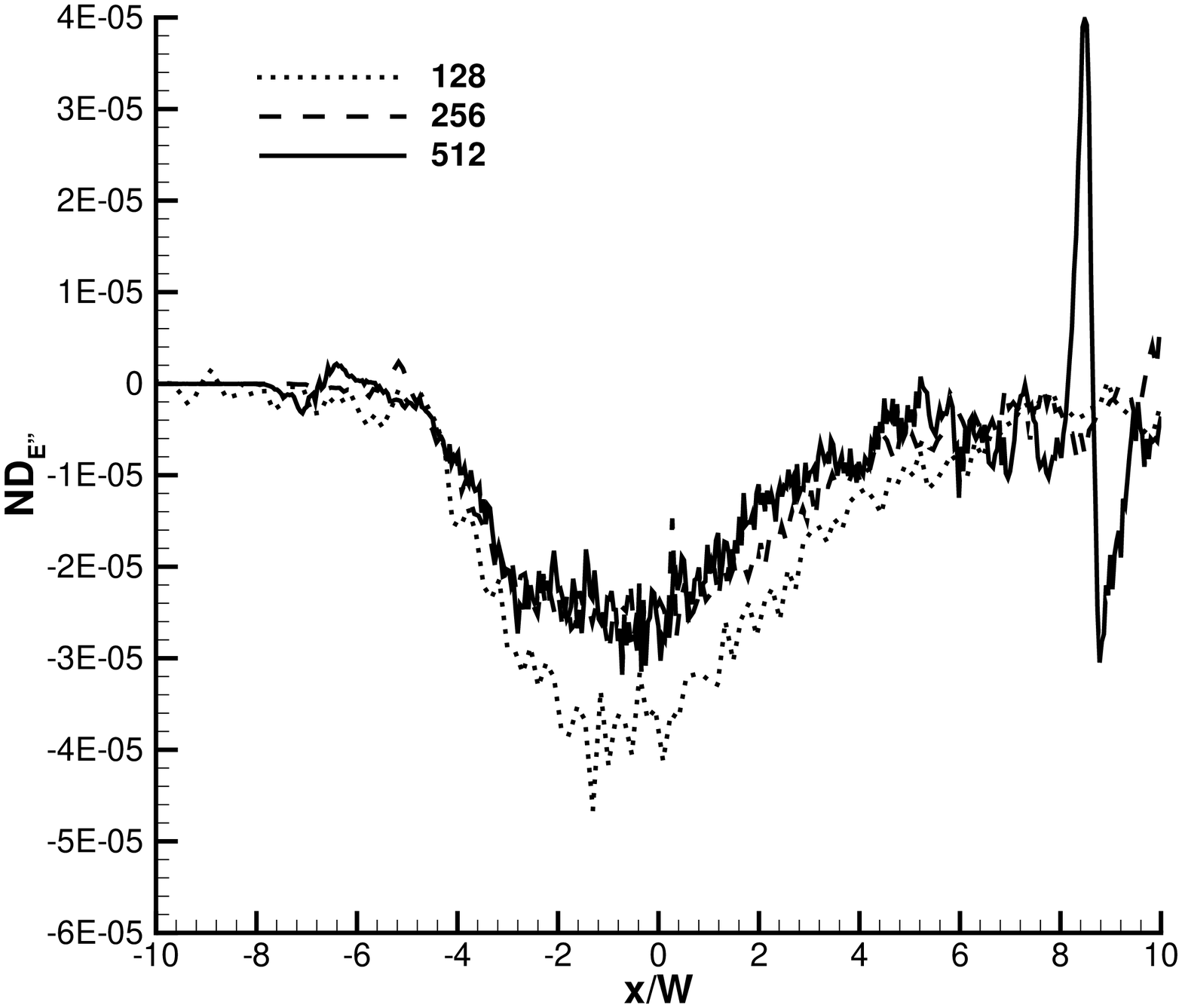}}
\subfigure[\hspace{0.1cm} Numerical dissipation in the $\widetilde{m''_1}$  transport equation]{\includegraphics[width=0.49\textwidth]{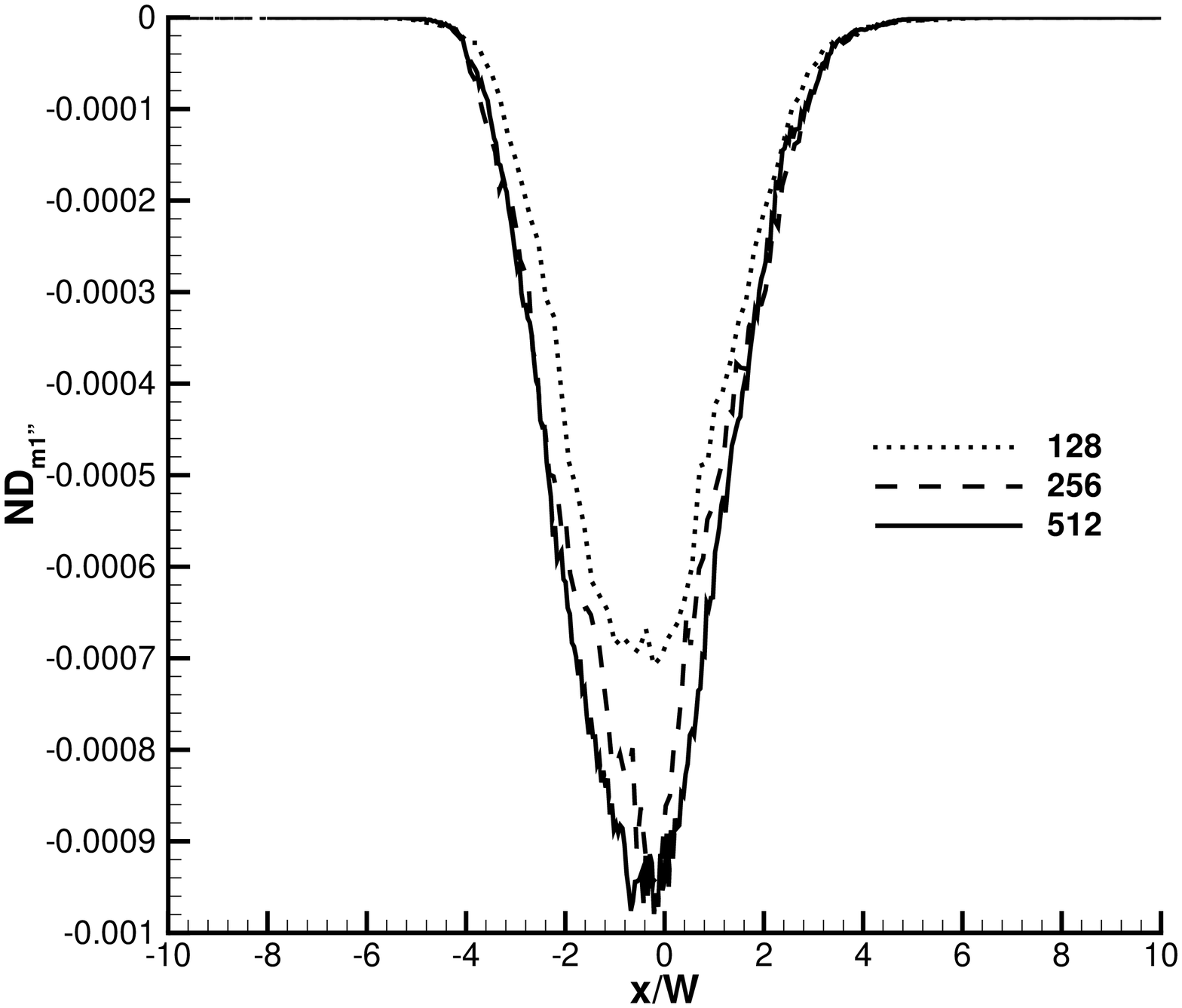}}
\caption{Convergence of non-dimensional numerical dissipation for Flamenco estimated by the residual in the turbulent transport balances, (a) $\tilde{u}$, (b) $\widetilde{m_1}$, (c) $\widetilde{E''}$ and (d) $\widetilde{m''_1}$.  \label{ndconv}}
\end{centering}
\end{figure*}

\begin{figure*}
\begin{centering}
\subfigure[\hspace{0.1cm} $\tilde{u}-u_{trans}$, $\tilde{v}$, $\tilde{w}$]{\includegraphics[width=0.49\textwidth]{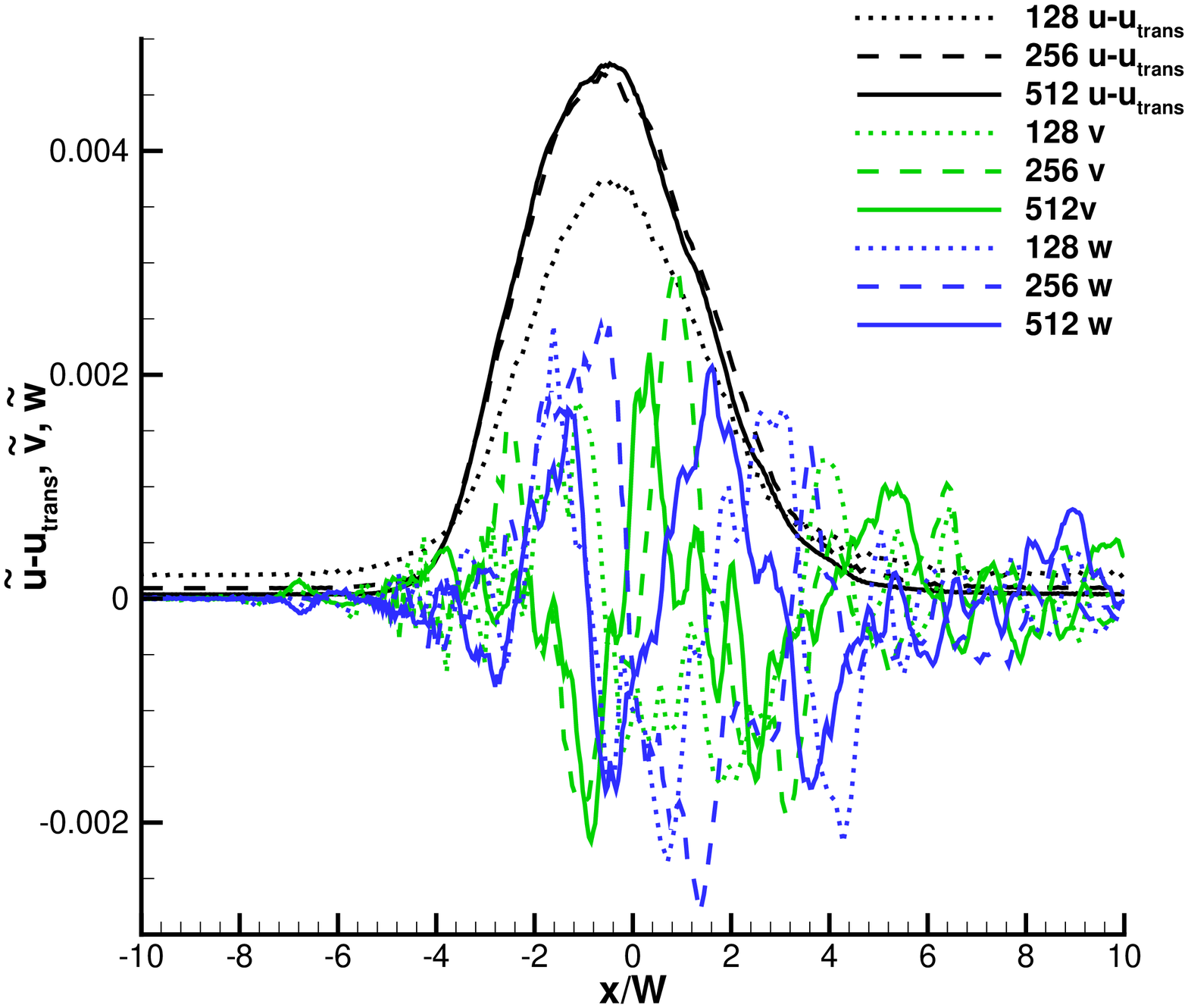}}
\subfigure[\hspace{0.1cm} $\widetilde{m_1}$ ]{\includegraphics[width=0.49\textwidth]{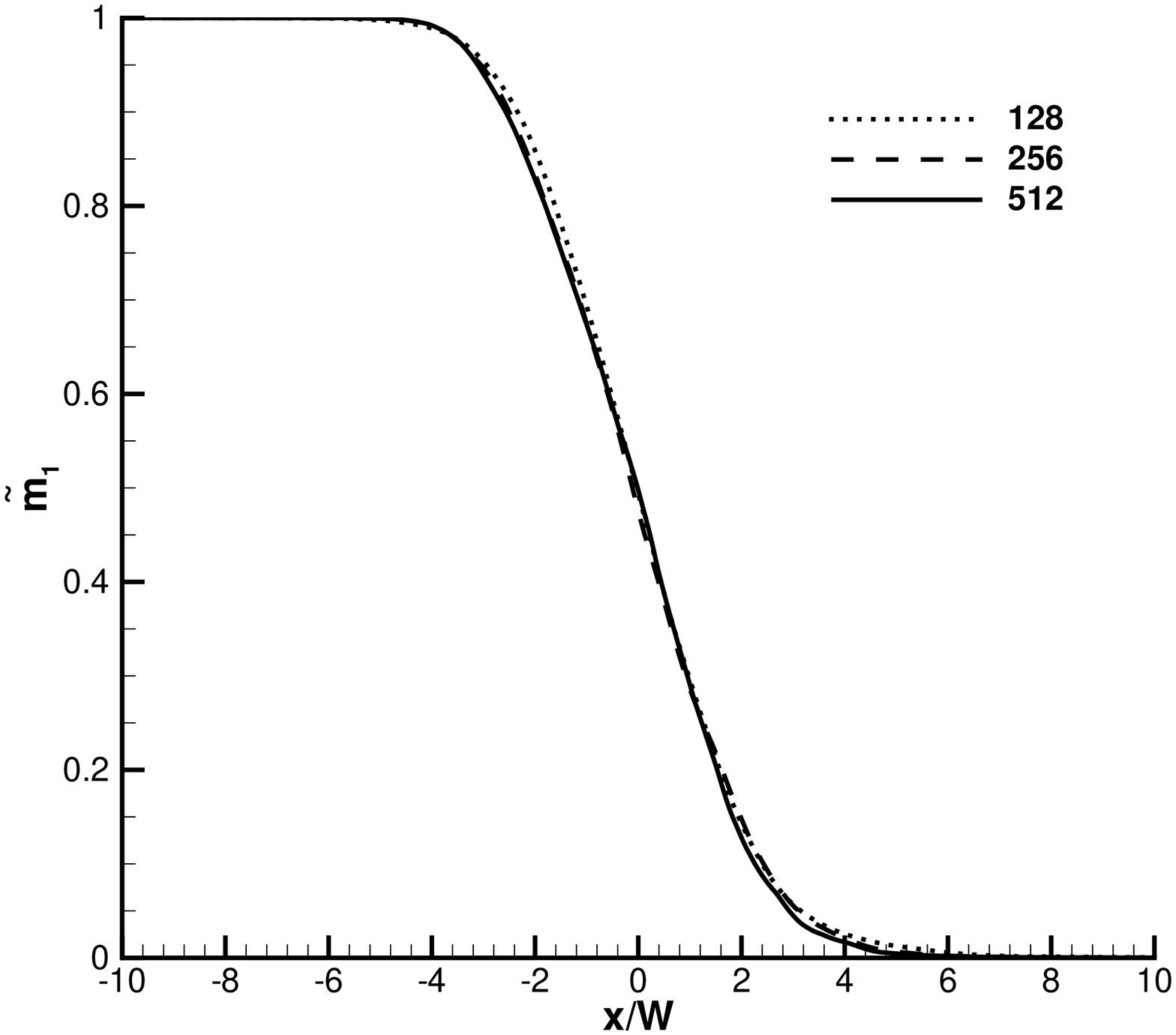}}
\subfigure[\hspace{0.1cm} $\widetilde{E''}$ \label{econv} ]{\includegraphics[width=0.49\textwidth]{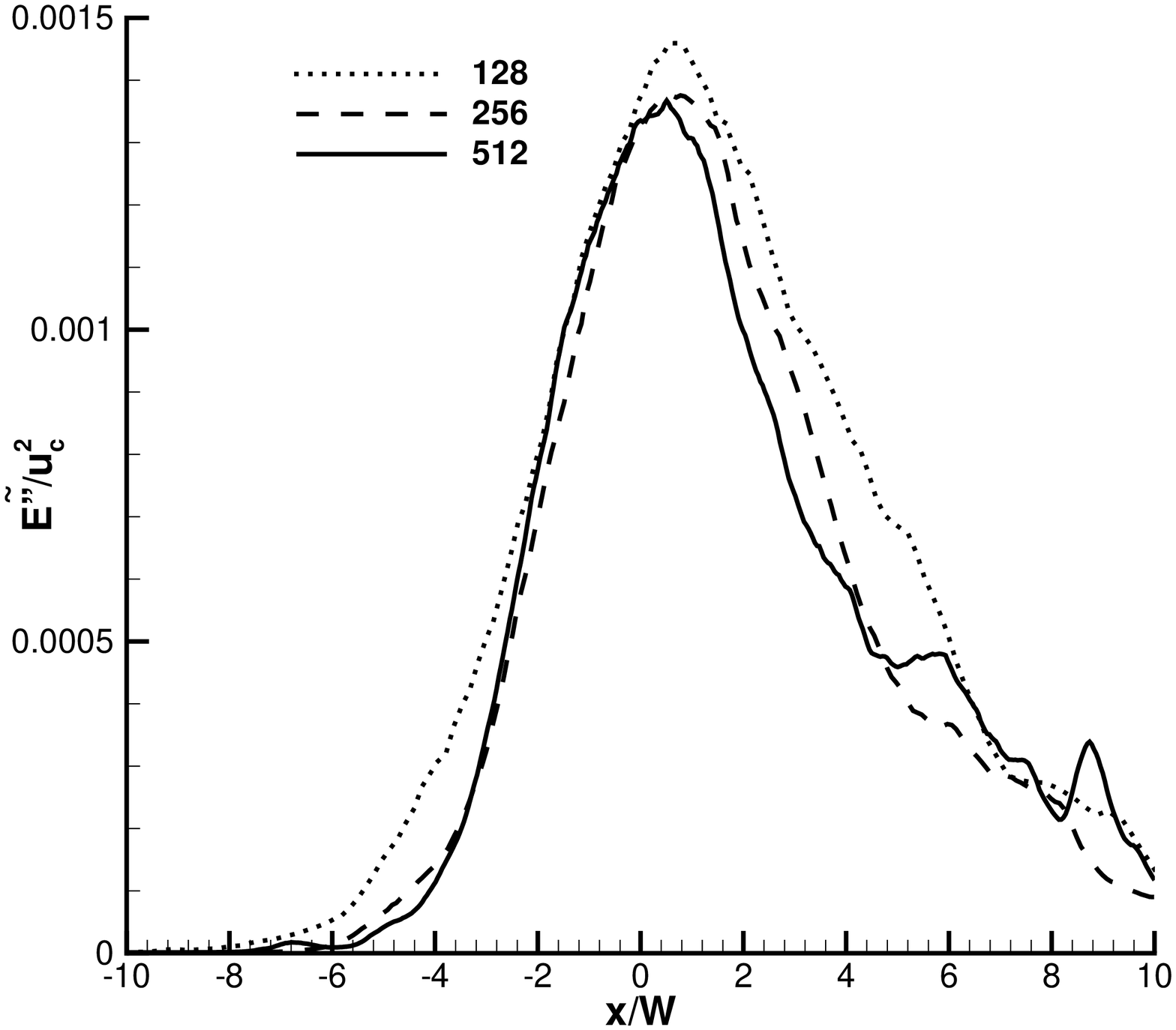}}
\subfigure[\hspace{0.1cm} $\widetilde{m''_1}$ ]{\includegraphics[width=0.49\textwidth]{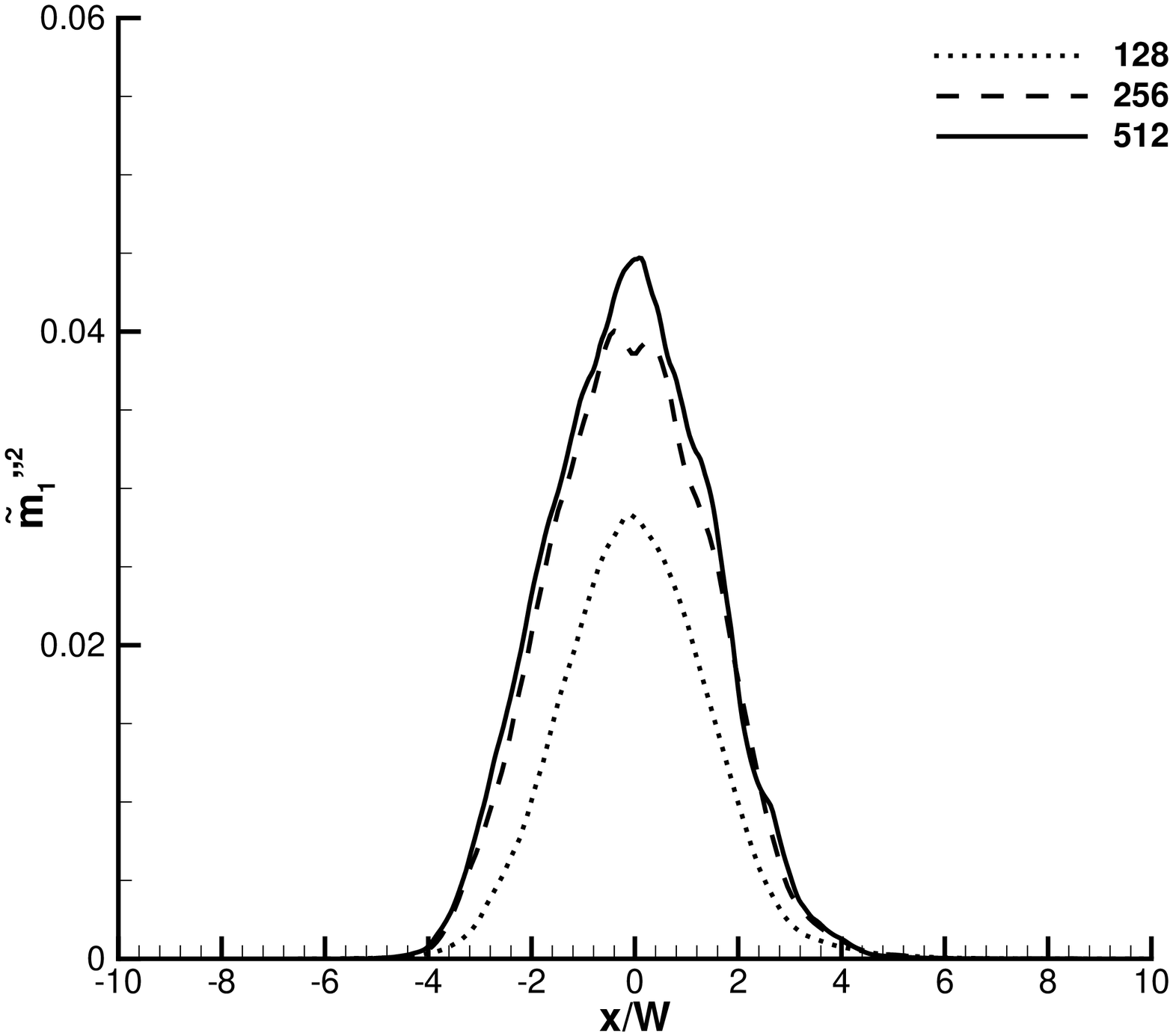}}
\caption{Convergence of mean properties at $1$ s for Flamenco, (a) $\tilde{u}-u_{trans}$, $\tilde{v}$, $\tilde{w}$, (b) $\widetilde{m_1}$, (c)$\widetilde{E''}$, and (d) $\widetilde{m''_1}$.\label{meanconv}}
\end{centering}
\end{figure*}

\begin{figure*}
\begin{centering}
\subfigure[\hspace{0.1cm} Terms in the $\tilde{u}$   transport equation]{\includegraphics[width=0.45\textwidth]{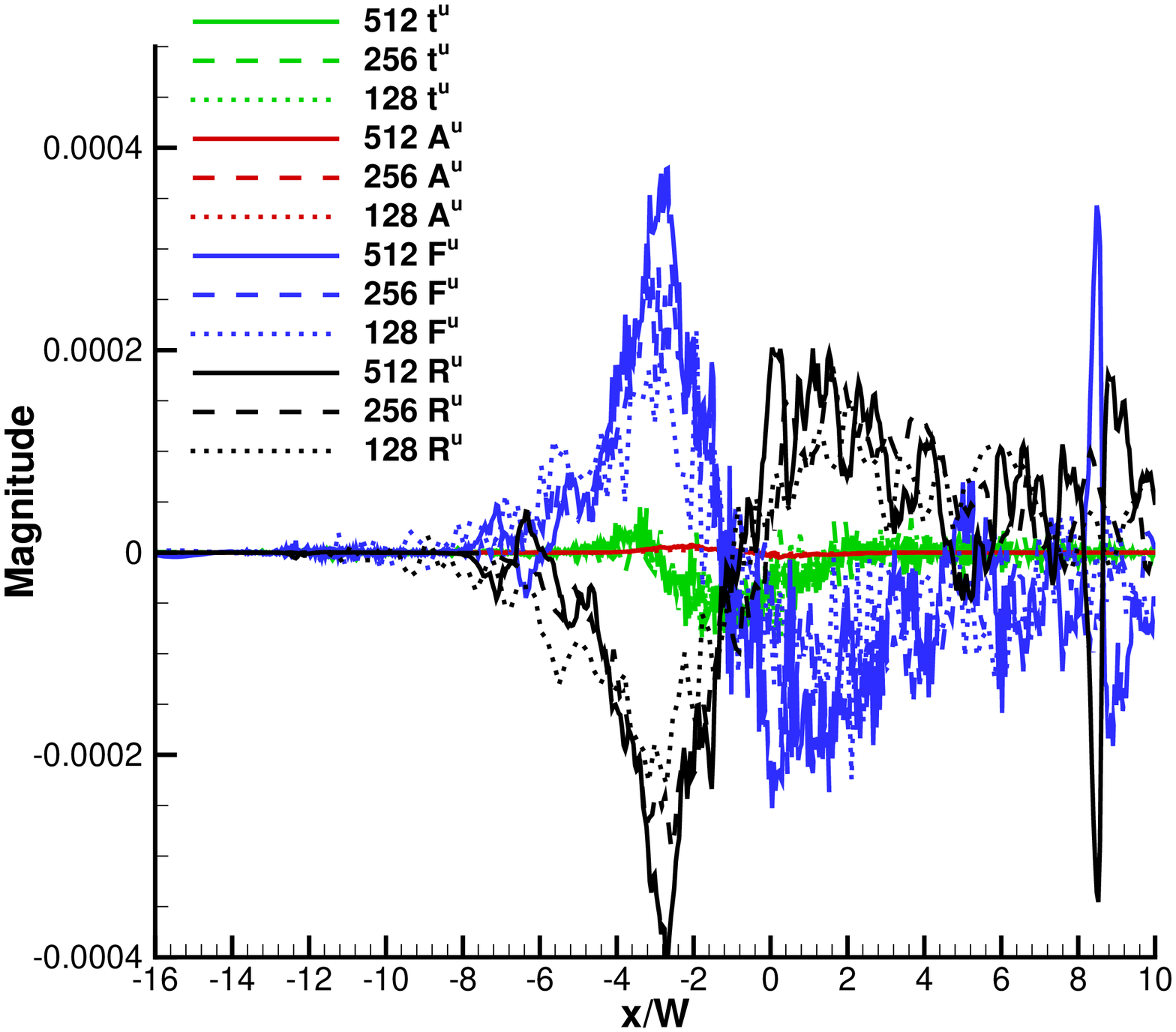}}
\subfigure[\hspace{0.1cm} Terms in the $\widetilde{m_1}$   transport equation]{\includegraphics[width=0.45\textwidth]{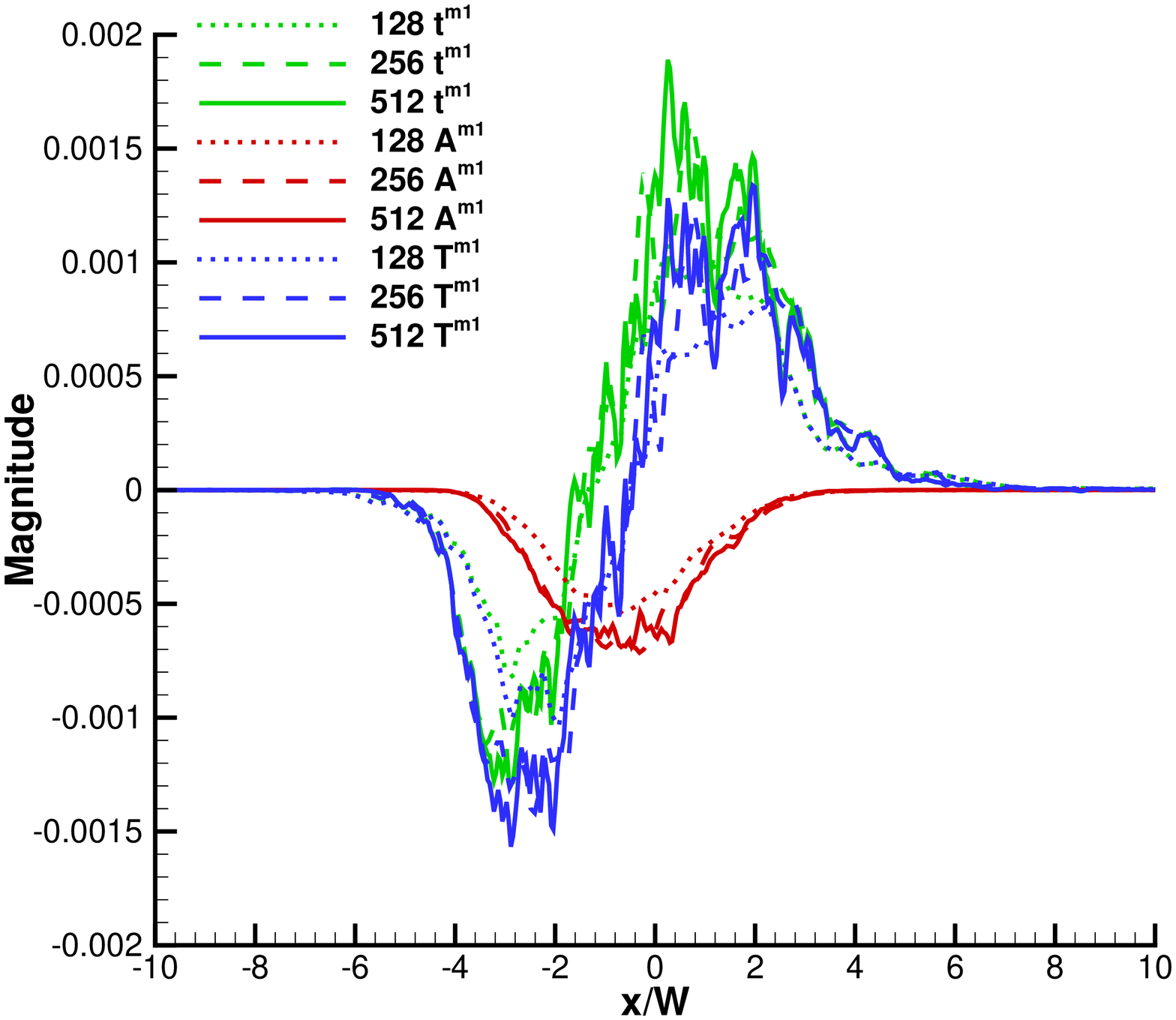}}
\subfigure[\hspace{0.1cm} Terms in the $\widetilde{E''} \label{etransconv}$   transport equation]{\includegraphics[width=0.45\textwidth]{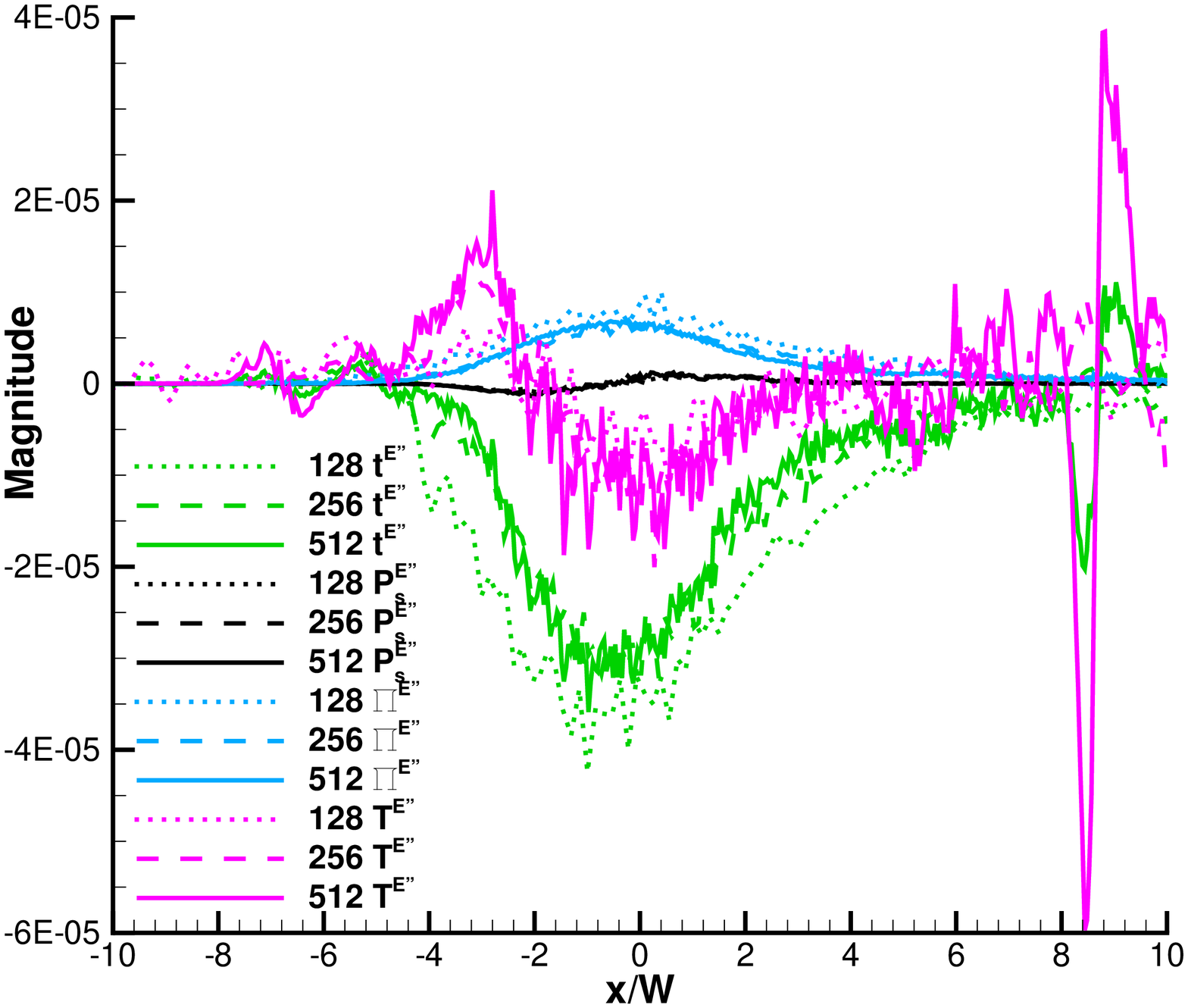}
\includegraphics[width=0.45\textwidth]{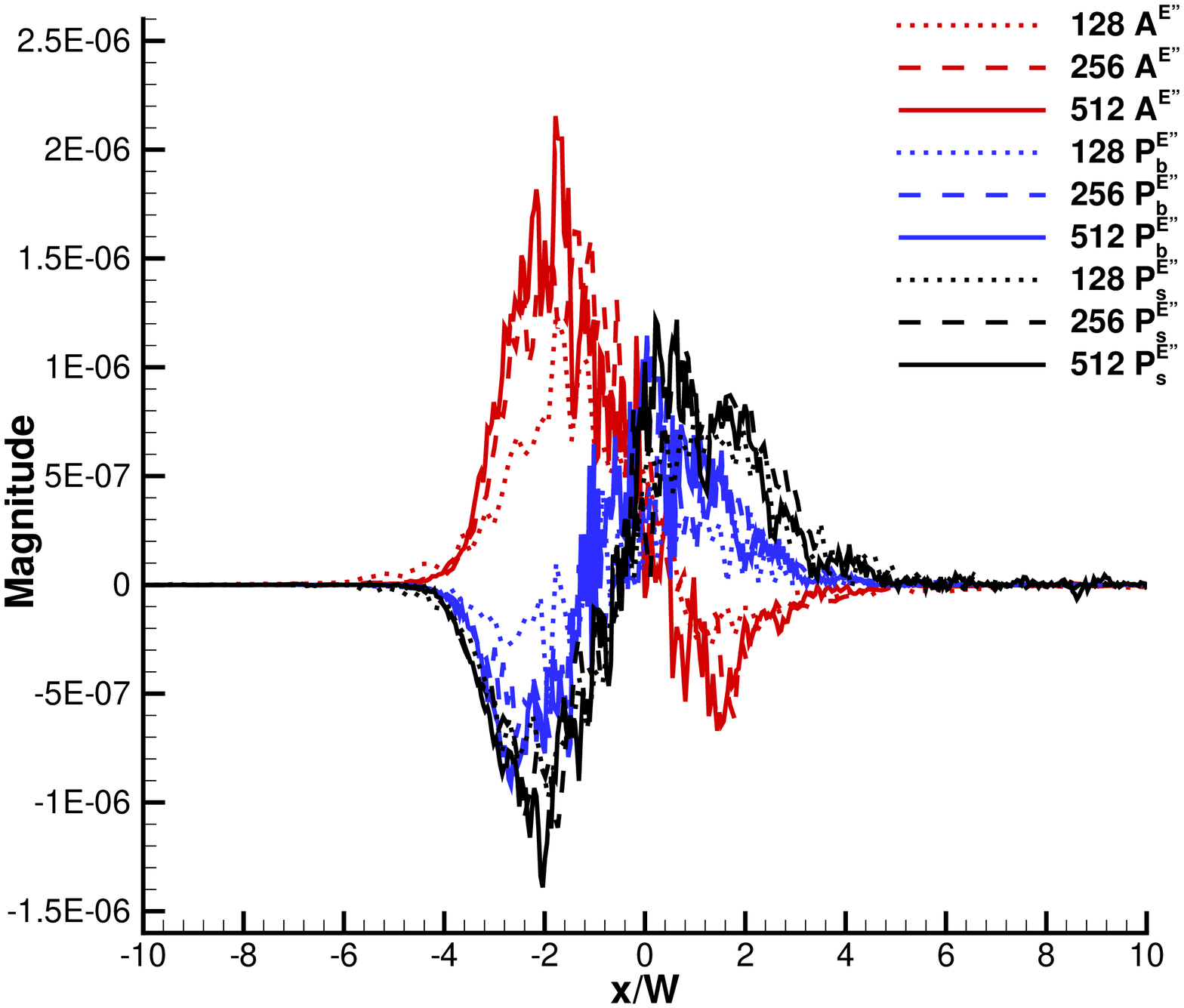}}
\subfigure[\hspace{0.1cm} Terms in the $\widetilde{m''_1}$   transport equation]{\includegraphics[width=0.45\textwidth]{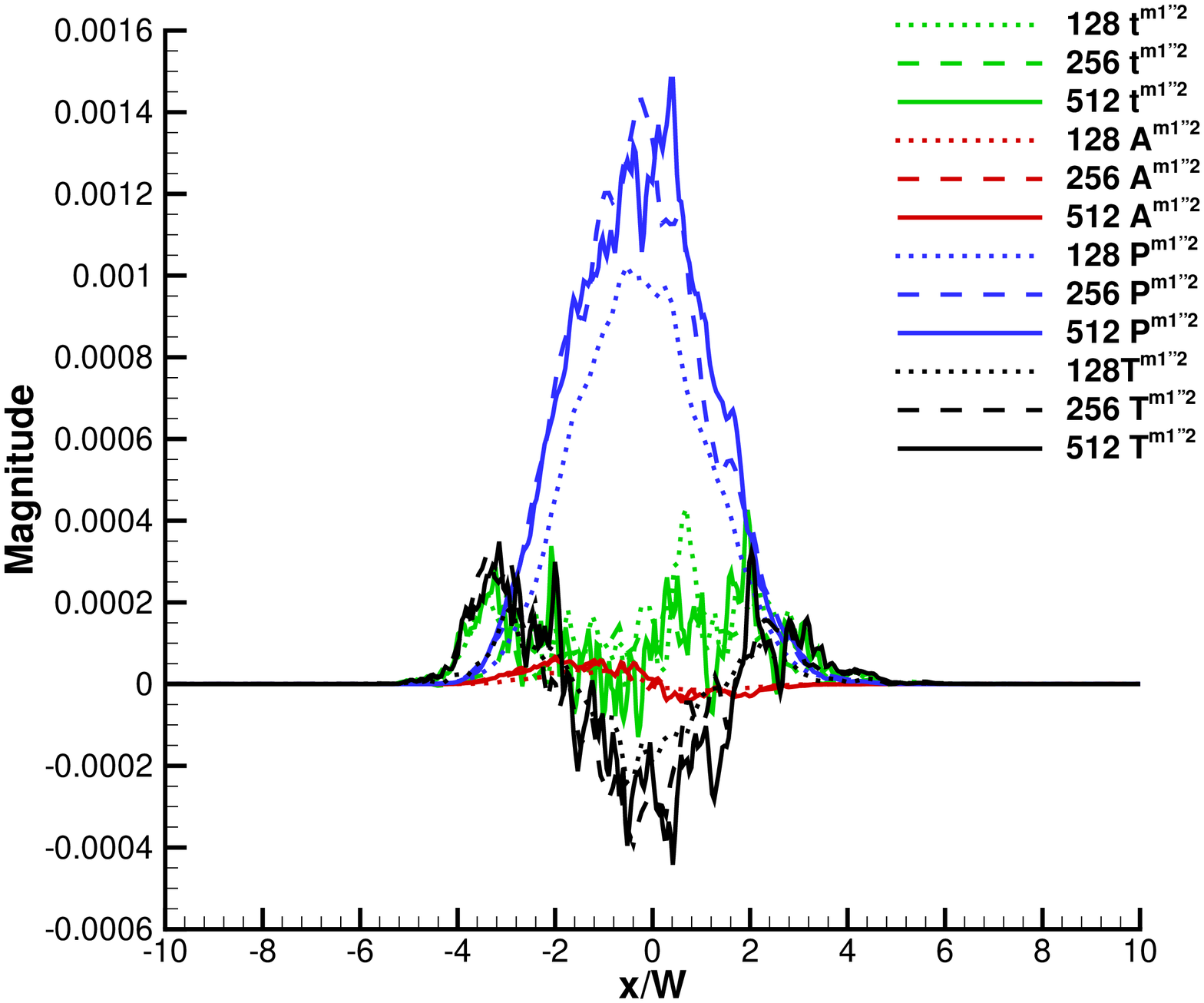}}
\caption{Convergence of individual terms in the (a) $\tilde{u}$ , (b) $\widetilde{m_1}$ , (c) $\widetilde{E''}$ , and (d) $\widetilde{m''_1}$  transport equations at $t=1$ s.\label{transpconv}}
\end{centering}
\end{figure*}

Due to the large number of individual terms computed, convergence is demonstrated here through examination of the Flamenco results alone, as detailed in Sec. \ref{conv}, using simulations ranging from $180\times 128^2$ to $720 \times 512^2$. Figure \ref{meanconv} shows the convergence of mean properties at $t=1$ s. The two highest grid levels are in very good agreement. Figure \ref{transpconv} shows the grid convergence of the individual terms in each transport equation for Flamenco. Again, the two highest resolutions agree well, particularly considering the higher-order nature of the metrics under examination. Finally, Fig. \ref{ndconv} shows the convergence of computed numerical dissipation for Flamenco. For the mean momentum and mass fraction, numerical dissipation is small and dominated by noise. For the mass fraction variance and kinetic energy, numerical dissipation is significant and the difference between the two highest grid resolutions is again small.

\bibliography{bibliography}

\begin{thebibliography}{56}%
\makeatletter
\providecommand \@ifxundefined [1]{%
 \@ifx{#1\undefined}
}%
\providecommand \@ifnum [1]{%
 \ifnum #1\expandafter \@firstoftwo
 \else \expandafter \@secondoftwo
 \fi
}%
\providecommand \@ifx [1]{%
 \ifx #1\expandafter \@firstoftwo
 \else \expandafter \@secondoftwo
 \fi
}%
\providecommand \natexlab [1]{#1}%
\providecommand \enquote  [1]{``#1''}%
\providecommand \bibnamefont  [1]{#1}%
\providecommand \bibfnamefont [1]{#1}%
\providecommand \citenamefont [1]{#1}%
\providecommand \href@noop [0]{\@secondoftwo}%
\providecommand \href [0]{\begingroup \@sanitize@url \@href}%
\providecommand \@href[1]{\@@startlink{#1}\@@href}%
\providecommand \@@href[1]{\endgroup#1\@@endlink}%
\providecommand \@sanitize@url [0]{\catcode `\\12\catcode `\$12\catcode
  `\&12\catcode `\#12\catcode `\^12\catcode `\_12\catcode `\%12\relax}%
\providecommand \@@startlink[1]{}%
\providecommand \@@endlink[0]{}%
\providecommand \url  [0]{\begingroup\@sanitize@url \@url }%
\providecommand \@url [1]{\endgroup\@href {#1}{\urlprefix }}%
\providecommand \urlprefix  [0]{URL }%
\providecommand \Eprint [0]{\href }%
\providecommand \doibase [0]{http://dx.doi.org/}%
\providecommand \selectlanguage [0]{\@gobble}%
\providecommand \bibinfo  [0]{\@secondoftwo}%
\providecommand \bibfield  [0]{\@secondoftwo}%
\providecommand \translation [1]{[#1]}%
\providecommand \BibitemOpen [0]{}%
\providecommand \bibitemStop [0]{}%
\providecommand \bibitemNoStop [0]{.\EOS\space}%
\providecommand \EOS [0]{\spacefactor3000\relax}%
\providecommand \BibitemShut  [1]{\csname bibitem#1\endcsname}%
\let\auto@bib@innerbib\@empty
\bibitem [{\citenamefont {Richtmyer}(1960)}]{Richtmyer1960}%
  \BibitemOpen
  \bibfield  {author} {\bibinfo {author} {\bibfnamefont {R.~D.}\ \bibnamefont
  {Richtmyer}},\ }\href@noop {} {\bibfield  {journal} {\bibinfo  {journal}
  {Comm. Pure Appl. Math.}\ }\textbf {\bibinfo {volume} {13}},\ \bibinfo
  {pages} {297} (\bibinfo {year} {1960})}\BibitemShut {NoStop}%
\bibitem [{\citenamefont {Meshkov}(1969)}]{Meshkov1969}%
  \BibitemOpen
  \bibfield  {author} {\bibinfo {author} {\bibfnamefont {E.~E.}\ \bibnamefont
  {Meshkov}},\ }\href@noop {} {\bibfield  {journal} {\bibinfo  {journal} {SSSR
  Mekh. Zhidk. Gaza.}\ }\textbf {\bibinfo {volume} {4}},\ \bibinfo {pages}
  {151} (\bibinfo {year} {1969})}\BibitemShut {NoStop}%
\bibitem [{\citenamefont {Clark~et al.}(2016)}]{Clark2016}%
  \BibitemOpen
  \bibfield  {author} {\bibinfo {author} {\bibfnamefont {D.~S.}\ \bibnamefont
  {Clark~et al.}},\ }\href@noop {} {\bibfield  {journal} {\bibinfo  {journal}
  {Physics of Plasmas}\ }\textbf {\bibinfo {volume} {23}},\ \bibinfo {pages}
  {056302} (\bibinfo {year} {2016})}\BibitemShut {NoStop}%
\bibitem [{\citenamefont {Burrows}(2000)}]{Burrows2000}%
  \BibitemOpen
  \bibfield  {author} {\bibinfo {author} {\bibfnamefont {A.}~\bibnamefont
  {Burrows}},\ }\href@noop {} {\bibfield  {journal} {\bibinfo  {journal}
  {Nature}\ }\textbf {\bibinfo {volume} {403}},\ \bibinfo {pages} {723}
  (\bibinfo {year} {2000})}\BibitemShut {NoStop}%
\bibitem [{\citenamefont {Yang}\ \emph {et~al.}(1993)\citenamefont {Yang},
  \citenamefont {Kubota},\ and\ \citenamefont {Zukoski}}]{yang1993}%
  \BibitemOpen
  \bibfield  {author} {\bibinfo {author} {\bibfnamefont {J.}~\bibnamefont
  {Yang}}, \bibinfo {author} {\bibfnamefont {T.}~\bibnamefont {Kubota}}, \ and\
  \bibinfo {author} {\bibfnamefont {E.~E.}\ \bibnamefont {Zukoski}},\
  }\href@noop {} {\bibfield  {journal} {\bibinfo  {journal} {AIAA journal}\
  }\textbf {\bibinfo {volume} {31}},\ \bibinfo {pages} {854} (\bibinfo {year}
  {1993})}\BibitemShut {NoStop}%
\bibitem [{\citenamefont {Zhou}(2017{\natexlab{a}})}]{Zhou2017a}%
  \BibitemOpen
  \bibfield  {author} {\bibinfo {author} {\bibfnamefont {Y.}~\bibnamefont
  {Zhou}},\ }\href@noop {} {\bibfield  {journal} {\bibinfo  {journal} {Physics
  Reports}\ }\textbf {\bibinfo {volume} {720-722}},\ \bibinfo {pages} {1 }
  (\bibinfo {year} {2017}{\natexlab{a}})}\BibitemShut {NoStop}%
\bibitem [{\citenamefont {Zhou}(2017{\natexlab{b}})}]{Zhou2017b}%
  \BibitemOpen
  \bibfield  {author} {\bibinfo {author} {\bibfnamefont {Y.}~\bibnamefont
  {Zhou}},\ }\href@noop {} {\bibfield  {journal} {\bibinfo  {journal} {Physics
  Reports}\ }\textbf {\bibinfo {volume} {723-725}},\ \bibinfo {pages} {1 }
  (\bibinfo {year} {2017}{\natexlab{b}})}\BibitemShut {NoStop}%
\bibitem [{\citenamefont {Barenblatt}\ \emph {et~al.}(1983)\citenamefont
  {Barenblatt}, \citenamefont {Looss},\ and\ \citenamefont
  {Joseph}}]{Barenblatt1983}%
  \BibitemOpen
  \bibfield  {author} {\bibinfo {author} {\bibfnamefont {G.~I.}\ \bibnamefont
  {Barenblatt}}, \bibinfo {author} {\bibfnamefont {G.}~\bibnamefont {Looss}}, \
  and\ \bibinfo {author} {\bibfnamefont {D.~D.}\ \bibnamefont {Joseph}},\
  }\href@noop {} {\emph {\bibinfo {title} {Nonlinear Dynamics and
  Turbulence}}}\ (\bibinfo  {publisher} {Pitman Publishing},\ \bibinfo {year}
  {1983})\BibitemShut {NoStop}%
\bibitem [{\citenamefont {Thornber}\ \emph {et~al.}(2010)\citenamefont
  {Thornber}, \citenamefont {Youngs}, \citenamefont {Drikakis},\ and\
  \citenamefont {Williams}}]{Thornber2010}%
  \BibitemOpen
  \bibfield  {author} {\bibinfo {author} {\bibfnamefont {B.}~\bibnamefont
  {Thornber}}, \bibinfo {author} {\bibfnamefont {D.}~\bibnamefont {Youngs}},
  \bibinfo {author} {\bibfnamefont {D.}~\bibnamefont {Drikakis}}, \ and\
  \bibinfo {author} {\bibfnamefont {R.~J.~R.}\ \bibnamefont {Williams}},\
  }\href@noop {} {\bibfield  {journal} {\bibinfo  {journal} {J. Fluid Mech.}\
  }\textbf {\bibinfo {volume} {654}},\ \bibinfo {pages} {99} (\bibinfo {year}
  {2010})}\BibitemShut {NoStop}%
\bibitem [{\citenamefont {Elbaz}\ and\ \citenamefont
  {Shvarts}(2018)}]{Elbaz2018}%
  \BibitemOpen
  \bibfield  {author} {\bibinfo {author} {\bibfnamefont {Y.}~\bibnamefont
  {Elbaz}}\ and\ \bibinfo {author} {\bibfnamefont {D.}~\bibnamefont
  {Shvarts}},\ }\href@noop {} {\bibfield  {journal} {\bibinfo  {journal}
  {Physics of Plasmas}\ }\textbf {\bibinfo {volume} {25}},\ \bibinfo {pages}
  {062126} (\bibinfo {year} {2018})}\BibitemShut {NoStop}%
\bibitem [{\citenamefont {Stefano}\ \emph {et~al.}(2015)\citenamefont
  {Stefano}, \citenamefont {Malamud}, \citenamefont {Kuranz}, \citenamefont
  {Klein},\ and\ \citenamefont {Drake}}]{DiStefano2015263}%
  \BibitemOpen
  \bibfield  {author} {\bibinfo {author} {\bibfnamefont {C.~D.}\ \bibnamefont
  {Stefano}}, \bibinfo {author} {\bibfnamefont {G.}~\bibnamefont {Malamud}},
  \bibinfo {author} {\bibfnamefont {C.}~\bibnamefont {Kuranz}}, \bibinfo
  {author} {\bibfnamefont {S.}~\bibnamefont {Klein}}, \ and\ \bibinfo {author}
  {\bibfnamefont {R.}~\bibnamefont {Drake}},\ }\href {\doibase
  http://dx.doi.org/10.1016/j.hedp.2015.09.001} {\bibfield  {journal} {\bibinfo
   {journal} {High Energy Density Physics}\ }\textbf {\bibinfo {volume} {17}},\
  \bibinfo {pages} {263} (\bibinfo {year} {2015})}\BibitemShut {NoStop}%
\bibitem [{\citenamefont {Flippo}\ \emph {et~al.}(2016)\citenamefont {Flippo},
  \citenamefont {Doss}, \citenamefont {Kline}, \citenamefont {Merritt},
  \citenamefont {Capelli}, \citenamefont {Cardenas}, \citenamefont {DeVolder},
  \citenamefont {Fierro}, \citenamefont {Huntington}, \citenamefont {Kot},
  \citenamefont {Loomis}, \citenamefont {MacLaren}, \citenamefont {Murphy},
  \citenamefont {Nagel}, \citenamefont {Perry}, \citenamefont {Randolph},
  \citenamefont {Rivera},\ and\ \citenamefont {Schmidt}}]{Flippo2016}%
  \BibitemOpen
  \bibfield  {author} {\bibinfo {author} {\bibfnamefont {K.~A.}\ \bibnamefont
  {Flippo}}, \bibinfo {author} {\bibfnamefont {F.~W.}\ \bibnamefont {Doss}},
  \bibinfo {author} {\bibfnamefont {J.~L.}\ \bibnamefont {Kline}}, \bibinfo
  {author} {\bibfnamefont {E.~C.}\ \bibnamefont {Merritt}}, \bibinfo {author}
  {\bibfnamefont {D.}~\bibnamefont {Capelli}}, \bibinfo {author} {\bibfnamefont
  {T.}~\bibnamefont {Cardenas}}, \bibinfo {author} {\bibfnamefont
  {B.}~\bibnamefont {DeVolder}}, \bibinfo {author} {\bibfnamefont
  {F.}~\bibnamefont {Fierro}}, \bibinfo {author} {\bibfnamefont {C.~M.}\
  \bibnamefont {Huntington}}, \bibinfo {author} {\bibfnamefont
  {L.}~\bibnamefont {Kot}}, \bibinfo {author} {\bibfnamefont {E.~N.}\
  \bibnamefont {Loomis}}, \bibinfo {author} {\bibfnamefont {S.~A.}\
  \bibnamefont {MacLaren}}, \bibinfo {author} {\bibfnamefont {T.~J.}\
  \bibnamefont {Murphy}}, \bibinfo {author} {\bibfnamefont {S.~R.}\
  \bibnamefont {Nagel}}, \bibinfo {author} {\bibfnamefont {T.~S.}\ \bibnamefont
  {Perry}}, \bibinfo {author} {\bibfnamefont {R.~B.}\ \bibnamefont {Randolph}},
  \bibinfo {author} {\bibfnamefont {G.}~\bibnamefont {Rivera}}, \ and\ \bibinfo
  {author} {\bibfnamefont {D.~W.}\ \bibnamefont {Schmidt}},\ }\href {\doibase
  10.1103/PhysRevLett.117.225001} {\bibfield  {journal} {\bibinfo  {journal}
  {Phys. Rev. Lett.}\ }\textbf {\bibinfo {volume} {117}},\ \bibinfo {pages}
  {225001} (\bibinfo {year} {2016})}\BibitemShut {NoStop}%
\bibitem [{\citenamefont {Krivets}\ \emph {et~al.}(2017)\citenamefont
  {Krivets}, \citenamefont {Ferguson},\ and\ \citenamefont
  {Jacobs}}]{Krivets2017}%
  \BibitemOpen
  \bibfield  {author} {\bibinfo {author} {\bibfnamefont {V.~V.}\ \bibnamefont
  {Krivets}}, \bibinfo {author} {\bibfnamefont {K.~J.}\ \bibnamefont
  {Ferguson}}, \ and\ \bibinfo {author} {\bibfnamefont {J.~W.}\ \bibnamefont
  {Jacobs}},\ }in\ \href@noop {} {\emph {\bibinfo {booktitle} {AIP Conference
  Proceedings}}},\ Vol.\ \bibinfo {volume} {1793}\ (\bibinfo {organization}
  {AIP Publishing},\ \bibinfo {year} {2017})\ p.\ \bibinfo {pages}
  {150003}\BibitemShut {NoStop}%
\bibitem [{\citenamefont {Soulard}\ \emph {et~al.}(2018)\citenamefont
  {Soulard}, \citenamefont {Guillois}, \citenamefont {Griffond}, \citenamefont
  {Sabelnikov},\ and\ \citenamefont {Simo\"ens}}]{Soulard2018}%
  \BibitemOpen
  \bibfield  {author} {\bibinfo {author} {\bibfnamefont {O.}~\bibnamefont
  {Soulard}}, \bibinfo {author} {\bibfnamefont {F.}~\bibnamefont {Guillois}},
  \bibinfo {author} {\bibfnamefont {J.}~\bibnamefont {Griffond}}, \bibinfo
  {author} {\bibfnamefont {V.}~\bibnamefont {Sabelnikov}}, \ and\ \bibinfo
  {author} {\bibfnamefont {S.}~\bibnamefont {Simo\"ens}},\ }\href {\doibase
  10.1103/PhysRevFluids.3.104603} {\bibfield  {journal} {\bibinfo  {journal}
  {Phys. Rev. Fluids}\ }\textbf {\bibinfo {volume} {3}},\ \bibinfo {pages}
  {104603} (\bibinfo {year} {2018})}\BibitemShut {NoStop}%
\bibitem [{\citenamefont {Schilling}\ and\ \citenamefont
  {Latini}(2010)}]{Schilling2010}%
  \BibitemOpen
  \bibfield  {author} {\bibinfo {author} {\bibfnamefont {O.}~\bibnamefont
  {Schilling}}\ and\ \bibinfo {author} {\bibfnamefont {M.}~\bibnamefont
  {Latini}},\ }\href@noop {} {\bibfield  {journal} {\bibinfo  {journal} {Acta
  Math. Sci.}\ }\textbf {\bibinfo {volume} {30B}},\ \bibinfo {pages} {595}
  (\bibinfo {year} {2010})}\BibitemShut {NoStop}%
\bibitem [{\citenamefont {Thornber}\ \emph {et~al.}(2017)\citenamefont
  {Thornber}, \citenamefont {Griffond}, \citenamefont {Poujade}, \citenamefont
  {Attal}, \citenamefont {Varshochi}, \citenamefont {Bigdelou}, \citenamefont
  {Ramaprabhu}, \citenamefont {Olson}, \citenamefont {Greenough}, \citenamefont
  {Zhou}, \citenamefont {Schilling}, \citenamefont {Garside}, \citenamefont
  {Williams}, \citenamefont {Batha}, \citenamefont {Kuchugov}, \citenamefont
  {Ladonkina}, \citenamefont {Tishkin}, \citenamefont {Zmitrenko},
  \citenamefont {Rozanov},\ and\ \citenamefont {Youngs}}]{Thornber2017}%
  \BibitemOpen
  \bibfield  {author} {\bibinfo {author} {\bibfnamefont {B.}~\bibnamefont
  {Thornber}}, \bibinfo {author} {\bibfnamefont {J.}~\bibnamefont {Griffond}},
  \bibinfo {author} {\bibfnamefont {O.}~\bibnamefont {Poujade}}, \bibinfo
  {author} {\bibfnamefont {N.}~\bibnamefont {Attal}}, \bibinfo {author}
  {\bibfnamefont {H.}~\bibnamefont {Varshochi}}, \bibinfo {author}
  {\bibfnamefont {P.}~\bibnamefont {Bigdelou}}, \bibinfo {author}
  {\bibfnamefont {P.}~\bibnamefont {Ramaprabhu}}, \bibinfo {author}
  {\bibfnamefont {B.}~\bibnamefont {Olson}}, \bibinfo {author} {\bibfnamefont
  {J.}~\bibnamefont {Greenough}}, \bibinfo {author} {\bibfnamefont
  {Y.}~\bibnamefont {Zhou}}, \bibinfo {author} {\bibfnamefont {O.}~\bibnamefont
  {Schilling}}, \bibinfo {author} {\bibfnamefont {K.~A.}\ \bibnamefont
  {Garside}}, \bibinfo {author} {\bibfnamefont {R.~J.~R.}\ \bibnamefont
  {Williams}}, \bibinfo {author} {\bibfnamefont {C.~A.}\ \bibnamefont {Batha}},
  \bibinfo {author} {\bibfnamefont {P.~A.}\ \bibnamefont {Kuchugov}}, \bibinfo
  {author} {\bibfnamefont {M.~E.}\ \bibnamefont {Ladonkina}}, \bibinfo {author}
  {\bibfnamefont {V.~F.}\ \bibnamefont {Tishkin}}, \bibinfo {author}
  {\bibfnamefont {N.~V.}\ \bibnamefont {Zmitrenko}}, \bibinfo {author}
  {\bibfnamefont {V.~B.}\ \bibnamefont {Rozanov}}, \ and\ \bibinfo {author}
  {\bibfnamefont {D.~L.}\ \bibnamefont {Youngs}},\ }\href {\doibase
  10.1063/1.4993464} {\bibfield  {journal} {\bibinfo  {journal} {Physics of
  Fluids}\ }\textbf {\bibinfo {volume} {29}},\ \bibinfo {pages} {105107}
  (\bibinfo {year} {2017})}\BibitemShut {NoStop}%
\bibitem [{\citenamefont {Favre}(1958)}]{Favre1958}%
  \BibitemOpen
  \bibfield  {author} {\bibinfo {author} {\bibfnamefont {A.}~\bibnamefont
  {Favre}},\ }\href@noop {} {\bibfield  {journal} {\bibinfo  {journal} {C.R.
  Acd. Sci. Paris}\ }\textbf {\bibinfo {volume} {273}},\ \bibinfo {pages}
  {1289} (\bibinfo {year} {1958})}\BibitemShut {NoStop}%
\bibitem [{\citenamefont {Chassaing}(2001)}]{Chassaing2001}%
  \BibitemOpen
  \bibfield  {author} {\bibinfo {author} {\bibfnamefont {P.}~\bibnamefont
  {Chassaing}},\ }\href {\doibase 10.1023/A:1013533322651} {\bibfield
  {journal} {\bibinfo  {journal} {Flow, Turbulence and Combustion}\ }\textbf
  {\bibinfo {volume} {66}},\ \bibinfo {pages} {293} (\bibinfo {year}
  {2001})}\BibitemShut {NoStop}%
\bibitem [{\citenamefont {Dimonte}\ and\ \citenamefont
  {Tipton}(2006)}]{Dimonte2006}%
  \BibitemOpen
  \bibfield  {author} {\bibinfo {author} {\bibfnamefont {G.}~\bibnamefont
  {Dimonte}}\ and\ \bibinfo {author} {\bibfnamefont {R.}~\bibnamefont
  {Tipton}},\ }\href@noop {} {\bibfield  {journal} {\bibinfo  {journal} {Phys.
  Fluids}\ }\textbf {\bibinfo {volume} {18}},\ \bibinfo {pages} {085101}
  (\bibinfo {year} {2006})}\BibitemShut {NoStop}%
\bibitem [{\citenamefont {Gauthier}\ and\ \citenamefont
  {Bonnet}(1990)}]{Gauthier1990}%
  \BibitemOpen
  \bibfield  {author} {\bibinfo {author} {\bibfnamefont {S.}~\bibnamefont
  {Gauthier}}\ and\ \bibinfo {author} {\bibfnamefont {M.}~\bibnamefont
  {Bonnet}},\ }\href@noop {} {\bibfield  {journal} {\bibinfo  {journal} {Phys.
  Fluids A}\ }\textbf {\bibinfo {volume} {2}},\ \bibinfo {pages} {1685}
  (\bibinfo {year} {1990})}\BibitemShut {NoStop}%
\bibitem [{\citenamefont {Mor{\'a}n-L{\'o}pez}\ and\ \citenamefont
  {Schilling}(2013)}]{Moran2013}%
  \BibitemOpen
  \bibfield  {author} {\bibinfo {author} {\bibfnamefont {J.~T.}\ \bibnamefont
  {Mor{\'a}n-L{\'o}pez}}\ and\ \bibinfo {author} {\bibfnamefont
  {O.}~\bibnamefont {Schilling}},\ }\href@noop {} {\bibfield  {journal}
  {\bibinfo  {journal} {High Energy Density Physics}\ }\textbf {\bibinfo
  {volume} {9}},\ \bibinfo {pages} {112} (\bibinfo {year} {2013})}\BibitemShut
  {NoStop}%
\bibitem [{\citenamefont {Besnard}\ \emph {et~al.}(1992)\citenamefont
  {Besnard}, \citenamefont {Harlow}, \citenamefont {Rauenzahn},\ and\
  \citenamefont {Zemach}}]{Besnard1992}%
  \BibitemOpen
  \bibfield  {author} {\bibinfo {author} {\bibfnamefont {D.}~\bibnamefont
  {Besnard}}, \bibinfo {author} {\bibfnamefont {F.}~\bibnamefont {Harlow}},
  \bibinfo {author} {\bibfnamefont {R.}~\bibnamefont {Rauenzahn}}, \ and\
  \bibinfo {author} {\bibfnamefont {C.}~\bibnamefont {Zemach}},\ }\href@noop {}
  {\emph {\bibinfo {title} {Turbulence transport equations for variable-density
  turbulence and their relationship to two-field models}}},\ \bibinfo {type}
  {Tech. Rep.}\ (\bibinfo  {institution} {Los Alamos},\ \bibinfo {year}
  {1992})\BibitemShut {NoStop}%
\bibitem [{\citenamefont {Banerjee}\ \emph {et~al.}(2010)\citenamefont
  {Banerjee}, \citenamefont {Gore},\ and\ \citenamefont
  {Andrews}}]{Banerjee2010}%
  \BibitemOpen
  \bibfield  {author} {\bibinfo {author} {\bibfnamefont {A.}~\bibnamefont
  {Banerjee}}, \bibinfo {author} {\bibfnamefont {R.}~\bibnamefont {Gore}}, \
  and\ \bibinfo {author} {\bibfnamefont {M.}~\bibnamefont {Andrews}},\
  }\href@noop {} {\bibfield  {journal} {\bibinfo  {journal} {Phys. Rev. E}\
  }\textbf {\bibinfo {volume} {82}},\ \bibinfo {pages} {046309} (\bibinfo
  {year} {2010})}\BibitemShut {NoStop}%
\bibitem [{\citenamefont {Stalsberg-Zarling}\ and\ \citenamefont
  {Gore}(2011)}]{Stalsberg2011}%
  \BibitemOpen
  \bibfield  {author} {\bibinfo {author} {\bibfnamefont {K.}~\bibnamefont
  {Stalsberg-Zarling}}\ and\ \bibinfo {author} {\bibfnamefont {R.}~\bibnamefont
  {Gore}},\ }\href@noop {} {\emph {\bibinfo {title} {The {BHR2} turbulence
  model Incompressible isotropic decay, {R}ayleigh-{T}aylor,
  {K}elvin-{H}elmholtz and homogeneous variable density turbulence}}},\
  \bibinfo {type} {Tech. Rep.}\ (\bibinfo  {institution} {Los Alamos},\
  \bibinfo {year} {2011})\BibitemShut {NoStop}%
\bibitem [{\citenamefont {Schwarzkopf}\ \emph {et~al.}(2011)\citenamefont
  {Schwarzkopf}, \citenamefont {Livescu}, \citenamefont {Gore}, \citenamefont
  {Rauenzahn},\ and\ \citenamefont
  {Ristorcelli}}]{doi:10.1080/14685248.2011.633084}%
  \BibitemOpen
  \bibfield  {author} {\bibinfo {author} {\bibfnamefont {J.~D.}\ \bibnamefont
  {Schwarzkopf}}, \bibinfo {author} {\bibfnamefont {D.}~\bibnamefont
  {Livescu}}, \bibinfo {author} {\bibfnamefont {R.~A.}\ \bibnamefont {Gore}},
  \bibinfo {author} {\bibfnamefont {R.~M.}\ \bibnamefont {Rauenzahn}}, \ and\
  \bibinfo {author} {\bibfnamefont {J.~R.}\ \bibnamefont {Ristorcelli}},\
  }\href@noop {} {\bibfield  {journal} {\bibinfo  {journal} {Journal of
  Turbulence}\ }\textbf {\bibinfo {volume} {12}},\ \bibinfo {pages} {N49}
  (\bibinfo {year} {2011})}\BibitemShut {NoStop}%
\bibitem [{\citenamefont {Schwarzkopf}\ \emph {et~al.}(2016)\citenamefont
  {Schwarzkopf}, \citenamefont {Livescu}, \citenamefont {Baltzer},
  \citenamefont {Gore},\ and\ \citenamefont
  {Ristorcelli}}]{schwarzkopf2016two}%
  \BibitemOpen
  \bibfield  {author} {\bibinfo {author} {\bibfnamefont {J.~D.}\ \bibnamefont
  {Schwarzkopf}}, \bibinfo {author} {\bibfnamefont {D.}~\bibnamefont
  {Livescu}}, \bibinfo {author} {\bibfnamefont {J.~R.}\ \bibnamefont
  {Baltzer}}, \bibinfo {author} {\bibfnamefont {R.~A.}\ \bibnamefont {Gore}}, \
  and\ \bibinfo {author} {\bibfnamefont {J.}~\bibnamefont {Ristorcelli}},\
  }\href@noop {} {\bibfield  {journal} {\bibinfo  {journal} {Flow, Turbulence
  and Combustion}\ }\textbf {\bibinfo {volume} {96}},\ \bibinfo {pages} {1}
  (\bibinfo {year} {2016})}\BibitemShut {NoStop}%
\bibitem [{\citenamefont {Gr\'egoire}\ \emph {et~al.}(2005)\citenamefont
  {Gr\'egoire}, \citenamefont {Souffland},\ and\ \citenamefont
  {Gauthier}}]{Gregoire05}%
  \BibitemOpen
  \bibfield  {author} {\bibinfo {author} {\bibfnamefont {O.}~\bibnamefont
  {Gr\'egoire}}, \bibinfo {author} {\bibfnamefont {D.}~\bibnamefont
  {Souffland}}, \ and\ \bibinfo {author} {\bibfnamefont {S.}~\bibnamefont
  {Gauthier}},\ }\href@noop {} {\bibfield  {journal} {\bibinfo  {journal}
  {Journal of Turbulence}\ }\textbf {\bibinfo {volume} {6}},\ \bibinfo {pages}
  {1} (\bibinfo {year} {2005})}\BibitemShut {NoStop}%
\bibitem [{\citenamefont {Souffland}\ \emph {et~al.}(2014)\citenamefont
  {Souffland}, \citenamefont {Soulard},\ and\ \citenamefont
  {Griffond}}]{Souffland14}%
  \BibitemOpen
  \bibfield  {author} {\bibinfo {author} {\bibfnamefont {D.}~\bibnamefont
  {Souffland}}, \bibinfo {author} {\bibfnamefont {O.}~\bibnamefont {Soulard}},
  \ and\ \bibinfo {author} {\bibfnamefont {J.}~\bibnamefont {Griffond}},\
  }\href@noop {} {\bibfield  {journal} {\bibinfo  {journal} {ASME Journal of
  Fluids Engineering}\ }\textbf {\bibinfo {volume} {136}},\ \bibinfo {pages}
  {091102} (\bibinfo {year} {2014})}\BibitemShut {NoStop}%
\bibitem [{\citenamefont {Youngs}(1994)}]{Youngs1994}%
  \BibitemOpen
  \bibfield  {author} {\bibinfo {author} {\bibfnamefont {D.~L.}\ \bibnamefont
  {Youngs}},\ }\href@noop {} {\bibfield  {journal} {\bibinfo  {journal} {Laser
  Part. Beams}\ }\textbf {\bibinfo {volume} {12}},\ \bibinfo {pages} {725}
  (\bibinfo {year} {1994})}\BibitemShut {NoStop}%
\bibitem [{\citenamefont {Llor}(2005)}]{Llor2005}%
  \BibitemOpen
  \bibfield  {author} {\bibinfo {author} {\bibfnamefont {A.}~\bibnamefont
  {Llor}},\ }\href@noop {} {\emph {\bibinfo {title} {Statistical Hydrodynamic
  Models for Developing Mixing Instability Flows}}}\ (\bibinfo  {publisher}
  {Springer, Berlin,},\ \bibinfo {year} {2005})\BibitemShut {NoStop}%
\bibitem [{\citenamefont {Llor}\ and\ \citenamefont
  {Bailly}(2003)}]{llor_bailly_2003}%
  \BibitemOpen
  \bibfield  {author} {\bibinfo {author} {\bibfnamefont {A.}~\bibnamefont
  {Llor}}\ and\ \bibinfo {author} {\bibfnamefont {P.}~\bibnamefont {Bailly}},\
  }\href {\doibase 10.1017/S0263034603213033} {\bibfield  {journal} {\bibinfo
  {journal} {Laser and Particle Beams}\ }\textbf {\bibinfo {volume} {21}},\
  \bibinfo {pages} {311–315} (\bibinfo {year} {2003})}\BibitemShut {NoStop}%
\bibitem [{\citenamefont {Leinov}\ \emph {et~al.}(2009)\citenamefont {Leinov},
  \citenamefont {Malamud}, \citenamefont {Elbaz}, \citenamefont {Levin},
  \citenamefont {Ben-Dor}, \citenamefont {Shvarts},\ and\ \citenamefont
  {Sadot}}]{Leinov2009}%
  \BibitemOpen
  \bibfield  {author} {\bibinfo {author} {\bibfnamefont {E.}~\bibnamefont
  {Leinov}}, \bibinfo {author} {\bibfnamefont {G.}~\bibnamefont {Malamud}},
  \bibinfo {author} {\bibfnamefont {Y.}~\bibnamefont {Elbaz}}, \bibinfo
  {author} {\bibfnamefont {L.}~\bibnamefont {Levin}}, \bibinfo {author}
  {\bibfnamefont {G.}~\bibnamefont {Ben-Dor}}, \bibinfo {author} {\bibfnamefont
  {D.}~\bibnamefont {Shvarts}}, \ and\ \bibinfo {author} {\bibfnamefont
  {O.}~\bibnamefont {Sadot}},\ }\href@noop {} {\bibfield  {journal} {\bibinfo
  {journal} {J. Fluid Mech.}\ }\textbf {\bibinfo {volume} {626}},\ \bibinfo
  {pages} {449} (\bibinfo {year} {2009})}\BibitemShut {NoStop}%
\bibitem [{\citenamefont {Malamud}\ \emph {et~al.}(2014)\citenamefont
  {Malamud}, \citenamefont {Leinov}, \citenamefont {Sadot}, \citenamefont
  {Elbaz}, \citenamefont {Ben-Dor},\ and\ \citenamefont
  {Shvarts}}]{Malamud2014}%
  \BibitemOpen
  \bibfield  {author} {\bibinfo {author} {\bibfnamefont {G.}~\bibnamefont
  {Malamud}}, \bibinfo {author} {\bibfnamefont {E.}~\bibnamefont {Leinov}},
  \bibinfo {author} {\bibfnamefont {O.}~\bibnamefont {Sadot}}, \bibinfo
  {author} {\bibfnamefont {Y.}~\bibnamefont {Elbaz}}, \bibinfo {author}
  {\bibfnamefont {G.}~\bibnamefont {Ben-Dor}}, \ and\ \bibinfo {author}
  {\bibfnamefont {D.}~\bibnamefont {Shvarts}},\ }\href@noop {} {\bibfield
  {journal} {\bibinfo  {journal} {Physics of Fluids}\ }\textbf {\bibinfo
  {volume} {26}},\ \bibinfo {pages} {084107} (\bibinfo {year}
  {2014})}\BibitemShut {NoStop}%
\bibitem [{\citenamefont {Mohaghar}\ \emph {et~al.}(2019)\citenamefont
  {Mohaghar}, \citenamefont {Carter}, \citenamefont {Pathikonda},\ and\
  \citenamefont {Ranjan}}]{mohaghar_carter_pathikonda_ranjan_2019}%
  \BibitemOpen
  \bibfield  {author} {\bibinfo {author} {\bibfnamefont {M.}~\bibnamefont
  {Mohaghar}}, \bibinfo {author} {\bibfnamefont {J.}~\bibnamefont {Carter}},
  \bibinfo {author} {\bibfnamefont {G.}~\bibnamefont {Pathikonda}}, \ and\
  \bibinfo {author} {\bibfnamefont {D.}~\bibnamefont {Ranjan}},\ }\href
  {\doibase 10.1017/jfm.2019.330} {\bibfield  {journal} {\bibinfo  {journal}
  {Journal of Fluid Mechanics}\ }\textbf {\bibinfo {volume} {871}},\ \bibinfo
  {pages} {595–635} (\bibinfo {year} {2019})}\BibitemShut {NoStop}%
\bibitem [{\citenamefont {Weber}\ \emph {et~al.}(2014)\citenamefont {Weber},
  \citenamefont {Haehn}, \citenamefont {Oakley}, \citenamefont {Rothamer},\
  and\ \citenamefont {Bonazza}}]{weber_haehn_oakley_rothamer_bonazza_2014}%
  \BibitemOpen
  \bibfield  {author} {\bibinfo {author} {\bibfnamefont {C.}~\bibnamefont
  {Weber}}, \bibinfo {author} {\bibfnamefont {N.}~\bibnamefont {Haehn}},
  \bibinfo {author} {\bibfnamefont {J.}~\bibnamefont {Oakley}}, \bibinfo
  {author} {\bibfnamefont {D.}~\bibnamefont {Rothamer}}, \ and\ \bibinfo
  {author} {\bibfnamefont {R.}~\bibnamefont {Bonazza}},\ }\href {\doibase
  10.1017/jfm.2014.188} {\bibfield  {journal} {\bibinfo  {journal} {Journal of
  Fluid Mechanics}\ }\textbf {\bibinfo {volume} {748}},\ \bibinfo {pages}
  {457–487} (\bibinfo {year} {2014})}\BibitemShut {NoStop}%
\bibitem [{\citenamefont {Reese}\ \emph {et~al.}(2018)\citenamefont {Reese},
  \citenamefont {Ames}, \citenamefont {Noble}, \citenamefont {Oakley},
  \citenamefont {Rothamer},\ and\ \citenamefont
  {Bonazza}}]{reese_ames_noble_oakley_rothamer_bonazza_2018}%
  \BibitemOpen
  \bibfield  {author} {\bibinfo {author} {\bibfnamefont {D.~T.}\ \bibnamefont
  {Reese}}, \bibinfo {author} {\bibfnamefont {A.~M.}\ \bibnamefont {Ames}},
  \bibinfo {author} {\bibfnamefont {C.~D.}\ \bibnamefont {Noble}}, \bibinfo
  {author} {\bibfnamefont {J.~G.}\ \bibnamefont {Oakley}}, \bibinfo {author}
  {\bibfnamefont {D.~A.}\ \bibnamefont {Rothamer}}, \ and\ \bibinfo {author}
  {\bibfnamefont {R.}~\bibnamefont {Bonazza}},\ }\href {\doibase
  10.1017/jfm.2018.419} {\bibfield  {journal} {\bibinfo  {journal} {Journal of
  Fluid Mechanics}\ }\textbf {\bibinfo {volume} {849}},\ \bibinfo {pages}
  {541–575} (\bibinfo {year} {2018})}\BibitemShut {NoStop}%
\bibitem [{\citenamefont {Clark}\ \emph {et~al.}(2018)\citenamefont {Clark},
  \citenamefont {Kritcher}, \citenamefont {Yi}, \citenamefont {Zylstra},
  \citenamefont {Haan},\ and\ \citenamefont {Weber}}]{Clark2018}%
  \BibitemOpen
  \bibfield  {author} {\bibinfo {author} {\bibfnamefont {D.~S.}\ \bibnamefont
  {Clark}}, \bibinfo {author} {\bibfnamefont {A.~L.}\ \bibnamefont {Kritcher}},
  \bibinfo {author} {\bibfnamefont {S.~A.}\ \bibnamefont {Yi}}, \bibinfo
  {author} {\bibfnamefont {A.~B.}\ \bibnamefont {Zylstra}}, \bibinfo {author}
  {\bibfnamefont {S.~W.}\ \bibnamefont {Haan}}, \ and\ \bibinfo {author}
  {\bibfnamefont {C.~R.}\ \bibnamefont {Weber}},\ }\href@noop {} {\bibfield
  {journal} {\bibinfo  {journal} {Physics of Plasmas}\ }\textbf {\bibinfo
  {volume} {25}},\ \bibinfo {pages} {032703} (\bibinfo {year}
  {2018})}\BibitemShut {NoStop}%
\bibitem [{\citenamefont {Nagel}\ \emph {et~al.}(2017)\citenamefont {Nagel},
  \citenamefont {Raman}, \citenamefont {Huntington}, \citenamefont {MacLaren},
  \citenamefont {Wang}, \citenamefont {Barrios}, \citenamefont {Baumann},
  \citenamefont {Bender}, \citenamefont {Benedetti}, \citenamefont {Doane},
  \citenamefont {Felker}, \citenamefont {Fitzsimmons}, \citenamefont {Flippo},
  \citenamefont {Holder}, \citenamefont {Kaczala}, \citenamefont {Perry},
  \citenamefont {Seugling}, \citenamefont {Savage},\ and\ \citenamefont
  {Zhou}}]{Nagel2017}%
  \BibitemOpen
  \bibfield  {author} {\bibinfo {author} {\bibfnamefont {S.~R.}\ \bibnamefont
  {Nagel}}, \bibinfo {author} {\bibfnamefont {K.~S.}\ \bibnamefont {Raman}},
  \bibinfo {author} {\bibfnamefont {C.~M.}\ \bibnamefont {Huntington}},
  \bibinfo {author} {\bibfnamefont {S.~A.}\ \bibnamefont {MacLaren}}, \bibinfo
  {author} {\bibfnamefont {P.}~\bibnamefont {Wang}}, \bibinfo {author}
  {\bibfnamefont {M.~A.}\ \bibnamefont {Barrios}}, \bibinfo {author}
  {\bibfnamefont {T.}~\bibnamefont {Baumann}}, \bibinfo {author} {\bibfnamefont
  {J.~D.}\ \bibnamefont {Bender}}, \bibinfo {author} {\bibfnamefont {L.~R.}\
  \bibnamefont {Benedetti}}, \bibinfo {author} {\bibfnamefont {D.~M.}\
  \bibnamefont {Doane}}, \bibinfo {author} {\bibfnamefont {S.}~\bibnamefont
  {Felker}}, \bibinfo {author} {\bibfnamefont {P.}~\bibnamefont {Fitzsimmons}},
  \bibinfo {author} {\bibfnamefont {K.~A.}\ \bibnamefont {Flippo}}, \bibinfo
  {author} {\bibfnamefont {J.~P.}\ \bibnamefont {Holder}}, \bibinfo {author}
  {\bibfnamefont {D.~N.}\ \bibnamefont {Kaczala}}, \bibinfo {author}
  {\bibfnamefont {T.~S.}\ \bibnamefont {Perry}}, \bibinfo {author}
  {\bibfnamefont {R.~M.}\ \bibnamefont {Seugling}}, \bibinfo {author}
  {\bibfnamefont {L.}~\bibnamefont {Savage}}, \ and\ \bibinfo {author}
  {\bibfnamefont {Y.}~\bibnamefont {Zhou}},\ }\href@noop {} {\bibfield
  {journal} {\bibinfo  {journal} {Physics of Plasmas}\ }\textbf {\bibinfo
  {volume} {24}},\ \bibinfo {pages} {072704} (\bibinfo {year}
  {2017})}\BibitemShut {NoStop}%
\bibitem [{\citenamefont {Schilling}\ and\ \citenamefont
  {Mueschke}(2010)}]{Schilling2010b}%
  \BibitemOpen
  \bibfield  {author} {\bibinfo {author} {\bibfnamefont {O.}~\bibnamefont
  {Schilling}}\ and\ \bibinfo {author} {\bibfnamefont {N.~J.}\ \bibnamefont
  {Mueschke}},\ }\href {\doibase 10.1063/1.3484247} {\bibfield  {journal}
  {\bibinfo  {journal} {Physics of Fluids}\ }\textbf {\bibinfo {volume} {22}},\
  \bibinfo {pages} {105102} (\bibinfo {year} {2010})}\BibitemShut {NoStop}%
\bibitem [{\citenamefont {Duff}\ \emph {et~al.}(1962)\citenamefont {Duff},
  \citenamefont {Harlow},\ and\ \citenamefont {Hirt}}]{Duff1962}%
  \BibitemOpen
  \bibfield  {author} {\bibinfo {author} {\bibfnamefont {R.~E.}\ \bibnamefont
  {Duff}}, \bibinfo {author} {\bibfnamefont {F.~H.}\ \bibnamefont {Harlow}}, \
  and\ \bibinfo {author} {\bibfnamefont {C.~W.}\ \bibnamefont {Hirt}},\
  }\href@noop {} {\bibfield  {journal} {\bibinfo  {journal} {The Physics of
  Fluids}\ }\textbf {\bibinfo {volume} {5}},\ \bibinfo {pages} {417} (\bibinfo
  {year} {1962})}\BibitemShut {NoStop}%
\bibitem [{\citenamefont {Thornber}\ \emph {et~al.}(2008)\citenamefont
  {Thornber}, \citenamefont {Mosedale}, \citenamefont {Drikakis}, \citenamefont
  {Youngs},\ and\ \citenamefont {Williams}}]{Thornber2007c}%
  \BibitemOpen
  \bibfield  {author} {\bibinfo {author} {\bibfnamefont {B.}~\bibnamefont
  {Thornber}}, \bibinfo {author} {\bibfnamefont {A.}~\bibnamefont {Mosedale}},
  \bibinfo {author} {\bibfnamefont {D.}~\bibnamefont {Drikakis}}, \bibinfo
  {author} {\bibfnamefont {D.}~\bibnamefont {Youngs}}, \ and\ \bibinfo {author}
  {\bibfnamefont {R.}~\bibnamefont {Williams}},\ }\href@noop {} {\bibfield
  {journal} {\bibinfo  {journal} {J. Comput. Phys.}\ }\textbf {\bibinfo
  {volume} {227}},\ \bibinfo {pages} {4873} (\bibinfo {year}
  {2008})}\BibitemShut {NoStop}%
\bibitem [{\citenamefont {Garcia-Uceda~Juarez}\ \emph
  {et~al.}(2014)\citenamefont {Garcia-Uceda~Juarez}, \citenamefont {Raimo},
  \citenamefont {Shapiro},\ and\ \citenamefont {Thornber}}]{Garcia2014}%
  \BibitemOpen
  \bibfield  {author} {\bibinfo {author} {\bibfnamefont {A.}~\bibnamefont
  {Garcia-Uceda~Juarez}}, \bibinfo {author} {\bibfnamefont {A.}~\bibnamefont
  {Raimo}}, \bibinfo {author} {\bibfnamefont {E.}~\bibnamefont {Shapiro}}, \
  and\ \bibinfo {author} {\bibfnamefont {B.}~\bibnamefont {Thornber}},\
  }\href@noop {} {\bibfield  {journal} {\bibinfo  {journal} {AIAA J.}\ }\textbf
  {\bibinfo {volume} {52}},\ \bibinfo {pages} {2559} (\bibinfo {year}
  {2014})}\BibitemShut {NoStop}%
\bibitem [{\citenamefont {Fryxell}\ \emph {et~al.}(2000)\citenamefont
  {Fryxell}, \citenamefont {Olson}, \citenamefont {Ricker}, \citenamefont
  {Timmes}, \citenamefont {Zingale}, \citenamefont {Lamb}, \citenamefont
  {MacNeice}, \citenamefont {Rosner}, \citenamefont {Truran},\ and\
  \citenamefont {Tufo}}]{fryxell2000flash}%
  \BibitemOpen
  \bibfield  {author} {\bibinfo {author} {\bibfnamefont {B.}~\bibnamefont
  {Fryxell}}, \bibinfo {author} {\bibfnamefont {K.}~\bibnamefont {Olson}},
  \bibinfo {author} {\bibfnamefont {P.}~\bibnamefont {Ricker}}, \bibinfo
  {author} {\bibfnamefont {F.~X.}\ \bibnamefont {Timmes}}, \bibinfo {author}
  {\bibfnamefont {M.}~\bibnamefont {Zingale}}, \bibinfo {author} {\bibfnamefont
  {D.~Q.}\ \bibnamefont {Lamb}}, \bibinfo {author} {\bibfnamefont
  {P.}~\bibnamefont {MacNeice}}, \bibinfo {author} {\bibfnamefont
  {R.}~\bibnamefont {Rosner}}, \bibinfo {author} {\bibfnamefont {J.~W.}\
  \bibnamefont {Truran}}, \ and\ \bibinfo {author} {\bibfnamefont
  {H.}~\bibnamefont {Tufo}},\ }\href@noop {} {\bibfield  {journal} {\bibinfo
  {journal} {The Astrophys. J. Suppl.}\ }\textbf {\bibinfo {volume} {131}},\
  \bibinfo {pages} {273} (\bibinfo {year} {2000})}\BibitemShut {NoStop}%
\bibitem [{\citenamefont {Colella}\ and\ \citenamefont
  {Woodward}(1984)}]{colella1984piecewise}%
  \BibitemOpen
  \bibfield  {author} {\bibinfo {author} {\bibfnamefont {P.}~\bibnamefont
  {Colella}}\ and\ \bibinfo {author} {\bibfnamefont {P.~R.}\ \bibnamefont
  {Woodward}},\ }\href@noop {} {\bibfield  {journal} {\bibinfo  {journal} {J.
  Comput. Phys.}\ }\textbf {\bibinfo {volume} {54}},\ \bibinfo {pages} {174}
  (\bibinfo {year} {1984})}\BibitemShut {NoStop}%
\bibitem [{\citenamefont {van Leer}(1977)}]{van1977towards}%
  \BibitemOpen
  \bibfield  {author} {\bibinfo {author} {\bibfnamefont {B.}~\bibnamefont {van
  Leer}},\ }\href@noop {} {\bibfield  {journal} {\bibinfo  {journal} {J.
  Comput. Phys.}\ }\textbf {\bibinfo {volume} {23}},\ \bibinfo {pages} {263}
  (\bibinfo {year} {1977})}\BibitemShut {NoStop}%
\bibitem [{\citenamefont {LeVeque}(2002)}]{LeVeque02}%
  \BibitemOpen
  \bibfield  {author} {\bibinfo {author} {\bibfnamefont {R.}~\bibnamefont
  {LeVeque}},\ }\href@noop {} {\emph {\bibinfo {title} {Finite Volume Methods
  for Hyperbolic Problems}}},\ \bibinfo {series} {Cambridge Texts in Applied
  Mathematics}\ No.~\bibinfo {number} {31}\ (\bibinfo  {publisher} {Cambridge
  University Press},\ \bibinfo {year} {2002})\BibitemShut {NoStop}%
\bibitem [{\citenamefont {Daru}\ and\ \citenamefont {Tenaud}(2004)}]{Daru04}%
  \BibitemOpen
  \bibfield  {author} {\bibinfo {author} {\bibfnamefont {V.}~\bibnamefont
  {Daru}}\ and\ \bibinfo {author} {\bibfnamefont {C.}~\bibnamefont {Tenaud}},\
  }\href@noop {} {\bibfield  {journal} {\bibinfo  {journal} {J. Comp. Phys.}\
  }\textbf {\bibinfo {volume} {193}},\ \bibinfo {pages} {563} (\bibinfo {year}
  {2004})}\BibitemShut {NoStop}%
\bibitem [{\citenamefont {Grinstein}\ \emph {et~al.}(2007)\citenamefont
  {Grinstein}, \citenamefont {Margolin},\ and\ \citenamefont
  {Rider}}]{Grinstein2007}%
  \BibitemOpen
  \bibinfo {editor} {\bibfnamefont {F.~F.}\ \bibnamefont {Grinstein}}, \bibinfo
  {editor} {\bibfnamefont {L.~G.}\ \bibnamefont {Margolin}}, \ and\ \bibinfo
  {editor} {\bibfnamefont {W.~J.}\ \bibnamefont {Rider}},\ eds.,\ \href@noop {}
  {\emph {\bibinfo {title} {Implicit Large Eddy Simulation: Computing Turbulent
  Fluid Dynamics}}}\ (\bibinfo  {publisher} {Cambridge University Press},\
  \bibinfo {address} {Cambridge},\ \bibinfo {year} {2007})\BibitemShut
  {NoStop}%
\bibitem [{\citenamefont {Youngs}(1991)}]{Youngs1991}%
  \BibitemOpen
  \bibfield  {author} {\bibinfo {author} {\bibfnamefont {D.~L.}\ \bibnamefont
  {Youngs}},\ }\href@noop {} {\bibfield  {journal} {\bibinfo  {journal} {Phys.
  Fluids A}\ }\textbf {\bibinfo {volume} {3}},\ \bibinfo {pages} {1312}
  (\bibinfo {year} {1991})}\BibitemShut {NoStop}%
\bibitem [{Wil(2018)}]{Williams2018}%
  \BibitemOpen
  \href@noop {} {\bibfield  {journal} {\bibinfo  {journal} {Journal of
  Computational Physics}\ }\textbf {\bibinfo {volume} {374}},\ \bibinfo {pages}
  {413} (\bibinfo {year} {2018})}\BibitemShut {NoStop}%
\bibitem [{\citenamefont {Thornber}\ and\ \citenamefont
  {Zhou}(2012)}]{Thornber2012b}%
  \BibitemOpen
  \bibfield  {author} {\bibinfo {author} {\bibfnamefont {B.}~\bibnamefont
  {Thornber}}\ and\ \bibinfo {author} {\bibfnamefont {Y.}~\bibnamefont
  {Zhou}},\ }\href {\doibase 10.1103/PhysRevE.86.056302} {\bibfield  {journal}
  {\bibinfo  {journal} {Phys. Rev. E}\ }\textbf {\bibinfo {volume} {86}},\
  \bibinfo {pages} {056302} (\bibinfo {year} {2012})}\BibitemShut {NoStop}%
\bibitem [{\citenamefont {Thornber}(2016)}]{Thornber2016}%
  \BibitemOpen
  \bibfield  {author} {\bibinfo {author} {\bibfnamefont {B.}~\bibnamefont
  {Thornber}},\ }\href@noop {} {\bibfield  {journal} {\bibinfo  {journal}
  {Phys. Fluids}\ }\textbf {\bibinfo {volume} {28}},\ \bibinfo {pages} {045106}
  (\bibinfo {year} {2016})}\BibitemShut {NoStop}%
\bibitem [{\citenamefont {Thornber}\ \emph {et~al.}(2012)\citenamefont
  {Thornber}, \citenamefont {Drikakis}, \citenamefont {Youngs},\ and\
  \citenamefont {Williams}}]{Thornber2012}%
  \BibitemOpen
  \bibfield  {author} {\bibinfo {author} {\bibfnamefont {B.}~\bibnamefont
  {Thornber}}, \bibinfo {author} {\bibfnamefont {D.}~\bibnamefont {Drikakis}},
  \bibinfo {author} {\bibfnamefont {D.}~\bibnamefont {Youngs}}, \ and\ \bibinfo
  {author} {\bibfnamefont {R.~J.~R.}\ \bibnamefont {Williams}},\ }\href@noop {}
  {\bibfield  {journal} {\bibinfo  {journal} {J. Turbul.}\ }\textbf {\bibinfo
  {volume} {13}},\ \bibinfo {pages} {1} (\bibinfo {year} {2012})}\BibitemShut
  {NoStop}%
\bibitem [{\citenamefont {Cook}\ and\ \citenamefont {Zhou}(2002)}]{Cook2002}%
  \BibitemOpen
  \bibfield  {author} {\bibinfo {author} {\bibfnamefont {A.~W.}\ \bibnamefont
  {Cook}}\ and\ \bibinfo {author} {\bibfnamefont {Y.}~\bibnamefont {Zhou}},\
  }\href@noop {} {\bibfield  {journal} {\bibinfo  {journal} {Phys. Rev. E}\
  }\textbf {\bibinfo {volume} {66}},\ \bibinfo {pages} {026312} (\bibinfo
  {year} {2002})}\BibitemShut {NoStop}%
\bibitem [{\citenamefont {Sarkar}(1992)}]{Sarkar1992}%
  \BibitemOpen
  \bibfield  {author} {\bibinfo {author} {\bibfnamefont {S.}~\bibnamefont
  {Sarkar}},\ }\href@noop {} {\bibfield  {journal} {\bibinfo  {journal} {Phys.
  Fluids A}\ }\textbf {\bibinfo {volume} {4}},\ \bibinfo {pages} {2674}
  (\bibinfo {year} {1992})}\BibitemShut {NoStop}%
\bibitem [{\citenamefont {Ristorcelli}(1997)}]{ristorcelli_1997}%
  \BibitemOpen
  \bibfield  {author} {\bibinfo {author} {\bibfnamefont {J.}~\bibnamefont
  {Ristorcelli}},\ }\href {\doibase 10.1017/S0022112097006083} {\bibfield
  {journal} {\bibinfo  {journal} {J. Fluid Mech.}\ }\textbf {\bibinfo {volume}
  {347}},\ \bibinfo {pages} {37} (\bibinfo {year} {1997})}\BibitemShut
  {NoStop}%
\end{thebibliography}%

\end{document}